\documentclass[a4paper,11pt]{article}

\usepackage{jheppub}
\usepackage[table]{xcolor}
\usepackage{slashbox}
\allowdisplaybreaks
\usepackage{graphicx}
\usepackage{epstopdf}
\usepackage{amsthm}

\newcommand{\be}{\begin{equation}}
	\newcommand{\ee}{\end{equation}}
\newcommand{\bea}{\begin{eqnarray}}
	\newcommand{\eea}{\end{eqnarray}}
\newcommand{\nn}{\nonumber}
\newcommand{\bdm}{\begin{displaymath}}
	\newcommand{\edm}{\end{displaymath}}

\newtheorem{definition}{Definition}
\newtheorem{conjecture}{Conjecture}
\newtheorem*{conjecture*}{Conjecture}

\title{Completing the Fifth PN Precision Frontier via 
	the \\EFT of Spinning Gravitating Objects}

\author[a]{Mich\`ele Levi,}
\author[b]{Zhewei Yin}

\affiliation[a]{Mathematical Institute, University of Oxford, \\
	Woodstock Road, Oxford OX2 6GG, United Kingdom}
\affiliation[b]{Department of Physics and Astronomy, Uppsala University,\\
	 Box 516, 75108 Uppsala, Sweden}

\emailAdd{levi@maths.ox.ac.uk}
\emailAdd{zhewei.yin@physics.uu.se}

\abstract{
We put forward a broader picture of the effective theory of a spinning particle within the EFT of 
spinning gravitating objects, through which we derive and establish the new precision frontier at the 
fifth PN ($5$PN) order. This frontier includes higher-spin sectors, quadratic and quartic in the spin, 
which both display novel physical features, due to the extension of the effective theory beyond linear 
order in the curvature. The quadratic-in-spin sectors give rise to a new tidal effect, and the 
quartic-in-spin sectors exhibit a new multipolar deformation. 
We then generalize the concept of tidal operators and of spin-induced multipolar operators, and make 
conjectures on the numerical values of their Wilson 
coefficients, and on the effective point-particle action of Kerr black holes. 
We confirm the generalized actions for generic compact binaries of the NLO quartic-in-spin sectors 
which were derived via the extension of the EFT of gravitating spinning objects. We first present the 
corresponding interaction potentials and general Hamiltonians, which consist of $12$ distinct sectors, 
with a new one due to the new multipolar deformation. These Hamiltonians give the full physical 
information on the binary system, which mostly gets lost in higher-spin sectors, when going to the 
aligned-spins configuration. Moreover these general Hamiltonians uniquely allow us to find the complete 
Poincar\'e algebra at the $5$PN order with spins, including the third subleading quadratic-in-spin 
sectors. We derive consequent observables for GW applications. Finally, to make contact with the 
scattering problem, we also derive the extrapolated scattering angles for aligned spins. Our completion 
of the Poincar\'e algebra provides the strongest validation of our most comprehensive new results, and 
thus that the $5$PN order has now been established as the new precision frontier.
}

\begin{document}
	

\preprint{UUITP-54/22} 
	
\maketitle
	
\flushbottom

\section{Introduction}
\label{intro}

Second-generation gravitational-wave (GW) experiments, currently including Advanced LIGO  \cite{LIGOScientific:2014pky}, Advanced 
Virgo \cite{VIRGO:2014yos}, and KAGRA \cite{KAGRA:2020tym}, have brought about one of the greatest breakthroughs of the century in 
physics: Measurements of GWs emitted from the inspirals and mergers of two black holes (BHs) \cite{Abbott:2016blz}, neutron stars 
(NSs) \cite{TheLIGOScientific:2017qsa}, or even of mixed BH-NS binaries \cite{LIGOScientific:2021qlt}. In just a few years the 
influx of GW data has been rapidly growing \cite{LIGOScientific:2018mvr,LIGOScientific:2020ibl,LIGOScientific:2021djp}, ushering in 
an exciting era of accelerated theoretical progress. The modelling of such GW signals is rooted in the analytic post-Newtonian (PN) 
approximation of General Relativity (GR) \cite{Blanchet:2013haa}, as the motion in sources of such signals is characterized by 
non-relativistic (NR) velocities. The effective-one-body (EOB) approach \cite{Buonanno:1998gg} builds on PN theory to generate theoretical 
waveforms across the entirety of GW signals. In particular, the progress in recent years on the precision frontier of the conservative 
dynamics of generic compact binaries has been spectacular. 

The state of the art is presented in table \ref{pnfrontier}, with an eye towards the next precision frontier.
In the point-mass sector, which occupies the first row of table \ref{pnfrontier}, the fifth PN ($5$PN) order has been tackled via 
traditional GR methods \cite{Bini:2019nra,Bini:2020wpo,Bini:2020uiq}, and the effective field theory (EFT) approach 
\cite{Goldberger:2004jt,Blumlein:2020pyo}, and work on the $6$PN order is underway in both methodologies 
\cite{Bini:2020nsb,Blumlein:2020znm,Blumlein:2021txj}. The $5$PN state of the art in the point-mass sector constitutes a unique 
milestone, due to the first appearance of finite-size effects in this sector, and thus of UV physics taking place at the short 
scales of the individual compact objects. 
In the spin-orbit sector, which occupies the second row of table \ref{pnfrontier}, 
finite-size effects are also postponed by $5$ PN orders to the high $6.5$PN accuracy, similar to the point-mass sector.
The spin-orbit sector has been fully completed and verified at the $4.5$PN order \cite{Levi:2020kvb,Kim:2022pou,Levi:2022dqm}, 
followed by a similar derivation in \cite{Mandal:2022nty}, via the EFT of spinning gravitating objects and the 
\texttt{EFTofPNG} public code \cite{Levi:2015msa,Levi:2017kzq}, and was also approached via traditional GR methods 
following \cite{Bini:2019nra,Bini:2020wpo} in 
\cite{Antonelli:2020aeb,Antonelli:2020ybz}. 

\begin{table}[t]
\begin{center}
\begin{tabular}{|l|l|l|l||l|l|l|l}
\hline
\backslashbox{\quad\boldmath{$l$}}{\boldmath{$n$}} & (N\boldmath{$^{0}$})LO
& N\boldmath{$^{(1)}$}LO & \boldmath{N$^2$LO}
& \boldmath{N$^3$LO} & \boldmath{N$^4$LO} & \boldmath{N$^5$LO} & \boldmath{N$^6$LO} 
\\
\hline
\boldmath{S$^0$} 
& \cellcolor{green!25} 0 
& \cellcolor{green!25} 1 
& \cellcolor{green!25} 2 
& 
3 & 
4 & 
\textcolor{gray}{5} & \textcolor{gray}{6}
\\
\hline
\boldmath{S$^1$} 
& \cellcolor{green!25} 1.5 
& \cellcolor{green!25} 2.5 
& \cellcolor{green!25} 3.5 
& 
\textbf{4.5} & \textcolor{gray}{5.5} & &
\\
\hline
\boldmath{S$^2$} 
& \cellcolor{green!25} 2 
& \cellcolor{green!25} 3 
& \cellcolor{green!25} 4 
& 
\textbf{5} & \textcolor{gray}{6} & &
\\
\hline
\boldmath{S$^3$} 
& \cellcolor{green!25} 3.5 
& \cellcolor{green!25} \textbf{4.5} 
& \cellcolor{green!25} \textcolor{gray}{5.5} 
&  &  &  &
\\
\hline
\boldmath{S$^4$} 
& \cellcolor{green!25} 4 
& \cellcolor{green!25} \textbf{5} 
& \cellcolor{green!25} \textcolor{gray}{6} 
& & & & \\
\hline
\boldmath{S$^5$} 
& \cellcolor{green!25} 5.5 
& & & & & & \\
\hline
\boldmath{S$^6$} 
& \cellcolor{green!25} 6 
& &  &  &  &  & \\
\end{tabular}
\caption{The complete state of the art of PN gravity for the conservative dynamics of generic compact 
binaries, with an eye 
towards the next precision frontier at the $6$PN order. 
Using the PN formula we introduced in \cite{Levi:2019kgk}, each correction enters at the order $n + l + 
\text{Parity}(l)/2$, where $n$ is highest $n$-loop and $l$ highest spin-induced multipole contained in 
each of the sectors, and the parity is $0$ or $1$ for even or odd $l$, respectively.
The new precision frontier at the $4.5$ and $5$PN orders with spins (in boldface) has been uniquely completed via the EFT of 
spinning gravitating objects, and has not been fully derived nor verified via independent methods.
Most of the sectors at the $5.5$ and $6$PN orders (in gray) are still unknown.
The green area of the table is explained in the text below.}
\label{pnfrontier}
\end{center}
\end{table}

This picture for finite-size effects dramatically changes once higher-spin sectors are considered. To begin with, the leading 
finite-size effect shows up already at the 2PN order \cite{Barker:1975ae}, due to the spin-induced quadrupole. Such spin-induced 
multipolar deformations then show up at all subsequent PN orders, as captured uniquely via the EFT of spinning gravitating objects, 
that introduced the relevant theory to all orders in spin \cite{Levi:2014gsa,Levi:2015msa}. In particular the work exclusively carried out via this framework 
\cite{Levi:2011eq,Levi:2014gsa,Levi:2015ixa,Levi:2016ofk}, which pioneered the general treatment of  higher-spin sectors, was key 
to the completion of the state of art at the $4$PN order \cite{Levi:2015uxa,Levi:2018nxp}. More recently, we completed the 
precision frontier at the $4.5$PN order, which consists only of sectors with spins, including the complete general Hamiltonians of 
the third subleading (N$^3$LO) spin-orbit sector \cite{Levi:2020kvb,Kim:2022bwv}, and the next-to-leading order (NLO) cubic-in-spin sectors 
\cite{Levi:2019kgk,Levi:2022dqm}, which were all verified via the full Poincar\'e algebra in \cite{Levi:2022dqm}. Moreover, at the $5$PN order, 
which we already noted is a milestone in the point-mass sector, we recently completed the N$^3$LO quadratic-in-spin sectors in 
\cite{Levi:2020uwu,Kim:2021rfj,Kim:2022bwv}, followed by \cite{Mandal:2022ufb} that used the same methodology and framework. 
To complete the $5$PN accuracy we also have had to obtain the NLO quartic-in-spin sectors \cite{Levi:2020lfn}.

These different sectors with spins at the $5$PN order required an extension of the effective action of a spinning particle from our
\cite{Levi:2015msa}, beyond linear order in the curvature. Similar to the milestone in the point-mass sector at this 
order, the extension of theory for spin uncovers yet another novel level of intricacy in the finite-size effects as of this order.
In the quadratic-in-spin sectors we discovered a new tidal effect \cite{Kim:2021rfj} that is unique to the presence of spin.
Yet, another type of new effect was featured first in \cite{Levi:2020lfn}. In section \ref{finaltheory} of this paper after we review the 
EFT of higher spin in gravity, we discuss these new finite-size effects in a broader perspective, and generalize their different 
notions. In section \ref{lagsandhams} we derive the reduced interaction potentials and general Hamiltonians of the NLO quartic-in-spin sectors, 
resulting from the generalized actions computed via EFT in \cite{Levi:2020lfn}, which consist of no less than $12$ subsectors. In 
section \ref{poincareupto5pn} we proceed to validate our general Hamiltonians across all sectors with spins at the $5$PN order via a 
derivation of the full Poincar\'e algebra. In that we find the global Poincar\'e invariants of the binary system, and at the same time carry out a 
highly non-trivial consistency check for the general Hamiltonians in arbitrary reference frames. Notably, the N$^3$LO 
sectors, particularly those that are quadratic in the spin, required a significant scaling of the approach to solve for the 
Poincar\'e algebra.

After having thus established the new precision frontier at the $5$PN order, we derive in section \ref{spechamsnobs} the consequent 
simplified Hamiltonians in restricted kinematic configurations, and then the GW observables in the form of gauge-invariant relations 
among the binding energy, the angular momentum, and the emitted frequency. Furthermore, to make contact with the scattering problem, we also 
derive the extension to scattering angles, which is relevant only in the restricted aligned-spins configuration. Yet, while studying higher-spin sectors generally provides invaluable 
input on QCD and gravity theories, the restriction to the aligned-spins configuration in these 
sectors entails a growing loss of information on the physics of precessing binaries with spins of arbitrary orientations, which have a clear observational signature on the 
gravitational waveforms. For this reason it is especially crucial in higher-spin sectors to obtain general Hamiltonians, rather than limited scattering angles. 

In the scattering problem in a weak-field approximation or so-called post-Minkowskian (PM) expansion, 
the scattering angle for BHs with aligned spins was first approached in \cite{Guevara:2018wpp} for the 
NLO quartic-in-spin sectors. NLO PM quartic-in-spin Hamiltonians in the center-of-mass (COM) frame for 
BHs, and for generic compact binaries, were then presented in \cite{Chen:2021qkk}, and 
\cite{Bern:2022kto}, respectively. 
However, some basic clarifications on such scattering-amplitudes approaches are in place. 
As far as application to GW measurements from binary mergers is concerned, velocities remain NR in most 
if not all of the GW signal. 
As long as the binary is bound, it is only meaningful to obtain precision corrections in the PN 
expansion: 
\be
n\text{PN} \sim \sum_{i=0}^{n+1} v^{2i}G^{n+1-i},
\ee 
with the speed of light $c\equiv1$. As can be easily seen, this requires for each $n$PN order a 
computation that includes a piece that is of order $G^{n+1}$. Then at the brief time that the binary is 
no longer bound, the weak-field approximation also breaks down, and does not hold any more. 

For these reasons, an $m$PM weak-field correction, which is an expansion to the order of $G^m$ (even if 
to all orders in the velocity) is useful phenomenologically only 
for an $(m-1)$PN correction, and only when it can be meaningfully transformed from the scattering to 
the bound problem. 
Such a transformation to the bound problem faces an obstacle at and beyond the N$^2$LO, 
or the $2$-loop level, essentially due to radiation-reaction effects that kick in at these orders in 
gravity, and for which a solution is currently unknown. 
At the same time, for higher-spin sectors of S$^l$, which map to scattering of massive particles 
with higher spin $s=l/2$, the $4$-particle amplitude of massive higher-spin particles and gravitons is 
imperative beyond tree level, namely as of the NLO, and this so-called ``Compton amplitude'' is as yet
murky for $l\ge5$. 

As can be illustrated from table \ref{pnfrontier} by the green area, these $2$ limitations on high loop 
and higher spin sectors, combine to restrict the applicability of current scattering-amplitudes 
approaches to the GW precision frontier. In fact already as of the $3$PN order, sectors that are 
required to complete a certain PN accuracy, are no longer within the reach of these approaches. 
Moreover, it is timely now to tackle sectors at the $5.5$ and $6$PN
orders, whereas resolving the Compton amplitude that is only needed as of the NLO quintic-in-spin 
sectors at the $6.5$PN order, is not experimentally meaningful, while none of the difficult sectors at 
the $5.5$ and $6$PN orders have been tackled yet. 

In contrast, our free-standing QFT-based approach efficiently uses QFT methods without the irrelevant 
quantum degrees of freedom (DOFs), has been directly suited to the bound problem, and thus readily gets at the 
necessary results for GW measurements. Our self-contained EFT approach has no limited reach to 
any sector, and provides the most general comprehensive results for GW measurements of generic compact 
binaries. For these very reasons our EFT approach is also key as a guide to such diverse 
particle-amplitudes methods in tackling the related gravitational scattering problem.

\section{Effective Field Theory of Higher Spin in Gravity}
\label{finaltheory}

At the $5$PN order we encounter a leap in the intricate array of finite-size effects for rotating compact objects. Such finite-size effects which 
provide unique input on unknown physics in small scales, further multiply and diversify at this order. In order to capture all that, we need to go step by 
step, from the minimal coupling of spin to gravity, to leading spin-induced multipolar deformations, which are the leading finite-size effects of rotating objects, 
and then further to include tidal and subleading spin-induced multipolar deformations. 

The minimal coupling of an extended relativistic spin already 
presents the first challenge which we resolved in the EFT of spinning gravitating objects \cite{Levi:2010zu,Levi:2015msa} via what we coined as 
``spin-gauge invariance''. 
Second, for the leading spin-induced multipolar deformations of generic compact objects we introduced an infinite tower of operators to all orders in spin 
\cite{Levi:2014gsa,Levi:2015msa}. Implementing these two formal developments at higher loop and spin orders already requires a thoughtful treatment 
\cite{Levi:2014gsa,Levi:2015msa,Levi:2015ixa,Levi:2019kgk}. Yet, for the sectors at the $5$PN order, one needs to further extend the theory to include couplings 
beyond linear in the curvature, namely tidal 
and subleading spin-induced multipolar deformations \cite{Levi:2020uwu,Levi:2020lfn,Kim:2021rfj,Kim:2022bwv}. There are also additional subtleties 
due to the implementation of spin-gauge invariance in the leading spin-induced multipolar deformations \cite{Levi:2019kgk,Levi:2020lfn}. Here we will review 
these formal developments, and further elaborate on the more advanced effective theory beyond linear in the curvature. 

We recall that our computations for the compact binary inspiral start at the orbital scale from a two-particle system in weak gravity. To 
arrive at that, we have invoked new DOFs localized on two worldlines for each object that capture all of the small-scale physics which 
we are suppressing at this stage \cite{Goldberger:2004jt}:    
\be 
\label{fullorbit}
S_{\text{eff}}=S_{\text{gr}}[g_{\mu\nu}]+\sum_{a=1}^{2}S_{\text{pp}}(\lambda_a).
\ee
In addition to the purely gravitational action of the weak-field modes $S_{\text{gr}}$, we should then prescribe the infinite point-particle 
action for a spinning particle, $S_{\text{pp}}$, interacting with weak-field gravity. $\lambda_a$ parametrizes the worldline of each object.

\subsection{EFT of Spinning Gravitating Objects}
\label{startpoint}

We start with the initial generic point-particle action we can write for a spinning object as 
\cite{Hanson:1974qy,Bailey:1975fe,Porto:2005ac,Levi:2010zu,Levi:2015msa}:
\begin{align}
\label{spprotoldwcount}
S_{\text{pp}}\left[g_{\mu\nu},y^\mu,e^{\mu}_{A}\right]
= \int d \lambda \left[ \underbrace{-m \sqrt{u^2}}_{0\text{PN}}
\underbrace{- \frac{1}{2} S_{\mu\nu}\Omega^{\mu\nu}}_{1.5\text{PN}}
\underbrace{+ L_{\text{SI}}\left[u^{\mu}, S_{\mu\nu}, g_{\mu\nu}\left(y^\mu\right)\right]}_{2\text{PN}}\right],
\end{align} 
with $y^\mu$ and $e^{\mu}_{A}$, the coordinate and tetrad, respectively, as the new worldline DOFs. 
$u^\mu\equiv dy^\mu/d\lambda$ is then the $4$-velocity, and 
$\Omega^{\mu\nu}(\lambda)\equiv e^\mu_A\frac{De^{A\nu}}{D\lambda}$ the generalized angular velocity. 
The conjugate of the latter is spin, $S_{\mu\nu}(\lambda)$, which is featured explicitly in the 
effective action, and in fact accounts for all the non-minimal coupling of the rotating object to 
gravity, $L_{\text{SI}}$. We noted for each term in eq.~\eqref{spprotoldwcount} the PN order, in which 
it starts to play a role according to PN power-counting.

\subsubsection{Spin-Gauge Invariance}
\label{sscsucks}

The minimal coupling as it appears in eq.~\eqref{spprotoldwcount} actually takes into account only the 
basic symmetries that play a role when we consider only position DOFs, as in the point-mass sector \cite{Levi:2015msa}. 
Once we also consider rotation, $SO(3)$ invariance plays a key role in the construction of all parts of the 
spinning-particle action \cite{Levi:2015msa}. Our spin-gauge invariance then arises naturally as its 
consequent gauge redundancy, which is concealed in the form of eq.~\eqref{spprotoldwcount} 
\cite{Levi:2015msa}. In our formulation of the effective action of a spinning particle 
\cite{Levi:2015msa}, we made this hidden gauge invariance manifest, by deriving from 
the minimal coupling of spin to gravity, the more general form:
\begin{align}
\label{wigneriskinematic}
\frac{1}{2} S_{\mu\nu} \Omega^{\mu\nu} &= 
\frac{1}{2} \hat{S}_{\mu\nu} \hat{\Omega}^{\mu\nu}
+ \frac{\hat{S}^{\mu\nu} p_{\nu}}{p^2} \frac{D p_{\mu}}{D \sigma},
\end{align}
with generic rotational variables at this time, related to a general gauge of the spin (or "SSC").
Notably there is an extra term that shows up, which is kinematic. This term affects all orders in spin 
in all sectors with spin, but is always already fixed from (the dynamics of) lower perturbative orders. Therefore 
we have kept it expressed in terms of only worldline DOFs, since at each new order it always contains 
no new dynamics involving the gravitational field, which therefore does not have to be integrated out 
of it again. 
Our formulation then differed from Yee and Bander \cite{Yee:1993ya}, and later \cite{Porto:2008jj} that 
followed suit, who added an ad-hoc term with curvature/field dependence, which could not capture correctly
this significant general kinematic effect for a relativistic rotating object. 

In the EFT of spinning gravitating objects we exploit the spin-gauge invariance, and switch to a 
generalized canonical gauge that we formulated in \cite{Levi:2015msa}, which enables to land directly 
on spin variables that satisfy the canonical $SO(3)$ Poisson brackets (or commutation relations) 
\cite{Levi:2008nh,Levi:2015msa}.
In order to switch to a generic gauge for the spin also in the non-minimal couplings, one then needs to 
use the relation:
\begin{equation}
\label{projecttorest}
S_{\mu\nu} = \hat{S}_{\mu\nu} - \frac{\hat{S}_{\mu\rho} p^{\rho} p_{\nu}}{p^2}
+ \frac{\hat{S}_{\nu\rho} p^{\rho} p_{\mu}}{p^2}.
\end{equation}

\subsubsection{Tower of Higher-Spin Couplings}
\label{spindeform}

To begin with we construct the non-minimal coupling of the action with the analogue of the 
Pauli-Lubanski vector, $S^{\mu}$, which is orthogonal to the linear 
momentum \cite{Levi:2014gsa,Levi:2015msa}. This naturally leads to definite-parity $SO(3)$ tensors for 
the spin-induced multipoles, in a construction that is guided mainly by the symmetries of parity and 
$SO(3)$ invariance \cite{Levi:2014gsa,Levi:2015msa}. Using the complete set of symmetries spelled out 
in \cite{Levi:2015msa}, and with the definite-parity curvature components, $E_{\mu\nu}$ and 
$B_{\mu\nu}$, our rigorous analysis gave rise to an infinite tower of leading non-minimal couplings 
to all orders in spin \cite{Levi:2014gsa,Levi:2015msa}: 
\begin{align} 
\label{spintower}
L_{\text{NMC}}=&\sum_{n=1}^{\infty} \frac{\left(-1\right)^n}{\left(2n\right)!}
\frac{C_{ES^{2n}}}{m^{2n-1}} D_{\mu_{2n}}\cdots D_{\mu_3}
\frac{E_{\mu_1\mu_2}}{\sqrt{u^2}} S^{\mu_1}S^{\mu_2}\cdots 
S^{\mu_{2n-1}}S^{\mu_{2n}}\nn\\
&+\sum_{n=1}^{\infty} \frac{\left(-1\right)^n}{\left(2n+1\right)!}
\frac{C_{BS^{2n+1}}}{m^{2n}} 
D_{\mu_{2n+1}}\cdots D_{\mu_3}\frac{B_{\mu_1\mu_2}}{\sqrt{u^2}} 
S^{\mu_1}S^{\mu_2}\cdots 
S^{\mu_{2n-1}}S^{\mu_{2n}}S^{\mu_{2n+1}}.
\end{align}
This new set of operators represents spin-induced multipolar deformations, and are preceded by new 
Wilson coefficients, that match what would be called in traditional GR ``multipolar deformation 
parameters''. 
From this infinite series we need the $3$ leading terms for the quartic-in-spin sectors 
\cite{Levi:2014gsa,Levi:2015msa}:
\begin{align} 
\label{tos4}
L_{\text{NMC(R)}}^{\le5\text{PN}} =& 
\underbrace{-\frac{C_{ES^2}}{2m} \frac{E_{\mu\nu}}{\sqrt{u^2}} S^{\mu} S^{\nu}}_{2\text{PN}} 
\underbrace{-\frac{C_{BS^3}}{6m^2}\frac{D_\lambda B_{\mu\nu}}{\sqrt{u^2}}
	S^{\mu} S^{\nu} S^{\lambda}}_{3.5\text{PN}}
\underbrace{+\frac{C_{ES^4}}{24m^3} \frac{D_{\kappa}D_{\lambda}E_{\mu\nu}}{\sqrt{u^2}} 
S^\mu S^\nu S^\lambda S^\kappa}_{4\text{PN}}, 
\end{align}
with the spin-induced quadrupole \cite{Barker:1975ae,Porto:2005ac}, octupole, and hexadecapole 
\cite{Levi:2014gsa}, which enter at the PN orders noted, inferred from power-counting as detailed in 
\cite{Levi:2014gsa,Levi:2015msa,Levi:2018nxp}. Notably the leading relativistic corrections start only from the 
octupole \cite{Levi:2014gsa,Levi:2015msa,Levi:2018nxp}. 

The various approaches which have implemented scattering-amplitudes methods to study scattering of 
massive higher-spin particles, such as \cite{Guevara:2018wpp,Chen:2021qkk,Bern:2022kto}, have relied 
on the above theory, introduced in \cite{Levi:2014gsa,Levi:2015msa}. 
The $S^l$ couplings to gravity in eq.~\eqref{spintower} provided the $3$-point amplitudes for the 
scattering of massive particles of spin $s=l/2$, where $3$-point amplitudes are the critical building 
blocks used to derive any amplitude in such methods. 
Moreover, these approaches all used further input from implementation of the above theory for the case 
of BHs in traditional GR, e.g.~\cite{Guevara:2018wpp}, to critically guide their derivations.
The dependence of Guevara et al.~\cite{Guevara:2018wpp} on our worldline theory for higher-spin from 
\cite{Levi:2014gsa,Levi:2015msa} should be noted here in particular, since it was omitted in 
\cite{Guevara:2018wpp}.

At this point we can notice that the iterative substitution of the linear momentum in 
eqs.~\eqref{wigneriskinematic}, \eqref{projecttorest}, and \eqref{tos4}, becomes subtle at cubic- 
and quartic-in-spin orders \cite{Levi:2019kgk,Levi:2020lfn}. In the latter it can 
give rise to contributions that are already quadratic in the curvature (though still 
without further new Wilson coefficients) 
\cite{Levi:2020lfn}. In any case the $5$PN order necessitates an extension of the effective action 
of a spinning particle, to include operators that are beyond linear in the curvature 
\cite{Levi:2020uwu,Levi:2020lfn,Kim:2021rfj,Kim:2022bwv}.

\subsection{Going Beyond Linear in Curvature}
\label{beyondlincurv}

It was already noted in \cite{Goldberger:2004jt}, which treated the point-mass sector (without spin), 
that the effective action at the $5$PN order should be extended to quadratic order in the curvature. 
We consider here such an extension with eyes towards the next precision frontier -- covering up to the 
$6$PN order, where we restrict the discussion to strictly conservative operators. Indeed, in 
\cite{Levi:2020uwu,Levi:2020lfn,Kim:2021rfj,Kim:2022bwv} for the present $5$PN precision frontier, we 
approached such an extension of the non-minimal coupling to gravity for all sectors up to quartic 
order in the spin. Using the complete symmetries, and similar considerations spelled out in 
\cite{Levi:2015msa}, we can thus write schematically such an extension:
\begin{align} 
\label{nmcto6pn}
L_{\text{NMC(R$^2$)}}^{\le6\text{PN}}&
= \underbrace{C_{E^2} \frac{E_{\alpha\beta}E^{\alpha\beta}}{\sqrt{u^2}^{\,3}}
+ C_{E^2S^2} S^{\mu} S^{\nu} \frac{E_{\mu\alpha}E_{\nu}^{\,\alpha}}{\sqrt{u^2}^{\,3}}
+ C_{E^2S^4} S^{\mu} S^{\nu} S^{\kappa} S^{\rho} 
\frac{E_{\mu\nu}E_{\kappa\rho}}{\sqrt{u^2}^{\,3}}}_{5\text{PN}}
\nn\\&
\underbrace{+ C_{B^2} \frac{B_{\alpha\beta}B^{\alpha\beta}}{\sqrt{u^2}^{\,3}}
+ C_{B^2S^2} S^{\mu} S^{\nu} \frac{B_{\mu\alpha}B_{\nu}^{\alpha}}{\sqrt{u^2}^{\,3}}
+ C_{B^2S^4} S^{\mu} S^{\nu} S^{\kappa} S^{\rho}
\frac{B_{\mu\nu}B_{\kappa\rho}}{\sqrt{u^2}^{\,3}}}_{6\text{PN}},
\end{align}
and beyond the 6PN order:
\begin{align} 
\label{nmcabove6pn}
L_{\text{NMC(R$^2$)}}^{\le7\text{PN};\le S^4}&
= \underbrace{C_{\nabla EBS} S^{\mu} \frac{D_{\mu}E_{\alpha\beta}B^{\alpha\beta}}{\sqrt{u^2}^{\,3}}
+ C_{E\nabla BS} S^{\mu} \frac{E_{\alpha\beta}D_{\mu}B^{\alpha\beta}}{\sqrt{u^2}^{\,3}}}_{6.5\text{PN}}
\nn\\& 
\underbrace{+ C_{\nabla EBS^3} S^{\mu} S^{\nu} S^{\kappa} 
\frac{D_{\kappa}E_{\mu\alpha}B_{\nu}^{\alpha}}{\sqrt{u^2}^{\,3}}
+ C_{E\nabla BS^3} S^{\mu} S^{\nu} S^{\kappa}
\frac{E_{\mu\alpha}D_{\kappa}B_{\nu}^{\alpha}}{\sqrt{u^2}^{\,3}}}_{6.5\text{PN}}
\nn\\& 
\underbrace{+ C_{(\nabla E)^2} 
	\frac{D_{\mu}E_{\alpha\beta}D^{\mu}E^{\alpha\beta}}{\sqrt{u^2}^{\,3}}
	+ C_{(\nabla B)^2} 
	\frac{D_{\mu}B_{\alpha\beta}D^{\mu}B^{\alpha\beta}}{\sqrt{u^2}^{\,3}}}_{7\text{PN}}
\nn\\& 
\underbrace{+ C_{(\nabla E)^2S^2} S^{\mu} S^{\nu} 
\frac{D_{\mu}E_{\alpha\beta}D_{\nu}E^{\alpha\beta}}{\sqrt{u^2}^{\,3}}
+ C_{(\nabla B)^2S^2} S^{\mu} S^{\nu} 
\frac{D_{\mu}B_{\alpha\beta}D_{\nu}B^{\alpha\beta}}{\sqrt{u^2}^{\,3}}}_{7\text{PN}}
\nn\\& 
\underbrace{+ C_{(\nabla E)^2S^4} S^{\mu} S^{\nu} S^{\kappa} S^{\rho}
\frac{D_{\kappa}E_{\mu\alpha}D_{\rho}E_{\nu}^{\alpha}}{\sqrt{u^2}^{\,3}}
+ C_{(\nabla B)^2S^4} S^{\mu} S^{\nu} S^{\kappa} S^{\rho}
\frac{D_{\kappa}B_{\mu\alpha}D_{\rho}B_{\nu}^{\alpha}}{\sqrt{u^2}^{\,3}}}_{7\text{PN}}, 
\end{align}
where we noted the PN order in which the various new operators enter. Notice that since the indices of 
covariant derivatives and the curvature components are symmetric \cite{Levi:2015msa}, one can always 
write the contractions among two curvature components and their covariant derivatives, starting 
with the indices of curvature components. 
Notice that only operators quadratic- and quartic-in-spin enter up to the $6$PN order, 
whereas the linear and cubic-in-spin sectors both get such corrections only as of the $6.5$PN order.
Moreover, from eq.~\eqref{nmcabove6pn} it is also easy to realize that an extension to higher orders 
in the curvature (beyond quadratic) is not relevant even up to the $7$PN order. 

Thus at the present $5$PN order we only need to consider the first line of eq.~\eqref{nmcto6pn}, 
with $3$ leading operators. 
Note however that the $2$ first operators do not contribute to the NLO quartic-in-spin sectors, which 
are further studied in the present paper, due to their lower order in the spin: 
At the $5$PN order they can only contribute to the sector without spin, or to the quadratic-in-spin 
sectors, as detailed in \cite{Kim:2022bwv}. 
Yet even within these leading 
quadratic-in-curvature operators, we can already notice a clear distinction between the first two, 
where indices of the curvature components are contracted among themselves, and the operator that 
is quartic in the spin, where none of the curvature indices are contacted among 
themselves, similar to the operators in eq.~\eqref{spintower}. Furthermore, as we elaborated in 
\cite{Kim:2021rfj,Kim:2022bwv}, the first two operators in eq.~\eqref{nmcto6pn} both capture tidal 
effects. These observations then motivate us to make the following useful definitions:
\begin{definition} 
We call all operators whose indices of curvature components and their covariant derivatives have
contractions among themselves -- ``tidal operators''. 
We call the Wilson coefficients of such operators ``tidal coefficients'', or ``generalized Love 
numbers''. 	
\end{definition}	
\begin{definition} 
We call all operators, of which no indices of curvature components and their covariant derivatives 
are contracted among themselves -- ``spin-induced multipolar operators''. 
We call the Wilson coefficients of such operators ``spin-induced multipolar coefficients''. 	
\end{definition}

It is clear that the infinite set in the form of direct products of the leading linear-in-curvature 
basis of spin-induced multipolar-deformation operators in eq.~\eqref{spintower}, will appear as 
subleading spin-induced multipolar operators of higher orders in spin and curvature. 
The first of these subleading operators is for S$^4$, and is of the form of a direct product of two 
spin-induced quadrupole operators \cite{Levi:2020lfn}. It is easy to see that the next similar 
operator appears at the NLO S$^5$ sectors, and so on and so forth, as in:
\begin{align} 
\label{nmcsub}
L_{\text{NMC(R$^2$)}}^{\text{SI}}& \supset 
\quad \underbrace{C_{E^2S^4} S^{\mu} S^{\nu} S^{\kappa} S^{\rho} 
E_{\mu\nu}E_{\kappa\rho}}_{\text{NLO S$^4$}},
\quad
\underbrace{C_{EBS^5} S^{\mu} S^{\nu} S^{\kappa} S^{\rho} S^{\lambda} 
E_{\mu\nu}D_\lambda B_{\kappa\rho}}_{\text{NLO S$^5$}}, \, \ldots
\end{align}
Notice that though no ladder graphs are allowed in our EFT diagrammatic expansion, here we have an 
infinite tower of ladder-like operators (albeit, localized on the worldline also in time), as 
subleading corrections to the leading spin-induced multipolar ones in 
eq.~\eqref{spintower}. Incidentally, it would be inconsistent to include from eq.~\eqref{nmcsub} the 
operator for S$^5$, but not that for S$^4$, which is the case in \cite{Bern:2022kto}, where the latter 
was omitted. 

We can now notice the following interesting patterns for operators at the sectors S$^l$. 
For $l=0,1$, namely for the sectors dominated by minimal coupling to gravity, corrections to the leading 
effects enter at a shift of $5$ PN orders, and they are of the tidal type. 
For $l=2,3$, corrections to the leading effects, which include the spin-induced multipolar type, enter 
at a shift of $3$ PN orders, and they are also of the tidal type.
As of $l\ge4$, we can also tell from power counting, that corrections to the leading effects, which
include the spin-induced multipolar type, will always enter at a shift of $1$ PN order, but will also
include the new spin-induced multipolar type, from subleading spin-induced multipolar operators as in 
eq.~\eqref{nmcsub}. 
However, we can actually also see from power-counting that in fact, for $l\ge2$, tidal corrections to 
the leading effects, always enter at a shift of $3$ PN orders.   

With the two definitions above, we proceed to make two corresponding conjectures:	
\begin{conjecture}
\label{KerrLove}
All tidal coefficients (or generalized Love numbers) vanish for rotating black holes in Einstein's 
theory of gravity (in $4$ spacetime dimensions).
\end{conjecture} 
\begin{conjecture}
\label{Kerrdeformation}
All spin-induced multipolar coefficients, when rendered dimensionless, and normalized according to the 
symmetry of their operator, are of order unity for rotating black holes in Einstein's theory of 
gravity (in $4$ spacetime dimensions).
\end{conjecture} 

Such conjectures should of course be rigorously proved (or disproved) using matching computations of 
the EFT to full GR. 
Conjecture \ref{KerrLove} would follow from some extended ``Love symmetry'', see for example some 
seminal and recent surge in studies of basic Love numbers, namely the traditional ones -- from the 
point-mass sector (i.e.~simply of the Wilson coefficient of the first operator in 
eq.~\eqref{nmcto6pn}) in e.g.~\cite{Hinderer:2007mb,Damour:2009vw,Binnington:2009bb,Kol:2011vg}, and 
\cite{LeTiec:2020spy,LeTiec:2020bos,Chia:2020yla,Charalambous:2021mea,Charalambous:2021kcz,
	Castro:2021wyc,Charalambous:2022rre}, respectively.
As to the spin-induced multipolar coefficients in conjecture \ref{Kerrdeformation}, they can in 
general be fixed from studying leading observables, using only the linearized Kerr BH metric (namely 
of GR at $4$ spacetime dimensions), since the leading contributions from these operators would always 
arise from the linearized curvature components, for which only the linearized Kerr field is needed. 

Conjecture \ref{Kerrdeformation} then stipulates that the new spin-induced multipolar Wilson 
coefficient in eq.~\eqref{nmcto6pn} in the NLO quartic-in-spin sectors at the $5$PN order, after the 
proper normalization, satisfies that C$_{E^2S^4}$ is of order $1$ for rotating BHs in standard GR, 
as in:
\begin{align}
\label{submultipolS4}
L_{\text{S$^4$(R$^2$)}}
=  
\frac{C_{E^2S^4}}{4! m^3}\Bigg[\frac{S^{\mu_1} S^{\mu_2} S^{\mu_3} S^{\mu_4}} {\sqrt{u^2}^3} 
\Big( E_{\mu_1 \mu_2} E_{\mu_3 \mu_4} +  E_{\mu_1 \mu_3} E_{\mu_2 \mu_4}  +  E_{\mu_1 \mu_4} E_{\mu_2 
	\mu_3} \Big)\nn\\
-\frac{12}{7}\frac{S^2 S^{\mu_1} S^{\mu_2}} {\sqrt{u^2}^3}E_{\mu_1\nu}E_{\mu_2\nu}
+\frac{6}{35}\frac{S^4} {\sqrt{u^2}^3}E_{\mu\nu}E_{\mu\nu}\Bigg],
\end{align}
where here the Wilson coefficient was indeed rendered dimensionless, and normalized according to the 
symmetry of the operator, and the operator is written in terms of its fully symmteric and tracless 
rank-$4$ $SO(3)$ tensor components. Notice that the traces that are removed from the spin-induced  
multipolar operator constitute tidal deformations, and are accounted for in the respective point-mass 
and quadratic-in-spin operators/sectors, similar to what happened in the quadratic-in-spin sectors as 
detailed in \cite{Kim:2022bwv} and \cite{Mandal:2022ufb}. 
The Feynman rule due to this operator, recalling that all indices in our Feynman rules are Euclidean, 
reads:
\bea
\label{propFeyn}
\int dt \, \frac{C_{E^2S^4}}{8m^3}\Big[S^iS^jS^kS^l 
\phi,_{ij}\phi,_{kl}-\frac{4}{7}S^2S^iS^j\phi,_{ik}\phi,_{jk} 
+\frac{2}{35}S^4\phi,_{ij}\phi,_{ij}\Big], 
\eea
that is a two-graviton coupling of the KK scalar field as in \cite{Levi:2020lfn}.

The two conjectures above can be jointly stated as one conjecture for the general effective 
point-particle theory of Kerr BHs:
\begin{conjecture*}
\label{Kerrasparticle}
Rotating black holes in Einstein's theory of gravity (in $4$ spacetime dimensions) 
are captured by an effective point-particle theory, which contains an infinite set 
of only spin-induced multipolar-deformation operators, whose Wilson coefficients are 
of order unity, once their dimensions and symmetry factors are accounted for.
\end{conjecture*}

At this stage an additional point should be made regarding the construction of our EFT of a spinning 
particle, and of EFTs in general.
EFTs are devised to be an efficient and economic tool for precision computations. Therefore, when 
operators can be shifted to higher perturbative orders, using EOMs of the lower-order effective 
theory, namely when such operators are redundant, then they should indeed be removed at the level of 
the effective theory already (albeit, if left in the EFT, their onset in the observables will anyway 
be postponed to higher perturbative orders). According to this general EFT philosophy, we argued 
e.g.~in \cite{Levi:2015msa} for possible linear-in-curvature operators, that could be recast as 
quadratic in the curvature, and higher order in the spin, using lower-order EOMs, and thus are omitted 
from the effective theory.



\section{Effective Actions and General Hamiltonians}
\label{lagsandhams}

Equipped with the extended effective theory of a spinning particle as discussed above, the NLO 
quartic-in-spin interactions were evaluated through an EFT computation in \cite{Levi:2020lfn}.
The evaluation involved $28$ unique Feynman graphs \cite{Levi:2020lfn}.
The NLO quartic-in-spin generalized actions can then be summarized as follows \cite{Levi:2020lfn}:
\begin{equation}
L^{\text{NLO}}_{\text{S}^4}=
L^{\text{NLO}}_{\text{S}_1^2\text{S}_2^2}
+L^{\text{NLO}}_{\text{S}_1^3\text{S}_2}
+L^{\text{NLO}}_{\text{S}_1^4}+(1\leftrightarrow 2),
\end{equation}
with the simplest part being:
\begin{align}
L^{\text{NLO}}_{\text{S}_1^2\text{S}_2^2}&
=\frac{1}{2}C_{1(\text{ES}^2)}C_{2(\text{ES}^2)}\frac{G}{m_1m_2}\left(\frac{3L_{(1)}}{8r^5}
+\frac{3L_{(2)}}{4r^4}+\frac{L_{(3)}}{2r^3}\right)\nonumber\\*
&\quad-\frac{9}{2}C_{1(\text{ES}^2)}C_{2(\text{ES}^2)}\frac{G^2}{m_1r^6}L_{(4)}
+\frac{1}{2}C_{1(\text{ES}^2)}\frac{G^2}{m_1r^6}L_{(5)},
\end{align}
then:
\begin{align}
L^{\text{NLO}}_{\text{S}_1^3S_2}&=C_{1(\text{BS}^3)}\frac{G}{m_1^2}
\left(\frac{{L}_{[1]}}{4r^5}+\frac{{L}_{[2]}}{4r^4}+\frac{{L}_{[3]}}{12r^3}\right)
+C_{1(\text{BS}^3)}\frac{G^2}{m_1r^6}L_{[4]}+C_{1(\text{BS}^3)}\frac{m_2G^2}{m_1^2r^6}L_{[5]} \nonumber\\*
&\quad+C_{1(\text{ES}^2)}\frac{G^2}{m_1r^6}L_{[6]}
+\frac{3}{2}C_{1(\text{ES}^2)}\frac{G}{m_1^2 r^4}L_{[7]}
+3C_{1(\text{ES}^2)}\frac{m_2 G^2}{m_1^2 r^6}L_{[8]},
\end{align}
and finally:
\begin{align}
L^{\text{NLO}}_{\text{S}_1^4}&=
C_{1(\text{ES}^4)}\frac{ Gm_2}{m_1^3}\left(\frac{L_{\{1\}}}{16r^5}
+\frac{L_{\{2\}}}{8r^4}+\frac{L_{\{3\}}}{12r^3}\right)
-\frac{1}{8}C_{1(\text{ES}^4)}\frac{G^2m_2}{r^6m_1^2}L_{\{4\}}
\nonumber\\
&\quad
-\frac{3}{8}C_{1(\text{ES}^4)}\frac{G^2m_2^2}{r^6m_1^3}L_{\{5\}}
+C_{1(\text{BS}^3)}\frac{G^2m_2}{r^6m_1^2}L_{\{6\}}
-\frac{1}{8}C_{1(\text{ES}^2)}^2\frac{G^2m_2}{r^6m_1^2}L_{\{7\}} \nonumber\\*
&\quad +\frac{1}{2}C_{1(\text{BS}^3)}\frac{G m_2}{r^4m_1^3}L_{\{8\}}
+C_{1(\text{BS}^3)}\frac{G^2m_2^2}{r^6m_1^3}L_{\{9\}}
+\frac{C_{1(\text{E}^2\text{S}^4)}}{8}\frac{G^2m_2^2}{r^6m_1^3}L_{\{10\}},
\end{align}
where we confirm the various distinct pieces that make up the above as in \cite{Levi:2020lfn}. 
We also present these generalized actions in the supplementary material to this publication.

\subsection{Reduction of Generalized Actions}
\label{minisgood}

Though the generalized actions provided in \cite{Levi:2020lfn} already allow to directly derive the 
EOMs for both the position and spin variables, as we discussed in \cite{Levi:2014sba}, it is more 
useful to have at hand reduced actions, that contain only velocity and spin as the 
variables of highest time-derivative. The reduction involves formal redefinitions of both 
the position and rotational variables, that was introduced in \cite{Levi:2014sba}, and recently 
further extended in \cite{Kim:2022pou}. 
Table \ref{finalredefs} shows the build-up of redefinitions, that need to be applied in increasing PN 
order, up to the present NLO quartic-in-spin sectors, and here we will follow up on the derivations 
presented in \cite{Kim:2022pou,Kim:2022bwv,Levi:2022dqm}. 

\begin{table}[t]
\begin{center}
\begin{tabular}{|l|c|c|}
\hline
\backslashbox{\quad\boldmath{$l$}}{\boldmath{$n$}} & (N\boldmath{$^{0}$})LO
& N\boldmath{$^{(1)}$}LO 
\\
\hline
\boldmath{S$^0$} &  & 
\\
\hline
\boldmath{S$^1$} & + & ++ 
\\
\hline
\boldmath{S$^2$} &  & ++ 
\\
\hline
\boldmath{S$^3$} 
& 
\qquad + 
& 
++ 
\\
\hline
\boldmath{S$^4$} 
& 
\quad 
& 
++ 
\\
\hline
\end{tabular}
\caption{
$10$ sectors contribute through redefinitions to the present NLO quartic-in-spin sectors, $(1,4)$,
where in \cite{Levi:2019kgk} we introduced the notation, $(n,l)$, and the corresponding PN counting. 
"+" indicates that only position shifts or redefinitions of rotational variables need to be fixed, 
whereas "++" indicates that redefinitions for both position and rotational variables need to be 
fixed.}
\label{finalredefs}
\end{center}
\end{table}

First, we should note that as in all other higher-spin sectors starting from the NLO quadratic-in-spin 
\cite{Levi:2015msa}, 
position shifts need to be applied beyond linear order. As to the rotational variables in the present sectors no 
redefinitions are needed to be applied beyond linear order \cite{Kim:2022pou}. As table 
\ref{finalredefs} shows, we need to consider the redefinitions fixed from $6$ sectors, where they 
were provided for the $3$ sectors below cubic-in-spin in \cite{Kim:2022bwv,Kim:2022pou}, and for the 
$2$ cubic-in-spin sectors in \cite{Levi:2022dqm}. Thus here we are left to fix the redefinitions at 
the present NLO quartic-in-spin sectors, as captured in table \ref{nlos4redef}, following our 
conventions and notations from \cite{Kim:2022pou}. 

\begin{table}[t]
\begin{center}
\begin{tabular}{|l|c|c|c|c|c|}
\hline
\backslashbox{from}{\boldmath{to}} 
& (0P)N & LO S$^1$ & LO S$^2$ & LO S$^3$
\\
\hline
LO S$^1$ &  &  & $(\Delta \vec{x})^2$ & $\Delta \vec{x}$
\\
\hline
NLO S$^2$ &  & & $\Delta \vec{x}$ & 
\\
\hline
LO S$^3$ &  & & $\Delta \vec{S}$ &  
\\
\hline
NLO S$^3$ & & $\Delta \vec{x}$ & &  
\\
\hline
NLO S$^4$ & $\Delta \vec{x}$, $\Delta \vec{S}$ & & & 
\\
\hline
\end{tabular}
\caption{Contributions to the NLO quartic-in-spin sectors from position shifts and spin 
redefinitions in lower-order sectors.}
\label{nlos4redef}
\end{center}
\end{table}

Thus, we recall that the unreduced actions and redefinitions at the LO and NLO spin-orbit, NLO 
quadratic-in-spin, as well as LO and NLO cubic-in-spin sectors, are detailed in \cite{Kim:2022pou}, 
\cite{Kim:2022bwv}, and \cite{Levi:2022dqm}, respectively. We can now proceed to the new position shift 
that is fixed in the present sectors:
\bea
\left(\Delta \vec{x}_1\right)^{\text{NLO}}_{\text{S}^4} &=&- 	\frac{G C_{1\text{ES}^4} m_{2}}{8 m_{1}{}^4 r{}^4} \Big[ -5 S_{1}^2 \vec{n} ( \vec{S}_{1}\cdot\vec{n})^{2} + \vec{n} S_{1}^{4} + 2 \vec{S}_{1}\cdot\vec{n} S_{1}^2 \vec{S}_{1} \Big] \nn\\ 
&& - 	\frac{G C_{1\text{BS}^3} m_{2}}{m_{1}{}^4 r{}^4} \Big[ 5 S_{1}^2 \vec{n} ( \vec{S}_{1}\cdot\vec{n})^{2} - \vec{n} S_{1}^{4} + \vec{S}_{1}\cdot\vec{n} S_{1}^2 \vec{S}_{1} - 5 ( \vec{S}_{1}\cdot\vec{n})^{3} \vec{S}_{1} \Big] \nn\\ 
&& + 	\frac{3 G C_{1\text{ES}^2} m_{2}}{16 m_{1}{}^4 r{}^4} \Big[ 5 S_{1}^2 \vec{n} ( \vec{S}_{1}\cdot\vec{n})^{2} - \vec{n} S_{1}^{4} + \vec{S}_{1}\cdot\vec{n} S_{1}^2 \vec{S}_{1} - 5 ( \vec{S}_{1}\cdot\vec{n})^{3} \vec{S}_{1} \Big] \nn\\ 
&& + 	\frac{3 G C_{1\text{ES}^2} C_{2\text{ES}^2}}{2 m_{1}{}^2 m_{2} r{}^4} \Big[ 2 \vec{S}_{2}\cdot\vec{n} \vec{S}_{1}\cdot\vec{S}_{2} \vec{S}_{1} + \vec{S}_{1}\cdot\vec{n} S_{2}^2 \vec{S}_{1} - 5 \vec{S}_{1}\cdot\vec{n} ( \vec{S}_{2}\cdot\vec{n})^{2} \vec{S}_{1} \Big] \nn\\ 
&& - 	\frac{3 G}{8 m_{1}{}^3 r{}^4} \Big[ S_{1}^2 \vec{S}_{1}\cdot\vec{S}_{2} \vec{n} - 5 \vec{S}_{1}\cdot\vec{n} S_{1}^2 \vec{n} \vec{S}_{2}\cdot\vec{n} - 2 \vec{S}_{1}\cdot\vec{n} \vec{S}_{1}\cdot\vec{S}_{2} \vec{S}_{1} \nn\\ 
&& + 5 \vec{S}_{2}\cdot\vec{n} ( \vec{S}_{1}\cdot\vec{n})^{2} \vec{S}_{1} + \vec{S}_{1}\cdot\vec{n} S_{1}^2 \vec{S}_{2} \Big] + 	\frac{3 G C_{1\text{ES}^2}}{m_{1}{}^3 r{}^4} \Big[ S_{1}^2 \vec{S}_{1}\cdot\vec{S}_{2} \vec{n} \nn\\ 
&& - 5 \vec{S}_{1}\cdot\vec{n} S_{1}^2 \vec{n} \vec{S}_{2}\cdot\vec{n} - 2 \vec{S}_{1}\cdot\vec{n} \vec{S}_{1}\cdot\vec{S}_{2} \vec{S}_{1} + 5 \vec{S}_{2}\cdot\vec{n} ( \vec{S}_{1}\cdot\vec{n})^{2} \vec{S}_{1} + \vec{S}_{1}\cdot\vec{n} S_{1}^2 \vec{S}_{2} \Big] \nn\\ 
&& - 	\frac{3 G C_{2\text{ES}^2}}{16 m_{1}{}^2 m_{2} r{}^4} \Big[ S_{1}^2 S_{2}^2 \vec{n} - 5 S_{1}^2 \vec{n} ( \vec{S}_{2}\cdot\vec{n})^{2} - 2 \vec{S}_{2}\cdot\vec{n} \vec{S}_{1}\cdot\vec{S}_{2} \vec{S}_{1} - \vec{S}_{1}\cdot\vec{n} S_{2}^2 \vec{S}_{1} \nn\\ 
&& + 5 \vec{S}_{1}\cdot\vec{n} ( \vec{S}_{2}\cdot\vec{n})^{2} \vec{S}_{1} + 2 S_{1}^2 \vec{S}_{2}\cdot\vec{n} \vec{S}_{2} \Big] + 	\frac{G C_{1\text{BS}^3}}{2 m_{1}{}^3 r{}^4} \Big[ S_{1}^2 \vec{S}_{1}\cdot\vec{S}_{2} \vec{n} \nn\\ 
&& + 5 \vec{S}_{1}\cdot\vec{n} S_{1}^2 \vec{n} \vec{S}_{2}\cdot\vec{n} - 10 \vec{S}_{1}\cdot\vec{S}_{2} \vec{n} ( \vec{S}_{1}\cdot\vec{n})^{2} - S_{1}^2 \vec{S}_{2}\cdot\vec{n} \vec{S}_{1} + 4 \vec{S}_{1}\cdot\vec{n} \vec{S}_{1}\cdot\vec{S}_{2} \vec{S}_{1} \nn\\ 
&& - 7 \vec{S}_{1}\cdot\vec{n} S_{1}^2 \vec{S}_{2} + 10 ( \vec{S}_{1}\cdot\vec{n})^{3} \vec{S}_{2} \Big] + 	\frac{G C_{2\text{BS}^3}}{m_{1} m_{2}{}^2 r{}^4} \Big[ \vec{S}_{1}\cdot\vec{S}_{2} S_{2}^2 \vec{n} - 5 \vec{S}_{1}\cdot\vec{S}_{2} \vec{n} ( \vec{S}_{2}\cdot\vec{n})^{2} \nn\\ 
&& - \vec{S}_{1}\cdot\vec{n} S_{2}^2 \vec{S}_{2} + 5 \vec{S}_{1}\cdot\vec{n} ( \vec{S}_{2}\cdot\vec{n})^{2} \vec{S}_{2} \Big].
\eea
The new redefinitions of the spin in the present sectors are then fixed as:
\bea
\left(\omega^{ij}_1\right)^{\text{NLO}}_{\text{S}^4} =\left(\omega^{ij}_1\right)^{\text{NLO}}_{\text{S}_1^4}+\left(\omega^{ij}_1\right)^{\text{NLO}}_{\text{S}_1^3 \text{S}_2} +\left(\omega^{ij}_1\right)^{\text{NLO}}_{\text{S}_1^2 \text{S}_2^2}  +\left(\omega^{ij}_1\right)^{\text{NLO}}_{\text{S}_1 \text{S}_2^3}  - (i \leftrightarrow j),
\eea
where we present the explicit expressions in appendix \ref{newredefs} due to their large volume, and in 
machine-readable format in the supplementary material to this publication.

After the above reduction is done, we obtain the following NLO quartic-in-spin potentials, made up of 
no less than $12$ unique sectors:
\bea
\label{12actionbreakdown}
V^{\text{NLO}}_{\text{S}^4} &=&C_{1\text{ES}^2} V^{\text{NLO}}_{(\text{ES}_1^2 ) \text{S}_1^2}+C_{1\text{ES}^2}^2 V^{\text{NLO}}_{(\text{ES}_1^2 )^2} + C_{1\text{BS}^3} V^{\text{NLO}}_{(\text{BS}_1^3) \text{S}_1 } +C_{1\text{ES}^4} V^{\text{NLO}}_{\text{ES}_1^4} +C_{1\text{E}^2\text{S}^4} V^{\text{NLO}}_{\text{E}^2\text{S}_1^4}\nn\\
&& +V^{\text{NLO}}_{\text{S}_1^3 \text{S}_2}+C_{1\text{ES}^2} V^{\text{NLO}}_{(\text{ES}_1^2 ) \text{S}_1 \text{S}_2} + C_{1\text{ES}^2}^2 V^{\text{NLO}}_{C_{\text{ES}^2_1}^2 \text{S}_1^3 \text{S}_2 } + C_{1\text{BS}^3} V^{\text{NLO}}_{(\text{BS}_1^3) \text{S}_2 }\nn\\
&& +V^{\text{NLO}}_{\text{S}_1^2 \text{S}_2^2}+C_{1\text{ES}^2} V^{\text{NLO}}_{(\text{ES}_1^2 )  \text{S}_2^2} +C_{1\text{ES}^2} C_{2\text{ES}^2} V^{\text{NLO}}_{(\text{ES}_1^2 )(\text{ES}_2^2 ) }+  (1 \leftrightarrow 2),
\eea
where we present the explicit expressions in appendix \ref{neweffacts}, and in machine-readable format
in the supplementary material to this publication.

Notice that for generic compact binaries we have $2$ distinct sectors that are both proportional to 
the square of the quadrupolar deformation parameter, one in eq.~\eqref{quadsqtohexa}, and 
another unique contribution that is cubic in the individual spins in eq.~\eqref{quadsqtooctu}, which 
first emerged in the NLO cubic-in-spin sectors \cite{Levi:2022dqm}. Note also the new sector in 
eq.~\eqref{subhexa} due to the new subleading hexadecapole operator at this order from eq.~\eqref{submultipolS4}.


\subsection{General Hamiltonians}
\label{genhams}

The final actions that we obtain via the EFT of gravitating spinning objects contain position and spin 
variables that correspond to the generalized canonical gauge we formulated therein \cite{Levi:2015msa}. 
This enables to directly derive the general Hamiltonians via a Legendre transform only with respect to 
the position variables, similar to sectors without spin. For this derivation of the 
Hamiltonian all the sectors in table \ref{finalredefs} need to be taken into account consistently. 
It should be highlighted that we obtain the most general Hamiltonians in an arbitrary reference frame, 
and these in turn uniquely allow to study the global Poincar\'e invariants in phase space.

Due to the existence and uniqueness of the Poincar\'e algebra, the ability to find a closed solution 
for it, with some given derived Hamiltonian, provides a significant validation of the latter, as 
will be further discussed in section \ref{poincareupto5pn} below.
For phenomenological applications these general Hamiltonians can be gradually simplified, 
starting from their restriction to the COM frame, and then to further specialized configurations, in 
which one can obtain compact observables, see section \ref{spechamsnobs} below. The Hamiltonians can 
also be simplified into EOB models, which are crucial to generate GW templates.

Our general Hamiltonians for the NLO quartic-in-spin sectors, similar to the action potentials in 
eq.~\eqref{12actionbreakdown}, consist of $12$ unique sectors:  
\bea
\label{eq:hamcns4}
H^{\text{NLO}}_{\text{S}^4} &=&C_{1\text{ES}^2} H^{\text{NLO}}_{(\text{ES}_1^2 ) \text{S}_1^2} +C_{1\text{ES}^2}^2 H^{\text{NLO}}_{(\text{ES}_1^2 )^2} + C_{1\text{BS}^3} H^{\text{NLO}}_{(\text{BS}_1^3) \text{S}_1 } +C_{1\text{ES}^4} H^{\text{NLO}}_{\text{ES}_1^4} +C_{1\text{E}^2\text{S}^4} H^{\text{NLO}}_{\text{E}^2\text{S}_1^4}\nn\\
&& +H^{\text{NLO}}_{\text{S}_1^3 \text{S}_2}+C_{1\text{ES}^2} H^{\text{NLO}}_{(\text{ES}_1^2 ) \text{S}_1 \text{S}_2}  + C_{1\text{ES}^2}^2 H^{\text{NLO}}_{C_{\text{ES}^2_1}^2 \text{S}_1^3 \text{S}_2 } + C_{1\text{BS}^3} H^{\text{NLO}}_{(\text{BS}_1^3) \text{S}_2 }\nn\\
&& +H^{\text{NLO}}_{\text{S}_1^2 \text{S}_2^2}+C_{1\text{ES}^2} H^{\text{NLO}}_{(\text{ES}_1^2 )  \text{S}_2^2} +C_{1\text{ES}^2} C_{2\text{ES}^2} H^{\text{NLO}}_{(\text{ES}_1^2 )(\text{ES}_2^2 ) }+  (1 \leftrightarrow 2),
\eea
where we present the explicit expressions in appendix \ref{newgenhams}, and in machine-readable format
in the supplementary material to this publication.

\section{Poincar\'e Algebra with Spins at the $5$PN Order}
\label{poincareupto5pn}

In this section we study the global Poincar\'e symmetry of the binary system for 
all the sectors with spins at the $5$PN order. This symmetry is realized in phase space, 
where in sectors with spin variables, we use the generalized Poisson brackets 
\cite{Hanson:1974qy,Bel:1980}:
\bea
\{ f, g \} \equiv \{ f, g \}_x + \{ f, g \}_S,
\eea
with
\bea
\{ f, g \}_x &=& \sum_{I=1}^2 \left(\frac{\partial f}{\partial x_I} \cdot 
\frac{\partial g}{\partial p_I} 
- \frac{\partial f}{\partial p_I} \cdot \frac{\partial g}{\partial x_I} \right),\\
\{ f, g \}_S &=& \sum_{I=1}^2 S_I \times 
\frac{\partial f}{\partial S_I} \cdot \frac{\partial g}{\partial S_I}.
\eea
For our binary system we have:
\bea
\vec{P} = \vec{p}_1 + \vec{p}_2, \quad \vec{J} = 
\sum_{I=1}^2 \left( \vec{x}_I \times \vec{p}_I +  \vec{S}_I \right),
\eea
for $\vec{P}$, the total linear momentum, and $\vec{J}$, the total angular momentum. 
If our PN Hamiltonians for arbitrary reference frames are valid, then they should 
satisfy the following Poincar\'e algebra:
\begin{align}
\label{given}
\{P_i, P_j\} = \{P_i, H\} = \{J_i, H\}=0, 
\quad & \{J_i, J_j\} = \quad \epsilon_{ijk}J_k, 
\quad & \{J_i, P_j\} = \epsilon_{ijk}P_k, \\
\label{nontrivial}
\{G_i, P_j\} = \delta_{ij}H,
\quad \{G_i, H\} = P_i,
\quad & \{G_i, G_j\} = -\epsilon_{ijk}J_k, 
\quad & \{J_i, G_j\} = \epsilon_{ijk}G_k,
\end{align}
with $H$ for the Hamiltonian, and $\vec{G}$, the center-of-mass generator, related 
to the boost via $\vec{K}\equiv \vec{G}-t\vec{P}$. While the Poisson brackets in 
eq.~\eqref{given} are trivially satisfied, those in eq.~\eqref{nontrivial}, which 
involve $\vec{G}$, require careful consideration. Our task here is thus to construct 
the unique solution for $\vec{G}$ that will satisfy the latter equation. 

First, we construct $\vec{G}$ from the vectors $\vec{x}_I$, $\vec{p}_I$ and $\vec{S}_I$, 
such that the last Poisson brackets in eq.~\eqref{nontrivial} are automatically satisfied. 
Then, if we take the following form for $\vec{G}$ :
\bea
\vec{G} = h_1 \vec{x}_1 + h_2 \vec{x}_2 + \vec{Y},
\quad h_1 + h_2 = H,
\eea
where $h_I$ and $\vec{Y}$ satisfy:
\bea
\{ h_I, P_i \} = \{ Y_i, P_j \} = 0,
\eea
then 
\be
\{G_i, P_j\} = \delta_{ij}H.
\ee
Various considerations \cite{Levi:2022dqm} then lead to the more specific form:
\be
\vec{G} = H (\vec{x}_1 + \vec{x}_2)/2 + \vec{Y},
\ee
where $\vec{Y}$ is symmetric under the exchange of worldline labels 
$1 \leftrightarrow 2$, and depends on $x_I$ only through $\vec{n}$ and $r$. Thus 
we are left with the task of constructing and constraining $\vec{Y}$, such that 
$\vec{G}$ uniquely solves the non-trivial Poisson brackets:
\bea
\label{mainconstraint}
\{G_i, H\} & = & P_i.
\eea
This is since it turns out that if the closed form for $\vec{G}$ in flat spacetime 
\cite{Bel:1980}:
\bea
\vec{G}_{\text{flat}} = \sum_{I=1}^2 \left( \gamma_I m_I \vec{x}_I 
- \frac{\vec{S}_I \times \vec{p}_I}{m_I (1+ \gamma_I)}\right),
\label{comflat}
\eea
with $\gamma_I = \sqrt{1+ p_I^2/m_I^2}$, is also used to constrain the 
$\mathcal{O} (G_N^0)$ in $\vec{G}$, then the third Poisson brackets in 
eq.~\eqref{nontrivial} is also automatically satisfied by the solution
for $\vec{G}$.

Thus, in order to solve for $\vec{G}$ we first decompose it according to spin orders 
in each individual spin, PN orders, and dependence in Wilson coefficients. We 
then construct an ansatz for $\vec{Y}$ using the vectors 
$\vec{n}$, $\vec{p}_I$, $\vec{S}_I$. We use dimensional analysis and Euclidean 
covariance to constrain the ansatz, but note that due to the fact that in 
sectors with spin we have $4$ or even $5$ vectors to consider in $3$-dimensional 
space, each $4$ of them are necessarily linearly dependent, and thus our general 
ansatz will contain a certain redundancy. 

Following these considerations we proceed 
to solve for $\vec{G}$ at the sectors with spin that make up the new precision 
frontier at the $5$PN order, with the general Hamiltonians for the NLO quartic-in-spin 
sectors first presented in section \ref{genhams} above, and the general Hamiltonians 
for the N$^3$LO quadratic-in-spin sectors first presented in \cite{Kim:2022bwv}. 
For the latter, the general ansatz to solve for, contains an order of $\sim10^3$ free 
dimensionless coefficients, which means that the solution for the problem needed to be 
scaled significantly, even with respect to the most advanced Poincar\'e algebra at the 
$4$PN order that we presented in \cite{Levi:2016ofk}.

\subsection{NLO Quartic-in-Spin Sectors}
\label{poincares4}

We start by decomposing $H$ and $\vec{G}$ to all the sectors relevant to the solution for $\vec{G}$ 
at the NLO quartic-in-spin sectors. The Hamiltonian consists of:
\bea
H = H_{\text{N}} + H_{1\text{PN}} + H^{\text{LO}}_{\text{SO}} + H^{\text{LO}}_{\text{S}^2}
+ H^{\text{NLO}}_{\text{SO}} + H^{\text{NLO}}_{\text{S}^2} + H^{\text{LO}}_{\text{S}^3} 
+ H^{\text{LO}}_{\text{S}^4} + H^{\text{NLO}}_{\text{S}^3} + H^{\text{NLO}}_{\text{S}^4},\qquad
\eea
with:
\bea
H^{\text{LO}}_{\text{SO}} &=& H^{\text{LO}}_{\text{S}_1} + (1 \leftrightarrow 2), \qquad H^{\text{NLO}}_{\text{SO}} = H^{\text{NLO}}_{\text{S}_1} + (1 \leftrightarrow 2),\\
H^{\text{LO}}_{\text{S}^2} &=& C_{1 \text{ES}^2} H^{\text{LO}}_{\text{ES}_1^2} + H^{\text{LO}}_{\text{S}_1 \text{S}_2} + (1 \leftrightarrow 2),\\
H^{\text{NLO}}_{\text{S}^2} &=& H^{\text{NLO}}_{\text{S}_1^2} + C_{1 \text{ES}^2} H^{\text{NLO}}_{\text{ES}_1^2} + H^{\text{NLO}}_{\text{S}_1 \text{S}_2} + (1 \leftrightarrow 2),\\
H^{\text{LO}}_{\text{S}^3} &=& C_{1 \text{ES}^2} H^{\text{LO}}_{(\text{ES}_1^2) \text{S}_1} + C_{1 \text{BS}^3} H^{\text{LO}}_{\text{BS}_1^3}  + H^{\text{LO}}_{\text{S}_1^2 \text{S}_2} + C_{1 \text{ES}^2} H^{\text{LO}}_{(\text{ES}_1^2) \text{S}_2}+ (1 \leftrightarrow 2),\\
H^{\text{NLO}}_{\text{S}^3} &=& H^{\text{NLO}}_{\text{S}_1^3 } 
+ C_{1\text{ES}^2} H^{\text{NLO}}_{(\text{ES}_1^2 ) \text{S}_1} 
+ C_{1\text{ES}^2}^2 H^{\text{NLO}}_{C_{\text{ES}_1^2}^2 \text{S}_1^3 }
+ C_{1\text{BS}^3} H^{\text{NLO}}_{\text{BS}_1^3 } \nn \\
&& 
+ H^{\text{NLO}}_{\text{S}_1^2 S_2}  
+ C_{1\text{ES}^2} H^{\text{NLO}}_{(\text{ES}_1^2 ) \text{S}_2} 
	 +  (1 \leftrightarrow 2),\\
H^{\text{LO}}_{\text{S}^4} &=& C_{1 \text{ES}^4} H^{\text{NLO}}_{\text{ES}_1^4} + C_{1 \text{BS}^3} H^{\text{LO}}_{(\text{BS}_1^3) \text{S}_2}  + C_{1 \text{ES}^2} C_{2 \text{ES}^2} H^{\text{LO}}_{(\text{ES}_1^2) (\text{ES}_2^2)}+ (1 \leftrightarrow 2),\ 
\eea
and $H^{\text{NLO}}_{\text{S}^4}$ is given in eq.~\eqref{eq:hamcns4}.
For the COM generator we have:
\bea
\vec{G} = \vec{G}_{\text{N}} + \vec{G}_{1\text{PN}} + \vec{G}^{\text{LO}}_{\text{SO}} + \vec{G}^{\text{NLO}}_{\text{SO}} + \vec{G}^{\text{NLO}}_{\text{S}^2} + \vec{G}^{\text{NLO}}_{\text{S}^3} + \vec{G}^{\text{NLO}}_{\text{S}^4},
\eea
where we used that all generators of LO sectors with spin vanish beyond spin-orbit, and with:
\bea
\vec{G}^{\text{LO}}_{\text{SO}} &=& \vec{G}^{\text{LO}}_{\text{S}_1} + (1 \leftrightarrow 2), \qquad \vec{G}^{\text{NLO}}_{\text{SO}} = \vec{G}^{\text{NLO}}_{\text{S}_1} + (1 \leftrightarrow 2),\\
\vec{G}^{\text{NLO}}_{\text{S}^2} &=& \vec{G}^{\text{NLO}}_{\text{S}_1^2} + C_{1 \text{ES}^2} \vec{G}^{\text{NLO}}_{\text{ES}_1^2} + \vec{G}^{\text{NLO}}_{\text{S}_1 \text{S}_2} + (1 \leftrightarrow 2),\\
\vec{G}^{\text{NLO}}_{\text{S}^3} &=&\vec{G}^{\text{NLO}}_{\text{S}_1^3 } 
+C_{1\text{ES}^2} \vec{G}^{\text{NLO}}_{(\text{ES}_1^2 ) \text{S}_1} 
+ C_{1\text{ES}^2}^2 \vec{G}^{\text{NLO}}_{C_{\text{ES}^2_1}^2 \text{S}_1^3 }
+ C_{1\text{BS}^3} \vec{G}^{\text{NLO}}_{\text{BS}_1^3 } \nn\\
&&
+\vec{G}^{\text{NLO}}_{\text{S}_1^2 S_2} 
+ C_{1\text{ES}^2} \vec{G}^{\text{NLO}}_{(\text{ES}_1^2 ) \text{S}_2} 
+  (1 \leftrightarrow 2),
\eea
where we solved for the latter in \cite{Levi:2022dqm},
and now we should solve for:
\bea
\vec{G}^{\text{NLO}}_{\text{S}^4} &=& 
C_{1\text{ES}^2} \vec{G}^{\text{NLO}}_{(\text{ES}_1^2 ) \text{S}_1^2} 
+C_{1\text{ES}^2}^2 \vec{G}^{\text{NLO}}_{(\text{ES}_1^2 )^2} 
+ C_{1\text{BS}^3} \vec{G}^{\text{NLO}}_{(\text{BS}_1^3) \text{S}_1 } 
+C_{1\text{ES}^4} \vec{G}^{\text{NLO}}_{\text{ES}_1^4} 
+C_{1\text{E}^2\text{S}^4} \vec{G}^{\text{NLO}}_{\text{E}^2\text{S}_1^4}\nn\\
&& 
+\vec{G}^{\text{NLO}}_{\text{S}_1^3 \text{S}_2}
+ C_{1\text{ES}^2} \vec{G}^{\text{NLO}}_{(\text{ES}_1^2 ) \text{S}_1 \text{S}_2} 
+ C_{1\text{ES}^2}^2 \vec{G}^{\text{NLO}}_{C_{\text{ES}^2_1}^2 \text{S}_1^3 \text{S}_2 } 
+ C_{1\text{BS}^3} \vec{G}^{\text{NLO}}_{(\text{BS}_1^3) \text{S}_2 }\nn\\
&& +\vec{G}^{\text{NLO}}_{\text{S}_1^2 \text{S}_2^2}+C_{1\text{ES}^2} \vec{G}^{\text{NLO}}_{(\text{ES}_1^2 )  \text{S}_2^2} +C_{1\text{ES}^2} C_{2\text{ES}^2} \vec{G}^{\text{NLO}}_{(\text{ES}_1^2 )(\text{ES}_2^2 ) }+  (1 \leftrightarrow 2).
\eea

The Poisson brackets in eq.~\eqref{mainconstraint} then decouple to a set of independent equations to 
solve for the generators of each subsector according to the above decomposition. Let us then list this 
set of equations. 
We solve for $\vec{G}^{\text{NLO}}_{(\text{ES}_1^2 ) \text{S}_1^2}$ from:
\bea
0&=&\{ \vec{G}^{\text{NLO}}_{(\text{ES}_1^2 ) \text{S}_1^2},H_{\text{N}} \}_x + \{ \vec{G}_{\text{N}},H^{\text{NLO}}_{(\text{ES}_1^2 ) \text{S}_1^2} \}_x + \{ \vec{G}^{\text{LO}}_{\text{S}_1},H^{\text{LO}}_{(\text{ES}_1^2 ) \text{S}_1} \}_x  + \{ \vec{G}^{\text{NLO}}_{\text{S}_1^2},H^{\text{LO}}_{ \text{ES}_1^2} \}_x .
\eea
We solve for $\vec{G}^{\text{NLO}}_{(\text{ES}_1^2 )^2}$ from:
\bea
0&=&\{ \vec{G}^{\text{NLO}}_{(\text{ES}_1^2 )^2},H_{\text{N}} \}_x  + \{ \vec{G}_{\text{N}},H^{\text{NLO}}_{(\text{ES}_1^2 )^2} \}_x + \{ \vec{G}^{\text{NLO}}_{\text{ES}_1^2},H^{\text{LO}}_{ \text{ES}_1^2} \}_x.
\eea
We solve for $\vec{G}^{\text{NLO}}_{(\text{BS}_1^3) \text{S}_1 }$ from:
\bea
0&=&\{\vec{G}^{\text{NLO}}_{(\text{BS}_1^3) \text{S}_1 },H_{\text{N}} \}_x  + \{ \vec{G}_{\text{N}},H^{\text{NLO}}_{(\text{BS}_1^3 ) \text{S}_1} \}_x + \{ \vec{G}^{\text{LO}}_{\text{S}_1},H^{\text{LO}}_{\text{BS}_1^3} \}_x.
\eea
We solve for $ \vec{G}^{\text{NLO}}_{\text{ES}_1^4}$ from:
\bea
0&=&\{ \vec{G}^{\text{NLO}}_{\text{ES}_1^4},H_{\text{N}} \}_x + \{ \vec{G}_{\text{N}},H^{\text{NLO}}_{\text{ES}_1^4} \}_x +  \{ \vec{G}_{1\text{PN}},H^{\text{LO}}_{\text{ES}_1^4} \}_x +  \{ \vec{G}^{\text{LO}}_{\text{S}_1},H^{\text{LO}}_{\text{ES}_1^4} \}_S.
\eea
We solve for $ \vec{G}^{\text{NLO}}_{\text{E}^2\text{S}_1^4}$ from:
\bea
0&=&\{ \vec{G}^{\text{NLO}}_{\text{E}^2\text{S}_1^4},H_{\text{N}} \}_x + \{ \vec{G}_{\text{N}},H^{\text{NLO}}_{\text{E}^2\text{S}_1^4} \}_x.
\eea
We solve for $\vec{G}^{\text{NLO}}_{\text{S}_1^3 \text{S}_2}$ from:
\bea
0&=&\{ \vec{G}^{\text{NLO}}_{\text{S}_1^3 \text{S}_2},H_{\text{N}} \}_x + \{ \vec{G}_{\text{N}},H^{\text{NLO}}_{\text{S}_1^3 \text{S}_2} \}_x + \{ \vec{G}^{\text{LO}}_{\text{S}_1},H^{\text{LO}}_{\text{S}_1^2  \text{S}_2} \}_x + \{ \vec{G}^{\text{NLO}}_{\text{S}_1^2},2H^{\text{LO}}_{\text{S}_1  \text{S}_2} \}_x .
\eea
We solve for $ \vec{G}^{\text{NLO}}_{(\text{ES}_1^2 ) \text{S}_1 \text{S}_2} $ from:
\bea
0&=&\{ \vec{G}^{\text{NLO}}_{(\text{ES}_1^2 ) \text{S}_1 \text{S}_2},H_{\text{N}} \}_x + \{ \vec{G}_{\text{N}},H^{\text{NLO}}_{(\text{ES}_1^2 ) \text{S}_1 \text{S}_2} \}_x + \{ \vec{G}^{\text{LO}}_{\text{S}_2},H^{\text{LO}}_{(\text{ES}_1^2 ) \text{S}_1} \}_x + \{ \vec{G}^{\text{LO}}_{\text{S}_1},H^{\text{LO}}_{(\text{ES}_1^2 ) \text{S}_2} \}_x  \nn\\
&&+ \{ 2\vec{G}^{\text{NLO}}_{\text{S}_1  \text{S}_2},H^{\text{LO}}_{ \text{ES}_1^2} \}_x + \{ \vec{G}^{\text{NLO}}_{\text{ES}_1^2}, 2H^{\text{LO}}_{  \text{S}_1  \text{S}_2} \}_x .
\eea
We solve for $\vec{G}^{\text{NLO}}_{C_{\text{ES}^2_1}^2 \text{S}_1^3 \text{S}_2 }$ from:
\bea
0&=&\{ \vec{G}^{\text{NLO}}_{C_{\text{ES}^2_1}^2 \text{S}_1^3 \text{S}_2 } ,H_{\text{N}} \}_x + \{ \vec{G}_{\text{N}},H^{\text{NLO}}_{C_{\text{ES}^2_1}^2 \text{S}_1^3 \text{S}_2 } \}_x.
\eea
We solve for $ \vec{G}^{\text{NLO}}_{(\text{BS}_1^3) \text{S}_2 }$ from:
\bea
0&=&\{ \vec{G}^{\text{NLO}}_{(\text{BS}_1^3) \text{S}_2 },H_{\text{N}} \}_x + \{ \vec{G}_{\text{N}},H^{\text{NLO}}_{(\text{BS}_1^3 ) \text{S}_2} \}_x + \{ \vec{G}^{\text{LO}}_{\text{S}_2},H^{\text{LO}}_{\text{BS}_1^3} \}_x + \{ \vec{G}_{1\text{PN}},H^{\text{LO}}_{(\text{BS}_1^3 ) \text{S}_2} \}_x\nn\\
&&+ \{ \vec{G}^{\text{LO}}_{\text{S}_2},H^{\text{LO}}_{\text{BS}_1^3 \text{S}_2} \}_S + \{ \vec{G}^{\text{LO}}_{\text{S}_1},H^{\text{LO}}_{\text{BS}_1^3 \text{S}_2} \}_S .
\eea
We solve for $\vec{G}^{\text{NLO}}_{\text{S}_1^2 \text{S}_2^2}$ from:
\bea
0&=&\{ 2\vec{G}^{\text{NLO}}_{\text{S}_1^2 \text{S}_2^2},H_{\text{N}} \}_x + \{ \vec{G}_{\text{N}},2H^{\text{NLO}}_{\text{S}_1^2 \text{S}_2^2} \}_x + \{ \vec{G}^{\text{LO}}_{\text{S}_2},H^{\text{LO}}_{\text{S}_1^2  \text{S}_2} \}_x + \{ \vec{G}^{\text{LO}}_{\text{S}_1},H^{\text{LO}}_{\text{S}_1  \text{S}_2^2} \}_x \nn\\
&&+ \{2 \vec{G}^{\text{NLO}}_{\text{S}_1 \text{S}_2},2H^{\text{LO}}_{\text{S}_1  \text{S}_2} \}_x.
\eea
We solve for $ \vec{G}^{\text{NLO}}_{(\text{ES}_1^2 )  \text{S}_2^2}$ from:
\bea
0&=&\{ \vec{G}^{\text{NLO}}_{(\text{ES}_1^2 )  \text{S}_2^2},H_{\text{N}} \}_x + \{ \vec{G}_{\text{N}},H^{\text{NLO}}_{(\text{ES}_1^2 ) \text{S}_2^2} \}_x + \{ \vec{G}^{\text{LO}}_{\text{S}_2},H^{\text{LO}}_{(\text{ES}_1^2 ) \text{S}_2} \}_x  + \{ \vec{G}^{\text{NLO}}_{\text{S}_2^2},H^{\text{LO}}_{ \text{ES}_1^2} \}_x .
\eea
We solve for $  \vec{G}^{\text{NLO}}_{(\text{ES}_1^2 )(\text{ES}_2^2 ) }$ from:
\bea
0&=&\{ 2\vec{G}^{\text{NLO}}_{(\text{ES}_1^2 )(\text{ES}_2^2 ) },H_{\text{N}} \}_x + \{ \vec{G}_{\text{N}},2H^{\text{NLO}}_{(\text{ES}_1^2 ) (\text{ES}_2^2)} \}_x + \{ \vec{G}_{1\text{PN}},2H^{\text{LO}}_{(\text{ES}_1^2 ) (\text{ES}_2^2)} \}_x   + \{ \vec{G}^{\text{NLO}}_{\text{ES}_2^2},H^{\text{LO}}_{ \text{ES}_1^2} \}_x \nn\\
&&+ \{ \vec{G}^{\text{NLO}}_{\text{ES}_1^2},H^{\text{LO}}_{ \text{ES}_2^2} \}_x+ \{ \vec{G}^{\text{LO}}_{\text{S}_2},2H^{\text{LO}}_{(\text{ES}_1^2 ) (\text{ES}_2^2)} \}_S + \{ \vec{G}^{\text{LO}}_{\text{S}_1},2H^{\text{LO}}_{(\text{ES}_1^2 ) (\text{ES}_2^2)} \}_S.
\eea

We then write the solution of $\vec{G}^{\text{NLO}}_{\text{S}^4}$ as:
\bea
\vec{G}^{\text{NLO}}_{\text{S}^4} &=& 
H^{\text{LO}}_{\text{S}^4} \frac{ \vec{x}_1 + \vec{x}_2}{2}  + \vec{Y}^{\text{NLO}}_{\text{S}^4},
\eea
with:
\bea
\vec{Y}^{\text{NLO}}_{\text{S}^4 } &=&- 	\frac{G C_{1\text{ES}^4} m_{2}}{4 m_{1}{}^3 r{}^4} \Big[ 3 \vec{S}_{1}\cdot\vec{n} S_{1}^2 \vec{S}_{1} - 5 ( \vec{S}_{1}\cdot\vec{n})^{3} \vec{S}_{1} \Big] - 	\frac{G C_{1\text{BS}^3} m_{2}}{2 m_{1}{}^3 r{}^4} \Big[ 5 S_{1}^2 \vec{n} ( \vec{S}_{1}\cdot\vec{n})^{2} - \vec{n} S_{1}^{4} \nn\\ 
&& + \vec{S}_{1}\cdot\vec{n} S_{1}^2 \vec{S}_{1} - 5 ( \vec{S}_{1}\cdot\vec{n})^{3} \vec{S}_{1} \Big] + 	\frac{3 G C_{1\text{ES}^2} m_{2}}{4 m_{1}{}^3 r{}^4} \Big[ 5 S_{1}^2 \vec{n} ( \vec{S}_{1}\cdot\vec{n})^{2} - \vec{n} S_{1}^{4} + \vec{S}_{1}\cdot\vec{n} S_{1}^2 \vec{S}_{1} \nn\\ 
&& - 5 ( \vec{S}_{1}\cdot\vec{n})^{3} \vec{S}_{1} \Big] \nn\\
&&  - 	\frac{3 G}{2 m_{1}{}^2 r{}^4} \Big[ S_{1}^2 \vec{S}_{1}\cdot\vec{S}_{2} \vec{n} - 5 \vec{S}_{1}\cdot\vec{n} S_{1}^2 \vec{n} \vec{S}_{2}\cdot\vec{n} - 2 \vec{S}_{1}\cdot\vec{n} \vec{S}_{1}\cdot\vec{S}_{2} \vec{S}_{1} + 5 \vec{S}_{2}\cdot\vec{n} ( \vec{S}_{1}\cdot\vec{n})^{2} \vec{S}_{1} \nn\\ 
&& + \vec{S}_{1}\cdot\vec{n} S_{1}^2 \vec{S}_{2} \Big] - 	\frac{G C_{1\text{BS}^3}}{4 m_{1}{}^2 r{}^4} \Big[ 2 S_{1}^2 \vec{S}_{1}\cdot\vec{S}_{2} \vec{n} - 10 \vec{S}_{1}\cdot\vec{S}_{2} \vec{n} ( \vec{S}_{1}\cdot\vec{n})^{2} - S_{1}^2 \vec{S}_{2}\cdot\vec{n} \vec{S}_{1} \nn\\ 
&& + 2 \vec{S}_{1}\cdot\vec{n} \vec{S}_{1}\cdot\vec{S}_{2} \vec{S}_{1} + 5 \vec{S}_{2}\cdot\vec{n} ( \vec{S}_{1}\cdot\vec{n})^{2} \vec{S}_{1} + 3 \vec{S}_{1}\cdot\vec{n} S_{1}^2 \vec{S}_{2} - 5 ( \vec{S}_{1}\cdot\vec{n})^{3} \vec{S}_{2} \Big] \nn\\ 
&& + 	\frac{3 G C_{1\text{ES}^2}}{2 m_{1}{}^2 r{}^4} \Big[ S_{1}^2 \vec{S}_{1}\cdot\vec{S}_{2} \vec{n} - 5 \vec{S}_{1}\cdot\vec{S}_{2} \vec{n} ( \vec{S}_{1}\cdot\vec{n})^{2} + 2 \vec{S}_{1}\cdot\vec{n} \vec{S}_{1}\cdot\vec{S}_{2} \vec{S}_{1} - 3 \vec{S}_{1}\cdot\vec{n} S_{1}^2 \vec{S}_{2} \nn\\ 
&& + 5 ( \vec{S}_{1}\cdot\vec{n})^{3} \vec{S}_{2} \Big]  \nn\\
&& - 	\frac{3 G C_{1\text{ES}^2} C_{2\text{ES}^2}}{4 m_{1} m_{2} r{}^4} \Big[ 5 S_{1}^2 \vec{n} ( \vec{S}_{2}\cdot\vec{n})^{2} + 2 \vec{S}_{2}\cdot\vec{n} \vec{S}_{1}\cdot\vec{S}_{2} \vec{S}_{1} + 3 \vec{S}_{1}\cdot\vec{n} S_{2}^2 \vec{S}_{1} \nn\\ 
&& - 5 \vec{S}_{1}\cdot\vec{n} ( \vec{S}_{2}\cdot\vec{n})^{2} \vec{S}_{1} \Big] - 	\frac{3 G C_{1\text{ES}^2}}{4 m_{1} m_{2} r{}^4} \Big[ S_{1}^2 S_{2}^2 \vec{n} - 5 S_{2}^2 \vec{n} ( \vec{S}_{1}\cdot\vec{n})^{2} + 2 \vec{S}_{1}\cdot\vec{n} S_{2}^2 \vec{S}_{1} \nn\\ 
&& - S_{1}^2 \vec{S}_{2}\cdot\vec{n} \vec{S}_{2} - 2 \vec{S}_{1}\cdot\vec{n} \vec{S}_{1}\cdot\vec{S}_{2} \vec{S}_{2} + 5 \vec{S}_{2}\cdot\vec{n} ( \vec{S}_{1}\cdot\vec{n})^{2} \vec{S}_{2} \Big]   + \left(1 \leftrightarrow 2\right) .
\eea
Thus we found a solution for the Poincaré algebra of the NLO quartic-in-spin sectors, with our new 
general Hamiltonian from eq.~\eqref{eq:hamcns4}, and this provides significant confidence in the 
validity of these new results. 

\subsection{N$^3$LO Quadratic-in-Spin Sectors}
\label{poincares2n3lo}

To confirm the new precision frontier at the $5$PN order across all sectors, we proceed to solve for 
the Poincar\'e algebra of the N$^3$LO quadratic-in-spin sectors.
Again, we start by decomposing $H$ and $\vec{G}$ to all the sectors relevant to the solution for 
$\vec{G}$ at the N$^3$LO 
quadratic-in-spin sectors in question. The Hamiltonian consists of:
\bea
H&=& H_{\text{N}} + H_{1\text{PN}} + H^{\text{LO}}_{\text{SO}}  + H_{2\text{PN}} 
+ H^{\text{LO}}_{\text{S}^2}  + H^{\text{NLO}}_{\text{SO}} + H_{3\text{PN}} 
+ H^{\text{NLO}}_{\text{S}^2} \nn\\
&& + H^{\text{N}^2\text{LO}}_{\text{SO}} 
+ H^{\text{N}^2\text{LO}}_{\text{S}^2} + H^{\text{N}^3\text{LO}}_{\text{SO}} + H^{\text{N}^3\text{LO}}_{\text{S}^2}, 
\eea
with $H^{\text{N$^3$LO}}_{\text{SO}}$ taken from our \cite{Kim:2022pou}, and 
$H^{\text{N$^3$LO}}_{\text{S}^2}$ from our \cite{Kim:2022bwv}, and we have:
\bea
H^{\text{LO}}_{\text{SO}} &=& H^{\text{LO}}_{\text{S}_1} + (1 \leftrightarrow 2), \qquad H^{\text{NLO}}_{\text{SO}} = H^{\text{NLO}}_{\text{S}_1} + (1 \leftrightarrow 2),\\
H^{\text{N}^2\text{LO}}_{\text{SO}} &=& H^{\text{N}^2\text{LO}}_{\text{S}_1} + (1 \leftrightarrow 2), \qquad H^{\text{N}^3\text{LO}}_{\text{SO}} = H^{\text{N}^3\text{LO}}_{\text{S}_1} + (1 \leftrightarrow 2),\\
H^{\text{LO}}_{\text{S}^2} &=& C_{1 \text{ES}^2} H^{\text{LO}}_{\text{ES}_1^2} + H^{\text{LO}}_{\text{S}_1 \text{S}_2} + (1 \leftrightarrow 2),\\
H^{\text{NLO}}_{\text{S}^2} &=& H^{\text{NLO}}_{\text{S}_1^2} + C_{1 \text{ES}^2} H^{\text{NLO}}_{\text{ES}_1^2} + H^{\text{NLO}}_{\text{S}_1 \text{S}_2} + (1 \leftrightarrow 2),\\
H^{\text{N}^2\text{LO}}_{\text{S}^2} &=& H^{\text{N}^2\text{LO}}_{\text{S}_1^2} + C_{1 \text{ES}^2} H^{\text{N}^2\text{LO}}_{\text{ES}_1^2} + H^{\text{N}^2\text{LO}}_{\text{S}_1 \text{S}_2} + (1 \leftrightarrow 2),\\
H^{\text{N}^3\text{LO}}_{\text{S}^2} &=& H^{\text{N}^3\text{LO}}_{\text{S}_1^2} + C_{1 \text{ES}^2} H^{\text{N}^3\text{LO}}_{\text{ES}_1^2} + C_{1 \text{E}^2\text{S}^2} H^{\text{N}^3\text{LO}}_{\text{E}^2\text{S}_1^2} + H^{\text{N}^3\text{LO}}_{\text{S}_1 \text{S}_2} + (1 \leftrightarrow 2).
\eea
For the COM generator we have:
\bea
\vec{G}&=& \vec{G}_{\text{N}} + \vec{G}_{1\text{PN}} + \vec{G}^{\text{LO}}_{\text{SO}} 
+ \vec{G}_{2\text{PN}} + \vec{G}^{\text{NLO}}_{\text{SO}} + \vec{G}_{3\text{PN}} 
+ \vec{G}^{\text{NLO}}_{\text{S}^2}
+ \vec{G}^{\text{N}^2\text{LO}}_{\text{SO}} + \vec{G}^{\text{N}^2\text{LO}}_{\text{S}^2} \nn\\
&&+ \vec{G}^{\text{N}^3\text{LO}}_{\text{SO}} + \vec{G}^{\text{N}^3\text{LO}}_{\text{S}^2},
\eea
where we used again that all generators of LO sectors with spin vanish beyond spin-orbit, and with:
\bea
\vec{G}^{\text{LO}}_{\text{SO}} &=& \vec{G}^{\text{LO}}_{\text{S}_1} + (1 \leftrightarrow 2), \qquad \vec{G}^{\text{NLO}}_{\text{SO}} = \vec{G}^{\text{NLO}}_{\text{S}_1} + (1 \leftrightarrow 2),\\ 
\vec{G}^{\text{N}^2\text{LO}}_{\text{SO}} &=& \vec{G}^{\text{N}^2\text{LO}}_{\text{S}_1} + (1 \leftrightarrow 2), \quad
\vec{G}^{\text{N}^3\text{LO}}_{\text{SO}} = \vec{G}^{\text{N}^3\text{LO}}_{\text{S}_1} + (1 \leftrightarrow 2),\\
\vec{G}^{\text{NLO}}_{\text{S}^2} &=& \vec{G}^{\text{NLO}}_{\text{S}_1^2} + C_{1 \text{ES}^2} \vec{G}^{\text{NLO}}_{\text{ES}_1^2} + \vec{G}^{\text{NLO}}_{\text{S}_1 \text{S}_2} + (1 \leftrightarrow 2),\\
\vec{G}^{\text{N}^2\text{LO}}_{\text{S}^2} &=& \vec{G}^{\text{N}^2\text{LO}}_{\text{S}_1^2} + C_{1 \text{ES}^2} \vec{G}^{\text{N}^2\text{LO}}_{\text{ES}_1^2} + \vec{G}^{\text{N}^2\text{LO}}_{\text{S}_1 \text{S}_2} + (1 \leftrightarrow 2),
\eea
where we solved for $\vec{G}^{\text{N}^3\text{LO}}_{\text{SO}}$ in \cite{Levi:2022dqm},
and now we should solve for:
\bea
\vec{G}^{\text{N}^3\text{LO}}_{\text{S}^2} &=& \vec{G}^{\text{N}^3\text{LO}}_{\text{S}_1^2} + C_{1 \text{ES}^2} \vec{G}^{\text{N}^3\text{LO}}_{\text{ES}_1^2} + C_{1 \text{E}^2\text{S}^2} \vec{G}^{\text{N}^3\text{LO}}_{\text{E}^2\text{S}_1^2} + \vec{G}^{\text{N}^3\text{LO}}_{\text{S}_1 \text{S}_2} + (1 \leftrightarrow 2).
\eea

Let us then list the set of decoupled equations to solve for the generators of each sector. 
We solve for $ \vec{G}^{\text{N}^3\text{LO}}_{\text{S}_1^2}$ from:
\bea
0&=& \{  \vec{G}^{\text{N}^3\text{LO}}_{\text{S}_1^2} , H_{\text{N}} \}_x + \{  \vec{G}_{\text{N}} , H^{\text{N}^3\text{LO}}_{\text{S}_1^2} \}_x  + \{  \vec{G}^{\text{LO}}_{\text{S}_1} , H^{\text{N}^2\text{LO}}_{\text{S}_1} \}_x + \{  \vec{G}_{1\text{PN}} , H^{\text{N}^2\text{LO}}_{\text{S}_1^2} \}_x  \nn\\
&&+ \{  \vec{G}^{\text{NLO}}_{\text{S}_1} , H^{\text{NLO}}_{\text{S}_1} \}_x + \{  \vec{G}^{\text{NLO}}_{\text{S}_1^2} , H^{\text{N}^2\text{LO}}_{2\text{PN}} \}_x + \{  \vec{G}^{\text{N}^2\text{LO}}_{2\text{PN}} , H^{\text{NLO}}_{\text{S}_1^2} \}_x + \{  \vec{G}^{\text{N}^2\text{LO}}_{\text{S}_1} , H^{\text{LO}}_{\text{S}_1} \}_x  \nn\\
&&+ \{  \vec{G}^{\text{N}^2\text{LO}}_{\text{S}_1^2} , H_{1\text{PN}} \}_x + \{  \vec{G}^{\text{LO}}_{\text{S}_1} , H^{\text{N}^2\text{LO}}_{\text{S}_1^2} \}_S + \{  \vec{G}^{\text{NLO}}_{\text{S}_1} , H^{\text{NLO}}_{\text{S}_1^2} \}_S + \{  \vec{G}^{\text{NLO}}_{\text{S}_1^2} , H^{\text{NLO}}_{\text{S}_1} \}_S \nn\\
&&+ \{  \vec{G}^{\text{N}^2\text{LO}}_{\text{S}_1} , H^{\text{LO}}_{\text{S}_1^2} \}_S + \{  \vec{G}^{\text{LO}}_{\text{S}_1^2} , H^{\text{N}^2\text{LO}}_{\text{S}_1}  \}_S .
\eea
We solve for $ \vec{G}^{\text{N}^3\text{LO}}_{\text{ES}_1^2}$ from:
\bea
0&=& \{  \vec{G}^{\text{N}^3\text{LO}}_{\text{ES}_1^2} , H_{\text{N}} \}_x + \{  \vec{G}_{\text{N}} , H^{\text{N}^3\text{LO}}_{\text{ES}_1^2} \}_x + \{  \vec{G}_{1\text{PN}} , H^{\text{N}^2\text{LO}}_{\text{ES}_1^2} \}_x + \{  \vec{G}^{\text{NLO}}_{\text{ES}_1^2} , H^{\text{N}^2\text{LO}}_{2\text{PN}} \}_x \nn\\
&& + \{  \vec{G}^{\text{N}^2\text{LO}}_{2\text{PN}} , H^{\text{NLO}}_{\text{ES}_1^2} \}_x + \{  \vec{G}^{\text{N}^2\text{LO}}_{\text{ES}_1^2} , H_{1\text{PN}} \}_x  + \{  \vec{G}^{\text{N}^3\text{LO}}_{3\text{PN}} , H^{\text{LO}}_{\text{ES}_1^2} \}_x  + \{  \vec{G}^{\text{LO}}_{\text{S}_1} , H^{\text{N}^2\text{LO}}_{\text{ES}_1^2} \}_S \nn\\
&&+ \{  \vec{G}^{\text{NLO}}_{\text{S}_1} , H^{\text{NLO}}_{\text{ES}_1^2} \}_S + \{  \vec{G}^{\text{NLO}}_{\text{ES}_1^2} , H^{\text{NLO}}_{\text{S}_1} \}_S + \{  \vec{G}^{\text{N}^2\text{LO}}_{\text{S}_1} , H^{\text{LO}}_{\text{ES}_1^2} \}_S + \{  \vec{G}^{\text{LO}}_{\text{ES}_1^2} , H^{\text{N}^2\text{LO}}_{\text{S}_1}  \}_S . \qquad \qquad
\eea
We solve for $ \vec{G}^{\text{N}^3\text{LO}}_{\text{E}^2\text{S}_1^2}$ from:
\bea
0&=& \{  \vec{G}^{\text{N}^3\text{LO}}_{\text{E}^2\text{S}_1^2} , H_{\text{N}} \}_x + \{   \vec{G}_{\text{N}},  H^{\text{N}^3\text{LO}}_{\text{E}^2\text{S}_1^2}\}_x.
\eea
We solve for $ \vec{G}^{\text{N}^3\text{LO}}_{\text{S}_1 \text{S}_2}$ from:
\bea
0&=& \{  \vec{G}^{\text{N}^3\text{LO}}_{\text{S}_1 \text{S}_2} , H_{\text{N}} \}_x + \{  \vec{G}_{\text{N}} , H^{\text{N}^3\text{LO}}_{\text{S}_1 \text{S}_2} \}_x + \{  \vec{G}^{\text{LO}}_{\text{S}_2} , H^{\text{N}^2\text{LO}}_{ \text{S}_1} \}_x + \{  \vec{G}_{1\text{PN}} , H^{\text{N}^2\text{LO}}_{\text{S}_1 \text{S}_2} \}_x\nn\\
&&+ \{  \vec{G}^{\text{NLO}}_{\text{S}_2} , H^{\text{NLO}}_{ \text{S}_1} \}_x + \{  \vec{G}^{\text{NLO}}_{\text{S}_1 \text{S}_2} , H^{\text{N}^2\text{LO}}_{ 2\text{PN}} \}_x + \{  \vec{G}^{\text{N}^2\text{LO}}_{2\text{PN}} , H^{\text{NLO}}_{\text{S}_1 \text{S}_2} \}_x + \{  \vec{G}^{\text{N}^2\text{LO}}_{\text{S}_2} , H^{\text{LO}}_{ \text{S}_1} \}_x\nn\\ 
&&+ \{  \vec{G}^{\text{N}^2\text{LO}}_{\text{S}_1 \text{S}_2} , H_{ 1\text{PN}} \}_x + \{  \vec{G}^{\text{N}^3\text{LO}}_{\text{PN}} , H^{\text{LO}}_{\text{S}_1 \text{S}_2} \}_x + \{  \vec{G}^{\text{LO}}_{\text{S}_2} , 2H^{\text{N}^2\text{LO}}_{\text{S}_1 \text{S}_2} \}_S + \{  \vec{G}^{\text{NLO}}_{\text{S}_2} , 2H^{\text{NLO}}_{\text{S}_1 \text{S}_2} \}_S\nn\\
&&+ \{  2\vec{G}^{\text{NLO}}_{\text{S}_1 \text{S}_2} , H^{\text{NLO}}_{\text{S}_2} \}_S + \{  \vec{G}^{\text{N}^2\text{LO}}_{\text{S}_2} , 2H^{\text{LO}}_{\text{S}_1 \text{S}_2} \}_S + \{  2\vec{G}^{\text{N}^2\text{LO}}_{\text{S}_1 \text{S}_2} , H^{\text{LO}}_{\text{S}_2} \}_S + \left(1 \leftrightarrow 2 \right).
\eea

We then write the solution of $\vec{G}^{\text{N}^3\text{LO}}_{\text{S}^2}$ as:
\bea
\vec{G}^{\text{N}^3\text{LO}}_{\text{S}^2} &=& H^{\text{N}^2\text{LO}}_{\text{S}^2} \frac{ \vec{x}_1 + \vec{x}_2}{2} +  \left( \vec{Y}^{\text{N}^3\text{LO}}_{\text{S}_1^2 }  + C_{1\text{ES}^2} \vec{Y}^{\text{N}^3\text{LO}}_{\text{ES}_1^2} +  \vec{Y}^{\text{N}^3\text{LO}}_{\text{S}_1 \text{S}_2 } + \left(1 \leftrightarrow 2 \right) \right),
\eea
where we present the explicit expressions in appendix \ref{comgenn3los2}.

Thus, we solved for the Poincar\'e algebra of the N$^3$LO quadratic-in-spin sectors, 
with our full general Hamiltonians first presented in \cite{Kim:2022bwv}, and this provides 
significant confidence in the validity of these results.
It should be highlighted that though an agreement with the consequent observables in specific 
restricted configurations, which we first provided in \cite{Kim:2021rfj}, was found later in 
\cite{Mandal:2022ufb}, the Poincar\'e algebra which we solved for here, provides the most stringent 
check on the most comprehensive Hamiltonian results. 
This then completes the Poincar\'e algebra across all sectors with spins at the new precision 
frontier at the $5$PN order.  

\section{Specialized Hamiltonians and Observables}
\label{spechamsnobs}

Obtaining Hamiltonians is essential to derive EOMs, EOB models, or as we have shown, to study the global 
Poincar\'e algebra of the binary system. However, Hamiltonians are also critical to get sensible binding 
energies, which are key to GW applications. To that end, we gradually simplify our 
general Hamiltonians from section \ref{genhams}, by considering specific kinematic configurations, and 
eventually removing all gauge-dependence to get physical observables.
First, we define some binary-mass conventions:
\begin{align}
m & \equiv m_1+m_2, 
\quad q \equiv m_1/m_2,  \\
\mu \equiv m_1 m_2 / m, 
\quad	\nu & \equiv m_1 m_2/m^2 = \mu/m = q/(1+q)^2,
\end{align}
where $q$ and $\nu$ are the dimensionless mass ratio and symmetric mass-ratio, respectively, that are 
used to express the final observables. All the variables are also converted to dimensionless ones, 
denoted with a tilde, where $Gm$ and $\mu$ are the length and mass units, respectively. 
The general Hamiltonians are then specified to the COM frame, reducing to a single momentum vector, 
$\vec{p} \equiv \vec{p}_1 = - \vec{p}_2$, with which the orbital angular momentum is defined,
$\vec{L} \equiv r \vec{n} \times \vec{p}$. 

The NLO quartic-in-spin Hamiltonians in the COM frame are then:
\bea
\tilde{H}^{\text{NLO}}_{\text{S}^4} &=&
C_{1\text{ES}^2} \tilde{H}^{\text{NLO}}_{(\text{ES}_1^2 ) \text{S}_1^2} 
+C_{1\text{ES}^2}^2 \tilde{H}^{\text{NLO}}_{(\text{ES}_1^2 )^2} 
+ C_{1\text{BS}^3} \tilde{H}^{\text{NLO}}_{(\text{BS}_1^3) \text{S}_1 } 
+C_{1\text{ES}^4} \tilde{H}^{\text{NLO}}_{\text{ES}_1^4} 
+C_{1\text{E}^2\text{S}^4} \tilde{H}^{\text{NLO}}_{\text{E}^2\text{S}_1^4}\nn\\
&& +\tilde{H}^{\text{NLO}}_{\text{S}_1^3 \text{S}_2}
+C_{1\text{ES}^2} \tilde{H}^{\text{NLO}}_{(\text{ES}_1^2 ) \text{S}_1 \text{S}_2} 
+ C_{1\text{ES}^2}^2 \tilde{H}^{\text{NLO}}_{C_{\text{ES}^2_1}^2 \text{S}_1^3 \text{S}_2 } 
+ C_{1\text{BS}^3} \tilde{H}^{\text{NLO}}_{(\text{BS}_1^3) \text{S}_2 }\nn\\
&& +\tilde{H}^{\text{NLO}}_{\text{S}_1^2 \text{S}_2^2}
+C_{1\text{ES}^2} \tilde{H}^{\text{NLO}}_{(\text{ES}_1^2 )  \text{S}_2^2} 
+C_{1\text{ES}^2} C_{2\text{ES}^2} \tilde{H}^{\text{NLO}}_{(\text{ES}_1^2 )(\text{ES}_2^2 ) }
+  (1 \leftrightarrow 2),
\eea
where we present the explicit expressions in appendix \ref{newcomhams}.
Very few scattering-amplitudes approaches to the scattering problem also present 
analogous Hamiltonians for the bound problem, though only for low loop orders, as explained in  
section \ref{intro} and illustrated in table \ref{pnfrontier}, due to radiation-reaction effects that 
kick in as of the N$^2$LO. In any case, those analogous Hamiltonians are already restricted to the COM 
frame \cite{Chen:2021qkk,Bern:2022kto}.

As noted in \cite{Levi:2022dqm}, \cite{Chen:2021qkk} and \cite{Bern:2022kto} differ between them 
in several ways already at the level of their physical scattering amplitudes, which underlie their 
subsequent COM Hamiltonians for BHs, and for generic compact objects, respectively.
The results in \cite{Chen:2021qkk} have been known to be discrepant with ours as of our verified
cubic-in-spin orders, which was already clarified in \cite{Levi:2022dqm}, and at the present 
quartic-in-spin sectors.

As was also clarified in \cite{Levi:2022dqm}, the COM Hamiltonians presented in \cite{Bern:2022kto} 
for generic compact binaries, have been known to be discrepant even for BHs, with our results as of 
the LO cubic-in-spin sectors in \cite{Levi:2014gsa}, which have been well-verified along the years. 
This discrepancy is due to singularities that show up as of cubic order in spin on \cite{Bern:2022kto}.
At the NLO the COM Hamiltonians in \cite{Bern:2022kto} for the quartic-in-spin sectors exhibit a 
growing discrepancy, including similar singularities even for BHs. 
Moreover, at the NLO quartic-in-spin sectors the work in \cite{Bern:2022kto} contained contributions 
with $3$ claimed new Wilson coefficients, $H_2$, $H_3$, and $H_4$, which they stipulated in their 
formulation. Such extra free parameters violate spin-gauge invariance \cite{Levi:2015msa}, and are 
also absent and discrepant with other corresponding physical scattering amplitudes, e.g.~in 
\cite{Chen:2021qkk}. 
Moreover, there was an inconsistent omission in \cite{Bern:2022kto} of the new subleading spin-induced 
hexadecapular deformation operator, as noted in section \ref{finaltheory}, and thus the consequent  
contribution is also missing in their results.

A further simplification is obtained by constraining the spins to be aligned with the orbital angular 
momentum, thus also requiring $\vec{S}_a \cdot \vec{n} = \vec{S}_a \cdot \vec{p}=0$. 
However, this simplification is notably inappropriate at higher-spin sectors, as it entails a growing 
loss of information on the physics of the system -- the higher in spin the sectors are -- as of quadratic order in 
the spin. For the present quartic-in-spin sectors, the aligned-spins simplification leads to the 
following total Hamiltonian: 
\bea
\label{tothamaligned}
\tilde{H}^{\text{NLO}}_{\text{S}^4} &=&\frac{\nu^2 \tilde{S}_1^4 }{\tilde{r}^6} \left[ C_{1(\text{ES}^2)} \left( 
\frac{3}{2}-\frac{3 \nu }{2} -\frac{3 \nu }{4} \frac{\tilde{L}^2}{\tilde{r}}  + \tilde{p}_r^2 \tilde{r} \left( \frac{45}{16}-\frac{21 \nu }{8} \right) \right. \right.\nn\\
&&\left. + \frac{1}{ \nu q} \left( -\frac{3 \nu ^2}{2}+\frac{9 \nu }{2}-\frac{3}{2} + \frac{\tilde{L}^2}{\tilde{r}} \left(\frac{3 \nu }{4}-\frac{3 \nu ^2}{2} \right) + \tilde{p}_r^2 \tilde{r} \left( -\frac{39 \nu ^2}{16}+\frac{33 \nu }{4}-\frac{45}{16} \right) \right) \right)\nn\\
&&+C_{1(\text{ES}^2)}^2 \left( -\frac{\nu }{8}  + \frac{1}{ q} \left( \frac{1}{8}-\frac{\nu }{8} \right) \right)
+ C_{1(\text{BS}^3)} \left( 
-\frac{\nu }{2}-\frac{3}{2} + \frac{\tilde{L}^2}{\tilde{r}}  \left( \frac{3 \nu }{2}-\frac{3}{2} \right) \right.\nn\\
&&+ \tilde{p}_r^2 \tilde{r} \left( \frac{3 \nu }{2}-\frac{3}{2} \right)   + \frac{1}{ \nu q} \left( -\frac{\nu ^2}{2}-\frac{5 \nu }{2}+\frac{3}{2} + \frac{\tilde{L}^2}{\tilde{r}} \left( \frac{3 \nu ^2}{2}-\frac{9 \nu }{2}+\frac{3}{2} \right) \right.\nn\\
&&\left.\left.+ \tilde{p}_r^2 \tilde{r} \left( \frac{3 \nu ^2}{2}-\frac{9 \nu }{2}+\frac{3}{2} \right) \right) \right)
+ C_{1(\text{ES}^4)} \left( 
\frac{5 \nu }{2}-\frac{23}{8} + \frac{\tilde{L}^2}{\tilde{r}}  \left( \frac{3 \nu }{16}+\frac{9}{16} \right) \right.\nn\\
&&+ \tilde{p}_r^2 \tilde{r} \left( \frac{9 \nu }{8}+\frac{9}{16} \right)   + \frac{1}{ \nu q} \left( \frac{5 \nu ^2}{2}-\frac{33 \nu }{4}+\frac{23}{8} + \frac{\tilde{L}^2}{\tilde{r}} \left( \frac{3 \nu ^2}{8}+\frac{15 \nu }{16}-\frac{9}{16} \right) \right.\nn\\
&&\left.\left.\left.+ \tilde{p}_r^2 \tilde{r} \left( \frac{9 \nu ^2}{4}-\frac{9}{16} \right) \right) \right)
+\frac{27}{35}C_{1(\text{E}^2\text{S}^4)} \left( \frac{1}{8}-\frac{\nu }{8}  + \frac{1}{\nu q} \left( -\frac{\nu ^2}{8}+\frac{3\nu }{8}-\frac{1}{8} \right) \right)\right]\nn\\
&&+\frac{\nu^3 \tilde{S}_1^3 \tilde{S}_2 }{\tilde{r}^6} \left[ 3 + \frac{45}{8} \tilde{p}_r^2 \tilde{r} + \frac{1}{\nu q} \left(3 \nu -3 + \frac{\tilde{L}^2}{\tilde{r}} \left(\frac{3 \nu }{2} \right) + \tilde{p}_r^2 \tilde{r} \left( \frac{21 \nu }{4}-\frac{45}{8} \right) \right) \right.\nn\\
&&+ C_{1(\text{ES}^2)} \left( 
\frac{5}{2} -\frac{9}{2} \frac{\tilde{L}^2}{\tilde{r}}  -\frac{9}{2} \tilde{p}_r^2 \tilde{r}    + \frac{1}{ \nu q} \left(\frac{5 \nu }{2}+6 + \frac{\tilde{L}^2}{\tilde{r}} \left( \frac{9}{2}-6 \nu \right) \right.\right.\nn\\
&&\left.\left.+ \tilde{p}_r^2 \tilde{r} \left( \frac{9}{2}-\frac{33 \nu }{8} \right) \right) \right)
+ C_{1(\text{BS}^3)} \left( 
-6 + \frac{3}{2} \frac{\tilde{L}^2}{\tilde{r}}  + \frac{3}{2} \tilde{p}_r^2 \tilde{r}    \right.\nn\\
&&\left.\left.+ \frac{1}{ \nu q} \left( \frac{23}{2}-6 \nu + \frac{\tilde{L}^2}{\tilde{r}} \left( \frac{3 \nu }{4}-\frac{13}{4} \right) + \tilde{p}_r^2 \tilde{r} \left( \frac{7}{4}-3 \nu \right) \right) \right) \right]\nn\\
&&+\frac{\nu^2 \tilde{S}_1^2 \tilde{S}_2^2 }{2\tilde{r}^6} \left[ 3 -3 \nu \frac{\tilde{L}^2}{\tilde{r}}+ \frac{3 \nu }{4} \tilde{p}_r^2 \tilde{r} 
+ 2 C_{1(\text{ES}^2)} \left( 
4 \nu +3 + \frac{\tilde{L}^2}{\tilde{r}} \left( \frac{9}{2}-\frac{15 \nu }{4} \right)  \right.\right.\nn\\
&&\left.+ \tilde{p}_r^2 \tilde{r} \left( \frac{27}{16}-\frac{15 \nu }{8} \right)  + \frac{\nu}{  q} \left(4 -\frac{9}{2} \frac{\tilde{L}^2}{\tilde{r}} -\frac{27}{16} \tilde{p}_r^2 \tilde{r}  \right) \right)\nn\\
&&\left.+ C_{1(\text{ES}^2)}  C_{2(\text{ES}^2)} \left( 
9 +  \frac{\tilde{L}^2}{\tilde{r}} \left( -\frac{9 \nu }{8}-\frac{33}{8} \right)  + \tilde{p}_r^2 \tilde{r}  \left( -\frac{27 \nu }{4}-\frac{3}{8} \right) \right)   \right] + (1 \leftrightarrow 2).\nn\\
\eea
Due to the aforementioned loss of information in the aligned-spins configuration, we see that similar 
to what happens at the NLO cubic-in-sectors, the unique sector in the potentials and in 
eq.~\eqref{squadeform}, that is proportional to both $C_{1\text{ES}^2}^2$ and $S_2$, vanishes here, and 
thus it is absent in all the common observables that assume this simplified restriction.

Finally, in the long quasi-circular inspiral phase, it is reasonable to assume the circular-orbit 
condition, $p_r \equiv \vec{p} \cdot \vec{n}=0 \Rightarrow p^2=p_r^2+L^2/r^2\to L^2/r^2$. Subjecting our aligned-spins Hamiltonian to this condition then gives rise to:
\bea
\label{radtotham}
\tilde{H}^{\text{NLO}}_{\text{S}^4} &=&\frac{\nu^2 \tilde{S}_1^4 }{\tilde{r}^6} \left[ C_{1(\text{ES}^2)} \left( 
\frac{3}{2}-\frac{3 \nu }{2} -\frac{3 \nu }{4} \frac{\tilde{L}^2}{\tilde{r}}    + \frac{1}{ \nu q} \left( -\frac{3 \nu ^2}{2}+\frac{9 \nu }{2}-\frac{3}{2} + \frac{\tilde{L}^2}{\tilde{r}} \left(\frac{3 \nu }{4}-\frac{3 \nu ^2}{2} \right)  \right) \right) \right.\nn\\
&&+C_{1(\text{ES}^2)}^2 \left( -\frac{\nu }{8}  + \frac{1}{ q} \left( \frac{1}{8}-\frac{\nu }{8} \right) \right)
+ C_{1(\text{BS}^3)} \left( 
-\frac{\nu }{2}-\frac{3}{2} + \frac{\tilde{L}^2}{\tilde{r}}  \left( \frac{3 \nu }{2}-\frac{3}{2} \right) \right.\nn\\
&&\left.  + \frac{1}{ \nu q} \left( -\frac{\nu ^2}{2}-\frac{5 \nu }{2}+\frac{3}{2} + \frac{\tilde{L}^2}{\tilde{r}} \left( \frac{3 \nu ^2}{2}-\frac{9 \nu }{2}+\frac{3}{2} \right)  \right) \right)\nn\\
&&+ C_{1(\text{ES}^4)} \left( 
\frac{5 \nu }{2}-\frac{23}{8} + \frac{\tilde{L}^2}{\tilde{r}}  \left( \frac{3 \nu }{16}+\frac{9}{16} \right)    + \frac{1}{ \nu q} \left( \frac{5 \nu ^2}{2}-\frac{33 \nu }{4}+\frac{23}{8} \right.\right.\nn\\
&&\left.\left.\left.+ \frac{\tilde{L}^2}{\tilde{r}} \left( \frac{3 \nu ^2}{8}+\frac{15 \nu }{16}-\frac{9}{16} \right)  \right) \right)
+\frac{27}{35}C_{1(\text{E}^2\text{S}^4)} \left( \frac{1}{8}-\frac{\nu }{8}  + \frac{1}{\nu q} \left( -\frac{\nu ^2}{8}+\frac{3\nu }{8}-\frac{1}{8} \right) \right)\right]\nn\\
&&+\frac{\nu^3 \tilde{S}_1^3 \tilde{S}_2 }{\tilde{r}^6} \left[ 3 + \frac{1}{\nu q} \left(3 \nu -3 + \frac{\tilde{L}^2}{\tilde{r}} \left(\frac{3 \nu }{2} \right)  \right) 
+ C_{1(\text{ES}^2)} \left( 
\frac{5}{2} -\frac{9}{2} \frac{\tilde{L}^2}{\tilde{r}}     + \frac{1}{ \nu q} \left(\frac{5 \nu }{2}+6 \right.\right.\right.\nn\\
&&\left.\left.\left. + \frac{\tilde{L}^2}{\tilde{r}} \left( \frac{9}{2}-6 \nu \right) \right) \right)
+ C_{1(\text{BS}^3)} \left( 
-6 + \frac{3}{2} \frac{\tilde{L}^2}{\tilde{r}}      + \frac{1}{ \nu q} \left( \frac{23}{2}-6 \nu + \frac{\tilde{L}^2}{\tilde{r}} \left( \frac{3 \nu }{4}-\frac{13}{4} \right)  \right) \right) \right]\nn\\
&&+\frac{\nu^2 \tilde{S}_1^2 \tilde{S}_2^2 }{2\tilde{r}^6} \left[ 3 -3 \nu \frac{\tilde{L}^2}{\tilde{r}}
+ 2 C_{1(\text{ES}^2)} \left( 
4 \nu +3 + \frac{\tilde{L}^2}{\tilde{r}} \left( \frac{9}{2}-\frac{15 \nu }{4} \right)   + \frac{\nu}{  q} \left(4 -\frac{9}{2} \frac{\tilde{L}^2}{\tilde{r}}   \right) \right) \right.\nn\\
&&\left.+ C_{1(\text{ES}^2)}  C_{2(\text{ES}^2)} \left( 
9 +  \frac{\tilde{L}^2}{\tilde{r}} \left( -\frac{9 \nu }{8}-\frac{33}{8} \right)   \right)   \right] + (1 \leftrightarrow 2).
\eea

\subsection{Gravitational-Wave Observables}
\label{gwobs}

Now we can define the binding energies associated with the simplified Hamiltonians above as 
$e\equiv\tilde{H}$, and relate them to the emitted frequencies of GWs measured in the LIGO, Virgo, or 
KAGRA experiments. To that end, we need to supplement the circular-orbit condition of constant 
coordinate-separation with Hamilton's equation,
$\dot{p}_r=-\partial\tilde{H}(\tilde{r},\tilde{L})/\partial\tilde{r}=0$, which then removes completely 
the coordinate dependence from the simplified Hamiltonian in eq.~\eqref{radtotham}, and relates, to begin 
with, the binding energy with the angular momentum:
\bea
(e)^{\text{NLO}}_{\text{S}^4} (\tilde{L}) &=& \frac{\nu^2 \tilde{S}_1^4}{\tilde{L}^{12}}\left[ \left( \frac{285 \nu }{4}+\frac{483}{8} \right)  C_{1\text{ES}^2} + \left( \frac{85 \nu }{16}+\frac{417}{16} \right)  C_{1\text{ES}^2}^2 \right.\nn\\
&&+ \left( \frac{17 \nu }{2}+\frac{39}{2} \right)  C_{1 \text{BS}^3} + \left( \frac{43 \nu }{16}+\frac{83}{16} \right) C_{1\text{ES}^4} + \left(\frac{1}{8}-\frac{\nu }{8} \right) \frac{27}{35}C_{1\text{E}^2\text{S}^4} \nn\\
&&+ \frac{1}{\nu q} \left( \left( \frac{423 \nu ^2}{8}+\frac{99 \nu }{2}-\frac{483}{8} \right) C_{1\text{ES}^2} + \left( \frac{11 \nu ^2}{2}+\frac{749 \nu }{16}-\frac{417}{16} \right) C_{1\text{ES}^2}^2 \right.\nn\\
&&+\left( \frac{17 \nu ^2}{2}+\frac{61 \nu }{2}-\frac{39}{2} \right) C_{1 \text{BS}^3} + \left( \frac{23 \nu ^2}{8}+\frac{123 \nu }{16}-\frac{83}{16} \right) C_{1\text{ES}^4} \nn\\
&&\left.\left.+ \left( -\frac{\nu ^2}{8}+\frac{3\nu }{8}-\frac{1}{8} \right) \frac{27}{35}C_{1\text{E}^2\text{S}^4} \right)  \right] +\frac{\nu^3 \tilde{S}_1^3 \tilde{S}_2 }{\tilde{L}^{12}}\left[ -\frac{717}{4} -35  C_{1\text{ES}^2} +3 C_{1 \text{BS}^3} \right.\nn\\
&&\left.+ \frac{1}{ \nu q} \left( -\frac{285 \nu }{2}-\frac{483}{4} + \left( -\frac{145 \nu }{2}-\frac{1455}{4} \right) C_{1\text{ES}^2}   + \left( \frac{9 \nu }{4}-\frac{207}{4} \right) C_{1 \text{BS}^3} \right) \right]\nn\\
&& +\frac{\nu^2 \tilde{S}_1^2 \tilde{S}_2^2 }{2\tilde{L}^{12}}  \left[ -\frac{297 \nu }{4}-\frac{2025}{4} + \left( -\frac{383 \nu }{2}-\frac{951}{4} \right) C_{1\text{ES}^2} \right.\nn\\
&&\left.+  \left( -\frac{3 \nu }{2}-\frac{195}{2} \right) C_{1\text{ES}^2} C_{2\text{ES}^2} -\frac{913}{4} \frac{\nu}{q} C_{1\text{ES}^2}  \right]+ (1 \leftrightarrow 2).
\eea
To relate to the gauge-invariant frequency, we invoke Hamilton's equation for the orbital phase, 
$d\phi/d\tilde{t}\equiv\tilde{\omega}
=\partial\tilde{H}(\tilde{r},\tilde{L})/\partial\tilde{L}=0$, and define the PN parameter, $x \equiv \tilde{\omega}^{2/3}$. From this we get the angular momentum as a function of GW frequency:
\bea
\frac{1}{\tilde{L}^2 } & \supset & \nu^2 x^6 \tilde{S}_1^4\left[ \left( 30 \nu +\frac{381}{2} \right)  C_{1\text{ES}^2} + \left( \frac{99}{4}-\frac{125 \nu }{4} \right)  C_{1\text{ES}^2}^2 + \left( -7 \nu -69 \right)  C_{1 \text{BS}^3}\right.\nn\\
&& + \left( \frac{61 \nu }{4}-\frac{59}{4} \right) C_{1\text{ES}^4} + \left( 1- \nu \right) \frac{27}{35}C_{1\text{E}^2\text{S}^4} + \frac{1}{\nu q} \left( \left(\frac{61 \nu ^2}{2}+351 \nu -\frac{381}{2} \right) C_{1\text{ES}^2} \right.\nn\\
&&+ \left( -\frac{47 \nu ^2}{2}+\frac{323 \nu }{4}-\frac{99}{4} \right) C_{1\text{ES}^2}^2  +\left( -7 \nu ^2-131 \nu +69 \right) C_{1 \text{BS}^3} \nn\\
&&\left.\left.+ \left( \frac{21 \nu ^2}{2}-\frac{179 \nu }{4}+\frac{59}{4} \right) C_{1\text{ES}^4} + \left( -\nu^2 +3\nu -1 \right) \frac{27}{35}C_{1\text{E}^2\text{S}^4} \right)  \right] \nn\\
&&+ \nu^3 x^6 \tilde{S}_1^3 \tilde{S}_2 \left[-59 +105  C_{1\text{ES}^2} -51 C_{1 \text{BS}^3} \phantom{\frac{1}{\nu q}}\right.\nn\\
&&\left.+ \frac{1}{ \nu q} \left( -60 \nu -381 + \left(75 \nu -255 \right) C_{1\text{ES}^2}   + \left( 111-32 \nu \right) C_{1 \text{BS}^3} \right)\right]\nn\\
&& + \frac{1}{2} \nu^2 x^6 \tilde{S}_1^2 \tilde{S}_2^2   \left[ -29 \nu -825 + \left( 33-2\nu \right) C_{1\text{ES}^2} +  \left( 13 \nu +5 \right) C_{1\text{ES}^2} C_{2\text{ES}^2}\right.\nn\\
&&\left. - \frac{\nu}{q} C_{1\text{ES}^2}  \right]+ (1 \leftrightarrow 2).
\eea

Finally, we can express the binding energy as a function of the GW frequency by combining the former two 
relations:
\bea
(e)^{\text{NLO}}_{\text{S}^4} (x) &=& \nu^2 x^6 \tilde{S}_1^4\left[ \left( \frac{9 \nu }{2}-9 \right)  C_{1\text{ES}^2} + \left( \frac{135 \nu }{16}-\frac{45}{16} \right)  C_{1\text{ES}^2}^2 + \left( 9-3 \nu \right)  C_{1 \text{BS}^3}\right.\nn\\
&& + \left( \frac{21}{16}-\frac{99 \nu }{16} \right) C_{1\text{ES}^4} + \left( \frac{3\nu }{8}-\frac{3}{8} \right) \frac{27}{35}C_{1\text{E}^2\text{S}^4} + \frac{1}{\nu q} \left( \left( 6 \nu ^2-\frac{45 \nu }{2}+9 \right) C_{1\text{ES}^2} \right.\nn\\
&&+ \left( \frac{33 \nu ^2}{4}-\frac{225 \nu }{16}+\frac{45}{16} \right) C_{1\text{ES}^2}^2  +\left( -3 \nu ^2+21 \nu -9 \right) C_{1 \text{BS}^3} \nn\\
&&\left.\left.+ \left( -\frac{39 \nu ^2}{8}+\frac{141 \nu }{16}-\frac{21}{16} \right) C_{1\text{ES}^4} + \left( \frac{3\nu ^2}{8}-\frac{9 \nu }{8}+\frac{3}{8} \right) \frac{27}{35}C_{1\text{E}^2\text{S}^4} \right)  \right] \nn\\
&&+ \nu^3 x^6 \tilde{S}_1^3 \tilde{S}_2 \left[ -6 -\frac{21}{2}  C_{1\text{ES}^2} + \frac{27}{2} C_{1 \text{BS}^3} \right.\nn\\
&&\left.+ \frac{1}{ \nu q} \left( 18-9 \nu + \left( -\frac{27 \nu }{4}-\frac{45}{4} \right) C_{1\text{ES}^2}   + \left( \frac{33 \nu }{4}-\frac{9}{4} \right) C_{1 \text{BS}^3} \right)\right]\nn\\
&& + \frac{1}{2} \nu^2 x^6 \tilde{S}_1^2 \tilde{S}_2^2   \left[ \frac{27 \nu }{4}+\frac{81}{4} + \left( 3\nu -36 \right) C_{1\text{ES}^2} +  \left( \frac{33}{2}-\frac{15 \nu }{2} \right) C_{1\text{ES}^2} C_{2\text{ES}^2}\right.\nn\\
&&\left. +6 \frac{\nu}{q} C_{1\text{ES}^2}  \right]+ (1 \leftrightarrow 2).
\eea


\subsection{Extension to Scattering Angles}
\label{scattering}

In a weak gravitational field $2$-to-$2$ scattering events can be studied in a perturbative expansion 
in $G$, within the so-called post-Minkowskian (PM) approximation. 
In this setup the scattering angle for aligned spins in the COM frame can be derived, and only in low 
loop orders, as we explained in section \ref{intro}, it can then be linked to an extrapolated quantity 
associated with the NR binary inspiral in some overlap with the PN approximation. At this point it 
should be highlighted that
recently a unique novel approach was put forward in \cite{Edison:2022cdu}, which bypasses the need to 
link between scattering and inspiral, and is applicable to any loop order. This approach is formulated 
directly in the bound problem, and uses amplitudes methods to reach unprecedented 
state-of-the-art results in PN theory \cite{Edison:2022cdu}.

In the present sectors, we can extend our PN aligned-spins Hamiltonian in eq.~\eqref{tothamaligned} to 
a scattering energy function, and using general integration considerations \cite{Damour:1988mr}, we can 
then compute here such extrapolated scattering angles as we detailed in \cite{Kim:2022bwv}. We then only 
need to truncate the expansion in $G$ according to the loop order in the PM approximation, which in 
this case is only one loop. Using identical notation to \cite{Kim:2022bwv}:
\bea
p_{\infty} = \frac{m_1 m_2}{E} \sqrt{\gamma^2 - 1},\quad E
= \sqrt{m_1^2 + m_2^2 + 2 m_1 m_2 \gamma}, \quad \gamma 
= \frac{1}{\sqrt{1- v_\infty^2}},
\eea
and:
\bea
\tilde{b} = \frac{v_\infty^2 }{ Gm }b, 
\qquad \tilde{v} = \frac{v_\infty}{c}, 
\qquad \tilde{a}_i = \frac{ S_i }{b m_i c},
\qquad \Gamma = \frac{E}{mc^2} = \sqrt{1+2\nu (\gamma - 1) },
\eea
we then write the scattering angles in the NLO quartic-in-spin sectors as:
\bea
\theta_{\text{S}^4}  &=&  \theta_{\text{S}_1^4}  +\theta_{\text{S}_1^3 \text{S}_2}  +\theta_{\text{S}_1^2 \text{S}_2^2}+  (1 \leftrightarrow 2),
\eea
with
\bea
\frac{\theta_{\text{S}_1^4}}{\Gamma} &=& \tilde{a}_1^4 \left[\frac{1}{\tilde{b}} C_{1\text{ES}^4} (2 + 2\tilde{v}^2) + \frac{\pi}{\tilde{b}^2} \left( -\frac{75 \nu }{16} \tilde{v}^2 C_{1\text{ES}^2} +\left(\frac{15}{16}+\left(\frac{105 \nu }{64}+\frac{105}{32}\right) \tilde{v}^2\right) C_{1\text{ES}^2}^2\right.\right.\nn\\
&&+\left(15-\frac{45 \nu }{8}\right) \tilde{v}^2 C_{1\text{BS}^3} +\left(\frac{45}{16}+\left(\frac{315}{32}-\frac{75 \nu }{16}\right) \tilde{v}^2\right) C_{1\text{ES}^4}+\frac{81}{448} \nu  \tilde{v}^2  C_{1\text{E}^2\text{S}^4}\nn\\
&&\left.\left. + \frac{\nu}{q} \tilde{v}^2 \left(-\frac{75  }{16} C_{1\text{ES}^2} +\frac{105}{64} C_{1\text{ES}^2}^2 -\frac{45}{8} C_{1\text{BS}^3} -\frac{75}{16}C_{1\text{ES}^4} +\frac{81}{448} C_{1\text{E}^2\text{S}^4} \right)  \right)
\right],\nn\\
\eea
\bea
\frac{\theta_{\text{S}_1^3 \text{S}_2}}{\Gamma}&=&   \tilde{a}_1^3 \tilde{a}_2 \left[ \frac{1}{\tilde{b}} C_{1\text{BS}^3} \left( 8 +8 \tilde{v}^2\right)+\frac{\pi}{\tilde{b}^2} \left(  -\frac{75 \nu }{8}\tilde{v}^2 +\left(\frac{15}{4} + \left(\frac{885}{16}-\frac{195 \nu }{16}\right) \tilde{v}^2 \right)C_{1\text{ES}^2} \right.\right.\nn\\
&&+\left(\frac{45}{4}+\left(\frac{735}{16}-\frac{75 \nu }{16}\right) \tilde{v}^2\right)C_{1\text{BS}^3} \nn\\
&&\left.\left.+ \frac{\nu}{q} \tilde{v}^2\left( -\frac{75}{8}-\frac{195}{16}C_{1\text{ES}^2} -\frac{75}{16}C_{1\text{BS}^3} \right) \right)
\right],
\eea
\bea
\frac{\theta_{\text{S}_1^2 \text{S}_2^2}}{\Gamma}&=&\frac{1}{2}  \tilde{a}_1^2 \tilde{a}_2^2 \left[\frac{1}{\tilde{b}} C_{1\text{ES}^2} C_{2\text{ES}^2}  \left(12 +12\tilde{v}^2 \right)+\frac{\pi}{\tilde{b}^2} \left( \frac{15}{4} + \frac{165}{16} \tilde{v}^2 \right.\right.\nn\\
&&+ 2\left(\frac{315 \nu }{16}+\frac{375}{16}\right) \tilde{v}^2 C_{1\text{ES}^2}+\left( \frac{75}{4}+\frac{225 }{4}\tilde{v}^2\right)C_{1\text{ES}^2}C_{2\text{ES}^2} \nn\\
&&\left.\left. + \frac{315}{8} \frac{\nu }{q} \tilde{v}^2 C_{1\text{ES}^2}\right)
\right].
\eea
When all leading spin-induced multipolar Wilson coefficients are set to unity, as stipulated in 
\cite{Levi:2015msa} for BHs, namely $C_{\text{ES}^2} = C_{\text{BS}^3} = C_{\text{ES}^4}= 1$, and 
depending on our conjecture \ref{Kerrdeformation} in section \ref{finaltheory}, and the related 
matching of the new Wilson coefficient $C_{\text{E$^2$S$^4$}}$ for BHs from full GR or real-world 
data, then our scattering angles may agree or not with those presented for BHs in 
\cite{Guevara:2018wpp}. 
In any case, as noted in \cite{Levi:2022dqm} and in section \ref{finaltheory}, the derivations 
and results in Guevara et al.~\cite{Guevara:2018wpp} are inherently dependent on our free-standing 
framework, as they relied on our worldline spin theory and results introduced in 
\cite{Levi:2014gsa,Levi:2015msa}, which was omitted in \cite{Guevara:2018wpp}. 

At this point it is important to reiterate that such limited observables provide but a very partial 
physical picture of the system, and even more so for higher-spin sectors, where the full information is 
found at the most general Hamiltonians first presented in section \ref{genhams} above. This further 
highlights the indispensable comprehensive confirmation of the general Hamiltonians provided through 
the Poincar\'e algebra that we found in section \ref{poincareupto5pn} above.

\section{Conclusions}
\label{closingthebusiness}

We put forward a broader picture of the effective theory of a spinning particle, and in 
particular of Kerr BHs, within the EFT of spinning gravitating objects \cite{Levi:2015msa}. We also 
fully derived and established the new precision frontier at the $5$PN order for GW measurements from 
inspirals and mergers of generic compact binaries. The $5$PN precision frontier includes higher-spin 
sectors, quadratic and quartic in the spin, which both display novel physical effects, originating from 
the extension of the effective theory beyond linear order in the curvature. In the third subleading 
quadratic-in-spin sectors there is a new tidal effect, and in the quartic-in-spin sectors there is a 
new spin-induced multipolar effect. Using these observations, and with eyes towards the next precision 
frontier at the $6$PN order, we generalized the concept of tidal and of spin-induced 
multipolar operators, and made conjectures on the numerical values of their Wilson coefficients for 
rotating BHs in GR in $4$ spacetime dimensions. 

We then confirmed the generalized actions for generic 
compact binaries of the complete NLO quartic-in-spin sectors, which were computed via the extension of 
the EFT of gravitating spinning objects in \cite{Levi:2020lfn}. 
We derived for the first time the NLO quartic-in spin interaction potentials, that consist of no less 
than $12$ distinct sectors, with a new one due to the new spin-induced multipolar operator 
that is quadratic in the curvature. We also derived for the first time the corresponding general 
Hamiltonians in an arbitrary reference frame. These Hamiltonians give the full physical information on 
the binary system, which mostly gets lost in higher-spin sectors, when going to common observables 
which assume an aligned-spins configuration. This is since generic spin orientations of the rotating 
components of the binary have an observational signature in the gravitational waveforms.  

Moreover, the general Hamiltonians obtained exclusively via our framework uniquely enable to find the 
global Poincar\'e algebra, which we carried out successfully in all the sectors with spins that 
contribute at the present $5$PN order. It should be noted that in order to accomplish that, we needed 
to significantly scale our approach to the solution of the Poincar\'e algebra. Thus we solved
for the full Poincar\'e algebra of the third subleading quadratic-in-spin Hamiltonians first 
presented in \cite{Kim:2022bwv}, and of the NLO quartic-in-spin sectors computed in 
\cite{Levi:2020lfn}. This solution of the Poincar\'e algebra also provides the most stringent 
consistency check to the validity of our new comprehensive state-of-the-art results. 

We proceeded to derive simplified Hamiltonians in restricted kinematic configurations, starting with 
the COM Hamiltonians. We then derived observables for GW applications, namely gauge-invariant 
relations among binding energies, angular momentum, and GW frequency.
Furthermore, to make contact with the scattering problem, we also derived the extrapolated scattering 
angles, relevant only in the aligned-spins configuration. We could then specify our consequent 
scattering angles to the scattering of BHs, by fixing all leading spin-induced multipolar Wilson 
coefficients to $1$, as we prescribed in \cite{Levi:2015msa}; and depending on our conjecture 
\ref{Kerrdeformation} in section \ref{finaltheory}, and the related matching of the new Wilson 
coefficient in the present NLO quartic-in-spin sectors, $C_{\text{E$^2$S$^4$}}$, for BHs from full GR 
or real-world data. 
In any case, as noted in \cite{Levi:2022dqm} and in section \ref{finaltheory}, the derivations 
and results in the few overlapping scattering-amplitudes studies are inherently dependent on our 
free-standing framework, as they relied on our worldline spin theory and prior results introduced in 
\cite{Levi:2014gsa,Levi:2015msa}. We reiterate however, that such limited scattering observables can 
provide but very partial physical input on the system, especially at higher-spin sectors. The solution 
of the Poincar\'e algebra on the other hand provides the strongest confirmation that the $5$PN order 
has now been established as the new precision frontier.

\acknowledgments
ML has been supported by the Science and Technology Facilities Council (STFC) 
Rutherford Grant ST/V003895/2 ``\textit{Harnessing QFT for Gravity}'', 
and by the Mathematical Institute University of Oxford. 
ZY is supported by the Knut and Alice Wallenberg Foundation under grants 
KAW 2018.0116 and KAW 2018.0162.

\appendix

\section{Redefinition of Rotational Variables}
\label{newredefs}

As noted in section \ref{minisgood}, the new redefinitions of the spin in the present sectors are  
fixed as:
\bea
\left(\omega^{ij}_1\right)^{\text{NLO}}_{\text{S}^4} =\left(\omega^{ij}_1\right)^{\text{NLO}}_{\text{S}_1^4}+\left(\omega^{ij}_1\right)^{\text{NLO}}_{\text{S}_1^3 \text{S}_2} +\left(\omega^{ij}_1\right)^{\text{NLO}}_{\text{S}_1^2 \text{S}_2^2}  +\left(\omega^{ij}_1\right)^{\text{NLO}}_{\text{S}_1 \text{S}_2^3}  - (i \leftrightarrow j),
\eea
where 
\bea
\left(\omega^{ij}_1\right)^{\text{NLO}}_{\text{S}_1^4}&=&- 	\frac{G C_{1\text{BS}^3} m_{2}}{4 m_{1}{}^3 r{}^4} \Big[ 20 S_{1}^2 S_{1}^{ki} n^k \vec{v}_{1}\cdot\vec{n} n^j + 2 S_{1}^2 S_{1}^{ki} v_{1}^k n^j - 2 S_{1}^2 S_{1}^{kj} n^k v_{1}^i \nn\\ 
&& + 8 S_{1}^2 S_{1}^{ij} \vec{v}_{1}\cdot\vec{n} - 10 S_{1}^{ki} v_{1}^k n^j ( \vec{S}_{1}\cdot\vec{n})^{2} + 5 S_{1}^{ij} \vec{v}_{1}\cdot\vec{n} ( \vec{S}_{1}\cdot\vec{n})^{2} \nn\\ 
&& - 30 \vec{S}_{1}\cdot\vec{n} \vec{S}_{1}\cdot\vec{v}_{1} n^j S_{1}^{ki} n^k - 13 \vec{S}_{1}\cdot\vec{n} \vec{S}_{1}\cdot\vec{v}_{1} S_{1}^{ij} \Big] \nn\\ 
&& + 	\frac{3 G C_{1\text{ES}^2} m_{2}}{8 m_{1}{}^3 r{}^4} \Big[ 20 S_{1}^2 S_{1}^{ki} n^k \vec{v}_{1}\cdot\vec{n} n^j + 5 S_{1}^2 S_{1}^{ki} v_{1}^k n^j - S_{1}^2 S_{1}^{kj} n^k v_{1}^i \nn\\ 
&& + 9 S_{1}^2 S_{1}^{ij} \vec{v}_{1}\cdot\vec{n} - 15 S_{1}^{ki} v_{1}^k n^j ( \vec{S}_{1}\cdot\vec{n})^{2} + 5 S_{1}^{kj} n^k v_{1}^i ( \vec{S}_{1}\cdot\vec{n})^{2} \nn\\ 
&& - 20 \vec{S}_{1}\cdot\vec{n} \vec{S}_{1}\cdot\vec{v}_{1} n^j S_{1}^{ki} n^k - 9 \vec{S}_{1}\cdot\vec{n} \vec{S}_{1}\cdot\vec{v}_{1} S_{1}^{ij} \Big] \nn\\ 
&& + 	\frac{G C_{1\text{ES}^4} m_{2}}{8 m_{1}{}^3 r{}^4} \Big[ 10 S_{1}^2 S_{1}^{ki} n^k \big( \vec{v}_{1}\cdot\vec{n} n^j - 4 \vec{v}_{2}\cdot\vec{n} n^j \big) - 2 S_{1}^2 S_{1}^{ki} v_{1}^k n^j \nn\\ 
&& - 4 S_{1}^2 S_{1}^{ki} v_{2}^k n^j + 2 S_{1}^2 S_{1}^{kj} n^k \big( v_{1}^i + 2 v_{2}^i \big) + S_{1}^2 S_{1}^{ij} \big( \vec{v}_{1}\cdot\vec{n} - 19 \vec{v}_{2}\cdot\vec{n} \big) \nn\\ 
&& + 70 S_{1}^{ki} n^k \vec{v}_{2}\cdot\vec{n} n^j ( \vec{S}_{1}\cdot\vec{n})^{2} + 10 S_{1}^{ki} v_{2}^k n^j ( \vec{S}_{1}\cdot\vec{n})^{2} + 5 S_{1}^{ij} \big( \vec{v}_{1}\cdot\vec{n} \nn\\ 
&& + 4 \vec{v}_{2}\cdot\vec{n} \big) ( \vec{S}_{1}\cdot\vec{n})^{2} + 30 \vec{S}_{1}\cdot\vec{n} \vec{S}_{1}\cdot\vec{v}_{2} n^j S_{1}^{ki} n^k - 2 \vec{S}_{1}\cdot\vec{n} \vec{S}_{1}\cdot\vec{v}_{1} S_{1}^{ij} \nn\\ 
&& + 11 \vec{S}_{1}\cdot\vec{n} \vec{S}_{1}\cdot\vec{v}_{2} S_{1}^{ij} \Big],
\eea
\bea
\left(\omega^{ij}_1\right)^{\text{NLO}}_{\text{S}_1^3 \text{S}_2}&=& - 	\frac{G C_{1\text{BS}^3}}{8 m_{1}{}^2 r{}^4} \Big[ 20 \vec{S}_{1}\cdot\vec{S}_{2} S_{1}^{ki} n^k \big( 4 \vec{v}_{1}\cdot\vec{n} n^j - 3 \vec{v}_{2}\cdot\vec{n} n^j \big) - 10 S_{1}^2 S_{2}^{ki} n^k \big( 2 \vec{v}_{1}\cdot\vec{n} n^j \nn\\ 
&& - 5 \vec{v}_{2}\cdot\vec{n} n^j \big) - 4 \vec{S}_{1}\cdot\vec{v}_{2} S_{1}^{ki} S_{2}^k n^j - 16 \vec{S}_{1}\cdot\vec{S}_{2} S_{1}^{ki} v_{1}^k n^j + 4 S_{1}^2 S_{2}^{ki} v_{1}^k n^j \nn\\ 
&& + 24 \vec{S}_{1}\cdot\vec{S}_{2} S_{1}^{ki} v_{2}^k n^j - 10 S_{1}^2 S_{2}^{ki} v_{2}^k n^j + 4 \vec{S}_{1}\cdot\vec{S}_{2} S_{1}^{kj} n^k \big( 4 v_{1}^i - 5 v_{2}^i \big) \nn\\ 
&& - 2 S_{1}^2 S_{2}^{kj} n^k \big( 14 v_{1}^i - 11 v_{2}^i \big) + 4 S_{1}^i S_{1}^{kj} S_{2}^k \vec{v}_{2}\cdot\vec{n} + 4 \vec{S}_{1}\cdot\vec{S}_{2} S_{1}^{ij} \big( 5 \vec{v}_{1}\cdot\vec{n} - 2 \vec{v}_{2}\cdot\vec{n} \big) \nn\\ 
&& + 4 S_{1}^2 S_{2}^{ij} \big( \vec{v}_{1}\cdot\vec{n} + 3 \vec{v}_{2}\cdot\vec{n} \big) - 210 S_{2}^{ki} n^k \vec{v}_{2}\cdot\vec{n} n^j ( \vec{S}_{1}\cdot\vec{n})^{2} + 20 S_{2}^{ki} v_{2}^k n^j ( \vec{S}_{1}\cdot\vec{n})^{2} \nn\\ 
&& + 10 S_{2}^{ij} \big( 2 \vec{v}_{1}\cdot\vec{n} - 11 \vec{v}_{2}\cdot\vec{n} \big) ( \vec{S}_{1}\cdot\vec{n})^{2} - 120 \vec{S}_{1}\cdot\vec{n} \vec{S}_{2}\cdot\vec{v}_{1} n^j S_{1}^{ki} n^k \nn\\ 
&& + 40 \vec{S}_{2}\cdot\vec{n} \vec{S}_{1}\cdot\vec{v}_{2} n^j S_{1}^{ki} n^k + 90 \vec{S}_{1}\cdot\vec{n} \vec{S}_{2}\cdot\vec{v}_{2} n^j S_{1}^{ki} n^k - 20 \vec{S}_{1}\cdot\vec{n} \vec{S}_{1}\cdot\vec{v}_{2} n^j S_{2}^{ki} n^k \nn\\ 
&& - 10 \vec{S}_{1}\cdot\vec{n} \vec{S}_{2}\cdot\vec{n} n^j S_{1}^{ki} v_{2}^k - 4 \vec{S}_{2}\cdot\vec{n} S_{1}^i S_{1}^{kj} v_{2}^k + 4 \vec{S}_{1}\cdot\vec{n} S_{1}^i S_{2}^{kj} v_{2}^k \nn\\ 
&& - 5 \vec{S}_{1}\cdot\vec{n} \vec{S}_{2}\cdot\vec{n} \big( 4 \vec{v}_{1}\cdot\vec{n} - 9 \vec{v}_{2}\cdot\vec{n} \big) S_{1}^{ij} + 4 \vec{S}_{2}\cdot\vec{n} \vec{S}_{1}\cdot\vec{v}_{1} S_{1}^{ij} - 32 \vec{S}_{1}\cdot\vec{n} \vec{S}_{2}\cdot\vec{v}_{1} S_{1}^{ij} \nn\\ 
&& + 14 \vec{S}_{2}\cdot\vec{n} \vec{S}_{1}\cdot\vec{v}_{2} S_{1}^{ij} + 23 \vec{S}_{1}\cdot\vec{n} \vec{S}_{2}\cdot\vec{v}_{2} S_{1}^{ij} - 8 \vec{S}_{1}\cdot\vec{n} \vec{S}_{1}\cdot\vec{v}_{1} S_{2}^{ij} \Big] \nn\\ 
&& - 	\frac{3 G C_{1\text{ES}^2}}{4 m_{1}{}^2 r{}^4} \Big[ 3 \vec{S}_{1}\times\vec{S}_{2}\cdot\vec{v}_{1} S_{1}^i n^j + 20 \vec{S}_{1}\cdot\vec{S}_{2} S_{1}^{ki} n^k \vec{v}_{1}\cdot\vec{n} n^j - 3 \vec{S}_{1}\cdot\vec{v}_{1} S_{1}^{ki} S_{2}^k n^j \nn\\ 
&& - 3 \vec{S}_{1}\cdot\vec{S}_{2} S_{1}^{ki} v_{1}^k n^j - 3 S_{1}^2 S_{2}^{ki} v_{1}^k n^j + 2 \vec{S}_{1}\cdot\vec{S}_{2} S_{1}^{kj} n^k v_{1}^i - 4 S_{1}^2 S_{2}^{kj} n^k v_{1}^i \nn\\ 
&& - 4 \vec{S}_{1}\cdot\vec{n} S_{1}^{kj} S_{2}^k v_{1}^i + 4 S_{1}^i S_{1}^{kj} S_{2}^k \vec{v}_{1}\cdot\vec{n} + 2 \vec{S}_{1}\cdot\vec{S}_{2} S_{1}^{ij} \vec{v}_{1}\cdot\vec{n} + 3 S_{1}^2 S_{2}^{ij} \vec{v}_{1}\cdot\vec{n} \nn\\ 
&& - 10 \vec{S}_{2}\cdot\vec{n} \vec{S}_{1}\times\vec{n}\cdot\vec{v}_{1} n^j S_{1}^i + 20 \vec{S}_{1}\cdot\vec{n} \vec{S}_{1}\times\vec{n}\cdot\vec{v}_{1} n^j S_{2}^i \nn\\ 
&& - 20 \vec{S}_{1}\cdot\vec{n} \vec{S}_{2}\cdot\vec{v}_{1} n^j S_{1}^{ki} n^k - 10 \vec{S}_{1}\cdot\vec{n} \vec{S}_{1}\cdot\vec{v}_{1} n^j S_{2}^{ki} n^k - 10 \vec{S}_{1}\times\vec{n}\cdot\vec{v}_{1} S_{2}^i S_{1}^j \nn\\ 
&& - 4 \vec{S}_{2}\cdot\vec{v}_{1} S_{1}^i S_{1}^{kj} n^k + 4 \vec{S}_{1}\cdot\vec{v}_{1} S_{1}^i S_{2}^{kj} n^k + 5 \vec{S}_{1}\cdot\vec{n} \vec{S}_{2}\cdot\vec{n} \vec{v}_{1}\cdot\vec{n} S_{1}^{ij} \nn\\ 
&& - 3 \vec{S}_{1}\cdot\vec{n} \vec{S}_{2}\cdot\vec{v}_{1} S_{1}^{ij} - 7 \vec{S}_{1}\cdot\vec{n} \vec{S}_{1}\cdot\vec{v}_{1} S_{2}^{ij} \Big] - 	\frac{3 G}{8 m_{1}{}^2 r{}^4} \Big[ 2 \vec{S}_{1}\times\vec{S}_{2}\cdot\vec{v}_{1} S_{1}^i n^j \nn\\ 
&& - 20 S_{1}^2 S_{2}^{ki} n^k \vec{v}_{1}\cdot\vec{n} n^j - \vec{S}_{1}\cdot\vec{v}_{1} S_{1}^{ki} S_{2}^k n^j - 13 \vec{S}_{1}\cdot\vec{S}_{2} S_{1}^{ki} v_{1}^k n^j - S_{1}^2 S_{2}^{ki} v_{1}^k n^j \nn\\ 
&& - \vec{S}_{1}\times\vec{n}\cdot\vec{S}_{2} S_{1}^j v_{1}^i + 5 \vec{S}_{1}\cdot\vec{S}_{2} S_{1}^{kj} n^k v_{1}^i - 3 S_{1}^2 S_{2}^{kj} n^k v_{1}^i - \vec{S}_{1}\cdot\vec{n} S_{1}^{kj} S_{2}^k v_{1}^i \nn\\ 
&& - S_{1}^i S_{1}^{kj} S_{2}^k \vec{v}_{1}\cdot\vec{n} - 5 \vec{S}_{1}\cdot\vec{S}_{2} S_{1}^{ij} \vec{v}_{1}\cdot\vec{n} - 7 S_{1}^2 S_{2}^{ij} \vec{v}_{1}\cdot\vec{n} + 20 \vec{S}_{1}\cdot\vec{n} \vec{S}_{1}\cdot\vec{v}_{1} n^j S_{2}^{ki} n^k \nn\\ 
&& + 30 \vec{S}_{1}\cdot\vec{n} \vec{S}_{2}\cdot\vec{n} n^j S_{1}^{ki} v_{1}^k - 10 \vec{S}_{1}\cdot\vec{n} \vec{S}_{2}\cdot\vec{n} v_{1}^i S_{1}^{kj} n^k - 2 \vec{S}_{2}\cdot\vec{v}_{1} S_{1}^i S_{1}^{kj} n^k \nn\\ 
&& + 2 \vec{S}_{1}\cdot\vec{v}_{1} S_{1}^i S_{2}^{kj} n^k + 3 \vec{S}_{2}\cdot\vec{n} S_{1}^i S_{1}^{kj} v_{1}^k - 3 \vec{S}_{1}\cdot\vec{n} S_{1}^i S_{2}^{kj} v_{1}^k + 5 \vec{S}_{1}\cdot\vec{n} \vec{S}_{2}\cdot\vec{v}_{1} S_{1}^{ij} \nn\\ 
&& + 7 \vec{S}_{1}\cdot\vec{n} \vec{S}_{1}\cdot\vec{v}_{1} S_{2}^{ij} \Big]\nn\\
&&- 	\frac{G C_{1\text{BS}^3}}{24 m_{1}{}^2 r{}^3} \Big[ 6 \vec{S}_{1}\cdot\dot{\vec{S}}_{2} S_{1}^{ki} n^k n^j + 9 S_{1}^2 \dot{S}_{2}^{ki} n^k n^j - 45 \dot{S}_{2}^{ki} n^k n^j ( \vec{S}_{1}\cdot\vec{n})^{2} \nn\\ 
&& + 9 \vec{S}_{1}\cdot\vec{n} \dot{\vec{S}}_{2}\cdot\vec{n} S_{1}^{ij} + 2 \vec{S}_{1}\cdot\dot{\vec{S}}_{2} S_{1}^{ij} + 4 S_{1}^2 \dot{S}_{2}^{ij} - 21 ( \vec{S}_{1}\cdot\vec{n})^{2} \dot{S}_{2}^{ij} \Big],
\eea
\bea
\left(\omega^{ij}_1\right)^{\text{NLO}}_{\text{S}_1^2 \text{S}_2^2}&=&- 	\frac{3 G C_{2\text{ES}^2}}{8 m_{1} m_{2} r{}^4} \Big[ \vec{S}_{2}\cdot\vec{v}_{1} S_{1}^{ki} S_{2}^k n^j + 2 S_{2}^2 S_{1}^{ki} v_{1}^k n^j + \vec{S}_{1}\cdot\vec{S}_{2} S_{2}^{ki} v_{1}^k n^j \nn\\ 
&& - \vec{S}_{1}\times\vec{n}\cdot\vec{S}_{2} S_{2}^j v_{1}^i + 2 S_{2}^2 S_{1}^{kj} n^k v_{1}^i - \vec{S}_{1}\cdot\vec{S}_{2} S_{2}^{kj} n^k v_{1}^i + \vec{S}_{2}\cdot\vec{n} S_{1}^{kj} S_{2}^k v_{1}^i \nn\\ 
&& + S_{2}^i S_{1}^{kj} S_{2}^k \vec{v}_{1}\cdot\vec{n} + \vec{S}_{1}\cdot\vec{S}_{2} S_{2}^{ij} \vec{v}_{1}\cdot\vec{n} - 5 S_{1}^{ki} v_{1}^k n^j ( \vec{S}_{2}\cdot\vec{n})^{2} - 5 S_{1}^{kj} n^k v_{1}^i ( \vec{S}_{2}\cdot\vec{n})^{2} \nn\\ 
&& - \vec{S}_{2}\cdot\vec{n} S_{2}^i S_{1}^{kj} v_{1}^k + \vec{S}_{1}\cdot\vec{n} S_{2}^i S_{2}^{kj} v_{1}^k - \vec{S}_{1}\cdot\vec{n} \vec{S}_{2}\cdot\vec{v}_{1} S_{2}^{ij} \Big] \nn\\ 
&& + 	\frac{3 G C_{1\text{ES}^2}}{4 m_{1} m_{2} r{}^4} \Big[ 2 \vec{S}_{1}\times\vec{S}_{2}\cdot\vec{v}_{2} S_{2}^i n^j + 10 S_{2}^2 S_{1}^{ki} n^k \vec{v}_{2}\cdot\vec{n} n^j + 3 \vec{S}_{2}\cdot\vec{v}_{2} S_{1}^{ki} S_{2}^k n^j \nn\\ 
&& - 3 S_{2}^2 S_{1}^{ki} v_{2}^k n^j - \vec{S}_{1}\cdot\vec{S}_{2} S_{2}^{ki} v_{2}^k n^j + 3 S_{2}^2 S_{1}^{kj} n^k v_{2}^i - 3 \vec{S}_{1}\cdot\vec{S}_{2} S_{2}^{kj} n^k v_{2}^i \nn\\ 
&& - 3 \vec{S}_{2}\cdot\vec{n} S_{1}^{kj} S_{2}^k v_{2}^i + 4 S_{2}^i S_{1}^{kj} S_{2}^k \vec{v}_{2}\cdot\vec{n} + 2 \vec{S}_{1}\cdot\vec{S}_{2} S_{2}^{ij} \vec{v}_{2}\cdot\vec{n} \nn\\ 
&& - 10 \vec{S}_{2}\cdot\vec{n} \vec{S}_{2}\times\vec{n}\cdot\vec{v}_{2} n^j S_{1}^i + 20 \vec{S}_{1}\cdot\vec{n} \vec{S}_{2}\times\vec{n}\cdot\vec{v}_{2} n^j S_{2}^i \nn\\ 
&& - 10 \vec{S}_{2}\cdot\vec{n} \vec{S}_{2}\cdot\vec{v}_{2} n^j S_{1}^{ki} n^k - 10 \vec{S}_{2}\times\vec{n}\cdot\vec{v}_{2} S_{2}^i S_{1}^j - 3 \vec{S}_{2}\cdot\vec{v}_{2} S_{2}^i S_{1}^{kj} n^k \nn\\ 
&& + 3 \vec{S}_{1}\cdot\vec{v}_{2} S_{2}^i S_{2}^{kj} n^k - \vec{S}_{2}\cdot\vec{n} S_{2}^i S_{1}^{kj} v_{2}^k + \vec{S}_{1}\cdot\vec{n} S_{2}^i S_{2}^{kj} v_{2}^k - 2 \vec{S}_{2}\cdot\vec{n} \vec{S}_{1}\cdot\vec{v}_{2} S_{2}^{ij} \Big] \nn\\ 
&& + 	\frac{3 G C_{1\text{ES}^2} C_{2\text{ES}^2}}{16 m_{1} m_{2} r{}^4} \Big[ 20 S_{2}^2 S_{1}^{ki} n^k \vec{v}_{2}\cdot\vec{n} n^j - 40 \vec{S}_{1}\cdot\vec{S}_{2} S_{2}^{ki} n^k \vec{v}_{2}\cdot\vec{n} n^j \nn\\ 
&& + 8 S_{2}^2 S_{1}^{ki} v_{1}^k n^j + 16 \vec{S}_{1}\cdot\vec{S}_{2} S_{2}^{ki} v_{1}^k n^j - 20 S_{2}^2 S_{1}^{ki} v_{2}^k n^j - 24 \vec{S}_{1}\cdot\vec{S}_{2} S_{2}^{ki} v_{2}^k n^j \nn\\ 
&& - 4 S_{2}^2 S_{1}^{kj} n^k \big( 2 v_{1}^i - 5 v_{2}^i \big) - 8 \vec{S}_{1}\cdot\vec{S}_{2} S_{2}^{kj} n^k v_{2}^i + 8 S_{2}^2 S_{1}^{ij} \big( \vec{v}_{1}\cdot\vec{n} - \vec{v}_{2}\cdot\vec{n} \big) \nn\\ 
&& + 8 \vec{S}_{1}\cdot\vec{S}_{2} S_{2}^{ij} \big( \vec{v}_{1}\cdot\vec{n} - 3 \vec{v}_{2}\cdot\vec{n} \big) + 140 S_{1}^{ki} n^k \vec{v}_{2}\cdot\vec{n} n^j ( \vec{S}_{2}\cdot\vec{n})^{2} \nn\\ 
&& - 20 S_{1}^{ki} v_{1}^k n^j ( \vec{S}_{2}\cdot\vec{n})^{2} + 30 S_{1}^{ki} v_{2}^k n^j ( \vec{S}_{2}\cdot\vec{n})^{2} + 10 S_{1}^{kj} n^k \big( 2 v_{1}^i - 3 v_{2}^i \big) ( \vec{S}_{2}\cdot\vec{n})^{2} \nn\\ 
&& - 10 S_{1}^{ij} \big( 2 \vec{v}_{1}\cdot\vec{n} - 7 \vec{v}_{2}\cdot\vec{n} \big) ( \vec{S}_{2}\cdot\vec{n})^{2} - 20 \vec{S}_{2}\cdot\vec{n} \vec{S}_{2}\cdot\vec{v}_{2} n^j S_{1}^{ki} n^k \nn\\ 
&& - 20 \vec{S}_{2}\cdot\vec{n} \vec{S}_{1}\cdot\vec{v}_{1} n^j S_{2}^{ki} n^k + 30 \vec{S}_{2}\cdot\vec{n} \vec{S}_{1}\cdot\vec{v}_{2} n^j S_{2}^{ki} n^k - 20 \vec{S}_{1}\cdot\vec{n} \vec{S}_{2}\cdot\vec{v}_{2} n^j S_{2}^{ki} n^k \nn\\ 
&& - 20 \vec{S}_{1}\cdot\vec{n} \vec{S}_{2}\cdot\vec{n} n^j S_{2}^{ki} v_{1}^k + 30 \vec{S}_{1}\cdot\vec{n} \vec{S}_{2}\cdot\vec{n} n^j S_{2}^{ki} v_{2}^k + 2 \vec{S}_{2}\cdot\vec{n} \vec{S}_{2}\cdot\vec{v}_{2} S_{1}^{ij} \nn\\ 
&& - 5 \vec{S}_{1}\cdot\vec{n} \vec{S}_{2}\cdot\vec{n} \big( 2 \vec{v}_{1}\cdot\vec{n} + \vec{v}_{2}\cdot\vec{n} \big) S_{2}^{ij} - 2 \vec{S}_{2}\cdot\vec{n} \vec{S}_{1}\cdot\vec{v}_{1} S_{2}^{ij} + 3 \vec{S}_{2}\cdot\vec{n} \vec{S}_{1}\cdot\vec{v}_{2} S_{2}^{ij} \nn\\ 
&& - 6 \vec{S}_{1}\cdot\vec{n} \vec{S}_{2}\cdot\vec{v}_{2} S_{2}^{ij} \Big]\nn\\
&&- 	\frac{G C_{1\text{ES}^2} C_{2\text{ES}^2}}{8 m_{1} m_{2} r{}^3} \Big[ 42 \vec{S}_{2}\cdot\dot{\vec{S}}_{2} S_{1}^{ki} n^k n^j - 6 \vec{S}_{1}\cdot\dot{\vec{S}}_{2} S_{2}^{ki} n^k n^j - 6 \vec{S}_{1}\cdot\vec{S}_{2} \dot{S}_{2}^{ki} n^k n^j \nn\\ 
&& - 30 \vec{S}_{2}\cdot\vec{n} \dot{\vec{S}}_{2}\cdot\vec{n} n^j S_{1}^{ki} n^k - 18 \vec{S}_{2}\cdot\vec{n} \dot{\vec{S}}_{2}\cdot\vec{n} S_{1}^{ij} + 16 \vec{S}_{2}\cdot\dot{\vec{S}}_{2} S_{1}^{ij} \nn\\ 
&& - 3 \vec{S}_{1}\cdot\vec{n} \dot{\vec{S}}_{2}\cdot\vec{n} S_{2}^{ij} - 3 \vec{S}_{1}\cdot\vec{n} \vec{S}_{2}\cdot\vec{n} \dot{S}_{2}^{ij} \Big],
\eea
\bea
\left(\omega^{ij}_1\right)^{\text{NLO}}_{\text{S}_1 \text{S}_2^3}&=&- 	\frac{3 G}{4 m_{2}{}^2 r{}^4} \Big[ 10 S_{2}^2 S_{2}^{ki} n^k \vec{v}_{2}\cdot\vec{n} n^j - 2 S_{2}^2 S_{2}^{ki} v_{2}^k n^j + 3 S_{2}^2 S_{2}^{ij} \vec{v}_{2}\cdot\vec{n} \nn\\ 
&& - 10 \vec{S}_{2}\cdot\vec{n} \vec{S}_{2}\cdot\vec{v}_{2} n^j S_{2}^{ki} n^k - 3 \vec{S}_{2}\cdot\vec{n} \vec{S}_{2}\cdot\vec{v}_{2} S_{2}^{ij} \Big] + 	\frac{3 G C_{2\text{ES}^2}}{2 m_{2}{}^2 r{}^4} \Big[ 3 S_{2}^2 S_{2}^{ki} v_{2}^k n^j \nn\\ 
&& + S_{2}^2 S_{2}^{ij} \vec{v}_{2}\cdot\vec{n} - 5 S_{2}^{ki} v_{2}^k n^j ( \vec{S}_{2}\cdot\vec{n})^{2} - \vec{S}_{2}\cdot\vec{n} \vec{S}_{2}\cdot\vec{v}_{2} S_{2}^{ij} \Big] \nn\\ 
&& - 	\frac{G C_{2\text{BS}^3}}{8 m_{2}{}^2 r{}^4} \Big[ 30 S_{2}^2 S_{2}^{ki} n^k \vec{v}_{2}\cdot\vec{n} n^j - 6 S_{2}^2 S_{2}^{ki} v_{2}^k n^j - 2 S_{2}^2 S_{2}^{kj} n^k \big( 4 v_{1}^i - 3 v_{2}^i \big) \nn\\ 
&& + 8 S_{2}^2 S_{2}^{ij} \vec{v}_{2}\cdot\vec{n} - 70 S_{2}^{ki} n^k \vec{v}_{2}\cdot\vec{n} n^j ( \vec{S}_{2}\cdot\vec{n})^{2} + 10 S_{2}^{ki} v_{2}^k n^j ( \vec{S}_{2}\cdot\vec{n})^{2} \nn\\ 
&& + 5 S_{2}^{ij} \big( 4 \vec{v}_{1}\cdot\vec{n} - 5 \vec{v}_{2}\cdot\vec{n} \big) ( \vec{S}_{2}\cdot\vec{n})^{2} - 40 \vec{S}_{2}\cdot\vec{n} \vec{S}_{2}\cdot\vec{v}_{1} n^j S_{2}^{ki} n^k \nn\\ 
&& + 30 \vec{S}_{2}\cdot\vec{n} \vec{S}_{2}\cdot\vec{v}_{2} n^j S_{2}^{ki} n^k - 20 \vec{S}_{2}\cdot\vec{n} \vec{S}_{2}\cdot\vec{v}_{1} S_{2}^{ij} + 13 \vec{S}_{2}\cdot\vec{n} \vec{S}_{2}\cdot\vec{v}_{2} S_{2}^{ij} \Big]\nn\\
&&- 	\frac{G C_{2\text{BS}^3}}{24 m_{2}{}^2 r{}^3} \Big[ 18 \vec{S}_{2}\cdot\dot{\vec{S}}_{2} S_{2}^{ki} n^k n^j + 9 S_{2}^2 \dot{S}_{2}^{ki} n^k n^j - 45 \dot{S}_{2}^{ki} n^k n^j ( \vec{S}_{2}\cdot\vec{n})^{2} \nn\\ 
&& + 3 \vec{S}_{2}\cdot\vec{n} \dot{\vec{S}}_{2}\cdot\vec{n} S_{2}^{ij} + 8 \vec{S}_{2}\cdot\dot{\vec{S}}_{2} S_{2}^{ij} + 4 S_{2}^2 \dot{S}_{2}^{ij} - 21 ( \vec{S}_{2}\cdot\vec{n})^{2} \dot{S}_{2}^{ij} \Big].
\eea

\section{Effective Actions}
\label{neweffacts}

After the reduction in section \ref{minisgood} is done, we obtain the NLO quartic-in-spin potentials, 
made up of no less than $12$ unique sectors:
\bea
\label{12actionbreakdownapp}
V^{\text{NLO}}_{\text{S}^4} &=&C_{1\text{ES}^2} V^{\text{NLO}}_{(\text{ES}_1^2 ) \text{S}_1^2}+C_{1\text{ES}^2}^2 V^{\text{NLO}}_{(\text{ES}_1^2 )^2} + C_{1\text{BS}^3} V^{\text{NLO}}_{(\text{BS}_1^3) \text{S}_1 } +C_{1\text{ES}^4} V^{\text{NLO}}_{\text{ES}_1^4} +C_{1\text{E}^2\text{S}^4} V^{\text{NLO}}_{\text{E}^2\text{S}_1^4}\nn\\
&& +V^{\text{NLO}}_{\text{S}_1^3 \text{S}_2}+C_{1\text{ES}^2} V^{\text{NLO}}_{(\text{ES}_1^2 ) \text{S}_1 \text{S}_2} + C_{1\text{ES}^2}^2 V^{\text{NLO}}_{C_{\text{ES}^2_1}^2 \text{S}_1^3 \text{S}_2 } + C_{1\text{BS}^3} V^{\text{NLO}}_{(\text{BS}_1^3) \text{S}_2 }\nn\\
&& +V^{\text{NLO}}_{\text{S}_1^2 \text{S}_2^2}+C_{1\text{ES}^2} V^{\text{NLO}}_{(\text{ES}_1^2 )  \text{S}_2^2} +C_{1\text{ES}^2} C_{2\text{ES}^2} V^{\text{NLO}}_{(\text{ES}_1^2 )(\text{ES}_2^2 ) }+  (1 \leftrightarrow 2),
\eea
where 
\bea
V^{\text{NLO}}_{(\text{ES}_1^2 ) \text{S}_1^2 } &=& - 	\frac{3 G m_{2}}{16 m_{1}{}^3 r{}^5} \Big[ 10 \vec{S}_{1}\cdot\vec{n} S_{1}^2 \big( 3 \vec{v}_{1}\cdot\vec{n} + 2 \vec{v}_{2}\cdot\vec{n} \big) \vec{S}_{1}\cdot\vec{v}_{1} - 40 \vec{S}_{1}\cdot\vec{n} S_{1}^2 \vec{v}_{1}\cdot\vec{n} \vec{S}_{1}\cdot\vec{v}_{2} \nn\\ 
&& - 4 S_{1}^2 \vec{S}_{1}\cdot\vec{v}_{1} \vec{S}_{1}\cdot\vec{v}_{2} - 5 S_{1}^2 \big( 3 v_{1}^2 + 4 \vec{v}_{1}\cdot\vec{v}_{2} -28 \vec{v}_{1}\cdot\vec{n} \vec{v}_{2}\cdot\vec{n} + 21 ( \vec{v}_{1}\cdot\vec{n})^{2} \big) ( \vec{S}_{1}\cdot\vec{n})^{2} \nn\\ 
&& + 70 \vec{S}_{1}\cdot\vec{v}_{1} \big( \vec{v}_{1}\cdot\vec{n} - 2 \vec{v}_{2}\cdot\vec{n} \big) ( \vec{S}_{1}\cdot\vec{n})^{3} + 35 v_{1}^2 ( \vec{S}_{1}\cdot\vec{n})^{4} + 60 \vec{S}_{1}\cdot\vec{v}_{1} \vec{S}_{1}\cdot\vec{v}_{2} ( \vec{S}_{1}\cdot\vec{n})^{2} \nn\\ 
&& - 30 ( \vec{S}_{1}\cdot\vec{n})^{2} ( \vec{S}_{1}\cdot\vec{v}_{1})^{2} + \big( 4 \vec{v}_{1}\cdot\vec{v}_{2} -20 \vec{v}_{1}\cdot\vec{n} \vec{v}_{2}\cdot\vec{n} + 15 ( \vec{v}_{1}\cdot\vec{n})^{2} \big) S_{1}^{4} \Big]\nn\\ && - 	\frac{3 G^2 m_{2}{}^2}{8 m_{1}{}^3 r{}^6} \Big[ 5 ( \vec{S}_{1}\cdot\vec{n})^{4} - 6 S_{1}^2 ( \vec{S}_{1}\cdot\vec{n})^{2} + S_{1}^{4} \Big],
\eea
\bea
\label{quadsqtohexa}
V^{\text{NLO}}_{(\text{ES}_1^2 )^2 }&=&- 	\frac{3 G^2 m_{2}{}^2}{2 m_{1}{}^3 r{}^6} \Big[ 5 ( \vec{S}_{1}\cdot\vec{n})^{4} - 3 S_{1}^2 ( \vec{S}_{1}\cdot\vec{n})^{2} \Big] + 	\frac{G^2 m_{2}}{8 m_{1}{}^2 r{}^6} \Big[ 9 ( \vec{S}_{1}\cdot\vec{n})^{4} - 6 S_{1}^2 ( \vec{S}_{1}\cdot\vec{n})^{2} + S_{1}^{4} \Big],\nn\\
\eea
\bea
V^{\text{NLO}}_{(\text{BS}_1^3 ) \text{S}_1 } &=&  	\frac{G m_{2}}{2 m_{1}{}^3 r{}^5} \Big[ 15 \vec{S}_{1}\cdot\vec{n} S_{1}^2 \vec{v}_{1}\cdot\vec{n} \vec{S}_{1}\cdot\vec{v}_{1} - 15 \vec{S}_{1}\cdot\vec{n} S_{1}^2 \vec{v}_{1}\cdot\vec{n} \vec{S}_{1}\cdot\vec{v}_{2} + 3 S_{1}^2 \vec{S}_{1}\cdot\vec{v}_{1} \vec{S}_{1}\cdot\vec{v}_{2} \nn\\ 
&& - 30 S_{1}^2 \big( v_{1}^2 - \vec{v}_{1}\cdot\vec{v}_{2} \big) ( \vec{S}_{1}\cdot\vec{n})^{2} - 35 \vec{S}_{1}\cdot\vec{v}_{1} \vec{v}_{1}\cdot\vec{n} ( \vec{S}_{1}\cdot\vec{n})^{3} + 35 \vec{S}_{1}\cdot\vec{v}_{2} \vec{v}_{1}\cdot\vec{n} ( \vec{S}_{1}\cdot\vec{n})^{3} \nn\\ 
&& + 35 \big( v_{1}^2 - \vec{v}_{1}\cdot\vec{v}_{2} \big) ( \vec{S}_{1}\cdot\vec{n})^{4} - 15 \vec{S}_{1}\cdot\vec{v}_{1} \vec{S}_{1}\cdot\vec{v}_{2} ( \vec{S}_{1}\cdot\vec{n})^{2} + 15 ( \vec{S}_{1}\cdot\vec{n})^{2} ( \vec{S}_{1}\cdot\vec{v}_{1})^{2} \nn\\ 
&& - 3 S_{1}^2 ( \vec{S}_{1}\cdot\vec{v}_{1})^{2} + 3 \big( v_{1}^2 - \vec{v}_{1}\cdot\vec{v}_{2} \big) S_{1}^{4} \Big]\nn\\ && +  	\frac{G^2 m_{2}}{m_{1}{}^2 r{}^6} \Big[ 5 ( \vec{S}_{1}\cdot\vec{n})^{4} - 3 S_{1}^2 ( \vec{S}_{1}\cdot\vec{n})^{2} \Big],
\eea
\bea
V^{\text{NLO}}_{\text{ES}_1^4 } &=& - 	\frac{G m_{2}}{16 m_{1}{}^3 r{}^5} \Big[ 60 \vec{S}_{1}\cdot\vec{n} S_{1}^2 \big( \vec{v}_{1}\cdot\vec{n} - \vec{v}_{2}\cdot\vec{n} \big) \vec{S}_{1}\cdot\vec{v}_{1} - 60 \vec{S}_{1}\cdot\vec{n} S_{1}^2 \vec{v}_{1}\cdot\vec{n} \vec{S}_{1}\cdot\vec{v}_{2} \nn\\ 
&& + 12 S_{1}^2 \vec{S}_{1}\cdot\vec{v}_{1} \vec{S}_{1}\cdot\vec{v}_{2} - 30 S_{1}^2 \big( 3 v_{1}^2 - 7 \vec{v}_{1}\cdot\vec{v}_{2} + 3 v_{2}^2 -7 \vec{v}_{1}\cdot\vec{n} \vec{v}_{2}\cdot\vec{n} \big) ( \vec{S}_{1}\cdot\vec{n})^{2} \nn\\ 
&& - 140 \vec{S}_{1}\cdot\vec{v}_{1} \big( \vec{v}_{1}\cdot\vec{n} - \vec{v}_{2}\cdot\vec{n} \big) ( \vec{S}_{1}\cdot\vec{n})^{3} + 140 \vec{S}_{1}\cdot\vec{v}_{2} \vec{v}_{1}\cdot\vec{n} ( \vec{S}_{1}\cdot\vec{n})^{3} + 35 \big( 3 v_{1}^2 \nn\\ 
&& - 7 \vec{v}_{1}\cdot\vec{v}_{2} + 3 v_{2}^2 -9 \vec{v}_{1}\cdot\vec{n} \vec{v}_{2}\cdot\vec{n} \big) ( \vec{S}_{1}\cdot\vec{n})^{4} - 60 \vec{S}_{1}\cdot\vec{v}_{1} \vec{S}_{1}\cdot\vec{v}_{2} ( \vec{S}_{1}\cdot\vec{n})^{2} \nn\\ 
&& + 60 ( \vec{S}_{1}\cdot\vec{n})^{2} ( \vec{S}_{1}\cdot\vec{v}_{1})^{2} - 12 S_{1}^2 ( \vec{S}_{1}\cdot\vec{v}_{1})^{2} + 3 \big( 3 v_{1}^2 - 7 \vec{v}_{1}\cdot\vec{v}_{2} \nn\\ 
&& + 3 v_{2}^2 -5 \vec{v}_{1}\cdot\vec{n} \vec{v}_{2}\cdot\vec{n} \big) S_{1}^{4} \Big]\nn\\ && +  	\frac{G^2 m_{2}}{8 m_{1}{}^2 r{}^6} \Big[ 35 ( \vec{S}_{1}\cdot\vec{n})^{4} - 30 S_{1}^2 ( \vec{S}_{1}\cdot\vec{n})^{2} + 3 S_{1}^{4} \Big] + 	\frac{G^2 m_{2}{}^2}{8 m_{1}{}^3 r{}^6} \Big[ 285 ( \vec{S}_{1}\cdot\vec{n})^{4} \nn\\ 
&& - 240 S_{1}^2 ( \vec{S}_{1}\cdot\vec{n})^{2} + 23 S_{1}^{4} \Big],
\eea
\bea
\label{subhexa}
V^{\text{NLO}}_{\text{E}^2\text{S}_1^4 } &=& - 	\frac{G^2 m_{2}{}^2}{8 m_{1}{}^3 r{}^6} 
\Big[ 9( \vec{S}_{1}\cdot\vec{n})^{4} - \frac{54}{7} S_{1}^2 ( \vec{S}_{1}\cdot\vec{n})^{2} + \frac{27}{35} S_{1}^{4} \Big],
\eea
\bea
V^{\text{NLO}}_{\text{S}_1^3 \text{S}_2}&=&- 	\frac{3 G}{8 m_{1}{}^2 r{}^5} \Big[ S_{1}^2 \vec{S}_{1}\cdot\vec{S}_{2} \big( 4 \vec{v}_{1}\cdot\vec{v}_{2} -20 \vec{v}_{1}\cdot\vec{n} \vec{v}_{2}\cdot\vec{n} + 15 ( \vec{v}_{1}\cdot\vec{n})^{2} \big) \nn\\ 
&& - 5 \vec{S}_{1}\cdot\vec{n} S_{1}^2 \big( 4 \vec{v}_{1}\cdot\vec{v}_{2} -28 \vec{v}_{1}\cdot\vec{n} \vec{v}_{2}\cdot\vec{n} + 21 ( \vec{v}_{1}\cdot\vec{n})^{2} \big) \vec{S}_{2}\cdot\vec{n} + 20 S_{1}^2 \vec{S}_{2}\cdot\vec{n} \vec{v}_{1}\cdot\vec{n} \vec{S}_{1}\cdot\vec{v}_{1} \nn\\ 
&& - 20 \vec{S}_{1}\cdot\vec{n} \vec{S}_{1}\cdot\vec{S}_{2} \big( \vec{v}_{1}\cdot\vec{n} - 2 \vec{v}_{2}\cdot\vec{n} \big) \vec{S}_{1}\cdot\vec{v}_{1} + 10 \vec{S}_{1}\cdot\vec{n} S_{1}^2 \big( 3 \vec{v}_{1}\cdot\vec{n} - 2 \vec{v}_{2}\cdot\vec{n} \big) \vec{S}_{2}\cdot\vec{v}_{1} \nn\\ 
&& - 4 S_{1}^2 \vec{S}_{1}\cdot\vec{v}_{1} \vec{S}_{2}\cdot\vec{v}_{1} - 20 S_{1}^2 \vec{S}_{2}\cdot\vec{n} \vec{v}_{1}\cdot\vec{n} \vec{S}_{1}\cdot\vec{v}_{2} + 40 \vec{S}_{1}\cdot\vec{n} \vec{S}_{2}\cdot\vec{n} \vec{S}_{1}\cdot\vec{v}_{1} \vec{S}_{1}\cdot\vec{v}_{2} \nn\\ 
&& - 8 \vec{S}_{1}\cdot\vec{S}_{2} \vec{S}_{1}\cdot\vec{v}_{1} \vec{S}_{1}\cdot\vec{v}_{2} + 4 S_{1}^2 \vec{S}_{2}\cdot\vec{v}_{1} \vec{S}_{1}\cdot\vec{v}_{2} - 20 \vec{S}_{1}\cdot\vec{n} S_{1}^2 \vec{v}_{1}\cdot\vec{n} \vec{S}_{2}\cdot\vec{v}_{2} \nn\\ 
&& - 15 \vec{S}_{1}\cdot\vec{S}_{2} v_{1}^2 ( \vec{S}_{1}\cdot\vec{n})^{2} + 70 \vec{S}_{2}\cdot\vec{n} \vec{S}_{1}\cdot\vec{v}_{1} \big( \vec{v}_{1}\cdot\vec{n} - 2 \vec{v}_{2}\cdot\vec{n} \big) ( \vec{S}_{1}\cdot\vec{n})^{2} \nn\\ 
&& + 35 \vec{S}_{2}\cdot\vec{n} v_{1}^2 ( \vec{S}_{1}\cdot\vec{n})^{3} - 10 \vec{S}_{1}\cdot\vec{v}_{1} \vec{S}_{2}\cdot\vec{v}_{1} ( \vec{S}_{1}\cdot\vec{n})^{2} + 20 \vec{S}_{1}\cdot\vec{v}_{1} \vec{S}_{2}\cdot\vec{v}_{2} ( \vec{S}_{1}\cdot\vec{n})^{2} \nn\\ 
&& - 20 \vec{S}_{1}\cdot\vec{n} \vec{S}_{2}\cdot\vec{n} ( \vec{S}_{1}\cdot\vec{v}_{1})^{2} + 4 \vec{S}_{1}\cdot\vec{S}_{2} ( \vec{S}_{1}\cdot\vec{v}_{1})^{2} \Big]\nn\\ && +  	\frac{3 G^2 m_{2}}{4 m_{1}{}^2 r{}^6} \Big[ 4 \vec{S}_{1}\cdot\vec{n} S_{1}^2 \vec{S}_{2}\cdot\vec{n} - S_{1}^2 \vec{S}_{1}\cdot\vec{S}_{2} - 5 \vec{S}_{2}\cdot\vec{n} ( \vec{S}_{1}\cdot\vec{n})^{3} + 2 \vec{S}_{1}\cdot\vec{S}_{2} ( \vec{S}_{1}\cdot\vec{n})^{2} \Big],\nn\\
\eea
\bea
V^{\text{NLO}}_{(\text{ES}_1^2 ) \text{S}_1 \text{S}_2} &=& 	\frac{3 G}{8 m_{1}{}^2 r{}^5} \Big[ S_{1}^2 \vec{S}_{1}\cdot\vec{S}_{2} \big( 12 v_{1}^2 - 8 \vec{v}_{1}\cdot\vec{v}_{2} -5 \vec{v}_{1}\cdot\vec{n} \vec{v}_{2}\cdot\vec{n} \big) - 15 \vec{S}_{1}\cdot\vec{n} S_{1}^2 \big( 4 v_{1}^2 \nn\\ 
&& - 3 \vec{v}_{1}\cdot\vec{v}_{2} \big) \vec{S}_{2}\cdot\vec{n} + 20 S_{1}^2 \vec{S}_{2}\cdot\vec{n} \vec{v}_{1}\cdot\vec{n} \vec{S}_{1}\cdot\vec{v}_{1} + 10 \vec{S}_{1}\cdot\vec{n} \vec{S}_{1}\cdot\vec{S}_{2} \big( 4 \vec{v}_{1}\cdot\vec{n} \nn\\ 
&& - \vec{v}_{2}\cdot\vec{n} \big) \vec{S}_{1}\cdot\vec{v}_{1} + 15 \vec{S}_{1}\cdot\vec{n} S_{1}^2 \vec{v}_{2}\cdot\vec{n} \vec{S}_{2}\cdot\vec{v}_{1} - 4 S_{1}^2 \vec{S}_{1}\cdot\vec{v}_{1} \vec{S}_{2}\cdot\vec{v}_{1} \nn\\ 
&& - 15 S_{1}^2 \vec{S}_{2}\cdot\vec{n} \vec{v}_{1}\cdot\vec{n} \vec{S}_{1}\cdot\vec{v}_{2} - 40 \vec{S}_{1}\cdot\vec{n} \vec{S}_{1}\cdot\vec{S}_{2} \vec{v}_{1}\cdot\vec{n} \vec{S}_{1}\cdot\vec{v}_{2} \nn\\ 
&& - 30 \vec{S}_{1}\cdot\vec{n} \vec{S}_{2}\cdot\vec{n} \vec{S}_{1}\cdot\vec{v}_{1} \vec{S}_{1}\cdot\vec{v}_{2} + 8 \vec{S}_{1}\cdot\vec{S}_{2} \vec{S}_{1}\cdot\vec{v}_{1} \vec{S}_{1}\cdot\vec{v}_{2} - 5 \vec{S}_{1}\cdot\vec{S}_{2} \big( 12 v_{1}^2 \nn\\ 
&& - 8 \vec{v}_{1}\cdot\vec{v}_{2} -7 \vec{v}_{1}\cdot\vec{n} \vec{v}_{2}\cdot\vec{n} \big) ( \vec{S}_{1}\cdot\vec{n})^{2} - 140 \vec{S}_{2}\cdot\vec{n} \vec{S}_{1}\cdot\vec{v}_{1} \vec{v}_{1}\cdot\vec{n} ( \vec{S}_{1}\cdot\vec{n})^{2} \nn\\ 
&& + 105 \vec{S}_{2}\cdot\vec{n} \vec{S}_{1}\cdot\vec{v}_{2} \vec{v}_{1}\cdot\vec{n} ( \vec{S}_{1}\cdot\vec{n})^{2} + 35 \vec{S}_{2}\cdot\vec{n} \big( 4 v_{1}^2 - 3 \vec{v}_{1}\cdot\vec{v}_{2} \big) ( \vec{S}_{1}\cdot\vec{n})^{3} \nn\\ 
&& - 35 \vec{S}_{2}\cdot\vec{v}_{1} \vec{v}_{2}\cdot\vec{n} ( \vec{S}_{1}\cdot\vec{n})^{3} + 20 \vec{S}_{1}\cdot\vec{v}_{1} \vec{S}_{2}\cdot\vec{v}_{1} ( \vec{S}_{1}\cdot\vec{n})^{2} + 40 \vec{S}_{1}\cdot\vec{n} \vec{S}_{2}\cdot\vec{n} ( \vec{S}_{1}\cdot\vec{v}_{1})^{2} \nn\\ 
&& - 8 \vec{S}_{1}\cdot\vec{S}_{2} ( \vec{S}_{1}\cdot\vec{v}_{1})^{2} \Big]\nn\\ && +  	\frac{3 G^2 m_{2}}{m_{1}{}^2 r{}^6} \Big[ S_{1}^2 \vec{S}_{1}\cdot\vec{S}_{2} - 3 \vec{S}_{1}\cdot\vec{S}_{2} ( \vec{S}_{1}\cdot\vec{n})^{2} \Big] - 	\frac{G^2}{m_{1} r{}^6} \Big[ 15 \vec{S}_{1}\cdot\vec{n} S_{1}^2 \vec{S}_{2}\cdot\vec{n} - 4 S_{1}^2 \vec{S}_{1}\cdot\vec{S}_{2} \nn\\ 
&& - 39 \vec{S}_{2}\cdot\vec{n} ( \vec{S}_{1}\cdot\vec{n})^{3} + 18 \vec{S}_{1}\cdot\vec{S}_{2} ( \vec{S}_{1}\cdot\vec{n})^{2} \Big],
\eea
\bea
\label{quadsqtooctu}
V^{\text{NLO}}_{C_{\text{ES}^2_1}^2 \text{S}_1^3 \text{S}_2 } &=&  	\frac{9 G^2 m_{2}}{m_{1}{}^2 r{}^6} \Big[ \vec{S}_{2}\cdot\vec{n} ( \vec{S}_{1}\cdot\vec{n})^{3} - \vec{S}_{1}\cdot\vec{S}_{2} ( \vec{S}_{1}\cdot\vec{n})^{2} \Big] ,
\eea
\bea
V^{\text{NLO}}_{(\text{BS}_1^3) \text{S}_2 }&=&- 	\frac{G}{4 m_{1}{}^2 r{}^5} \Big[ S_{1}^2 \vec{S}_{1}\cdot\vec{S}_{2} \big( 13 v_{1}^2 - 23 \vec{v}_{1}\cdot\vec{v}_{2} + 7 v_{2}^2 + 25 \vec{v}_{1}\cdot\vec{n} \vec{v}_{2}\cdot\vec{n} - 20 ( \vec{v}_{1}\cdot\vec{n})^{2} \nn\\ 
&& - 20 ( \vec{v}_{2}\cdot\vec{n})^{2} \big) - 15 \vec{S}_{1}\cdot\vec{n} S_{1}^2 \big( 3 v_{1}^2 - 5 \vec{v}_{1}\cdot\vec{v}_{2} + v_{2}^2 -7 \vec{v}_{1}\cdot\vec{n} \vec{v}_{2}\cdot\vec{n} \big) \vec{S}_{2}\cdot\vec{n} \nn\\ 
&& + 25 S_{1}^2 \vec{S}_{2}\cdot\vec{n} \big( \vec{v}_{1}\cdot\vec{n} - \vec{v}_{2}\cdot\vec{n} \big) \vec{S}_{1}\cdot\vec{v}_{1} - 30 \vec{S}_{1}\cdot\vec{n} \vec{S}_{1}\cdot\vec{S}_{2} \big( \vec{v}_{1}\cdot\vec{n} - \vec{v}_{2}\cdot\vec{n} \big) \vec{S}_{1}\cdot\vec{v}_{1} \nn\\ 
&& + 15 \vec{S}_{1}\cdot\vec{n} S_{1}^2 \big( 4 \vec{v}_{1}\cdot\vec{n} - 3 \vec{v}_{2}\cdot\vec{n} \big) \vec{S}_{2}\cdot\vec{v}_{1} - 17 S_{1}^2 \vec{S}_{1}\cdot\vec{v}_{1} \vec{S}_{2}\cdot\vec{v}_{1} - 5 S_{1}^2 \vec{S}_{2}\cdot\vec{n} \big( 7 \vec{v}_{1}\cdot\vec{n} \nn\\ 
&& - 4 \vec{v}_{2}\cdot\vec{n} \big) \vec{S}_{1}\cdot\vec{v}_{2} + 10 \vec{S}_{1}\cdot\vec{n} \vec{S}_{1}\cdot\vec{S}_{2} \big( \vec{v}_{1}\cdot\vec{n} - 4 \vec{v}_{2}\cdot\vec{n} \big) \vec{S}_{1}\cdot\vec{v}_{2} \nn\\ 
&& - 90 \vec{S}_{1}\cdot\vec{n} \vec{S}_{2}\cdot\vec{n} \vec{S}_{1}\cdot\vec{v}_{1} \vec{S}_{1}\cdot\vec{v}_{2} + 2 \vec{S}_{1}\cdot\vec{S}_{2} \vec{S}_{1}\cdot\vec{v}_{1} \vec{S}_{1}\cdot\vec{v}_{2} + 13 S_{1}^2 \vec{S}_{2}\cdot\vec{v}_{1} \vec{S}_{1}\cdot\vec{v}_{2} \nn\\ 
&& - 15 \vec{S}_{1}\cdot\vec{n} S_{1}^2 \big( 3 \vec{v}_{1}\cdot\vec{n} - \vec{v}_{2}\cdot\vec{n} \big) \vec{S}_{2}\cdot\vec{v}_{2} + 11 S_{1}^2 \vec{S}_{1}\cdot\vec{v}_{1} \vec{S}_{2}\cdot\vec{v}_{2} - 7 S_{1}^2 \vec{S}_{1}\cdot\vec{v}_{2} \vec{S}_{2}\cdot\vec{v}_{2} \nn\\ 
&& - 5 \vec{S}_{1}\cdot\vec{S}_{2} \big( 13 v_{1}^2 - 23 \vec{v}_{1}\cdot\vec{v}_{2} + 7 v_{2}^2 + 35 \vec{v}_{1}\cdot\vec{n} \vec{v}_{2}\cdot\vec{n} - 28 ( \vec{v}_{1}\cdot\vec{n})^{2} \nn\\ 
&& - 28 ( \vec{v}_{2}\cdot\vec{n})^{2} \big) ( \vec{S}_{1}\cdot\vec{n})^{2} - 175 \vec{S}_{2}\cdot\vec{n} \vec{S}_{1}\cdot\vec{v}_{1} \big( \vec{v}_{1}\cdot\vec{n} - \vec{v}_{2}\cdot\vec{n} \big) ( \vec{S}_{1}\cdot\vec{n})^{2} \nn\\ 
&& + 35 \vec{S}_{2}\cdot\vec{n} \vec{S}_{1}\cdot\vec{v}_{2} \big( 7 \vec{v}_{1}\cdot\vec{n} - 4 \vec{v}_{2}\cdot\vec{n} \big) ( \vec{S}_{1}\cdot\vec{n})^{2} + 35 \vec{S}_{2}\cdot\vec{n} \big( 3 v_{1}^2 - 5 \vec{v}_{1}\cdot\vec{v}_{2} \nn\\ 
&& + v_{2}^2 -9 \vec{v}_{1}\cdot\vec{n} \vec{v}_{2}\cdot\vec{n} \big) ( \vec{S}_{1}\cdot\vec{n})^{3} - 35 \vec{S}_{2}\cdot\vec{v}_{1} \big( 4 \vec{v}_{1}\cdot\vec{n} - 3 \vec{v}_{2}\cdot\vec{n} \big) ( \vec{S}_{1}\cdot\vec{n})^{3} \nn\\ 
&& + 35 \vec{S}_{2}\cdot\vec{v}_{2} \big( 3 \vec{v}_{1}\cdot\vec{n} - \vec{v}_{2}\cdot\vec{n} \big) ( \vec{S}_{1}\cdot\vec{n})^{3} + 85 \vec{S}_{1}\cdot\vec{v}_{1} \vec{S}_{2}\cdot\vec{v}_{1} ( \vec{S}_{1}\cdot\vec{n})^{2} \nn\\ 
&& - 65 \vec{S}_{2}\cdot\vec{v}_{1} \vec{S}_{1}\cdot\vec{v}_{2} ( \vec{S}_{1}\cdot\vec{n})^{2} - 55 \vec{S}_{1}\cdot\vec{v}_{1} \vec{S}_{2}\cdot\vec{v}_{2} ( \vec{S}_{1}\cdot\vec{n})^{2} + 35 \vec{S}_{1}\cdot\vec{v}_{2} \vec{S}_{2}\cdot\vec{v}_{2} ( \vec{S}_{1}\cdot\vec{n})^{2} \nn\\ 
&& + 50 \vec{S}_{1}\cdot\vec{n} \vec{S}_{2}\cdot\vec{n} ( \vec{S}_{1}\cdot\vec{v}_{1})^{2} - 2 \vec{S}_{1}\cdot\vec{S}_{2} ( \vec{S}_{1}\cdot\vec{v}_{1})^{2} + 40 \vec{S}_{1}\cdot\vec{n} \vec{S}_{2}\cdot\vec{n} ( \vec{S}_{1}\cdot\vec{v}_{2})^{2} \Big]\nn\\ && - 	\frac{G^2}{m_{1} r{}^6} \Big[ 19 \vec{S}_{1}\cdot\vec{n} S_{1}^2 \vec{S}_{2}\cdot\vec{n} - 4 S_{1}^2 \vec{S}_{1}\cdot\vec{S}_{2} - 45 \vec{S}_{2}\cdot\vec{n} ( \vec{S}_{1}\cdot\vec{n})^{3} + 20 \vec{S}_{1}\cdot\vec{S}_{2} ( \vec{S}_{1}\cdot\vec{n})^{2} \Big] \nn\\ 
&& - 	\frac{G^2 m_{2}}{2 m_{1}{}^2 r{}^6} \Big[ 105 \vec{S}_{1}\cdot\vec{n} S_{1}^2 \vec{S}_{2}\cdot\vec{n} - 19 S_{1}^2 \vec{S}_{1}\cdot\vec{S}_{2} - 240 \vec{S}_{2}\cdot\vec{n} ( \vec{S}_{1}\cdot\vec{n})^{3} \nn\\ 
&& + 96 \vec{S}_{1}\cdot\vec{S}_{2} ( \vec{S}_{1}\cdot\vec{n})^{2} \Big],
\eea
\bea
V^{\text{NLO}}_{\text{S}_1^2 \text{S}_2^2}&=&- 	\frac{3 G}{8 m_{1} m_{2} r{}^5} \Big[ S_{1}^2 S_{2}^2 \big( 9 \vec{v}_{1}\cdot\vec{v}_{2} -10 \vec{v}_{1}\cdot\vec{n} \vec{v}_{2}\cdot\vec{n} \big) \nn\\ 
&& + 5 \vec{S}_{1}\cdot\vec{n} \vec{S}_{2}\cdot\vec{n} \big( 9 \vec{v}_{1}\cdot\vec{v}_{2} -7 \vec{v}_{1}\cdot\vec{n} \vec{v}_{2}\cdot\vec{n} \big) \vec{S}_{1}\cdot\vec{S}_{2} - 10 \vec{S}_{2}\cdot\vec{n} \vec{S}_{1}\cdot\vec{S}_{2} \vec{v}_{2}\cdot\vec{n} \vec{S}_{1}\cdot\vec{v}_{1} \nn\\ 
&& + 20 \vec{S}_{1}\cdot\vec{n} S_{2}^2 \vec{v}_{2}\cdot\vec{n} \vec{S}_{1}\cdot\vec{v}_{1} + 10 S_{1}^2 \vec{S}_{2}\cdot\vec{n} \vec{v}_{2}\cdot\vec{n} \vec{S}_{2}\cdot\vec{v}_{1} - 30 \vec{S}_{1}\cdot\vec{n} \vec{S}_{1}\cdot\vec{S}_{2} \vec{v}_{2}\cdot\vec{n} \vec{S}_{2}\cdot\vec{v}_{1} \nn\\ 
&& - 18 S_{2}^2 \vec{S}_{1}\cdot\vec{v}_{1} \vec{S}_{1}\cdot\vec{v}_{2} - 30 \vec{S}_{1}\cdot\vec{n} \vec{S}_{2}\cdot\vec{n} \vec{S}_{2}\cdot\vec{v}_{1} \vec{S}_{1}\cdot\vec{v}_{2} + 13 \vec{S}_{1}\cdot\vec{S}_{2} \vec{S}_{2}\cdot\vec{v}_{1} \vec{S}_{1}\cdot\vec{v}_{2} \nn\\ 
&& - 10 \vec{S}_{1}\cdot\vec{n} \vec{S}_{2}\cdot\vec{n} \vec{S}_{1}\cdot\vec{v}_{1} \vec{S}_{2}\cdot\vec{v}_{2} + 9 \vec{S}_{1}\cdot\vec{S}_{2} \vec{S}_{1}\cdot\vec{v}_{1} \vec{S}_{2}\cdot\vec{v}_{2} \nn\\ 
&& + 70 \vec{S}_{2}\cdot\vec{n} \vec{S}_{2}\cdot\vec{v}_{1} \vec{v}_{2}\cdot\vec{n} ( \vec{S}_{1}\cdot\vec{n})^{2} - 10 S_{1}^2 \vec{v}_{1}\cdot\vec{v}_{2} ( \vec{S}_{2}\cdot\vec{n})^{2} - 35 \vec{v}_{1}\cdot\vec{v}_{2} ( \vec{S}_{1}\cdot\vec{n})^{2} ( \vec{S}_{2}\cdot\vec{n})^{2} \nn\\ 
&& - \big( 13 \vec{v}_{1}\cdot\vec{v}_{2} -15 \vec{v}_{1}\cdot\vec{n} \vec{v}_{2}\cdot\vec{n} \big) ( \vec{S}_{1}\cdot\vec{S}_{2})^{2} + 10 \vec{S}_{1}\cdot\vec{v}_{1} \vec{S}_{1}\cdot\vec{v}_{2} ( \vec{S}_{2}\cdot\vec{n})^{2} \Big]\nn\\ && - 	\frac{6 G^2}{m_{1} r{}^6} \Big[ 5 \vec{S}_{1}\cdot\vec{n} \vec{S}_{2}\cdot\vec{n} \vec{S}_{1}\cdot\vec{S}_{2} - 5 ( \vec{S}_{1}\cdot\vec{n})^{2} ( \vec{S}_{2}\cdot\vec{n})^{2} - ( \vec{S}_{1}\cdot\vec{S}_{2})^{2} \Big] \nn\\ 
&& - 	\frac{6 G^2}{m_{2} r{}^6} S_{1}^2 ( \vec{S}_{2}\cdot\vec{n})^{2} ,
\eea
\bea
V^{\text{NLO}}_{(\text{ES}_1^2 )  \text{S}_2^2}&=& - 	\frac{3 G}{16 m_{1} m_{2} r{}^5} \Big[ S_{1}^2 S_{2}^2 \big( 92 \vec{v}_{1}\cdot\vec{v}_{2} - 80 v_{2}^2 -100 \vec{v}_{1}\cdot\vec{n} \vec{v}_{2}\cdot\vec{n} + 85 ( \vec{v}_{2}\cdot\vec{n})^{2} \big) \nn\\ 
&& - 80 \vec{S}_{2}\cdot\vec{n} \vec{S}_{1}\cdot\vec{S}_{2} \vec{v}_{2}\cdot\vec{n} \vec{S}_{1}\cdot\vec{v}_{1} + 40 \vec{S}_{1}\cdot\vec{n} S_{2}^2 \vec{v}_{2}\cdot\vec{n} \vec{S}_{1}\cdot\vec{v}_{1} + 120 S_{1}^2 \vec{S}_{2}\cdot\vec{n} \vec{v}_{2}\cdot\vec{n} \vec{S}_{2}\cdot\vec{v}_{1} \nn\\ 
&& - 20 \vec{S}_{2}\cdot\vec{n} \vec{S}_{1}\cdot\vec{S}_{2} \big( 4 \vec{v}_{1}\cdot\vec{n} - 7 \vec{v}_{2}\cdot\vec{n} \big) \vec{S}_{1}\cdot\vec{v}_{2} + 40 \vec{S}_{1}\cdot\vec{n} S_{2}^2 \big( \vec{v}_{1}\cdot\vec{n} - 2 \vec{v}_{2}\cdot\vec{n} \big) \vec{S}_{1}\cdot\vec{v}_{2} \nn\\ 
&& - 72 S_{2}^2 \vec{S}_{1}\cdot\vec{v}_{1} \vec{S}_{1}\cdot\vec{v}_{2} + 64 \vec{S}_{1}\cdot\vec{S}_{2} \vec{S}_{2}\cdot\vec{v}_{1} \vec{S}_{1}\cdot\vec{v}_{2} + 10 S_{1}^2 \vec{S}_{2}\cdot\vec{n} \big( 10 \vec{v}_{1}\cdot\vec{n} \nn\\ 
&& - 19 \vec{v}_{2}\cdot\vec{n} \big) \vec{S}_{2}\cdot\vec{v}_{2} - 40 \vec{S}_{1}\cdot\vec{n} \vec{S}_{1}\cdot\vec{S}_{2} \big( \vec{v}_{1}\cdot\vec{n} - \vec{v}_{2}\cdot\vec{n} \big) \vec{S}_{2}\cdot\vec{v}_{2} \nn\\ 
&& - 40 \vec{S}_{1}\cdot\vec{n} \vec{S}_{2}\cdot\vec{n} \vec{S}_{1}\cdot\vec{v}_{1} \vec{S}_{2}\cdot\vec{v}_{2} + 72 \vec{S}_{1}\cdot\vec{S}_{2} \vec{S}_{1}\cdot\vec{v}_{1} \vec{S}_{2}\cdot\vec{v}_{2} - 92 S_{1}^2 \vec{S}_{2}\cdot\vec{v}_{1} \vec{S}_{2}\cdot\vec{v}_{2} \nn\\ 
&& + 40 \vec{S}_{1}\cdot\vec{n} \vec{S}_{2}\cdot\vec{n} \vec{S}_{1}\cdot\vec{v}_{2} \vec{S}_{2}\cdot\vec{v}_{2} - 120 \vec{S}_{1}\cdot\vec{S}_{2} \vec{S}_{1}\cdot\vec{v}_{2} \vec{S}_{2}\cdot\vec{v}_{2} - 5 S_{2}^2 \big( 28 \vec{v}_{1}\cdot\vec{v}_{2} \nn\\ 
&& - 24 v_{2}^2 -28 \vec{v}_{1}\cdot\vec{n} \vec{v}_{2}\cdot\vec{n} + 21 ( \vec{v}_{2}\cdot\vec{n})^{2} \big) ( \vec{S}_{1}\cdot\vec{n})^{2} - 280 \vec{S}_{2}\cdot\vec{n} \vec{S}_{2}\cdot\vec{v}_{1} \vec{v}_{2}\cdot\vec{n} ( \vec{S}_{1}\cdot\vec{n})^{2} \nn\\ 
&& - 70 \vec{S}_{2}\cdot\vec{n} \vec{S}_{2}\cdot\vec{v}_{2} \big( 2 \vec{v}_{1}\cdot\vec{n} - 5 \vec{v}_{2}\cdot\vec{n} \big) ( \vec{S}_{1}\cdot\vec{n})^{2} + 140 \vec{S}_{2}\cdot\vec{v}_{1} \vec{S}_{2}\cdot\vec{v}_{2} ( \vec{S}_{1}\cdot\vec{n})^{2} \nn\\ 
&& - 15 S_{1}^2 \big( 8 \vec{v}_{1}\cdot\vec{v}_{2} - 7 v_{2}^2 \big) ( \vec{S}_{2}\cdot\vec{n})^{2} + 35 \big( 8 \vec{v}_{1}\cdot\vec{v}_{2} - 7 v_{2}^2 \big) ( \vec{S}_{1}\cdot\vec{n})^{2} ( \vec{S}_{2}\cdot\vec{n})^{2} \nn\\ 
&& - 2 \big( 32 \vec{v}_{1}\cdot\vec{v}_{2} - 28 v_{2}^2 -40 \vec{v}_{1}\cdot\vec{n} \vec{v}_{2}\cdot\vec{n} + 35 ( \vec{v}_{2}\cdot\vec{n})^{2} \big) ( \vec{S}_{1}\cdot\vec{S}_{2})^{2} \nn\\ 
&& + 80 \vec{S}_{1}\cdot\vec{v}_{1} \vec{S}_{1}\cdot\vec{v}_{2} ( \vec{S}_{2}\cdot\vec{n})^{2} - 70 ( \vec{S}_{2}\cdot\vec{n})^{2} ( \vec{S}_{1}\cdot\vec{v}_{2})^{2} + 64 S_{2}^2 ( \vec{S}_{1}\cdot\vec{v}_{2})^{2} \nn\\ 
&& - 120 ( \vec{S}_{1}\cdot\vec{n})^{2} ( \vec{S}_{2}\cdot\vec{v}_{2})^{2} + 80 S_{1}^2 ( \vec{S}_{2}\cdot\vec{v}_{2})^{2} \Big]\nn\\ && +  	\frac{3 G^2}{8 m_{2} r{}^6} \Big[ 2 \vec{S}_{1}\cdot\vec{n} \vec{S}_{2}\cdot\vec{n} \vec{S}_{1}\cdot\vec{S}_{2} - S_{1}^2 S_{2}^2 + 3 S_{2}^2 ( \vec{S}_{1}\cdot\vec{n})^{2} - 5 ( \vec{S}_{1}\cdot\vec{n})^{2} ( \vec{S}_{2}\cdot\vec{n})^{2} \nn\\ 
&& + S_{1}^2 ( \vec{S}_{2}\cdot\vec{n})^{2} \Big] - 	\frac{G^2}{2 m_{1} r{}^6} \Big[ 30 \vec{S}_{1}\cdot\vec{n} \vec{S}_{2}\cdot\vec{n} \vec{S}_{1}\cdot\vec{S}_{2} - 5 S_{1}^2 S_{2}^2 + 12 S_{2}^2 ( \vec{S}_{1}\cdot\vec{n})^{2} \nn\\ 
&& - 75 ( \vec{S}_{1}\cdot\vec{n})^{2} ( \vec{S}_{2}\cdot\vec{n})^{2} + 15 S_{1}^2 ( \vec{S}_{2}\cdot\vec{n})^{2} + 3 ( \vec{S}_{1}\cdot\vec{S}_{2})^{2} \Big],
\eea
\bea
V^{\text{NLO}}_{(\text{ES}_1^2 )(\text{ES}_2^2 ) }&=&- 	\frac{3 G}{16 m_{1} m_{2} r{}^5} \Big[ S_{1}^2 S_{2}^2 \big( 10 v_{1}^2 - 11 \vec{v}_{1}\cdot\vec{v}_{2} + 15 \vec{v}_{1}\cdot\vec{n} \vec{v}_{2}\cdot\vec{n} - 20 ( \vec{v}_{1}\cdot\vec{n})^{2} \big) \nn\\ 
&& - 20 \vec{S}_{1}\cdot\vec{n} \vec{S}_{2}\cdot\vec{n} \big( 6 v_{1}^2 - 7 \vec{v}_{1}\cdot\vec{v}_{2} -7 \vec{v}_{1}\cdot\vec{n} \vec{v}_{2}\cdot\vec{n} \big) \vec{S}_{1}\cdot\vec{S}_{2} + 40 \vec{S}_{2}\cdot\vec{n} \vec{S}_{1}\cdot\vec{S}_{2} \big( \vec{v}_{1}\cdot\vec{n} \nn\\ 
&& - \vec{v}_{2}\cdot\vec{n} \big) \vec{S}_{1}\cdot\vec{v}_{1} + 20 \vec{S}_{1}\cdot\vec{n} S_{2}^2 \big( \vec{v}_{1}\cdot\vec{n} + \vec{v}_{2}\cdot\vec{n} \big) \vec{S}_{1}\cdot\vec{v}_{1} - 20 S_{1}^2 \vec{S}_{2}\cdot\vec{n} \big( 4 \vec{v}_{1}\cdot\vec{n} \nn\\ 
&& - \vec{v}_{2}\cdot\vec{n} \big) \vec{S}_{2}\cdot\vec{v}_{1} - 40 \vec{S}_{1}\cdot\vec{n} \vec{S}_{1}\cdot\vec{S}_{2} \vec{v}_{2}\cdot\vec{n} \vec{S}_{2}\cdot\vec{v}_{1} + 40 \vec{S}_{1}\cdot\vec{n} \vec{S}_{2}\cdot\vec{n} \vec{S}_{1}\cdot\vec{v}_{1} \vec{S}_{2}\cdot\vec{v}_{1} \nn\\ 
&& - 8 \vec{S}_{1}\cdot\vec{S}_{2} \vec{S}_{1}\cdot\vec{v}_{1} \vec{S}_{2}\cdot\vec{v}_{1} - 4 S_{2}^2 \vec{S}_{1}\cdot\vec{v}_{1} \vec{S}_{1}\cdot\vec{v}_{2} - 20 \vec{S}_{1}\cdot\vec{n} \vec{S}_{2}\cdot\vec{n} \vec{S}_{2}\cdot\vec{v}_{1} \vec{S}_{1}\cdot\vec{v}_{2} \nn\\ 
&& + 4 \vec{S}_{1}\cdot\vec{S}_{2} \vec{S}_{2}\cdot\vec{v}_{1} \vec{S}_{1}\cdot\vec{v}_{2} - 20 \vec{S}_{1}\cdot\vec{n} \vec{S}_{2}\cdot\vec{n} \vec{S}_{1}\cdot\vec{v}_{1} \vec{S}_{2}\cdot\vec{v}_{2} + 4 \vec{S}_{1}\cdot\vec{S}_{2} \vec{S}_{1}\cdot\vec{v}_{1} \vec{S}_{2}\cdot\vec{v}_{2} \nn\\ 
&& - 30 S_{2}^2 v_{1}^2 ( \vec{S}_{1}\cdot\vec{n})^{2} + 140 \vec{S}_{2}\cdot\vec{n} \vec{S}_{2}\cdot\vec{v}_{1} \vec{v}_{2}\cdot\vec{n} ( \vec{S}_{1}\cdot\vec{n})^{2} - 10 S_{1}^2 \big( 5 v_{1}^2 - 9 \vec{v}_{1}\cdot\vec{v}_{2} \nn\\ 
&& + 7 \vec{v}_{1}\cdot\vec{n} \vec{v}_{2}\cdot\vec{n} - 14 ( \vec{v}_{1}\cdot\vec{n})^{2} \big) ( \vec{S}_{2}\cdot\vec{n})^{2} - 140 \vec{S}_{1}\cdot\vec{n} \vec{S}_{1}\cdot\vec{v}_{1} \big( \vec{v}_{1}\cdot\vec{n} - \vec{v}_{2}\cdot\vec{n} \big) ( \vec{S}_{2}\cdot\vec{n})^{2} \nn\\ 
&& + 35 \big( 6 v_{1}^2 - 7 \vec{v}_{1}\cdot\vec{v}_{2} -9 \vec{v}_{1}\cdot\vec{n} \vec{v}_{2}\cdot\vec{n} \big) ( \vec{S}_{1}\cdot\vec{n})^{2} ( \vec{S}_{2}\cdot\vec{n})^{2} + 2 \big( 6 v_{1}^2 \nn\\ 
&& - 7 \vec{v}_{1}\cdot\vec{v}_{2} -5 \vec{v}_{1}\cdot\vec{n} \vec{v}_{2}\cdot\vec{n} \big) ( \vec{S}_{1}\cdot\vec{S}_{2})^{2} - 20 \vec{S}_{1}\cdot\vec{v}_{1} \vec{S}_{1}\cdot\vec{v}_{2} ( \vec{S}_{2}\cdot\vec{n})^{2} \nn\\ 
&& + 20 ( \vec{S}_{2}\cdot\vec{n})^{2} ( \vec{S}_{1}\cdot\vec{v}_{1})^{2} - 4 S_{2}^2 ( \vec{S}_{1}\cdot\vec{v}_{1})^{2} + 8 S_{1}^2 ( \vec{S}_{2}\cdot\vec{v}_{1})^{2} \Big]\nn\\ && - 	\frac{3 G^2}{2 m_{1} r{}^6} \Big[ 32 \vec{S}_{1}\cdot\vec{n} \vec{S}_{2}\cdot\vec{n} \vec{S}_{1}\cdot\vec{S}_{2} - 3 S_{1}^2 S_{2}^2 - 62 ( \vec{S}_{1}\cdot\vec{n})^{2} ( \vec{S}_{2}\cdot\vec{n})^{2} + 12 S_{1}^2 ( \vec{S}_{2}\cdot\vec{n})^{2} \nn\\ 
&& - 3 ( \vec{S}_{1}\cdot\vec{S}_{2})^{2} \Big] - 	\frac{15 G^2}{m_{2} r{}^6} S_{1}^2 ( \vec{S}_{2}\cdot\vec{n})^{2} .
\eea

\section{General Hamiltonians}
\label{newgenhams}

As noted in section \ref{genhams}, our general Hamiltonians for the NLO quartic-in-spin sectors, 
similar to the action potentials in eq.~\eqref{12actionbreakdown}, consist of $12$ unique sectors:  
\bea
\label{eq:hamcns4app}
H^{\text{NLO}}_{\text{S}^4} &=&C_{1\text{ES}^2} H^{\text{NLO}}_{(\text{ES}_1^2 ) \text{S}_1^2} +C_{1\text{ES}^2}^2 H^{\text{NLO}}_{(\text{ES}_1^2 )^2} + C_{1\text{BS}^3} H^{\text{NLO}}_{(\text{BS}_1^3) \text{S}_1 } +C_{1\text{ES}^4} H^{\text{NLO}}_{\text{ES}_1^4} +C_{1\text{E}^2\text{S}^4} H^{\text{NLO}}_{\text{E}^2\text{S}_1^4}\nn\\
&& +H^{\text{NLO}}_{\text{S}_1^3 \text{S}_2}+C_{1\text{ES}^2} H^{\text{NLO}}_{(\text{ES}_1^2 ) \text{S}_1 \text{S}_2}  + C_{1\text{ES}^2}^2 H^{\text{NLO}}_{C_{\text{ES}^2_1}^2 \text{S}_1^3 \text{S}_2 } + C_{1\text{BS}^3} H^{\text{NLO}}_{(\text{BS}_1^3) \text{S}_2 }\nn\\
&& +H^{\text{NLO}}_{\text{S}_1^2 \text{S}_2^2}+C_{1\text{ES}^2} H^{\text{NLO}}_{(\text{ES}_1^2 )  \text{S}_2^2} +C_{1\text{ES}^2} C_{2\text{ES}^2} H^{\text{NLO}}_{(\text{ES}_1^2 )(\text{ES}_2^2 ) }+  (1 \leftrightarrow 2),
\eea
with:
\bea
H^{\text{NLO}}_{(\text{ES}_1^2 ) \text{S}_1^2} &=&- 	\frac{3 G}{4 m_{1}{}^4 r{}^5} \Big[ 5 \vec{S}_{1}\cdot\vec{n} \vec{p}_{1}\cdot\vec{S}_{1} \vec{p}_{2}\cdot\vec{n} S_{1}^2 - 10 \vec{S}_{1}\cdot\vec{n} \vec{p}_{2}\cdot\vec{S}_{1} \vec{p}_{1}\cdot\vec{n} S_{1}^2 - \vec{p}_{1}\cdot\vec{S}_{1} \vec{p}_{2}\cdot\vec{S}_{1} S_{1}^2 \nn\\ 
&& - 5 S_{1}^2 \big( \vec{p}_{1}\cdot\vec{p}_{2} -7 \vec{p}_{1}\cdot\vec{n} \vec{p}_{2}\cdot\vec{n} \big) ( \vec{S}_{1}\cdot\vec{n})^{2} - 35 \vec{p}_{1}\cdot\vec{S}_{1} \vec{p}_{2}\cdot\vec{n} ( \vec{S}_{1}\cdot\vec{n})^{3} \nn\\ 
&& + 15 \vec{p}_{1}\cdot\vec{S}_{1} \vec{p}_{2}\cdot\vec{S}_{1} ( \vec{S}_{1}\cdot\vec{n})^{2} + \big( \vec{p}_{1}\cdot\vec{p}_{2} -5 \vec{p}_{1}\cdot\vec{n} \vec{p}_{2}\cdot\vec{n} \big) S_{1}^{4} \Big] \nn\\ 
&& - 	\frac{15 G m_{2}}{16 m_{1}{}^5 r{}^5} \Big[ 6 \vec{S}_{1}\cdot\vec{n} \vec{p}_{1}\cdot\vec{S}_{1} \vec{p}_{1}\cdot\vec{n} S_{1}^2 - 3 S_{1}^2 \big( p_{1}^2 + 7 ( \vec{p}_{1}\cdot\vec{n})^{2} \big) ( \vec{S}_{1}\cdot\vec{n})^{2} \nn\\ 
&& + 14 \vec{p}_{1}\cdot\vec{S}_{1} \vec{p}_{1}\cdot\vec{n} ( \vec{S}_{1}\cdot\vec{n})^{3} + 7 p_{1}^2 ( \vec{S}_{1}\cdot\vec{n})^{4} - 6 ( \vec{S}_{1}\cdot\vec{n})^{2} ( \vec{p}_{1}\cdot\vec{S}_{1})^{2} + 3 ( \vec{p}_{1}\cdot\vec{n})^{2} S_{1}^{4} \Big]\nn\\ && - 	\frac{3 G^2 m_{2}{}^2}{2 m_{1}{}^3 r{}^6} \Big[ 5 ( \vec{S}_{1}\cdot\vec{n})^{4} - 6 S_{1}^2 ( \vec{S}_{1}\cdot\vec{n})^{2} + S_{1}^{4} \Big],
\eea
\bea
H^{\text{NLO}}_{(\text{ES}_1^2 )^2}&=& - 	\frac{3 G^2 m_{2}{}^2}{2 m_{1}{}^3 r{}^6} \Big[ 5 ( \vec{S}_{1}\cdot\vec{n})^{4} - 3 S_{1}^2 ( \vec{S}_{1}\cdot\vec{n})^{2} \Big] + 	\frac{G^2 m_{2}}{8 m_{1}{}^2 r{}^6} \Big[ 9 ( \vec{S}_{1}\cdot\vec{n})^{4} - 6 S_{1}^2 ( \vec{S}_{1}\cdot\vec{n})^{2} + S_{1}^{4} \Big],\nn\\
\eea
\bea
H^{\text{NLO}}_{(\text{BS}_1^3) \text{S}_1 } &=& 	\frac{G m_{2}}{2 m_{1}{}^5 r{}^5} \Big[ 15 \vec{S}_{1}\cdot\vec{n} \vec{p}_{1}\cdot\vec{S}_{1} \vec{p}_{1}\cdot\vec{n} S_{1}^2 - 30 S_{1}^2 p_{1}^2 ( \vec{S}_{1}\cdot\vec{n})^{2} - 35 \vec{p}_{1}\cdot\vec{S}_{1} \vec{p}_{1}\cdot\vec{n} ( \vec{S}_{1}\cdot\vec{n})^{3} \nn\\ 
&& + 35 p_{1}^2 ( \vec{S}_{1}\cdot\vec{n})^{4} + 15 ( \vec{S}_{1}\cdot\vec{n})^{2} ( \vec{p}_{1}\cdot\vec{S}_{1})^{2} - 3 S_{1}^2 ( \vec{p}_{1}\cdot\vec{S}_{1})^{2} + 3 p_{1}^2 S_{1}^{4} \Big] \nn\\ 
&& - 	\frac{G}{2 m_{1}{}^4 r{}^5} \Big[ 15 \vec{S}_{1}\cdot\vec{n} \vec{p}_{2}\cdot\vec{S}_{1} \vec{p}_{1}\cdot\vec{n} S_{1}^2 - 3 \vec{p}_{1}\cdot\vec{S}_{1} \vec{p}_{2}\cdot\vec{S}_{1} S_{1}^2 - 30 S_{1}^2 \vec{p}_{1}\cdot\vec{p}_{2} ( \vec{S}_{1}\cdot\vec{n})^{2} \nn\\ 
&& - 35 \vec{p}_{2}\cdot\vec{S}_{1} \vec{p}_{1}\cdot\vec{n} ( \vec{S}_{1}\cdot\vec{n})^{3} + 35 \vec{p}_{1}\cdot\vec{p}_{2} ( \vec{S}_{1}\cdot\vec{n})^{4} + 15 \vec{p}_{1}\cdot\vec{S}_{1} \vec{p}_{2}\cdot\vec{S}_{1} ( \vec{S}_{1}\cdot\vec{n})^{2} \nn\\ 
&& + 3 \vec{p}_{1}\cdot\vec{p}_{2} S_{1}^{4} \Big]\nn\\ && +  	\frac{3 G^2 m_{2}{}^2}{2 m_{1}{}^3 r{}^6} \Big[ 5 ( \vec{S}_{1}\cdot\vec{n})^{4} - 6 S_{1}^2 ( \vec{S}_{1}\cdot\vec{n})^{2} + S_{1}^{4} \Big] + 	\frac{G^2 m_{2}}{m_{1}{}^2 r{}^6} \Big[ 15 ( \vec{S}_{1}\cdot\vec{n})^{4} - 15 S_{1}^2 ( \vec{S}_{1}\cdot\vec{n})^{2} \nn\\ 
&& + 2 S_{1}^{4} \Big],
\eea
\bea
H^{\text{NLO}}_{\text{ES}_1^4} &=& 	\frac{G}{16 m_{1}{}^4 r{}^5} \Big[ 60 \vec{S}_{1}\cdot\vec{n} \vec{p}_{1}\cdot\vec{S}_{1} \vec{p}_{2}\cdot\vec{n} S_{1}^2 + 60 \vec{S}_{1}\cdot\vec{n} \vec{p}_{2}\cdot\vec{S}_{1} \vec{p}_{1}\cdot\vec{n} S_{1}^2 - 12 \vec{p}_{1}\cdot\vec{S}_{1} \vec{p}_{2}\cdot\vec{S}_{1} S_{1}^2 \nn\\ 
&& - 210 S_{1}^2 \big( \vec{p}_{1}\cdot\vec{p}_{2} + \vec{p}_{1}\cdot\vec{n} \vec{p}_{2}\cdot\vec{n} \big) ( \vec{S}_{1}\cdot\vec{n})^{2} - 140 \vec{p}_{1}\cdot\vec{S}_{1} \vec{p}_{2}\cdot\vec{n} ( \vec{S}_{1}\cdot\vec{n})^{3} \nn\\ 
&& - 140 \vec{p}_{2}\cdot\vec{S}_{1} \vec{p}_{1}\cdot\vec{n} ( \vec{S}_{1}\cdot\vec{n})^{3} + 35 \big( 7 \vec{p}_{1}\cdot\vec{p}_{2} + 9 \vec{p}_{1}\cdot\vec{n} \vec{p}_{2}\cdot\vec{n} \big) ( \vec{S}_{1}\cdot\vec{n})^{4} \nn\\ 
&& + 60 \vec{p}_{1}\cdot\vec{S}_{1} \vec{p}_{2}\cdot\vec{S}_{1} ( \vec{S}_{1}\cdot\vec{n})^{2} + 3 \big( 7 \vec{p}_{1}\cdot\vec{p}_{2} + 5 \vec{p}_{1}\cdot\vec{n} \vec{p}_{2}\cdot\vec{n} \big) S_{1}^{4} \Big] \nn\\ 
&& - 	\frac{G m_{2}}{16 m_{1}{}^5 r{}^5} \Big[ 60 \vec{S}_{1}\cdot\vec{n} \vec{p}_{1}\cdot\vec{S}_{1} \vec{p}_{1}\cdot\vec{n} S_{1}^2 - 90 S_{1}^2 p_{1}^2 ( \vec{S}_{1}\cdot\vec{n})^{2} - 140 \vec{p}_{1}\cdot\vec{S}_{1} \vec{p}_{1}\cdot\vec{n} ( \vec{S}_{1}\cdot\vec{n})^{3} \nn\\ 
&& + 105 p_{1}^2 ( \vec{S}_{1}\cdot\vec{n})^{4} + 60 ( \vec{S}_{1}\cdot\vec{n})^{2} ( \vec{p}_{1}\cdot\vec{S}_{1})^{2} - 12 S_{1}^2 ( \vec{p}_{1}\cdot\vec{S}_{1})^{2} + 9 p_{1}^2 S_{1}^{4} \Big] \nn\\ 
&& + 	\frac{3 G}{16 m_{1}{}^3 m_{2} r{}^5} \Big[ 30 S_{1}^2 p_{2}^2 ( \vec{S}_{1}\cdot\vec{n})^{2} - 35 p_{2}^2 ( \vec{S}_{1}\cdot\vec{n})^{4} - 3 p_{2}^2 S_{1}^{4} \Big]\nn\\ && +  	\frac{G^2 m_{2}}{8 m_{1}{}^2 r{}^6} \Big[ 35 ( \vec{S}_{1}\cdot\vec{n})^{4} - 30 S_{1}^2 ( \vec{S}_{1}\cdot\vec{n})^{2} + 3 S_{1}^{4} \Big] + 	\frac{G^2 m_{2}{}^2}{8 m_{1}{}^3 r{}^6} \Big[ 285 ( \vec{S}_{1}\cdot\vec{n})^{4} \nn\\ 
&& - 240 S_{1}^2 ( \vec{S}_{1}\cdot\vec{n})^{2} + 23 S_{1}^{4} \Big],
\eea
\bea
\label{subhexaham}
H^{\text{NLO}}_{\text{E}^2\text{S}_1^4}  &=& - 	\frac{G^2 m_{2}{}^2}{8 m_{1}{}^3 r{}^6} 
\Big[ 9( \vec{S}_{1}\cdot\vec{n})^{4} - \frac{54}{7} S_{1}^2 ( \vec{S}_{1}\cdot\vec{n})^{2} + \frac{27}{35} S_{1}^{4} \Big],
\eea
\bea
H^{\text{NLO}}_{\text{S}_1^3 \text{S}_2} &=&- 	\frac{3 G}{8 m_{1}{}^4 r{}^5} \Big[ 15 S_{1}^2 \vec{S}_{1}\cdot\vec{S}_{2} ( \vec{p}_{1}\cdot\vec{n})^{2} - 105 \vec{S}_{1}\cdot\vec{n} S_{1}^2 ( \vec{p}_{1}\cdot\vec{n})^{2} \vec{S}_{2}\cdot\vec{n} \nn\\ 
&& + 20 \vec{p}_{1}\cdot\vec{S}_{1} S_{1}^2 \vec{p}_{1}\cdot\vec{n} \vec{S}_{2}\cdot\vec{n} + 30 \vec{S}_{1}\cdot\vec{n} S_{1}^2 \vec{p}_{1}\cdot\vec{n} \vec{p}_{1}\cdot\vec{S}_{2} - 4 \vec{p}_{1}\cdot\vec{S}_{1} S_{1}^2 \vec{p}_{1}\cdot\vec{S}_{2} \nn\\ 
&& - 20 \vec{S}_{1}\cdot\vec{n} \vec{p}_{1}\cdot\vec{S}_{1} \vec{p}_{1}\cdot\vec{n} \vec{S}_{1}\cdot\vec{S}_{2} + 70 \vec{p}_{1}\cdot\vec{S}_{1} \vec{S}_{2}\cdot\vec{n} \vec{p}_{1}\cdot\vec{n} ( \vec{S}_{1}\cdot\vec{n})^{2} \nn\\ 
&& - 15 \vec{S}_{1}\cdot\vec{S}_{2} p_{1}^2 ( \vec{S}_{1}\cdot\vec{n})^{2} + 35 \vec{S}_{2}\cdot\vec{n} p_{1}^2 ( \vec{S}_{1}\cdot\vec{n})^{3} - 20 \vec{S}_{1}\cdot\vec{n} \vec{S}_{2}\cdot\vec{n} ( \vec{p}_{1}\cdot\vec{S}_{1})^{2} \nn\\ 
&& - 10 \vec{p}_{1}\cdot\vec{S}_{1} \vec{p}_{1}\cdot\vec{S}_{2} ( \vec{S}_{1}\cdot\vec{n})^{2} + 4 \vec{S}_{1}\cdot\vec{S}_{2} ( \vec{p}_{1}\cdot\vec{S}_{1})^{2} \Big] \nn\\ 
&& - 	\frac{3 G}{2 m_{1}{}^3 m_{2} r{}^5} \Big[ S_{1}^2 \vec{S}_{1}\cdot\vec{S}_{2} \big( \vec{p}_{1}\cdot\vec{p}_{2} -5 \vec{p}_{1}\cdot\vec{n} \vec{p}_{2}\cdot\vec{n} \big) \nn\\ 
&& - 5 \vec{S}_{1}\cdot\vec{n} S_{1}^2 \big( \vec{p}_{1}\cdot\vec{p}_{2} -7 \vec{p}_{1}\cdot\vec{n} \vec{p}_{2}\cdot\vec{n} \big) \vec{S}_{2}\cdot\vec{n} - 5 \vec{p}_{2}\cdot\vec{S}_{1} S_{1}^2 \vec{p}_{1}\cdot\vec{n} \vec{S}_{2}\cdot\vec{n} \nn\\ 
&& + 10 \vec{S}_{1}\cdot\vec{n} \vec{p}_{1}\cdot\vec{S}_{1} \vec{p}_{2}\cdot\vec{S}_{1} \vec{S}_{2}\cdot\vec{n} - 5 \vec{S}_{1}\cdot\vec{n} S_{1}^2 \vec{p}_{2}\cdot\vec{n} \vec{p}_{1}\cdot\vec{S}_{2} + \vec{p}_{2}\cdot\vec{S}_{1} S_{1}^2 \vec{p}_{1}\cdot\vec{S}_{2} \nn\\ 
&& - 5 \vec{S}_{1}\cdot\vec{n} S_{1}^2 \vec{p}_{1}\cdot\vec{n} \vec{p}_{2}\cdot\vec{S}_{2} + 10 \vec{S}_{1}\cdot\vec{n} \vec{p}_{1}\cdot\vec{S}_{1} \vec{p}_{2}\cdot\vec{n} \vec{S}_{1}\cdot\vec{S}_{2} - 2 \vec{p}_{1}\cdot\vec{S}_{1} \vec{p}_{2}\cdot\vec{S}_{1} \vec{S}_{1}\cdot\vec{S}_{2} \nn\\ 
&& - 35 \vec{p}_{1}\cdot\vec{S}_{1} \vec{S}_{2}\cdot\vec{n} \vec{p}_{2}\cdot\vec{n} ( \vec{S}_{1}\cdot\vec{n})^{2} + 5 \vec{p}_{1}\cdot\vec{S}_{1} \vec{p}_{2}\cdot\vec{S}_{2} ( \vec{S}_{1}\cdot\vec{n})^{2} \Big]\nn\\ && +  	\frac{3 G^2 m_{2}}{m_{1}{}^2 r{}^6} \Big[ 4 \vec{S}_{1}\cdot\vec{n} S_{1}^2 \vec{S}_{2}\cdot\vec{n} - S_{1}^2 \vec{S}_{1}\cdot\vec{S}_{2} - 5 \vec{S}_{2}\cdot\vec{n} ( \vec{S}_{1}\cdot\vec{n})^{3} + 2 \vec{S}_{1}\cdot\vec{S}_{2} ( \vec{S}_{1}\cdot\vec{n})^{2} \Big],\nn\\
\eea
\bea
H^{\text{NLO}}_{(\text{ES}_1^2 ) \text{S}_1 \text{S}_2} &=& 	\frac{3 G}{2 m_{1}{}^4 r{}^5} \Big[ 3 S_{1}^2 \vec{S}_{1}\cdot\vec{S}_{2} p_{1}^2 - 15 \vec{S}_{1}\cdot\vec{n} S_{1}^2 p_{1}^2 \vec{S}_{2}\cdot\vec{n} + 5 \vec{p}_{1}\cdot\vec{S}_{1} S_{1}^2 \vec{p}_{1}\cdot\vec{n} \vec{S}_{2}\cdot\vec{n} \nn\\ 
&& - \vec{p}_{1}\cdot\vec{S}_{1} S_{1}^2 \vec{p}_{1}\cdot\vec{S}_{2} + 10 \vec{S}_{1}\cdot\vec{n} \vec{p}_{1}\cdot\vec{S}_{1} \vec{p}_{1}\cdot\vec{n} \vec{S}_{1}\cdot\vec{S}_{2} - 35 \vec{p}_{1}\cdot\vec{S}_{1} \vec{S}_{2}\cdot\vec{n} \vec{p}_{1}\cdot\vec{n} ( \vec{S}_{1}\cdot\vec{n})^{2} \nn\\ 
&& - 15 \vec{S}_{1}\cdot\vec{S}_{2} p_{1}^2 ( \vec{S}_{1}\cdot\vec{n})^{2} + 35 \vec{S}_{2}\cdot\vec{n} p_{1}^2 ( \vec{S}_{1}\cdot\vec{n})^{3} + 10 \vec{S}_{1}\cdot\vec{n} \vec{S}_{2}\cdot\vec{n} ( \vec{p}_{1}\cdot\vec{S}_{1})^{2} \nn\\ 
&& + 5 \vec{p}_{1}\cdot\vec{S}_{1} \vec{p}_{1}\cdot\vec{S}_{2} ( \vec{S}_{1}\cdot\vec{n})^{2} - 2 \vec{S}_{1}\cdot\vec{S}_{2} ( \vec{p}_{1}\cdot\vec{S}_{1})^{2} \Big] - 	\frac{3 G}{8 m_{1}{}^3 m_{2} r{}^5} \Big[ S_{1}^2 \vec{S}_{1}\cdot\vec{S}_{2} \big( 8 \vec{p}_{1}\cdot\vec{p}_{2} \nn\\ 
&& + 5 \vec{p}_{1}\cdot\vec{n} \vec{p}_{2}\cdot\vec{n} \big) - 45 \vec{S}_{1}\cdot\vec{n} S_{1}^2 \vec{p}_{1}\cdot\vec{p}_{2} \vec{S}_{2}\cdot\vec{n} + 15 \vec{p}_{2}\cdot\vec{S}_{1} S_{1}^2 \vec{p}_{1}\cdot\vec{n} \vec{S}_{2}\cdot\vec{n} \nn\\ 
&& + 30 \vec{S}_{1}\cdot\vec{n} \vec{p}_{1}\cdot\vec{S}_{1} \vec{p}_{2}\cdot\vec{S}_{1} \vec{S}_{2}\cdot\vec{n} - 15 \vec{S}_{1}\cdot\vec{n} S_{1}^2 \vec{p}_{2}\cdot\vec{n} \vec{p}_{1}\cdot\vec{S}_{2} \nn\\ 
&& + 10 \vec{S}_{1}\cdot\vec{n} \vec{p}_{1}\cdot\vec{S}_{1} \vec{p}_{2}\cdot\vec{n} \vec{S}_{1}\cdot\vec{S}_{2} + 40 \vec{S}_{1}\cdot\vec{n} \vec{p}_{2}\cdot\vec{S}_{1} \vec{p}_{1}\cdot\vec{n} \vec{S}_{1}\cdot\vec{S}_{2} - 8 \vec{p}_{1}\cdot\vec{S}_{1} \vec{p}_{2}\cdot\vec{S}_{1} \vec{S}_{1}\cdot\vec{S}_{2} \nn\\ 
&& - 105 \vec{p}_{2}\cdot\vec{S}_{1} \vec{S}_{2}\cdot\vec{n} \vec{p}_{1}\cdot\vec{n} ( \vec{S}_{1}\cdot\vec{n})^{2} - 5 \vec{S}_{1}\cdot\vec{S}_{2} \big( 8 \vec{p}_{1}\cdot\vec{p}_{2} + 7 \vec{p}_{1}\cdot\vec{n} \vec{p}_{2}\cdot\vec{n} \big) ( \vec{S}_{1}\cdot\vec{n})^{2} \nn\\ 
&& + 105 \vec{S}_{2}\cdot\vec{n} \vec{p}_{1}\cdot\vec{p}_{2} ( \vec{S}_{1}\cdot\vec{n})^{3} + 35 \vec{p}_{1}\cdot\vec{S}_{2} \vec{p}_{2}\cdot\vec{n} ( \vec{S}_{1}\cdot\vec{n})^{3} \Big]\nn\\ && - 	\frac{3 G^2 m_{2}}{m_{1}{}^2 r{}^6} \Big[ 4 \vec{S}_{1}\cdot\vec{n} S_{1}^2 \vec{S}_{2}\cdot\vec{n} - 2 S_{1}^2 \vec{S}_{1}\cdot\vec{S}_{2} - 5 \vec{S}_{2}\cdot\vec{n} ( \vec{S}_{1}\cdot\vec{n})^{3} + 5 \vec{S}_{1}\cdot\vec{S}_{2} ( \vec{S}_{1}\cdot\vec{n})^{2} \Big] \nn\\ 
&& - 	\frac{G^2}{2 m_{1} r{}^6} \Big[ 57 \vec{S}_{1}\cdot\vec{n} S_{1}^2 \vec{S}_{2}\cdot\vec{n} - 17 S_{1}^2 \vec{S}_{1}\cdot\vec{S}_{2} - 123 \vec{S}_{2}\cdot\vec{n} ( \vec{S}_{1}\cdot\vec{n})^{3} \nn\\ 
&& + 63 \vec{S}_{1}\cdot\vec{S}_{2} ( \vec{S}_{1}\cdot\vec{n})^{2} \Big],
\eea
\bea
H^{\text{NLO}}_{C_{\text{ES}^2_1}^2 \text{S}_1^3 \text{S}_2 } &=& 	\frac{9 G^2 m_{2}}{m_{1}{}^2 r{}^6} \Big[ \vec{S}_{2}\cdot\vec{n} ( \vec{S}_{1}\cdot\vec{n})^{3} - \vec{S}_{1}\cdot\vec{S}_{2} ( \vec{S}_{1}\cdot\vec{n})^{2} \Big] ,
\eea
\bea
H^{\text{NLO}}_{(\text{BS}_1^3) \text{S}_2 } &=&- 	\frac{G}{4 m_{1}{}^2 m_{2}{}^2 r{}^5} \Big[ S_{1}^2 \vec{S}_{1}\cdot\vec{S}_{2} \big( 7 p_{2}^2 -20 ( \vec{p}_{2}\cdot\vec{n})^{2} \big) - 15 \vec{S}_{1}\cdot\vec{n} S_{1}^2 p_{2}^2 \vec{S}_{2}\cdot\vec{n} \nn\\ 
&& + 20 \vec{p}_{2}\cdot\vec{S}_{1} S_{1}^2 \vec{p}_{2}\cdot\vec{n} \vec{S}_{2}\cdot\vec{n} + 15 \vec{S}_{1}\cdot\vec{n} S_{1}^2 \vec{p}_{2}\cdot\vec{n} \vec{p}_{2}\cdot\vec{S}_{2} - 7 \vec{p}_{2}\cdot\vec{S}_{1} S_{1}^2 \vec{p}_{2}\cdot\vec{S}_{2} \nn\\ 
&& - 40 \vec{S}_{1}\cdot\vec{n} \vec{p}_{2}\cdot\vec{S}_{1} \vec{p}_{2}\cdot\vec{n} \vec{S}_{1}\cdot\vec{S}_{2} - 140 \vec{p}_{2}\cdot\vec{S}_{1} \vec{S}_{2}\cdot\vec{n} \vec{p}_{2}\cdot\vec{n} ( \vec{S}_{1}\cdot\vec{n})^{2} \nn\\ 
&& - 35 \vec{S}_{1}\cdot\vec{S}_{2} \big( p_{2}^2 -4 ( \vec{p}_{2}\cdot\vec{n})^{2} \big) ( \vec{S}_{1}\cdot\vec{n})^{2} + 35 \vec{S}_{2}\cdot\vec{n} p_{2}^2 ( \vec{S}_{1}\cdot\vec{n})^{3} \nn\\ 
&& - 35 \vec{p}_{2}\cdot\vec{S}_{2} \vec{p}_{2}\cdot\vec{n} ( \vec{S}_{1}\cdot\vec{n})^{3} + 40 \vec{S}_{1}\cdot\vec{n} \vec{S}_{2}\cdot\vec{n} ( \vec{p}_{2}\cdot\vec{S}_{1})^{2} + 35 \vec{p}_{2}\cdot\vec{S}_{1} \vec{p}_{2}\cdot\vec{S}_{2} ( \vec{S}_{1}\cdot\vec{n})^{2} \Big] \nn\\ 
&& - 	\frac{G}{4 m_{1}{}^4 r{}^5} \Big[ S_{1}^2 \vec{S}_{1}\cdot\vec{S}_{2} \big( 13 p_{1}^2 -20 ( \vec{p}_{1}\cdot\vec{n})^{2} \big) - 45 \vec{S}_{1}\cdot\vec{n} S_{1}^2 p_{1}^2 \vec{S}_{2}\cdot\vec{n} \nn\\ 
&& + 25 \vec{p}_{1}\cdot\vec{S}_{1} S_{1}^2 \vec{p}_{1}\cdot\vec{n} \vec{S}_{2}\cdot\vec{n} + 60 \vec{S}_{1}\cdot\vec{n} S_{1}^2 \vec{p}_{1}\cdot\vec{n} \vec{p}_{1}\cdot\vec{S}_{2} - 17 \vec{p}_{1}\cdot\vec{S}_{1} S_{1}^2 \vec{p}_{1}\cdot\vec{S}_{2} \nn\\ 
&& - 30 \vec{S}_{1}\cdot\vec{n} \vec{p}_{1}\cdot\vec{S}_{1} \vec{p}_{1}\cdot\vec{n} \vec{S}_{1}\cdot\vec{S}_{2} - 175 \vec{p}_{1}\cdot\vec{S}_{1} \vec{S}_{2}\cdot\vec{n} \vec{p}_{1}\cdot\vec{n} ( \vec{S}_{1}\cdot\vec{n})^{2} \nn\\ 
&& - 5 \vec{S}_{1}\cdot\vec{S}_{2} \big( 13 p_{1}^2 -28 ( \vec{p}_{1}\cdot\vec{n})^{2} \big) ( \vec{S}_{1}\cdot\vec{n})^{2} + 105 \vec{S}_{2}\cdot\vec{n} p_{1}^2 ( \vec{S}_{1}\cdot\vec{n})^{3} \nn\\ 
&& - 140 \vec{p}_{1}\cdot\vec{S}_{2} \vec{p}_{1}\cdot\vec{n} ( \vec{S}_{1}\cdot\vec{n})^{3} + 50 \vec{S}_{1}\cdot\vec{n} \vec{S}_{2}\cdot\vec{n} ( \vec{p}_{1}\cdot\vec{S}_{1})^{2} + 85 \vec{p}_{1}\cdot\vec{S}_{1} \vec{p}_{1}\cdot\vec{S}_{2} ( \vec{S}_{1}\cdot\vec{n})^{2} \nn\\ 
&& - 2 \vec{S}_{1}\cdot\vec{S}_{2} ( \vec{p}_{1}\cdot\vec{S}_{1})^{2} \Big] + 	\frac{G}{4 m_{1}{}^3 m_{2} r{}^5} \Big[ S_{1}^2 \vec{S}_{1}\cdot\vec{S}_{2} \big( 23 \vec{p}_{1}\cdot\vec{p}_{2} -25 \vec{p}_{1}\cdot\vec{n} \vec{p}_{2}\cdot\vec{n} \big) \nn\\ 
&& - 15 \vec{S}_{1}\cdot\vec{n} S_{1}^2 \big( 5 \vec{p}_{1}\cdot\vec{p}_{2} + 7 \vec{p}_{1}\cdot\vec{n} \vec{p}_{2}\cdot\vec{n} \big) \vec{S}_{2}\cdot\vec{n} + 25 \vec{p}_{1}\cdot\vec{S}_{1} S_{1}^2 \vec{p}_{2}\cdot\vec{n} \vec{S}_{2}\cdot\vec{n} \nn\\ 
&& + 35 \vec{p}_{2}\cdot\vec{S}_{1} S_{1}^2 \vec{p}_{1}\cdot\vec{n} \vec{S}_{2}\cdot\vec{n} + 90 \vec{S}_{1}\cdot\vec{n} \vec{p}_{1}\cdot\vec{S}_{1} \vec{p}_{2}\cdot\vec{S}_{1} \vec{S}_{2}\cdot\vec{n} + 45 \vec{S}_{1}\cdot\vec{n} S_{1}^2 \vec{p}_{2}\cdot\vec{n} \vec{p}_{1}\cdot\vec{S}_{2} \nn\\ 
&& - 13 \vec{p}_{2}\cdot\vec{S}_{1} S_{1}^2 \vec{p}_{1}\cdot\vec{S}_{2} + 45 \vec{S}_{1}\cdot\vec{n} S_{1}^2 \vec{p}_{1}\cdot\vec{n} \vec{p}_{2}\cdot\vec{S}_{2} - 11 \vec{p}_{1}\cdot\vec{S}_{1} S_{1}^2 \vec{p}_{2}\cdot\vec{S}_{2} \nn\\ 
&& - 30 \vec{S}_{1}\cdot\vec{n} \vec{p}_{1}\cdot\vec{S}_{1} \vec{p}_{2}\cdot\vec{n} \vec{S}_{1}\cdot\vec{S}_{2} - 10 \vec{S}_{1}\cdot\vec{n} \vec{p}_{2}\cdot\vec{S}_{1} \vec{p}_{1}\cdot\vec{n} \vec{S}_{1}\cdot\vec{S}_{2} - 2 \vec{p}_{1}\cdot\vec{S}_{1} \vec{p}_{2}\cdot\vec{S}_{1} \vec{S}_{1}\cdot\vec{S}_{2} \nn\\ 
&& - 175 \vec{p}_{1}\cdot\vec{S}_{1} \vec{S}_{2}\cdot\vec{n} \vec{p}_{2}\cdot\vec{n} ( \vec{S}_{1}\cdot\vec{n})^{2} - 245 \vec{p}_{2}\cdot\vec{S}_{1} \vec{S}_{2}\cdot\vec{n} \vec{p}_{1}\cdot\vec{n} ( \vec{S}_{1}\cdot\vec{n})^{2} \nn\\ 
&& - 5 \vec{S}_{1}\cdot\vec{S}_{2} \big( 23 \vec{p}_{1}\cdot\vec{p}_{2} -35 \vec{p}_{1}\cdot\vec{n} \vec{p}_{2}\cdot\vec{n} \big) ( \vec{S}_{1}\cdot\vec{n})^{2} + 35 \vec{S}_{2}\cdot\vec{n} \big( 5 \vec{p}_{1}\cdot\vec{p}_{2} \nn\\ 
&& + 9 \vec{p}_{1}\cdot\vec{n} \vec{p}_{2}\cdot\vec{n} \big) ( \vec{S}_{1}\cdot\vec{n})^{3} - 105 \vec{p}_{1}\cdot\vec{S}_{2} \vec{p}_{2}\cdot\vec{n} ( \vec{S}_{1}\cdot\vec{n})^{3} - 105 \vec{p}_{2}\cdot\vec{S}_{2} \vec{p}_{1}\cdot\vec{n} ( \vec{S}_{1}\cdot\vec{n})^{3} \nn\\ 
&& + 65 \vec{p}_{2}\cdot\vec{S}_{1} \vec{p}_{1}\cdot\vec{S}_{2} ( \vec{S}_{1}\cdot\vec{n})^{2} + 55 \vec{p}_{1}\cdot\vec{S}_{1} \vec{p}_{2}\cdot\vec{S}_{2} ( \vec{S}_{1}\cdot\vec{n})^{2} \Big]\nn\\ && - 	\frac{G^2}{2 m_{1} r{}^6} \Big[ 41 \vec{S}_{1}\cdot\vec{n} S_{1}^2 \vec{S}_{2}\cdot\vec{n} - 11 S_{1}^2 \vec{S}_{1}\cdot\vec{S}_{2} - 105 \vec{S}_{2}\cdot\vec{n} ( \vec{S}_{1}\cdot\vec{n})^{3} \nn\\ 
&& + 55 \vec{S}_{1}\cdot\vec{S}_{2} ( \vec{S}_{1}\cdot\vec{n})^{2} \Big] - 	\frac{G^2 m_{2}}{2 m_{1}{}^2 r{}^6} \Big[ 109 \vec{S}_{1}\cdot\vec{n} S_{1}^2 \vec{S}_{2}\cdot\vec{n} - 23 S_{1}^2 \vec{S}_{1}\cdot\vec{S}_{2} \nn\\ 
&& - 260 \vec{S}_{2}\cdot\vec{n} ( \vec{S}_{1}\cdot\vec{n})^{3} + 116 \vec{S}_{1}\cdot\vec{S}_{2} ( \vec{S}_{1}\cdot\vec{n})^{2} \Big],
\eea
\bea 
H^{\text{NLO}}_{\text{S}_1^2 \text{S}_2^2}
&=& - 	\frac{3 G}{8 m_{1}{}^2 m_{2}{}^2 r{}^5} \Big[ S_{1}^2 S_{2}^2 \big( 9 \vec{p}_{1}\cdot\vec{p}_{2} -10 \vec{p}_{1}\cdot\vec{n} \vec{p}_{2}\cdot\vec{n} \big) + 10 S_{1}^2 \vec{S}_{2}\cdot\vec{n} \vec{p}_{2}\cdot\vec{n} \vec{p}_{1}\cdot\vec{S}_{2} \nn\\ 
&& - 30 \vec{S}_{1}\cdot\vec{n} \vec{p}_{2}\cdot\vec{S}_{1} \vec{S}_{2}\cdot\vec{n} \vec{p}_{1}\cdot\vec{S}_{2} + 20 S_{1}^2 \vec{S}_{2}\cdot\vec{n} \vec{p}_{1}\cdot\vec{n} \vec{p}_{2}\cdot\vec{S}_{2} \nn\\ 
&& - 10 \vec{S}_{1}\cdot\vec{n} \vec{p}_{1}\cdot\vec{S}_{1} \vec{S}_{2}\cdot\vec{n} \vec{p}_{2}\cdot\vec{S}_{2} - 18 S_{1}^2 \vec{p}_{1}\cdot\vec{S}_{2} \vec{p}_{2}\cdot\vec{S}_{2} \nn\\ 
&& + 5 \vec{S}_{1}\cdot\vec{n} \vec{S}_{2}\cdot\vec{n} \big( 9 \vec{p}_{1}\cdot\vec{p}_{2} -7 \vec{p}_{1}\cdot\vec{n} \vec{p}_{2}\cdot\vec{n} \big) \vec{S}_{1}\cdot\vec{S}_{2} - 10 \vec{p}_{1}\cdot\vec{S}_{1} \vec{S}_{2}\cdot\vec{n} \vec{p}_{2}\cdot\vec{n} \vec{S}_{1}\cdot\vec{S}_{2} \nn\\ 
&& - 30 \vec{p}_{2}\cdot\vec{S}_{1} \vec{S}_{2}\cdot\vec{n} \vec{p}_{1}\cdot\vec{n} \vec{S}_{1}\cdot\vec{S}_{2} + 13 \vec{p}_{2}\cdot\vec{S}_{1} \vec{p}_{1}\cdot\vec{S}_{2} \vec{S}_{1}\cdot\vec{S}_{2} + 9 \vec{p}_{1}\cdot\vec{S}_{1} \vec{p}_{2}\cdot\vec{S}_{2} \vec{S}_{1}\cdot\vec{S}_{2} \nn\\ 
&& + 70 \vec{S}_{1}\cdot\vec{n} \vec{p}_{2}\cdot\vec{S}_{1} \vec{p}_{1}\cdot\vec{n} ( \vec{S}_{2}\cdot\vec{n})^{2} - 10 S_{1}^2 \vec{p}_{1}\cdot\vec{p}_{2} ( \vec{S}_{2}\cdot\vec{n})^{2} - 35 \vec{p}_{1}\cdot\vec{p}_{2} ( \vec{S}_{1}\cdot\vec{n})^{2} ( \vec{S}_{2}\cdot\vec{n})^{2} \nn\\ 
&& + 10 \vec{p}_{1}\cdot\vec{S}_{1} \vec{p}_{2}\cdot\vec{S}_{1} ( \vec{S}_{2}\cdot\vec{n})^{2} - \big( 13 \vec{p}_{1}\cdot\vec{p}_{2} -15 \vec{p}_{1}\cdot\vec{n} \vec{p}_{2}\cdot\vec{n} \big) ( \vec{S}_{1}\cdot\vec{S}_{2})^{2} \Big]\nn\\ && - 	\frac{3 G^2}{m_{1} r{}^6} \Big[ 5 \vec{S}_{1}\cdot\vec{n} \vec{S}_{2}\cdot\vec{n} \vec{S}_{1}\cdot\vec{S}_{2} - 5 ( \vec{S}_{1}\cdot\vec{n})^{2} ( \vec{S}_{2}\cdot\vec{n})^{2} - ( \vec{S}_{1}\cdot\vec{S}_{2})^{2} \Big] \nn\\ 
&& - 	\frac{3 G^2}{m_{2} r{}^6} S_{1}^2 ( \vec{S}_{2}\cdot\vec{n})^{2} ,
\eea
\bea
H^{\text{NLO}}_{(\text{ES}_1^2 )  \text{S}_2^2} 
&=& - 	\frac{3 G}{4 m_{1}{}^2 m_{2}{}^2 r{}^5} \Big[ S_{1}^2 S_{2}^2 \big( 23 \vec{p}_{1}\cdot\vec{p}_{2} -25 \vec{p}_{1}\cdot\vec{n} \vec{p}_{2}\cdot\vec{n} \big) + 30 S_{1}^2 \vec{S}_{2}\cdot\vec{n} \vec{p}_{2}\cdot\vec{n} \vec{p}_{1}\cdot\vec{S}_{2} \nn\\ 
&& + 25 S_{1}^2 \vec{S}_{2}\cdot\vec{n} \vec{p}_{1}\cdot\vec{n} \vec{p}_{2}\cdot\vec{S}_{2} - 10 \vec{S}_{1}\cdot\vec{n} \vec{p}_{1}\cdot\vec{S}_{1} \vec{S}_{2}\cdot\vec{n} \vec{p}_{2}\cdot\vec{S}_{2} - 23 S_{1}^2 \vec{p}_{1}\cdot\vec{S}_{2} \vec{p}_{2}\cdot\vec{S}_{2} \nn\\ 
&& - 20 \vec{p}_{1}\cdot\vec{S}_{1} \vec{S}_{2}\cdot\vec{n} \vec{p}_{2}\cdot\vec{n} \vec{S}_{1}\cdot\vec{S}_{2} - 20 \vec{p}_{2}\cdot\vec{S}_{1} \vec{S}_{2}\cdot\vec{n} \vec{p}_{1}\cdot\vec{n} \vec{S}_{1}\cdot\vec{S}_{2} \nn\\ 
&& - 10 \vec{S}_{1}\cdot\vec{n} \vec{p}_{2}\cdot\vec{S}_{2} \vec{p}_{1}\cdot\vec{n} \vec{S}_{1}\cdot\vec{S}_{2} + 16 \vec{p}_{2}\cdot\vec{S}_{1} \vec{p}_{1}\cdot\vec{S}_{2} \vec{S}_{1}\cdot\vec{S}_{2} + 18 \vec{p}_{1}\cdot\vec{S}_{1} \vec{p}_{2}\cdot\vec{S}_{2} \vec{S}_{1}\cdot\vec{S}_{2} \nn\\ 
&& + 10 \vec{S}_{1}\cdot\vec{n} \vec{p}_{1}\cdot\vec{S}_{1} \vec{p}_{2}\cdot\vec{n} S_{2}^2 + 10 \vec{S}_{1}\cdot\vec{n} \vec{p}_{2}\cdot\vec{S}_{1} \vec{p}_{1}\cdot\vec{n} S_{2}^2 - 18 \vec{p}_{1}\cdot\vec{S}_{1} \vec{p}_{2}\cdot\vec{S}_{1} S_{2}^2 \nn\\ 
&& - 70 \vec{S}_{2}\cdot\vec{n} \vec{p}_{1}\cdot\vec{S}_{2} \vec{p}_{2}\cdot\vec{n} ( \vec{S}_{1}\cdot\vec{n})^{2} - 35 \vec{S}_{2}\cdot\vec{n} \vec{p}_{2}\cdot\vec{S}_{2} \vec{p}_{1}\cdot\vec{n} ( \vec{S}_{1}\cdot\vec{n})^{2} - 35 S_{2}^2 \big( \vec{p}_{1}\cdot\vec{p}_{2} \nn\\ 
&& - \vec{p}_{1}\cdot\vec{n} \vec{p}_{2}\cdot\vec{n} \big) ( \vec{S}_{1}\cdot\vec{n})^{2} + 35 \vec{p}_{1}\cdot\vec{S}_{2} \vec{p}_{2}\cdot\vec{S}_{2} ( \vec{S}_{1}\cdot\vec{n})^{2} - 30 S_{1}^2 \vec{p}_{1}\cdot\vec{p}_{2} ( \vec{S}_{2}\cdot\vec{n})^{2} \nn\\ 
&& + 70 \vec{p}_{1}\cdot\vec{p}_{2} ( \vec{S}_{1}\cdot\vec{n})^{2} ( \vec{S}_{2}\cdot\vec{n})^{2} + 20 \vec{p}_{1}\cdot\vec{S}_{1} \vec{p}_{2}\cdot\vec{S}_{1} ( \vec{S}_{2}\cdot\vec{n})^{2} \nn\\ 
&& - 4 \big( 4 \vec{p}_{1}\cdot\vec{p}_{2} -5 \vec{p}_{1}\cdot\vec{n} \vec{p}_{2}\cdot\vec{n} \big) ( \vec{S}_{1}\cdot\vec{S}_{2})^{2} \Big] + 	\frac{3 G}{16 m_{1} m_{2}{}^3 r{}^5} \Big[ 5 S_{1}^2 S_{2}^2 \big( 16 p_{2}^2 -17 ( \vec{p}_{2}\cdot\vec{n})^{2} \big) \nn\\ 
&& + 190 S_{1}^2 \vec{S}_{2}\cdot\vec{n} \vec{p}_{2}\cdot\vec{n} \vec{p}_{2}\cdot\vec{S}_{2} - 40 \vec{S}_{1}\cdot\vec{n} \vec{p}_{2}\cdot\vec{S}_{1} \vec{S}_{2}\cdot\vec{n} \vec{p}_{2}\cdot\vec{S}_{2} \nn\\ 
&& - 140 \vec{p}_{2}\cdot\vec{S}_{1} \vec{S}_{2}\cdot\vec{n} \vec{p}_{2}\cdot\vec{n} \vec{S}_{1}\cdot\vec{S}_{2} - 40 \vec{S}_{1}\cdot\vec{n} \vec{p}_{2}\cdot\vec{S}_{2} \vec{p}_{2}\cdot\vec{n} \vec{S}_{1}\cdot\vec{S}_{2} \nn\\ 
&& + 120 \vec{p}_{2}\cdot\vec{S}_{1} \vec{p}_{2}\cdot\vec{S}_{2} \vec{S}_{1}\cdot\vec{S}_{2} + 80 \vec{S}_{1}\cdot\vec{n} \vec{p}_{2}\cdot\vec{S}_{1} \vec{p}_{2}\cdot\vec{n} S_{2}^2 \nn\\ 
&& - 350 \vec{S}_{2}\cdot\vec{n} \vec{p}_{2}\cdot\vec{S}_{2} \vec{p}_{2}\cdot\vec{n} ( \vec{S}_{1}\cdot\vec{n})^{2} - 15 S_{2}^2 \big( 8 p_{2}^2 -7 ( \vec{p}_{2}\cdot\vec{n})^{2} \big) ( \vec{S}_{1}\cdot\vec{n})^{2} \nn\\ 
&& - 64 S_{2}^2 ( \vec{p}_{2}\cdot\vec{S}_{1})^{2} - 105 S_{1}^2 p_{2}^2 ( \vec{S}_{2}\cdot\vec{n})^{2} + 245 p_{2}^2 ( \vec{S}_{1}\cdot\vec{n})^{2} ( \vec{S}_{2}\cdot\vec{n})^{2} \nn\\ 
&& + 70 ( \vec{p}_{2}\cdot\vec{S}_{1})^{2} ( \vec{S}_{2}\cdot\vec{n})^{2} + 120 ( \vec{S}_{1}\cdot\vec{n})^{2} ( \vec{p}_{2}\cdot\vec{S}_{2})^{2} - 80 S_{1}^2 ( \vec{p}_{2}\cdot\vec{S}_{2})^{2} \nn\\ 
&& - 14 \big( 4 p_{2}^2 -5 ( \vec{p}_{2}\cdot\vec{n})^{2} \big) ( \vec{S}_{1}\cdot\vec{S}_{2})^{2} \Big]\nn\\ && - 	\frac{3 G^2}{m_{2} r{}^6} \Big[ 2 \vec{S}_{1}\cdot\vec{n} \vec{S}_{2}\cdot\vec{n} \vec{S}_{1}\cdot\vec{S}_{2} - S_{1}^2 S_{2}^2 + 3 S_{2}^2 ( \vec{S}_{1}\cdot\vec{n})^{2} - 5 ( \vec{S}_{1}\cdot\vec{n})^{2} ( \vec{S}_{2}\cdot\vec{n})^{2} \nn\\ 
&& + S_{1}^2 ( \vec{S}_{2}\cdot\vec{n})^{2} \Big] - 	\frac{G^2}{2 m_{1} r{}^6} \Big[ 54 \vec{S}_{1}\cdot\vec{n} \vec{S}_{2}\cdot\vec{n} \vec{S}_{1}\cdot\vec{S}_{2} - 17 S_{1}^2 S_{2}^2 + 48 S_{2}^2 ( \vec{S}_{1}\cdot\vec{n})^{2} \nn\\ 
&& - 135 ( \vec{S}_{1}\cdot\vec{n})^{2} ( \vec{S}_{2}\cdot\vec{n})^{2} + 27 S_{1}^2 ( \vec{S}_{2}\cdot\vec{n})^{2} + 3 ( \vec{S}_{1}\cdot\vec{S}_{2})^{2} \Big],
\eea
\bea
H^{\text{NLO}}_{(\text{ES}_1^2 )(\text{ES}_2^2 ) }&=& - 	\frac{3 G}{8 m_{1} m_{2}{}^3 r{}^5} \Big[ 10 S_{1}^2 \vec{S}_{2}\cdot\vec{n} \vec{p}_{2}\cdot\vec{n} \vec{p}_{2}\cdot\vec{S}_{2} - 15 S_{1}^2 p_{2}^2 ( \vec{S}_{2}\cdot\vec{n})^{2} - 2 S_{1}^2 ( \vec{p}_{2}\cdot\vec{S}_{2})^{2} \Big] \nn\\ 
&& + 	\frac{3 G}{16 m_{1}{}^2 m_{2}{}^2 r{}^5} \Big[ S_{1}^2 S_{2}^2 \big( 11 \vec{p}_{1}\cdot\vec{p}_{2} -15 \vec{p}_{1}\cdot\vec{n} \vec{p}_{2}\cdot\vec{n} \big) - 20 S_{1}^2 \vec{S}_{2}\cdot\vec{n} \vec{p}_{2}\cdot\vec{n} \vec{p}_{1}\cdot\vec{S}_{2} \nn\\ 
&& + 20 \vec{S}_{1}\cdot\vec{n} \vec{p}_{2}\cdot\vec{S}_{1} \vec{S}_{2}\cdot\vec{n} \vec{p}_{1}\cdot\vec{S}_{2} - 20 S_{1}^2 \vec{S}_{2}\cdot\vec{n} \vec{p}_{1}\cdot\vec{n} \vec{p}_{2}\cdot\vec{S}_{2} \nn\\ 
&& + 20 \vec{S}_{1}\cdot\vec{n} \vec{p}_{1}\cdot\vec{S}_{1} \vec{S}_{2}\cdot\vec{n} \vec{p}_{2}\cdot\vec{S}_{2} + 4 S_{1}^2 \vec{p}_{1}\cdot\vec{S}_{2} \vec{p}_{2}\cdot\vec{S}_{2} - 140 \vec{S}_{1}\cdot\vec{n} \vec{S}_{2}\cdot\vec{n} \big( \vec{p}_{1}\cdot\vec{p}_{2} \nn\\ 
&& + \vec{p}_{1}\cdot\vec{n} \vec{p}_{2}\cdot\vec{n} \big) \vec{S}_{1}\cdot\vec{S}_{2} + 40 \vec{p}_{1}\cdot\vec{S}_{1} \vec{S}_{2}\cdot\vec{n} \vec{p}_{2}\cdot\vec{n} \vec{S}_{1}\cdot\vec{S}_{2} + 40 \vec{p}_{2}\cdot\vec{S}_{1} \vec{S}_{2}\cdot\vec{n} \vec{p}_{1}\cdot\vec{n} \vec{S}_{1}\cdot\vec{S}_{2} \nn\\ 
&& - 4 \vec{p}_{2}\cdot\vec{S}_{1} \vec{p}_{1}\cdot\vec{S}_{2} \vec{S}_{1}\cdot\vec{S}_{2} - 4 \vec{p}_{1}\cdot\vec{S}_{1} \vec{p}_{2}\cdot\vec{S}_{2} \vec{S}_{1}\cdot\vec{S}_{2} - 140 \vec{S}_{1}\cdot\vec{n} \vec{p}_{1}\cdot\vec{S}_{1} \vec{p}_{2}\cdot\vec{n} ( \vec{S}_{2}\cdot\vec{n})^{2} \nn\\ 
&& - 140 \vec{S}_{1}\cdot\vec{n} \vec{p}_{2}\cdot\vec{S}_{1} \vec{p}_{1}\cdot\vec{n} ( \vec{S}_{2}\cdot\vec{n})^{2} - 10 S_{1}^2 \big( 9 \vec{p}_{1}\cdot\vec{p}_{2} -7 \vec{p}_{1}\cdot\vec{n} \vec{p}_{2}\cdot\vec{n} \big) ( \vec{S}_{2}\cdot\vec{n})^{2} \nn\\ 
&& + 35 \big( 7 \vec{p}_{1}\cdot\vec{p}_{2} + 9 \vec{p}_{1}\cdot\vec{n} \vec{p}_{2}\cdot\vec{n} \big) ( \vec{S}_{1}\cdot\vec{n})^{2} ( \vec{S}_{2}\cdot\vec{n})^{2} + 20 \vec{p}_{1}\cdot\vec{S}_{1} \vec{p}_{2}\cdot\vec{S}_{1} ( \vec{S}_{2}\cdot\vec{n})^{2} \nn\\ 
&& + 2 \big( 7 \vec{p}_{1}\cdot\vec{p}_{2} + 5 \vec{p}_{1}\cdot\vec{n} \vec{p}_{2}\cdot\vec{n} \big) ( \vec{S}_{1}\cdot\vec{S}_{2})^{2} \Big] - 	\frac{3 G}{8 m_{1}{}^3 m_{2} r{}^5} \Big[ 5 S_{1}^2 S_{2}^2 \big( p_{1}^2 -2 ( \vec{p}_{1}\cdot\vec{n})^{2} \big) \nn\\ 
&& - 40 S_{1}^2 \vec{S}_{2}\cdot\vec{n} \vec{p}_{1}\cdot\vec{n} \vec{p}_{1}\cdot\vec{S}_{2} + 20 \vec{S}_{1}\cdot\vec{n} \vec{p}_{1}\cdot\vec{S}_{1} \vec{S}_{2}\cdot\vec{n} \vec{p}_{1}\cdot\vec{S}_{2} - 60 \vec{S}_{1}\cdot\vec{n} \vec{S}_{2}\cdot\vec{n} p_{1}^2 \vec{S}_{1}\cdot\vec{S}_{2} \nn\\ 
&& + 20 \vec{p}_{1}\cdot\vec{S}_{1} \vec{S}_{2}\cdot\vec{n} \vec{p}_{1}\cdot\vec{n} \vec{S}_{1}\cdot\vec{S}_{2} - 4 \vec{p}_{1}\cdot\vec{S}_{1} \vec{p}_{1}\cdot\vec{S}_{2} \vec{S}_{1}\cdot\vec{S}_{2} \nn\\ 
&& - 70 \vec{S}_{1}\cdot\vec{n} \vec{p}_{1}\cdot\vec{S}_{1} \vec{p}_{1}\cdot\vec{n} ( \vec{S}_{2}\cdot\vec{n})^{2} - 5 S_{1}^2 \big( 5 p_{1}^2 -14 ( \vec{p}_{1}\cdot\vec{n})^{2} \big) ( \vec{S}_{2}\cdot\vec{n})^{2} \nn\\ 
&& + 105 p_{1}^2 ( \vec{S}_{1}\cdot\vec{n})^{2} ( \vec{S}_{2}\cdot\vec{n})^{2} + 10 ( \vec{p}_{1}\cdot\vec{S}_{1})^{2} ( \vec{S}_{2}\cdot\vec{n})^{2} + 4 S_{1}^2 ( \vec{p}_{1}\cdot\vec{S}_{2})^{2} + 6 p_{1}^2 ( \vec{S}_{1}\cdot\vec{S}_{2})^{2} \Big]\nn\\ && - 	\frac{3 G^2}{2 m_{1} r{}^6} \Big[ 32 \vec{S}_{1}\cdot\vec{n} \vec{S}_{2}\cdot\vec{n} \vec{S}_{1}\cdot\vec{S}_{2} - 3 S_{1}^2 S_{2}^2 - 62 ( \vec{S}_{1}\cdot\vec{n})^{2} ( \vec{S}_{2}\cdot\vec{n})^{2} + 12 S_{1}^2 ( \vec{S}_{2}\cdot\vec{n})^{2} \nn\\ 
&& - 3 ( \vec{S}_{1}\cdot\vec{S}_{2})^{2} \Big] - 	\frac{15 G^2}{m_{2} r{}^6} S_{1}^2 ( \vec{S}_{2}\cdot\vec{n})^{2} .
\eea

\section{COM Generator of the N$^3$LO Quadratic-in-Spin Sectors} 
\label{comgenn3los2}

As noted in section \ref{poincares2n3lo}, we write the solution of the COM generator of the N$^3$LO 
quadratic-in-spin sectors, $\vec{G}^{\text{N}^3\text{LO}}_{\text{S}^2}$, as:
\bea
\vec{G}^{\text{N}^3\text{LO}}_{\text{S}^2} &=& H^{\text{N}^2\text{LO}}_{\text{S}^2} \frac{ \vec{x}_1 + \vec{x}_2}{2} +  \left( \vec{Y}^{\text{N}^3\text{LO}}_{\text{S}_1^2 }  + C_{1\text{ES}^2} \vec{Y}^{\text{N}^3\text{LO}}_{\text{ES}_1^2} +  \vec{Y}^{\text{N}^3\text{LO}}_{\text{S}_1 \text{S}_2 } + \left(1 \leftrightarrow 2 \right) \right),
\eea
with:
\bea
\vec{Y}^{\text{N}^3\text{LO}}_{\text{S}_1^2 } &=&  - 	\frac{G m_{2}}{32 m_{1}{}^5 r{}^2} \Big[ \vec{S}_{1}\cdot\vec{n} \vec{p}_{1}\cdot\vec{S}_{1} \big( p_{1}^2 \vec{p}_{1} -9 \vec{p}_{1}\cdot\vec{n} p_{1}^2 \vec{n} \big) - 5 S_{1}^2 \big( 3 p_{1}^{4} \vec{n} \nn\\ 
&& + 4 \vec{p}_{1}\cdot\vec{n} p_{1}^2 \vec{p}_{1} -12 p_{1}^2 ( \vec{p}_{1}\cdot\vec{n})^{2} \vec{n} \big) - 2 \vec{S}_{1}\cdot\vec{n} \vec{S}_{1} p_{1}^{4} - 33 \vec{p}_{1}\cdot\vec{S}_{1} \vec{S}_{1} \vec{p}_{1}\cdot\vec{n} p_{1}^2 \nn\\ 
&& - 15 p_{1}^{4} \vec{n} ( \vec{S}_{1}\cdot\vec{n})^{2} + 3 \big( 5 p_{1}^2 \vec{n} + 18 \vec{p}_{1}\cdot\vec{n} \vec{p}_{1} -12 ( \vec{p}_{1}\cdot\vec{n})^{2} \vec{n} \big) ( \vec{p}_{1}\cdot\vec{S}_{1})^{2} \Big] \nn\\ 
&& - 	\frac{G}{32 m_{1}{}^4 r{}^2} \Big[ \vec{S}_{1}\cdot\vec{n} \vec{p}_{1}\cdot\vec{S}_{1} \big( 18 \vec{p}_{1}\cdot\vec{p}_{2} \vec{p}_{1} + 14 p_{1}^2 \vec{p}_{2} + 9 p_{1}^2 \vec{p}_{2}\cdot\vec{n} \vec{n} + 69 \vec{p}_{1}\cdot\vec{n} \vec{p}_{1}\cdot\vec{p}_{2} \vec{n} \nn\\ 
&& - 57 \vec{p}_{1}\cdot\vec{n} \vec{p}_{2}\cdot\vec{n} \vec{p}_{1} - 12 ( \vec{p}_{1}\cdot\vec{n})^{2} \vec{p}_{2} + 60 \vec{p}_{2}\cdot\vec{n} ( \vec{p}_{1}\cdot\vec{n})^{2} \vec{n} \big) \nn\\ 
&& - \vec{S}_{1}\cdot\vec{n} \vec{p}_{2}\cdot\vec{S}_{1} \big( 38 p_{1}^2 \vec{p}_{1} -63 \vec{p}_{1}\cdot\vec{n} p_{1}^2 \vec{n} + 3 ( \vec{p}_{1}\cdot\vec{n})^{2} \vec{p}_{1} \big) - \vec{p}_{1}\cdot\vec{S}_{1} \vec{p}_{2}\cdot\vec{S}_{1} \big( 11 p_{1}^2 \vec{n} \nn\\ 
&& + 99 \vec{p}_{1}\cdot\vec{n} \vec{p}_{1} + 15 ( \vec{p}_{1}\cdot\vec{n})^{2} \vec{n} \big) + S_{1}^2 \big( 99 p_{1}^2 \vec{p}_{1}\cdot\vec{p}_{2} \vec{n} + 46 p_{1}^2 \vec{p}_{2}\cdot\vec{n} \vec{p}_{1} + 11 \vec{p}_{1}\cdot\vec{n} \vec{p}_{1}\cdot\vec{p}_{2} \vec{p}_{1} \nn\\ 
&& - 9 \vec{p}_{1}\cdot\vec{n} p_{1}^2 \vec{p}_{2} -168 \vec{p}_{1}\cdot\vec{n} p_{1}^2 \vec{p}_{2}\cdot\vec{n} \vec{n} - 57 \vec{p}_{1}\cdot\vec{p}_{2} ( \vec{p}_{1}\cdot\vec{n})^{2} \vec{n} + 102 \vec{p}_{2}\cdot\vec{n} ( \vec{p}_{1}\cdot\vec{n})^{2} \vec{p}_{1} \nn\\ 
&& + 12 ( \vec{p}_{1}\cdot\vec{n})^{3} \vec{p}_{2} -60 \vec{p}_{2}\cdot\vec{n} ( \vec{p}_{1}\cdot\vec{n})^{3} \vec{n} \big) + \vec{S}_{1}\cdot\vec{n} \vec{S}_{1} \big( 10 p_{1}^2 \vec{p}_{1}\cdot\vec{p}_{2} + 9 \vec{p}_{1}\cdot\vec{n} p_{1}^2 \vec{p}_{2}\cdot\vec{n} \nn\\ 
&& - 3 \vec{p}_{1}\cdot\vec{p}_{2} ( \vec{p}_{1}\cdot\vec{n})^{2} \big) + \vec{p}_{1}\cdot\vec{S}_{1} \vec{S}_{1} \big( 22 p_{1}^2 \vec{p}_{2}\cdot\vec{n} + 67 \vec{p}_{1}\cdot\vec{n} \vec{p}_{1}\cdot\vec{p}_{2} -63 \vec{p}_{2}\cdot\vec{n} ( \vec{p}_{1}\cdot\vec{n})^{2} \big) \nn\\ 
&& + 3 \vec{p}_{2}\cdot\vec{S}_{1} \vec{S}_{1} \big( 5 \vec{p}_{1}\cdot\vec{n} p_{1}^2 + ( \vec{p}_{1}\cdot\vec{n})^{3} \big) - 3 \big( 20 p_{1}^2 \vec{p}_{1}\cdot\vec{p}_{2} \vec{n} + 5 p_{1}^2 \vec{p}_{2}\cdot\vec{n} \vec{p}_{1} \nn\\ 
&& - \vec{p}_{1}\cdot\vec{n} \vec{p}_{1}\cdot\vec{p}_{2} \vec{p}_{1} \big) ( \vec{S}_{1}\cdot\vec{n})^{2} - \big( 92 \vec{p}_{1}\cdot\vec{p}_{2} \vec{n} + 68 \vec{p}_{2}\cdot\vec{n} \vec{p}_{1} \nn\\ 
&& - 15 \vec{p}_{1}\cdot\vec{n} \vec{p}_{2} -183 \vec{p}_{1}\cdot\vec{n} \vec{p}_{2}\cdot\vec{n} \vec{n} \big) ( \vec{p}_{1}\cdot\vec{S}_{1})^{2} \Big] - 	\frac{G}{16 m_{1} m_{2}{}^3 r{}^2} \Big[ 2 \vec{S}_{1}\cdot\vec{n} \vec{p}_{2}\cdot\vec{S}_{1} \big( p_{2}^2 \vec{p}_{2} \nn\\ 
&& + 3 \vec{p}_{2}\cdot\vec{n} p_{2}^2 \vec{n} + 3 ( \vec{p}_{2}\cdot\vec{n})^{2} \vec{p}_{2} -15 ( \vec{p}_{2}\cdot\vec{n})^{3} \vec{n} \big) + S_{1}^2 \big( - p_{2}^{4} \vec{n} + 38 \vec{p}_{2}\cdot\vec{n} p_{2}^2 \vec{p}_{2} \nn\\ 
&& + 18 p_{2}^2 ( \vec{p}_{2}\cdot\vec{n})^{2} \vec{n} - 54 ( \vec{p}_{2}\cdot\vec{n})^{3} \vec{p}_{2} \big) + \vec{S}_{1}\cdot\vec{n} \vec{S}_{1} \big( p_{2}^{4} -18 p_{2}^2 ( \vec{p}_{2}\cdot\vec{n})^{2} \big) \nn\\ 
&& - 2 \vec{p}_{2}\cdot\vec{S}_{1} \vec{S}_{1} \big( 19 \vec{p}_{2}\cdot\vec{n} p_{2}^2 -27 ( \vec{p}_{2}\cdot\vec{n})^{3} \big) - 6 \big( \vec{p}_{2}\cdot\vec{n} p_{2}^2 \vec{p}_{2} -5 ( \vec{p}_{2}\cdot\vec{n})^{3} \vec{p}_{2} \big) ( \vec{S}_{1}\cdot\vec{n})^{2} \nn\\ 
&& - 2 \big( p_{2}^2 \vec{n} + 3 ( \vec{p}_{2}\cdot\vec{n})^{2} \vec{n} \big) ( \vec{p}_{2}\cdot\vec{S}_{1})^{2} \Big] + 	\frac{G}{32 m_{1}{}^3 m_{2} r{}^2} \Big[ \vec{S}_{1}\cdot\vec{n} \vec{p}_{1}\cdot\vec{S}_{1} \big( 23 p_{2}^2 \vec{p}_{1} \nn\\ 
&& - 22 \vec{p}_{1}\cdot\vec{p}_{2} \vec{p}_{2} -174 \vec{p}_{2}\cdot\vec{n} \vec{p}_{1}\cdot\vec{p}_{2} \vec{n} + 135 \vec{p}_{1}\cdot\vec{n} p_{2}^2 \vec{n} - 45 ( \vec{p}_{2}\cdot\vec{n})^{2} \vec{p}_{1} \nn\\ 
&& - 114 \vec{p}_{1}\cdot\vec{n} \vec{p}_{2}\cdot\vec{n} \vec{p}_{2} -135 \vec{p}_{1}\cdot\vec{n} ( \vec{p}_{2}\cdot\vec{n})^{2} \vec{n} \big) + 8 \vec{S}_{1}\cdot\vec{n} \vec{p}_{2}\cdot\vec{S}_{1} \big( \vec{p}_{1}\cdot\vec{p}_{2} \vec{p}_{1} + 12 p_{1}^2 \vec{p}_{2}\cdot\vec{n} \vec{n} \nn\\ 
&& - 3 \vec{p}_{1}\cdot\vec{n} \vec{p}_{2}\cdot\vec{n} \vec{p}_{1} \big) + 2 \vec{p}_{1}\cdot\vec{S}_{1} \vec{p}_{2}\cdot\vec{S}_{1} \big( 23 \vec{p}_{1}\cdot\vec{p}_{2} \vec{n} - 53 \vec{p}_{2}\cdot\vec{n} \vec{p}_{1} - 15 \vec{p}_{1}\cdot\vec{n} \vec{p}_{2} \nn\\ 
&& + 69 \vec{p}_{1}\cdot\vec{n} \vec{p}_{2}\cdot\vec{n} \vec{n} \big) + S_{1}^2 \big( 105 p_{1}^2 p_{2}^2 \vec{n} + 2 ( \vec{p}_{1}\cdot\vec{p}_{2})^{2} \vec{n} - 78 \vec{p}_{2}\cdot\vec{n} \vec{p}_{1}\cdot\vec{p}_{2} \vec{p}_{1} \nn\\ 
&& - 27 \vec{p}_{1}\cdot\vec{n} p_{2}^2 \vec{p}_{1} - 16 p_{1}^2 \vec{p}_{2}\cdot\vec{n} \vec{p}_{2} -48 \vec{p}_{1}\cdot\vec{n} \vec{p}_{2}\cdot\vec{n} \vec{p}_{1}\cdot\vec{p}_{2} \vec{n} - 45 p_{2}^2 ( \vec{p}_{1}\cdot\vec{n})^{2} \vec{n} \nn\\ 
&& - 174 p_{1}^2 ( \vec{p}_{2}\cdot\vec{n})^{2} \vec{n} + 93 \vec{p}_{1}\cdot\vec{n} ( \vec{p}_{2}\cdot\vec{n})^{2} \vec{p}_{1} + 162 \vec{p}_{2}\cdot\vec{n} ( \vec{p}_{1}\cdot\vec{n})^{2} \vec{p}_{2} \nn\\ 
&& + 45 ( \vec{p}_{1}\cdot\vec{n})^{2} ( \vec{p}_{2}\cdot\vec{n})^{2} \vec{n} \big) - 3 \vec{S}_{1}\cdot\vec{n} \vec{S}_{1} \big( 3 p_{1}^2 p_{2}^2 - 10 ( \vec{p}_{1}\cdot\vec{p}_{2})^{2} + 8 \vec{p}_{1}\cdot\vec{n} \vec{p}_{2}\cdot\vec{n} \vec{p}_{1}\cdot\vec{p}_{2} \nn\\ 
&& - 7 p_{2}^2 ( \vec{p}_{1}\cdot\vec{n})^{2} + 2 p_{1}^2 ( \vec{p}_{2}\cdot\vec{n})^{2} + 35 ( \vec{p}_{1}\cdot\vec{n})^{2} ( \vec{p}_{2}\cdot\vec{n})^{2} \big) + \vec{p}_{1}\cdot\vec{S}_{1} \vec{S}_{1} \big( 162 \vec{p}_{2}\cdot\vec{n} \vec{p}_{1}\cdot\vec{p}_{2} \nn\\ 
&& + 19 \vec{p}_{1}\cdot\vec{n} p_{2}^2 -51 \vec{p}_{1}\cdot\vec{n} ( \vec{p}_{2}\cdot\vec{n})^{2} \big) + 6 \vec{p}_{2}\cdot\vec{S}_{1} \vec{S}_{1} \big( 4 p_{1}^2 \vec{p}_{2}\cdot\vec{n} \nn\\ 
&& + 4 \vec{p}_{1}\cdot\vec{n} \vec{p}_{1}\cdot\vec{p}_{2} -15 \vec{p}_{2}\cdot\vec{n} ( \vec{p}_{1}\cdot\vec{n})^{2} \big) - 3 \big( 32 p_{1}^2 p_{2}^2 \vec{n} - 26 \vec{p}_{2}\cdot\vec{n} \vec{p}_{1}\cdot\vec{p}_{2} \vec{p}_{1} \nn\\ 
&& + 5 \vec{p}_{1}\cdot\vec{n} p_{2}^2 \vec{p}_{1} -40 p_{1}^2 ( \vec{p}_{2}\cdot\vec{n})^{2} \vec{n} - 25 \vec{p}_{1}\cdot\vec{n} ( \vec{p}_{2}\cdot\vec{n})^{2} \vec{p}_{1} \big) ( \vec{S}_{1}\cdot\vec{n})^{2} - \big( 111 p_{2}^2 \vec{n} \nn\\ 
&& - 14 \vec{p}_{2}\cdot\vec{n} \vec{p}_{2} -183 ( \vec{p}_{2}\cdot\vec{n})^{2} \vec{n} \big) ( \vec{p}_{1}\cdot\vec{S}_{1})^{2} - 2 \big( 24 p_{1}^2 \vec{n} + 5 \vec{p}_{1}\cdot\vec{n} \vec{p}_{1} \big) ( \vec{p}_{2}\cdot\vec{S}_{1})^{2} \Big] \nn\\ 
&& - 	\frac{G}{32 m_{1}{}^2 m_{2}{}^2 r{}^2} \Big[ 3 \vec{S}_{1}\cdot\vec{n} \vec{p}_{1}\cdot\vec{S}_{1} \big( 5 p_{2}^2 \vec{p}_{2} -35 \vec{p}_{2}\cdot\vec{n} p_{2}^2 \vec{n} - 3 ( \vec{p}_{2}\cdot\vec{n})^{2} \vec{p}_{2} + 25 ( \vec{p}_{2}\cdot\vec{n})^{3} \vec{n} \big) \nn\\ 
&& + 2 \vec{S}_{1}\cdot\vec{n} \vec{p}_{2}\cdot\vec{S}_{1} \big( 3 p_{2}^2 \vec{p}_{1} - 4 \vec{p}_{1}\cdot\vec{p}_{2} \vec{p}_{2} + 18 \vec{p}_{1}\cdot\vec{n} p_{2}^2 \vec{n} \nn\\ 
&& - 45 ( \vec{p}_{2}\cdot\vec{n})^{2} \vec{p}_{1} -90 \vec{p}_{1}\cdot\vec{n} ( \vec{p}_{2}\cdot\vec{n})^{2} \vec{n} \big) - \vec{p}_{1}\cdot\vec{S}_{1} \vec{p}_{2}\cdot\vec{S}_{1} \big( 41 p_{2}^2 \vec{n} \nn\\ 
&& + 90 \vec{p}_{2}\cdot\vec{n} \vec{p}_{2} -183 ( \vec{p}_{2}\cdot\vec{n})^{2} \vec{n} \big) + S_{1}^2 \big( 37 \vec{p}_{1}\cdot\vec{p}_{2} p_{2}^2 \vec{n} - 134 \vec{p}_{2}\cdot\vec{n} p_{2}^2 \vec{p}_{1} \nn\\ 
&& - 22 \vec{p}_{2}\cdot\vec{n} \vec{p}_{1}\cdot\vec{p}_{2} \vec{p}_{2} - 51 \vec{p}_{1}\cdot\vec{n} p_{2}^2 \vec{p}_{2} + 33 \vec{p}_{1}\cdot\vec{n} \vec{p}_{2}\cdot\vec{n} p_{2}^2 \vec{n} - 171 \vec{p}_{1}\cdot\vec{p}_{2} ( \vec{p}_{2}\cdot\vec{n})^{2} \vec{n} \nn\\ 
&& + 150 ( \vec{p}_{2}\cdot\vec{n})^{3} \vec{p}_{1} + 117 \vec{p}_{1}\cdot\vec{n} ( \vec{p}_{2}\cdot\vec{n})^{2} \vec{p}_{2} + 45 \vec{p}_{1}\cdot\vec{n} ( \vec{p}_{2}\cdot\vec{n})^{3} \vec{n} \big) \nn\\ 
&& + 3 \vec{S}_{1}\cdot\vec{n} \vec{S}_{1} \big( \vec{p}_{1}\cdot\vec{p}_{2} p_{2}^2 -3 \vec{p}_{1}\cdot\vec{n} \vec{p}_{2}\cdot\vec{n} p_{2}^2 + 9 \vec{p}_{1}\cdot\vec{p}_{2} ( \vec{p}_{2}\cdot\vec{n})^{2} -15 \vec{p}_{1}\cdot\vec{n} ( \vec{p}_{2}\cdot\vec{n})^{3} \big) \nn\\ 
&& + 2 \vec{p}_{1}\cdot\vec{S}_{1} \vec{S}_{1} \big( 67 \vec{p}_{2}\cdot\vec{n} p_{2}^2 -75 ( \vec{p}_{2}\cdot\vec{n})^{3} \big) + \vec{p}_{2}\cdot\vec{S}_{1} \vec{S}_{1} \big( 62 \vec{p}_{2}\cdot\vec{n} \vec{p}_{1}\cdot\vec{p}_{2} \nn\\ 
&& + 31 \vec{p}_{1}\cdot\vec{n} p_{2}^2 -57 \vec{p}_{1}\cdot\vec{n} ( \vec{p}_{2}\cdot\vec{n})^{2} \big) - 3 \big( 8 \vec{p}_{1}\cdot\vec{p}_{2} p_{2}^2 \vec{n} - 27 \vec{p}_{2}\cdot\vec{n} p_{2}^2 \vec{p}_{1} \nn\\ 
&& + 4 \vec{p}_{1}\cdot\vec{n} p_{2}^2 \vec{p}_{2} -40 \vec{p}_{1}\cdot\vec{p}_{2} ( \vec{p}_{2}\cdot\vec{n})^{2} \vec{n} + 25 ( \vec{p}_{2}\cdot\vec{n})^{3} \vec{p}_{1} - 20 \vec{p}_{1}\cdot\vec{n} ( \vec{p}_{2}\cdot\vec{n})^{2} \vec{p}_{2} \big) ( \vec{S}_{1}\cdot\vec{n})^{2} \nn\\ 
&& + 2 \big( 4 \vec{p}_{1}\cdot\vec{p}_{2} \vec{n} + 25 \vec{p}_{2}\cdot\vec{n} \vec{p}_{1} \big) ( \vec{p}_{2}\cdot\vec{S}_{1})^{2} \Big] - 	\frac{G^2 m_{2}{}^2}{32 m_{1}{}^3 r{}^3} \Big[ 2 \vec{S}_{1}\cdot\vec{n} \vec{p}_{1}\cdot\vec{S}_{1} \big( 26 \vec{p}_{1} \nn\\ 
&& + 167 \vec{p}_{1}\cdot\vec{n} \vec{n} \big) + S_{1}^2 \big( 205 p_{1}^2 \vec{n} - 60 \vec{p}_{1}\cdot\vec{n} \vec{p}_{1} -122 ( \vec{p}_{1}\cdot\vec{n})^{2} \vec{n} \big) \nn\\ 
&& - 2 \vec{S}_{1}\cdot\vec{n} \vec{S}_{1} \big( 27 p_{1}^2 -32 ( \vec{p}_{1}\cdot\vec{n})^{2} \big) + 60 \vec{p}_{1}\cdot\vec{S}_{1} \vec{S}_{1} \vec{p}_{1}\cdot\vec{n} - \big( 217 p_{1}^2 \vec{n} \nn\\ 
&& + 64 \vec{p}_{1}\cdot\vec{n} \vec{p}_{1} \big) ( \vec{S}_{1}\cdot\vec{n})^{2} - 198 \vec{n} ( \vec{p}_{1}\cdot\vec{S}_{1})^{2} \Big] - 	\frac{G^2 m_{2}}{48 m_{1}{}^2 r{}^3} \Big[ \vec{S}_{1}\cdot\vec{n} \vec{p}_{1}\cdot\vec{S}_{1} \big( 491 \vec{p}_{1} + 138 \vec{p}_{2} \nn\\ 
&& + 3604 \vec{p}_{1}\cdot\vec{n} \vec{n} + 432 \vec{p}_{2}\cdot\vec{n} \vec{n} \big) - 3 \vec{S}_{1}\cdot\vec{n} \vec{p}_{2}\cdot\vec{S}_{1} \big( 46 \vec{p}_{1} + 151 \vec{p}_{1}\cdot\vec{n} \vec{n} \big) \nn\\ 
&& + 462 \vec{p}_{1}\cdot\vec{S}_{1} \vec{p}_{2}\cdot\vec{S}_{1} \vec{n} + S_{1}^2 \big( 833 p_{1}^2 \vec{n} - 999 \vec{p}_{1}\cdot\vec{p}_{2} \vec{n} + 433 \vec{p}_{1}\cdot\vec{n} \vec{p}_{1} + 627 \vec{p}_{2}\cdot\vec{n} \vec{p}_{1} \nn\\ 
&& + 72 \vec{p}_{1}\cdot\vec{n} \vec{p}_{2} -477 \vec{p}_{1}\cdot\vec{n} \vec{p}_{2}\cdot\vec{n} \vec{n} - 1187 ( \vec{p}_{1}\cdot\vec{n})^{2} \vec{n} \big) - \vec{S}_{1}\cdot\vec{n} \vec{S}_{1} \big( 272 p_{1}^2 \nn\\ 
&& - 705 \vec{p}_{1}\cdot\vec{p}_{2} -792 \vec{p}_{1}\cdot\vec{n} \vec{p}_{2}\cdot\vec{n} - 920 ( \vec{p}_{1}\cdot\vec{n})^{2} \big) - 5 \vec{p}_{1}\cdot\vec{S}_{1} \vec{S}_{1} \big( 113 \vec{p}_{1}\cdot\vec{n} + 195 \vec{p}_{2}\cdot\vec{n} \big) \nn\\ 
&& + 3 \vec{p}_{2}\cdot\vec{S}_{1} \vec{S}_{1} \vec{p}_{1}\cdot\vec{n} - \big( 1157 p_{1}^2 \vec{n} - 546 \vec{p}_{1}\cdot\vec{p}_{2} \vec{n} + 1480 \vec{p}_{1}\cdot\vec{n} \vec{p}_{1} + 555 \vec{p}_{2}\cdot\vec{n} \vec{p}_{1} \nn\\ 
&& + 180 \vec{p}_{1}\cdot\vec{n} \vec{p}_{2} + 438 ( \vec{p}_{1}\cdot\vec{n})^{2} \vec{n} \big) ( \vec{S}_{1}\cdot\vec{n})^{2} - 1166 \vec{n} ( \vec{p}_{1}\cdot\vec{S}_{1})^{2} \Big] \nn\\ 
&& - 	\frac{G^2}{8 m_{2} r{}^3} \Big[ 2 \vec{S}_{1}\cdot\vec{n} \vec{p}_{2}\cdot\vec{S}_{1} \big( 61 \vec{p}_{2} + 10 \vec{p}_{2}\cdot\vec{n} \vec{n} \big) + S_{1}^2 \big( 75 p_{2}^2 \vec{n} - 146 \vec{p}_{2}\cdot\vec{n} \vec{p}_{2} \nn\\ 
&& + 45 ( \vec{p}_{2}\cdot\vec{n})^{2} \vec{n} \big) - 2 \vec{S}_{1}\cdot\vec{n} \vec{S}_{1} \big( 35 p_{2}^2 + 29 ( \vec{p}_{2}\cdot\vec{n})^{2} \big) + 128 \vec{p}_{2}\cdot\vec{S}_{1} \vec{S}_{1} \vec{p}_{2}\cdot\vec{n} - \big( 29 p_{2}^2 \vec{n} \nn\\ 
&& + 46 \vec{p}_{2}\cdot\vec{n} \vec{p}_{2} -3 ( \vec{p}_{2}\cdot\vec{n})^{2} \vec{n} \big) ( \vec{S}_{1}\cdot\vec{n})^{2} - 106 \vec{n} ( \vec{p}_{2}\cdot\vec{S}_{1})^{2} \Big] \nn\\ 
&& + 	\frac{G^2}{48 m_{1} r{}^3} \Big[ 2 \vec{S}_{1}\cdot\vec{n} \vec{p}_{1}\cdot\vec{S}_{1} \big( 200 \vec{p}_{2} + 541 \vec{p}_{2}\cdot\vec{n} \vec{n} \big) + 4 \vec{S}_{1}\cdot\vec{n} \vec{p}_{2}\cdot\vec{S}_{1} \big( 58 \vec{p}_{1} - 84 \vec{p}_{2} \nn\\ 
&& + 752 \vec{p}_{1}\cdot\vec{n} \vec{n} - 51 \vec{p}_{2}\cdot\vec{n} \vec{n} \big) - 1700 \vec{p}_{1}\cdot\vec{S}_{1} \vec{p}_{2}\cdot\vec{S}_{1} \vec{n} + S_{1}^2 \big( 1394 \vec{p}_{1}\cdot\vec{p}_{2} \vec{n} - 822 p_{2}^2 \vec{n} \nn\\ 
&& - 310 \vec{p}_{2}\cdot\vec{n} \vec{p}_{1} + 62 \vec{p}_{1}\cdot\vec{n} \vec{p}_{2} + 369 \vec{p}_{2}\cdot\vec{n} \vec{p}_{2} -902 \vec{p}_{1}\cdot\vec{n} \vec{p}_{2}\cdot\vec{n} \vec{n} + 585 ( \vec{p}_{2}\cdot\vec{n})^{2} \vec{n} \big) \nn\\ 
&& - \vec{S}_{1}\cdot\vec{n} \vec{S}_{1} \big( 716 \vec{p}_{1}\cdot\vec{p}_{2} - 840 p_{2}^2 -206 \vec{p}_{1}\cdot\vec{n} \vec{p}_{2}\cdot\vec{n} + 327 ( \vec{p}_{2}\cdot\vec{n})^{2} \big) \nn\\ 
&& + 538 \vec{p}_{1}\cdot\vec{S}_{1} \vec{S}_{1} \vec{p}_{2}\cdot\vec{n} - \vec{p}_{2}\cdot\vec{S}_{1} \vec{S}_{1} \big( 128 \vec{p}_{1}\cdot\vec{n} + 531 \vec{p}_{2}\cdot\vec{n} \big) - \big( 1208 \vec{p}_{1}\cdot\vec{p}_{2} \vec{n} - 147 p_{2}^2 \vec{n} \nn\\ 
&& - 406 \vec{p}_{2}\cdot\vec{n} \vec{p}_{1} + 1088 \vec{p}_{1}\cdot\vec{n} \vec{p}_{2} - 9 \vec{p}_{2}\cdot\vec{n} \vec{p}_{2} + 780 \vec{p}_{1}\cdot\vec{n} \vec{p}_{2}\cdot\vec{n} \vec{n} \nn\\ 
&& + 144 ( \vec{p}_{2}\cdot\vec{n})^{2} \vec{n} \big) ( \vec{S}_{1}\cdot\vec{n})^{2} + 414 \vec{n} ( \vec{p}_{2}\cdot\vec{S}_{1})^{2} \Big] + 	\frac{G^3 m_{2}{}^3}{2 m_{1} r{}^4} \Big[ 28 S_{1}^2 \vec{n} - 15 \vec{n} ( \vec{S}_{1}\cdot\vec{n})^{2} \nn\\ 
&& - 13 \vec{S}_{1}\cdot\vec{n} \vec{S}_{1} \Big] + 	\frac{3 G^3 m_{1} m_{2}}{245 r{}^4} \Big[ 1471 S_{1}^2 \vec{n} - 1860 \vec{n} ( \vec{S}_{1}\cdot\vec{n})^{2} + 72 \vec{S}_{1}\cdot\vec{n} \vec{S}_{1} \Big] \nn\\ 
&& + 	\frac{G^3 m_{2}{}^2}{64 r{}^4} \Big[ S_{1}^2 {(2452 - 45 \pi^2)} \vec{n} - 25 {(124 - 9 \pi^2)} \vec{n} ( \vec{S}_{1}\cdot\vec{n})^{2} \nn\\ 
&& - 2 {(188 + 45 \pi^2)} \vec{S}_{1}\cdot\vec{n} \vec{S}_{1} \Big] ,
\eea
\bea
\vec{Y}^{\text{N}^3\text{LO}}_{\text{ES}_1^2 } &=&  - 	\frac{5 G}{16 m_{1} m_{2}{}^3 r{}^2} \Big[ S_{1}^2 p_{2}^{4} \vec{n} - \vec{S}_{1}\cdot\vec{n} \vec{S}_{1} p_{2}^{4} \Big] + 	\frac{G}{8 m_{1}{}^2 m_{2}{}^2 r{}^2} \Big[ 2 \vec{S}_{1}\cdot\vec{n} \vec{p}_{1}\cdot\vec{S}_{1} p_{2}^2 \vec{p}_{2} \nn\\ 
&& - 2 \vec{S}_{1}\cdot\vec{n} \vec{p}_{2}\cdot\vec{S}_{1} p_{2}^2 \vec{p}_{1} + S_{1}^2 \big( 2 \vec{p}_{1}\cdot\vec{p}_{2} p_{2}^2 \vec{n} + 3 \vec{p}_{2}\cdot\vec{n} p_{2}^2 \vec{p}_{1} \nn\\ 
&& + \vec{p}_{1}\cdot\vec{n} p_{2}^2 \vec{p}_{2} -6 \vec{p}_{1}\cdot\vec{n} \vec{p}_{2}\cdot\vec{n} p_{2}^2 \vec{n} \big) - 2 \vec{S}_{1}\cdot\vec{n} \vec{S}_{1} \big( \vec{p}_{1}\cdot\vec{p}_{2} p_{2}^2 -3 \vec{p}_{1}\cdot\vec{n} \vec{p}_{2}\cdot\vec{n} p_{2}^2 \big) \nn\\ 
&& - 2 \vec{p}_{1}\cdot\vec{S}_{1} \vec{S}_{1} \vec{p}_{2}\cdot\vec{n} p_{2}^2 - 2 \vec{p}_{2}\cdot\vec{S}_{1} \vec{S}_{1} \vec{p}_{1}\cdot\vec{n} p_{2}^2 + 3 \big( \vec{p}_{2}\cdot\vec{n} p_{2}^2 \vec{p}_{1} - \vec{p}_{1}\cdot\vec{n} p_{2}^2 \vec{p}_{2} \big) ( \vec{S}_{1}\cdot\vec{n})^{2} \Big] \nn\\ 
&& + 	\frac{G m_{2}}{16 m_{1}{}^5 r{}^2} \Big[ 3 \vec{S}_{1}\cdot\vec{n} \vec{p}_{1}\cdot\vec{S}_{1} p_{1}^2 \vec{p}_{1} - S_{1}^2 \big( 5 p_{1}^{4} \vec{n} + 4 \vec{p}_{1}\cdot\vec{n} p_{1}^2 \vec{p}_{1} \big) + 5 \vec{S}_{1}\cdot\vec{n} \vec{S}_{1} p_{1}^{4} \nn\\ 
&& + 3 \vec{p}_{1}\cdot\vec{S}_{1} \vec{S}_{1} \vec{p}_{1}\cdot\vec{n} p_{1}^2 - 2 \vec{p}_{1}\cdot\vec{n} \vec{p}_{1} ( \vec{p}_{1}\cdot\vec{S}_{1})^{2} \Big] - 	\frac{G}{8 m_{1}{}^4 r{}^2} \Big[ \vec{S}_{1}\cdot\vec{n} \vec{p}_{1}\cdot\vec{S}_{1} \big( 6 \vec{p}_{1}\cdot\vec{p}_{2} \vec{p}_{1} \nn\\ 
&& - p_{1}^2 \vec{p}_{2} + 6 \vec{p}_{1}\cdot\vec{n} \vec{p}_{2}\cdot\vec{n} \vec{p}_{1} - 3 ( \vec{p}_{1}\cdot\vec{n})^{2} \vec{p}_{2} \big) + 2 \vec{S}_{1}\cdot\vec{n} \vec{p}_{2}\cdot\vec{S}_{1} p_{1}^2 \vec{p}_{1} \nn\\ 
&& - 2 \vec{p}_{1}\cdot\vec{S}_{1} \vec{p}_{2}\cdot\vec{S}_{1} \vec{p}_{1}\cdot\vec{n} \vec{p}_{1} - S_{1}^2 \big( 2 p_{1}^2 \vec{p}_{1}\cdot\vec{p}_{2} \vec{n} + 10 \vec{p}_{1}\cdot\vec{n} \vec{p}_{1}\cdot\vec{p}_{2} \vec{p}_{1} \nn\\ 
&& + 4 \vec{p}_{1}\cdot\vec{n} p_{1}^2 \vec{p}_{2} -6 \vec{p}_{1}\cdot\vec{n} p_{1}^2 \vec{p}_{2}\cdot\vec{n} \vec{n} + 9 \vec{p}_{2}\cdot\vec{n} ( \vec{p}_{1}\cdot\vec{n})^{2} \vec{p}_{1} - 3 ( \vec{p}_{1}\cdot\vec{n})^{3} \vec{p}_{2} \big) \nn\\ 
&& + 2 \vec{S}_{1}\cdot\vec{n} \vec{S}_{1} \big( p_{1}^2 \vec{p}_{1}\cdot\vec{p}_{2} -3 \vec{p}_{1}\cdot\vec{n} p_{1}^2 \vec{p}_{2}\cdot\vec{n} \big) + \vec{p}_{1}\cdot\vec{S}_{1} \vec{S}_{1} \big( p_{1}^2 \vec{p}_{2}\cdot\vec{n} + 6 \vec{p}_{1}\cdot\vec{n} \vec{p}_{1}\cdot\vec{p}_{2} \nn\\ 
&& + 3 \vec{p}_{2}\cdot\vec{n} ( \vec{p}_{1}\cdot\vec{n})^{2} \big) + 2 \vec{p}_{2}\cdot\vec{S}_{1} \vec{S}_{1} \vec{p}_{1}\cdot\vec{n} p_{1}^2 - 3 \big( p_{1}^2 \vec{p}_{2}\cdot\vec{n} \vec{p}_{1} - \vec{p}_{1}\cdot\vec{n} p_{1}^2 \vec{p}_{2} \big) ( \vec{S}_{1}\cdot\vec{n})^{2} \nn\\ 
&& - 2 \big( \vec{p}_{2}\cdot\vec{n} \vec{p}_{1} - \vec{p}_{1}\cdot\vec{n} \vec{p}_{2} \big) ( \vec{p}_{1}\cdot\vec{S}_{1})^{2} \Big] + 	\frac{G}{16 m_{1}{}^3 m_{2} r{}^2} \Big[ \vec{S}_{1}\cdot\vec{n} \vec{p}_{1}\cdot\vec{S}_{1} \big( 21 p_{2}^2 \vec{p}_{1} \nn\\ 
&& - 29 \vec{p}_{1}\cdot\vec{p}_{2} \vec{p}_{2} + 18 \vec{p}_{2}\cdot\vec{n} \vec{p}_{1}\cdot\vec{p}_{2} \vec{n} - 48 \vec{p}_{1}\cdot\vec{n} p_{2}^2 \vec{n} - 6 ( \vec{p}_{2}\cdot\vec{n})^{2} \vec{p}_{1} + 30 \vec{p}_{1}\cdot\vec{n} \vec{p}_{2}\cdot\vec{n} \vec{p}_{2} \big) \nn\\ 
&& - \vec{S}_{1}\cdot\vec{n} \vec{p}_{2}\cdot\vec{S}_{1} \big( 5 \vec{p}_{1}\cdot\vec{p}_{2} \vec{p}_{1} - 19 p_{1}^2 \vec{p}_{2} + 18 p_{1}^2 \vec{p}_{2}\cdot\vec{n} \vec{n} - 48 \vec{p}_{1}\cdot\vec{n} \vec{p}_{1}\cdot\vec{p}_{2} \vec{n} \nn\\ 
&& + 36 \vec{p}_{1}\cdot\vec{n} \vec{p}_{2}\cdot\vec{n} \vec{p}_{1} - 12 ( \vec{p}_{1}\cdot\vec{n})^{2} \vec{p}_{2} \big) + \vec{p}_{1}\cdot\vec{S}_{1} \vec{p}_{2}\cdot\vec{S}_{1} \big( 4 \vec{p}_{1}\cdot\vec{p}_{2} \vec{n} - 9 \vec{p}_{2}\cdot\vec{n} \vec{p}_{1} \nn\\ 
&& - 31 \vec{p}_{1}\cdot\vec{n} \vec{p}_{2} + 36 \vec{p}_{1}\cdot\vec{n} \vec{p}_{2}\cdot\vec{n} \vec{n} \big) + S_{1}^2 \big( 5 p_{1}^2 p_{2}^2 \vec{n} - 9 \vec{p}_{2}\cdot\vec{n} \vec{p}_{1}\cdot\vec{p}_{2} \vec{p}_{1} - 42 \vec{p}_{1}\cdot\vec{n} p_{2}^2 \vec{p}_{1} \nn\\ 
&& - 6 p_{1}^2 \vec{p}_{2}\cdot\vec{n} \vec{p}_{2} + 25 \vec{p}_{1}\cdot\vec{n} \vec{p}_{1}\cdot\vec{p}_{2} \vec{p}_{2} + 30 p_{2}^2 ( \vec{p}_{1}\cdot\vec{n})^{2} \vec{n} + 15 \vec{p}_{1}\cdot\vec{n} ( \vec{p}_{2}\cdot\vec{n})^{2} \vec{p}_{1} \nn\\ 
&& - 27 \vec{p}_{2}\cdot\vec{n} ( \vec{p}_{1}\cdot\vec{n})^{2} \vec{p}_{2} + 15 ( \vec{p}_{1}\cdot\vec{n})^{2} ( \vec{p}_{2}\cdot\vec{n})^{2} \vec{n} \big) - 5 \vec{S}_{1}\cdot\vec{n} \vec{S}_{1} \big( 4 p_{1}^2 p_{2}^2 - 3 ( \vec{p}_{1}\cdot\vec{p}_{2})^{2} \nn\\ 
&& + 6 \vec{p}_{1}\cdot\vec{n} \vec{p}_{2}\cdot\vec{n} \vec{p}_{1}\cdot\vec{p}_{2} + 3 p_{2}^2 ( \vec{p}_{1}\cdot\vec{n})^{2} - 3 p_{1}^2 ( \vec{p}_{2}\cdot\vec{n})^{2} + 3 ( \vec{p}_{1}\cdot\vec{n})^{2} ( \vec{p}_{2}\cdot\vec{n})^{2} \big) \nn\\ 
&& + \vec{p}_{1}\cdot\vec{S}_{1} \vec{S}_{1} \big( 7 \vec{p}_{2}\cdot\vec{n} \vec{p}_{1}\cdot\vec{p}_{2} + 39 \vec{p}_{1}\cdot\vec{n} p_{2}^2 -24 \vec{p}_{1}\cdot\vec{n} ( \vec{p}_{2}\cdot\vec{n})^{2} \big) + \vec{p}_{2}\cdot\vec{S}_{1} \vec{S}_{1} \big( 3 p_{1}^2 \vec{p}_{2}\cdot\vec{n} \nn\\ 
&& - 23 \vec{p}_{1}\cdot\vec{n} \vec{p}_{1}\cdot\vec{p}_{2} + 36 \vec{p}_{2}\cdot\vec{n} ( \vec{p}_{1}\cdot\vec{n})^{2} \big) + 3 \big( 5 p_{1}^2 p_{2}^2 \vec{n} - 5 ( \vec{p}_{1}\cdot\vec{p}_{2})^{2} \vec{n} - 2 \vec{p}_{2}\cdot\vec{n} \vec{p}_{1}\cdot\vec{p}_{2} \vec{p}_{1} \nn\\ 
&& + 7 \vec{p}_{1}\cdot\vec{n} p_{2}^2 \vec{p}_{1} - 5 p_{1}^2 \vec{p}_{2}\cdot\vec{n} \vec{p}_{2} -5 \vec{p}_{1}\cdot\vec{n} ( \vec{p}_{2}\cdot\vec{n})^{2} \vec{p}_{1} + 5 \vec{p}_{2}\cdot\vec{n} ( \vec{p}_{1}\cdot\vec{n})^{2} \vec{p}_{2} \big) ( \vec{S}_{1}\cdot\vec{n})^{2} \nn\\ 
&& - \big( 2 p_{2}^2 \vec{n} - 7 \vec{p}_{2}\cdot\vec{n} \vec{p}_{2} + 3 ( \vec{p}_{2}\cdot\vec{n})^{2} \vec{n} \big) ( \vec{p}_{1}\cdot\vec{S}_{1})^{2} - \big( 2 p_{1}^2 \vec{n} - 33 \vec{p}_{1}\cdot\vec{n} \vec{p}_{1} \nn\\ 
&& + 33 ( \vec{p}_{1}\cdot\vec{n})^{2} \vec{n} \big) ( \vec{p}_{2}\cdot\vec{S}_{1})^{2} \Big] - 	\frac{G^2 m_{2}}{48 m_{1}{}^2 r{}^3} \Big[ 2 \vec{S}_{1}\cdot\vec{n} \vec{p}_{1}\cdot\vec{S}_{1} \big( 323 \vec{p}_{1} - 46 \vec{p}_{2} -179 \vec{p}_{1}\cdot\vec{n} \vec{n} \nn\\ 
&& - 380 \vec{p}_{2}\cdot\vec{n} \vec{n} \big) + 2 \vec{S}_{1}\cdot\vec{n} \vec{p}_{2}\cdot\vec{S}_{1} \big( 497 \vec{p}_{1} -677 \vec{p}_{1}\cdot\vec{n} \vec{n} \big) + 136 \vec{p}_{1}\cdot\vec{S}_{1} \vec{p}_{2}\cdot\vec{S}_{1} \vec{n} \nn\\ 
&& + S_{1}^2 \big( 256 p_{1}^2 \vec{n} - 784 \vec{p}_{1}\cdot\vec{p}_{2} \vec{n} + 140 \vec{p}_{1}\cdot\vec{n} \vec{p}_{1} - 856 \vec{p}_{2}\cdot\vec{n} \vec{p}_{1} - 796 \vec{p}_{1}\cdot\vec{n} \vec{p}_{2} \nn\\ 
&& + 2992 \vec{p}_{1}\cdot\vec{n} \vec{p}_{2}\cdot\vec{n} \vec{n} + 227 ( \vec{p}_{1}\cdot\vec{n})^{2} \vec{n} \big) + \vec{S}_{1}\cdot\vec{n} \vec{S}_{1} \big( 407 p_{1}^2 \nn\\ 
&& + 1000 \vec{p}_{1}\cdot\vec{p}_{2} -2104 \vec{p}_{1}\cdot\vec{n} \vec{p}_{2}\cdot\vec{n} - 161 ( \vec{p}_{1}\cdot\vec{n})^{2} \big) + 4 \vec{p}_{1}\cdot\vec{S}_{1} \vec{S}_{1} \big( 19 \vec{p}_{1}\cdot\vec{n} \nn\\ 
&& + 142 \vec{p}_{2}\cdot\vec{n} \big) + 958 \vec{p}_{2}\cdot\vec{S}_{1} \vec{S}_{1} \vec{p}_{1}\cdot\vec{n} - 2 \big( 461 p_{1}^2 \vec{n} + 376 \vec{p}_{1}\cdot\vec{p}_{2} \vec{n} + 637 \vec{p}_{1}\cdot\vec{n} \vec{p}_{1} \nn\\ 
&& + 637 \vec{p}_{2}\cdot\vec{n} \vec{p}_{1} + 31 \vec{p}_{1}\cdot\vec{n} \vec{p}_{2} -573 \vec{p}_{1}\cdot\vec{n} \vec{p}_{2}\cdot\vec{n} \vec{n} - 255 ( \vec{p}_{1}\cdot\vec{n})^{2} \vec{n} \big) ( \vec{S}_{1}\cdot\vec{n})^{2} \nn\\ 
&& + 47 \vec{n} ( \vec{p}_{1}\cdot\vec{S}_{1})^{2} \Big] + 	\frac{G^2 m_{2}{}^2}{12 m_{1}{}^3 r{}^3} \Big[ \vec{S}_{1}\cdot\vec{n} \vec{p}_{1}\cdot\vec{S}_{1} \big( \vec{p}_{1} -70 \vec{p}_{1}\cdot\vec{n} \vec{n} \big) - 2 S_{1}^2 \big( 28 p_{1}^2 \vec{n} \nn\\ 
&& + 41 \vec{p}_{1}\cdot\vec{n} \vec{p}_{1} -31 ( \vec{p}_{1}\cdot\vec{n})^{2} \vec{n} \big) + \vec{S}_{1}\cdot\vec{n} \vec{S}_{1} \big( 23 p_{1}^2 -8 ( \vec{p}_{1}\cdot\vec{n})^{2} \big) + 67 \vec{p}_{1}\cdot\vec{S}_{1} \vec{S}_{1} \vec{p}_{1}\cdot\vec{n} \nn\\ 
&& + \big( 83 p_{1}^2 \vec{n} + 40 \vec{p}_{1}\cdot\vec{n} \vec{p}_{1} -12 ( \vec{p}_{1}\cdot\vec{n})^{2} \vec{n} \big) ( \vec{S}_{1}\cdot\vec{n})^{2} - 13 \vec{n} ( \vec{p}_{1}\cdot\vec{S}_{1})^{2} \Big] \nn\\ 
&& + 	\frac{G^2}{24 m_{2} r{}^3} \Big[ 4 \vec{S}_{1}\cdot\vec{n} \vec{p}_{2}\cdot\vec{S}_{1} \big( 7 \vec{p}_{2} - \vec{p}_{2}\cdot\vec{n} \vec{n} \big) - S_{1}^2 \big( 38 p_{2}^2 \vec{n} - 2 \vec{p}_{2}\cdot\vec{n} \vec{p}_{2} -17 ( \vec{p}_{2}\cdot\vec{n})^{2} \vec{n} \big) \nn\\ 
&& + 14 \vec{S}_{1}\cdot\vec{n} \vec{S}_{1} \big( 4 p_{2}^2 - ( \vec{p}_{2}\cdot\vec{n})^{2} \big) + 4 \vec{p}_{2}\cdot\vec{S}_{1} \vec{S}_{1} \vec{p}_{2}\cdot\vec{n} - \big( 76 p_{2}^2 \vec{n} \nn\\ 
&& + 62 \vec{p}_{2}\cdot\vec{n} \vec{p}_{2} -15 ( \vec{p}_{2}\cdot\vec{n})^{2} \vec{n} \big) ( \vec{S}_{1}\cdot\vec{n})^{2} + 2 \vec{n} ( \vec{p}_{2}\cdot\vec{S}_{1})^{2} \Big] \nn\\ 
&& + 	\frac{G^2}{48 m_{1} r{}^3} \Big[ 4 \vec{S}_{1}\cdot\vec{n} \vec{p}_{1}\cdot\vec{S}_{1} \big( 121 \vec{p}_{2} -37 \vec{p}_{2}\cdot\vec{n} \vec{n} \big) + 2 \vec{S}_{1}\cdot\vec{n} \vec{p}_{2}\cdot\vec{S}_{1} \big( 32 \vec{p}_{1} \nn\\ 
&& - 163 \vec{p}_{2} -86 \vec{p}_{1}\cdot\vec{n} \vec{n} - 842 \vec{p}_{2}\cdot\vec{n} \vec{n} \big) + 40 \vec{p}_{1}\cdot\vec{S}_{1} \vec{p}_{2}\cdot\vec{S}_{1} \vec{n} + S_{1}^2 \big( 332 \vec{p}_{1}\cdot\vec{p}_{2} \vec{n} - 2 p_{2}^2 \vec{n} \nn\\ 
&& + 20 \vec{p}_{2}\cdot\vec{n} \vec{p}_{1} + 128 \vec{p}_{1}\cdot\vec{n} \vec{p}_{2} + 236 \vec{p}_{2}\cdot\vec{n} \vec{p}_{2} + 172 \vec{p}_{1}\cdot\vec{n} \vec{p}_{2}\cdot\vec{n} \vec{n} + 1781 ( \vec{p}_{2}\cdot\vec{n})^{2} \vec{n} \big) \nn\\ 
&& + 4 \vec{S}_{1}\cdot\vec{n} \vec{S}_{1} \big( 70 \vec{p}_{1}\cdot\vec{p}_{2} + 110 p_{2}^2 -37 \vec{p}_{1}\cdot\vec{n} \vec{p}_{2}\cdot\vec{n} - 239 ( \vec{p}_{2}\cdot\vec{n})^{2} \big) + 16 \vec{p}_{1}\cdot\vec{S}_{1} \vec{S}_{1} \vec{p}_{2}\cdot\vec{n} \nn\\ 
&& + 4 \vec{p}_{2}\cdot\vec{S}_{1} \vec{S}_{1} \big( 13 \vec{p}_{1}\cdot\vec{n} + 241 \vec{p}_{2}\cdot\vec{n} \big) - \big( 740 \vec{p}_{1}\cdot\vec{p}_{2} \vec{n} + 1177 p_{2}^2 \vec{n} + 188 \vec{p}_{2}\cdot\vec{n} \vec{p}_{1} \nn\\ 
&& + 920 \vec{p}_{1}\cdot\vec{n} \vec{p}_{2} + 1550 \vec{p}_{2}\cdot\vec{n} \vec{p}_{2} -480 \vec{p}_{1}\cdot\vec{n} \vec{p}_{2}\cdot\vec{n} \vec{n} - 897 ( \vec{p}_{2}\cdot\vec{n})^{2} \vec{n} \big) ( \vec{S}_{1}\cdot\vec{n})^{2} \nn\\ 
&& + 179 \vec{n} ( \vec{p}_{2}\cdot\vec{S}_{1})^{2} \Big] - 	\frac{G^3 m_{2}{}^2}{8 r{}^4} \Big[ 165 S_{1}^2 \vec{n} - 181 \vec{n} ( \vec{S}_{1}\cdot\vec{n})^{2} - 132 \vec{S}_{1}\cdot\vec{n} \vec{S}_{1} \Big] \nn\\ 
&& + 	\frac{G^3 m_{2}{}^3}{4 m_{1} r{}^4} \Big[ 51 S_{1}^2 \vec{n} - 46 \vec{n} ( \vec{S}_{1}\cdot\vec{n})^{2} - 11 \vec{S}_{1}\cdot\vec{n} \vec{S}_{1} \Big] + 	\frac{G^3 m_{1} m_{2}}{28 r{}^4} \Big[ 66 S_{1}^2 \vec{n} \nn\\ 
&& - 197 \vec{n} ( \vec{S}_{1}\cdot\vec{n})^{2} + 69 \vec{S}_{1}\cdot\vec{n} \vec{S}_{1} \Big] ,
\eea
\bea
\vec{Y}^{\text{N}^3\text{LO}}_{\text{S}_1 \text{S}_2 } &=& - 	\frac{G}{16 m_{1}{}^4 r{}^2} \Big[ 6 \vec{S}_{1}\cdot\vec{n} \vec{S}_{2}\cdot\vec{n} \big( - p_{1}^{4} \vec{n} + \vec{p}_{1}\cdot\vec{n} p_{1}^2 \vec{p}_{1} -5 ( \vec{p}_{1}\cdot\vec{n})^{3} \vec{p}_{1} \big) \nn\\ 
&& - \vec{p}_{1}\cdot\vec{S}_{1} \vec{S}_{2}\cdot\vec{n} \big( 5 p_{1}^2 \vec{p}_{1} -6 \vec{p}_{1}\cdot\vec{n} p_{1}^2 \vec{n} + 18 ( \vec{p}_{1}\cdot\vec{n})^{2} \vec{p}_{1} \big) \nn\\ 
&& + 2 \vec{S}_{1}\cdot\vec{n} \vec{p}_{1}\cdot\vec{S}_{2} \big( 4 p_{1}^2 \vec{p}_{1} -3 \vec{p}_{1}\cdot\vec{n} p_{1}^2 \vec{n} + 15 ( \vec{p}_{1}\cdot\vec{n})^{3} \vec{n} \big) \nn\\ 
&& - 2 \vec{p}_{1}\cdot\vec{S}_{1} \vec{p}_{1}\cdot\vec{S}_{2} \big( p_{1}^2 \vec{n} -9 ( \vec{p}_{1}\cdot\vec{n})^{2} \vec{n} \big) - 2 \vec{S}_{1}\cdot\vec{S}_{2} \big( - p_{1}^{4} \vec{n} + \vec{p}_{1}\cdot\vec{n} p_{1}^2 \vec{p}_{1} \nn\\ 
&& + 12 p_{1}^2 ( \vec{p}_{1}\cdot\vec{n})^{2} \vec{n} - 15 ( \vec{p}_{1}\cdot\vec{n})^{3} \vec{p}_{1} \big) + \vec{S}_{2}\cdot\vec{n} \vec{S}_{1} \big( 7 p_{1}^{4} + 24 p_{1}^2 ( \vec{p}_{1}\cdot\vec{n})^{2} \big) \nn\\ 
&& - 6 \vec{p}_{1}\cdot\vec{S}_{2} \vec{S}_{1} \big( \vec{p}_{1}\cdot\vec{n} p_{1}^2 + 5 ( \vec{p}_{1}\cdot\vec{n})^{3} \big) - 7 \vec{S}_{1}\cdot\vec{n} \vec{S}_{2} p_{1}^{4} + 5 \vec{p}_{1}\cdot\vec{S}_{1} \vec{S}_{2} \vec{p}_{1}\cdot\vec{n} p_{1}^2 \Big] \nn\\ 
&& + 	\frac{G}{32 m_{1}{}^3 m_{2} r{}^2} \Big[ 3 \vec{S}_{1}\cdot\vec{n} \vec{S}_{2}\cdot\vec{n} \big( 11 p_{1}^2 \vec{p}_{1}\cdot\vec{p}_{2} \vec{n} + 12 p_{1}^2 \vec{p}_{2}\cdot\vec{n} \vec{p}_{1} + 49 \vec{p}_{1}\cdot\vec{n} \vec{p}_{1}\cdot\vec{p}_{2} \vec{p}_{1} \nn\\ 
&& - 43 \vec{p}_{1}\cdot\vec{n} p_{1}^2 \vec{p}_{2} + 40 \vec{p}_{1}\cdot\vec{p}_{2} ( \vec{p}_{1}\cdot\vec{n})^{2} \vec{n} + 20 \vec{p}_{2}\cdot\vec{n} ( \vec{p}_{1}\cdot\vec{n})^{2} \vec{p}_{1} - 15 ( \vec{p}_{1}\cdot\vec{n})^{3} \vec{p}_{2} \big) \nn\\ 
&& + \vec{p}_{1}\cdot\vec{S}_{1} \vec{S}_{2}\cdot\vec{n} \big( 32 \vec{p}_{1}\cdot\vec{p}_{2} \vec{p}_{1} - 14 p_{1}^2 \vec{p}_{2} -12 p_{1}^2 \vec{p}_{2}\cdot\vec{n} \vec{n} - 63 \vec{p}_{1}\cdot\vec{n} \vec{p}_{1}\cdot\vec{p}_{2} \vec{n} \nn\\ 
&& - 84 \vec{p}_{1}\cdot\vec{n} \vec{p}_{2}\cdot\vec{n} \vec{p}_{1} + 162 ( \vec{p}_{1}\cdot\vec{n})^{2} \vec{p}_{2} \big) \nn\\ 
&& + \vec{p}_{2}\cdot\vec{S}_{1} \vec{S}_{2}\cdot\vec{n} \big( 10 p_{1}^2 \vec{p}_{1} -138 ( \vec{p}_{1}\cdot\vec{n})^{2} \vec{p}_{1} -15 ( \vec{p}_{1}\cdot\vec{n})^{3} \vec{n} \big) - \vec{S}_{1}\cdot\vec{n} \vec{p}_{1}\cdot\vec{S}_{2} \big( 8 \vec{p}_{1}\cdot\vec{p}_{2} \vec{p}_{1} \nn\\ 
&& + 6 p_{1}^2 \vec{p}_{2} -27 p_{1}^2 \vec{p}_{2}\cdot\vec{n} \vec{n} + 60 \vec{p}_{1}\cdot\vec{n} \vec{p}_{1}\cdot\vec{p}_{2} \vec{n} + 111 \vec{p}_{1}\cdot\vec{n} \vec{p}_{2}\cdot\vec{n} \vec{p}_{1} - 168 ( \vec{p}_{1}\cdot\vec{n})^{2} \vec{p}_{2} \nn\\ 
&& + 180 \vec{p}_{2}\cdot\vec{n} ( \vec{p}_{1}\cdot\vec{n})^{2} \vec{n} \big) + \vec{p}_{1}\cdot\vec{S}_{1} \vec{p}_{1}\cdot\vec{S}_{2} \big( 8 \vec{p}_{1}\cdot\vec{p}_{2} \vec{n} + 16 \vec{p}_{2}\cdot\vec{n} \vec{p}_{1} - 145 \vec{p}_{1}\cdot\vec{n} \vec{p}_{2} \nn\\ 
&& + 27 \vec{p}_{1}\cdot\vec{n} \vec{p}_{2}\cdot\vec{n} \vec{n} \big) - \vec{p}_{2}\cdot\vec{S}_{1} \vec{p}_{1}\cdot\vec{S}_{2} \big( 24 p_{1}^2 \vec{n} - 77 \vec{p}_{1}\cdot\vec{n} \vec{p}_{1} -96 ( \vec{p}_{1}\cdot\vec{n})^{2} \vec{n} \big) \nn\\ 
&& + 4 \vec{S}_{1}\cdot\vec{n} \vec{p}_{2}\cdot\vec{S}_{2} \big( 2 p_{1}^2 \vec{p}_{1} -21 ( \vec{p}_{1}\cdot\vec{n})^{2} \vec{p}_{1} + 15 ( \vec{p}_{1}\cdot\vec{n})^{3} \vec{n} \big) + 60 \vec{p}_{1}\cdot\vec{S}_{1} \vec{p}_{2}\cdot\vec{S}_{2} \vec{p}_{1}\cdot\vec{n} \vec{p}_{1} \nn\\ 
&& + \vec{S}_{1}\cdot\vec{S}_{2} \big( 24 p_{1}^2 \vec{p}_{1}\cdot\vec{p}_{2} \vec{n} - 50 p_{1}^2 \vec{p}_{2}\cdot\vec{n} \vec{p}_{1} - 161 \vec{p}_{1}\cdot\vec{n} \vec{p}_{1}\cdot\vec{p}_{2} \vec{p}_{1} \nn\\ 
&& + 165 \vec{p}_{1}\cdot\vec{n} p_{1}^2 \vec{p}_{2} -96 \vec{p}_{1}\cdot\vec{p}_{2} ( \vec{p}_{1}\cdot\vec{n})^{2} \vec{n} + 222 \vec{p}_{2}\cdot\vec{n} ( \vec{p}_{1}\cdot\vec{n})^{2} \vec{p}_{1} - 159 ( \vec{p}_{1}\cdot\vec{n})^{3} \vec{p}_{2} \nn\\ 
&& + 15 \vec{p}_{2}\cdot\vec{n} ( \vec{p}_{1}\cdot\vec{n})^{3} \vec{n} \big) - 12 \vec{S}_{2}\cdot\vec{n} \vec{S}_{1} \big( 2 p_{1}^2 \vec{p}_{1}\cdot\vec{p}_{2} -4 \vec{p}_{1}\cdot\vec{n} p_{1}^2 \vec{p}_{2}\cdot\vec{n} - \vec{p}_{1}\cdot\vec{p}_{2} ( \vec{p}_{1}\cdot\vec{n})^{2} \big) \nn\\ 
&& + 4 \vec{p}_{1}\cdot\vec{S}_{2} \vec{S}_{1} \big( 2 p_{1}^2 \vec{p}_{2}\cdot\vec{n} + 23 \vec{p}_{1}\cdot\vec{n} \vec{p}_{1}\cdot\vec{p}_{2} -18 \vec{p}_{2}\cdot\vec{n} ( \vec{p}_{1}\cdot\vec{n})^{2} \big) \nn\\ 
&& - 12 \vec{p}_{2}\cdot\vec{S}_{2} \vec{S}_{1} \big( 5 \vec{p}_{1}\cdot\vec{n} p_{1}^2 -2 ( \vec{p}_{1}\cdot\vec{n})^{3} \big) - \vec{S}_{1}\cdot\vec{n} \vec{S}_{2} \big( 10 p_{1}^2 \vec{p}_{1}\cdot\vec{p}_{2} -45 \vec{p}_{1}\cdot\vec{n} p_{1}^2 \vec{p}_{2}\cdot\vec{n} \nn\\ 
&& + 36 \vec{p}_{1}\cdot\vec{p}_{2} ( \vec{p}_{1}\cdot\vec{n})^{2} + 15 \vec{p}_{2}\cdot\vec{n} ( \vec{p}_{1}\cdot\vec{n})^{3} \big) + \vec{p}_{1}\cdot\vec{S}_{1} \vec{S}_{2} \big( 22 p_{1}^2 \vec{p}_{2}\cdot\vec{n} \nn\\ 
&& + 45 \vec{p}_{1}\cdot\vec{n} \vec{p}_{1}\cdot\vec{p}_{2} -90 \vec{p}_{2}\cdot\vec{n} ( \vec{p}_{1}\cdot\vec{n})^{2} \big) - \vec{p}_{2}\cdot\vec{S}_{1} \vec{S}_{2} \big( 65 \vec{p}_{1}\cdot\vec{n} p_{1}^2 -87 ( \vec{p}_{1}\cdot\vec{n})^{3} \big) \Big] \nn\\ 
&& - 	\frac{G}{32 m_{1}{}^2 m_{2}{}^2 r{}^2} \Big[ 3 \vec{S}_{1}\cdot\vec{n} \vec{S}_{2}\cdot\vec{n} \big( 32 \vec{p}_{2}\cdot\vec{n} \vec{p}_{1}\cdot\vec{p}_{2} \vec{p}_{1} -40 p_{1}^2 ( \vec{p}_{2}\cdot\vec{n})^{2} \vec{n} \nn\\ 
&& + 5 \vec{p}_{1}\cdot\vec{n} ( \vec{p}_{2}\cdot\vec{n})^{2} \vec{p}_{1} \big) - 3 \vec{p}_{1}\cdot\vec{S}_{1} \vec{S}_{2}\cdot\vec{n} \big( 12 p_{2}^2 \vec{p}_{1} -18 \vec{p}_{2}\cdot\vec{n} \vec{p}_{1}\cdot\vec{p}_{2} \vec{n} + 4 \vec{p}_{1}\cdot\vec{n} p_{2}^2 \vec{n} \nn\\ 
&& + 5 ( \vec{p}_{2}\cdot\vec{n})^{2} \vec{p}_{1} -60 \vec{p}_{1}\cdot\vec{n} ( \vec{p}_{2}\cdot\vec{n})^{2} \vec{n} \big) + 3 \vec{p}_{2}\cdot\vec{S}_{1} \vec{S}_{2}\cdot\vec{n} \big( 40 \vec{p}_{1}\cdot\vec{p}_{2} \vec{p}_{1} -8 p_{1}^2 \vec{p}_{2}\cdot\vec{n} \vec{n} \nn\\ 
&& - 26 \vec{p}_{1}\cdot\vec{n} \vec{p}_{1}\cdot\vec{p}_{2} \vec{n} - 18 \vec{p}_{1}\cdot\vec{n} \vec{p}_{2}\cdot\vec{n} \vec{p}_{1} + 5 \vec{p}_{2}\cdot\vec{n} ( \vec{p}_{1}\cdot\vec{n})^{2} \vec{n} \big) \nn\\ 
&& + 2 \vec{S}_{1}\cdot\vec{n} \vec{p}_{1}\cdot\vec{S}_{2} \big( 17 p_{2}^2 \vec{p}_{1} -27 ( \vec{p}_{2}\cdot\vec{n})^{2} \vec{p}_{1} \big) + 3 \vec{p}_{1}\cdot\vec{S}_{1} \vec{p}_{1}\cdot\vec{S}_{2} \big( 6 p_{2}^2 \vec{n} -53 ( \vec{p}_{2}\cdot\vec{n})^{2} \vec{n} \big) \nn\\ 
&& + 16 \vec{p}_{2}\cdot\vec{S}_{1} \vec{p}_{1}\cdot\vec{S}_{2} \vec{p}_{2}\cdot\vec{n} \vec{p}_{1} + 44 \vec{p}_{1}\cdot\vec{S}_{1} \vec{p}_{2}\cdot\vec{S}_{2} \vec{p}_{2}\cdot\vec{n} \vec{p}_{1} - 8 \vec{p}_{2}\cdot\vec{S}_{1} \vec{p}_{2}\cdot\vec{S}_{2} \vec{p}_{1}\cdot\vec{n} \vec{p}_{1} \nn\\ 
&& - 3 \vec{S}_{1}\cdot\vec{S}_{2} \big( 64 \vec{p}_{2}\cdot\vec{n} \vec{p}_{1}\cdot\vec{p}_{2} \vec{p}_{1} - 2 \vec{p}_{1}\cdot\vec{n} p_{2}^2 \vec{p}_{1} -51 p_{1}^2 ( \vec{p}_{2}\cdot\vec{n})^{2} \vec{n} - 17 \vec{p}_{1}\cdot\vec{n} ( \vec{p}_{2}\cdot\vec{n})^{2} \vec{p}_{1} \big) \nn\\ 
&& + \vec{S}_{2}\cdot\vec{n} \vec{S}_{1} \big( 30 p_{1}^2 p_{2}^2 - 82 ( \vec{p}_{1}\cdot\vec{p}_{2})^{2} -48 \vec{p}_{1}\cdot\vec{n} \vec{p}_{2}\cdot\vec{n} \vec{p}_{1}\cdot\vec{p}_{2} + 30 p_{2}^2 ( \vec{p}_{1}\cdot\vec{n})^{2} \nn\\ 
&& - 3 p_{1}^2 ( \vec{p}_{2}\cdot\vec{n})^{2} -105 ( \vec{p}_{1}\cdot\vec{n})^{2} ( \vec{p}_{2}\cdot\vec{n})^{2} \big) + 18 \vec{p}_{2}\cdot\vec{S}_{2} \vec{S}_{1} \vec{p}_{1}\cdot\vec{n} \vec{p}_{1}\cdot\vec{p}_{2} \nn\\ 
&& + 28 \vec{p}_{1}\cdot\vec{S}_{1} \vec{S}_{2} \vec{p}_{1}\cdot\vec{n} p_{2}^2 + \vec{p}_{2}\cdot\vec{S}_{1} \vec{S}_{2} \big( 46 p_{1}^2 \vec{p}_{2}\cdot\vec{n} \nn\\ 
&& - 134 \vec{p}_{1}\cdot\vec{n} \vec{p}_{1}\cdot\vec{p}_{2} -69 \vec{p}_{2}\cdot\vec{n} ( \vec{p}_{1}\cdot\vec{n})^{2} \big) \Big] + 	\frac{G^2 m_{2}}{32 m_{1}{}^2 r{}^3} \Big[ 6 \vec{S}_{1}\cdot\vec{n} \vec{S}_{2}\cdot\vec{n} \big( 34 p_{1}^2 \vec{n} \nn\\ 
&& + 77 \vec{p}_{1}\cdot\vec{n} \vec{p}_{1} \big) + 6 \vec{p}_{1}\cdot\vec{S}_{1} \vec{S}_{2}\cdot\vec{n} \big( 53 \vec{p}_{1} -40 \vec{p}_{1}\cdot\vec{n} \vec{n} \big) - 2 \vec{S}_{1}\cdot\vec{n} \vec{p}_{1}\cdot\vec{S}_{2} \big( 280 \vec{p}_{1} \nn\\ 
&& + 163 \vec{p}_{1}\cdot\vec{n} \vec{n} \big) + 354 \vec{p}_{1}\cdot\vec{S}_{1} \vec{p}_{1}\cdot\vec{S}_{2} \vec{n} - \vec{S}_{1}\cdot\vec{S}_{2} \big( 378 p_{1}^2 \vec{n} \nn\\ 
&& + 301 \vec{p}_{1}\cdot\vec{n} \vec{p}_{1} -465 ( \vec{p}_{1}\cdot\vec{n})^{2} \vec{n} \big) - \vec{S}_{2}\cdot\vec{n} \vec{S}_{1} \big( 298 p_{1}^2 + 601 ( \vec{p}_{1}\cdot\vec{n})^{2} \big) \nn\\ 
&& + 861 \vec{p}_{1}\cdot\vec{S}_{2} \vec{S}_{1} \vec{p}_{1}\cdot\vec{n} + 680 \vec{S}_{1}\cdot\vec{n} \vec{S}_{2} p_{1}^2 - 640 \vec{p}_{1}\cdot\vec{S}_{1} \vec{S}_{2} \vec{p}_{1}\cdot\vec{n} \Big] \nn\\ 
&& - 	\frac{G^2}{96 m_{2} r{}^3} \Big[ 2 \vec{S}_{1}\cdot\vec{n} \vec{S}_{2}\cdot\vec{n} \big( 2524 \vec{p}_{1}\cdot\vec{p}_{2} \vec{n} - 425 \vec{p}_{2}\cdot\vec{n} \vec{p}_{1} + 174 \vec{p}_{1}\cdot\vec{n} \vec{p}_{2}\cdot\vec{n} \vec{n} \big) \nn\\ 
&& + 2 \vec{p}_{2}\cdot\vec{S}_{1} \vec{S}_{2}\cdot\vec{n} \big( 566 \vec{p}_{1} -6767 \vec{p}_{1}\cdot\vec{n} \vec{n} \big) + 4333 \vec{p}_{2}\cdot\vec{S}_{1} \vec{p}_{1}\cdot\vec{S}_{2} \vec{n} \nn\\ 
&& - 2 \vec{S}_{1}\cdot\vec{n} \vec{p}_{2}\cdot\vec{S}_{2} \big( 817 \vec{p}_{1} -910 \vec{p}_{1}\cdot\vec{n} \vec{n} \big) - 803 \vec{p}_{1}\cdot\vec{S}_{1} \vec{p}_{2}\cdot\vec{S}_{2} \vec{n} - \vec{S}_{1}\cdot\vec{S}_{2} \big( 2690 \vec{p}_{1}\cdot\vec{p}_{2} \vec{n} \nn\\ 
&& - 955 \vec{p}_{2}\cdot\vec{n} \vec{p}_{1} -6896 \vec{p}_{1}\cdot\vec{n} \vec{p}_{2}\cdot\vec{n} \vec{n} \big) + 2884 \vec{p}_{1}\cdot\vec{S}_{2} \vec{S}_{1} \vec{p}_{2}\cdot\vec{n} \nn\\ 
&& + \vec{S}_{1}\cdot\vec{n} \vec{S}_{2} \big( 3949 \vec{p}_{1}\cdot\vec{p}_{2} -1246 \vec{p}_{1}\cdot\vec{n} \vec{p}_{2}\cdot\vec{n} \big) - 2819 \vec{p}_{1}\cdot\vec{S}_{1} \vec{S}_{2} \vec{p}_{2}\cdot\vec{n} \Big] \nn\\ 
&& + 	\frac{G^2}{96 m_{1} r{}^3} \Big[ 2 \vec{S}_{1}\cdot\vec{n} \vec{S}_{2}\cdot\vec{n} \big( 3301 p_{1}^2 \vec{n} + 3176 \vec{p}_{1}\cdot\vec{n} \vec{p}_{1} + 1699 \vec{p}_{2}\cdot\vec{n} \vec{p}_{1} -357 ( \vec{p}_{1}\cdot\vec{n})^{2} \vec{n} \big) \nn\\ 
&& + \vec{p}_{1}\cdot\vec{S}_{1} \vec{S}_{2}\cdot\vec{n} \big( 1355 \vec{p}_{1} -13178 \vec{p}_{1}\cdot\vec{n} \vec{n} \big) - \vec{p}_{2}\cdot\vec{S}_{1} \vec{S}_{2}\cdot\vec{n} \big( 347 \vec{p}_{1} + 6730 \vec{p}_{1}\cdot\vec{n} \vec{n} \big) \nn\\ 
&& - \vec{S}_{1}\cdot\vec{n} \vec{p}_{1}\cdot\vec{S}_{2} \big( 3163 \vec{p}_{1} + 2426 \vec{p}_{1}\cdot\vec{n} \vec{n} \big) + 4142 \vec{p}_{1}\cdot\vec{S}_{1} \vec{p}_{1}\cdot\vec{S}_{2} \vec{n} \nn\\ 
&& + \vec{S}_{1}\cdot\vec{n} \vec{p}_{2}\cdot\vec{S}_{2} \big( 1015 \vec{p}_{1} + 1922 \vec{p}_{1}\cdot\vec{n} \vec{n} \big) - \vec{S}_{1}\cdot\vec{S}_{2} \big( 3632 p_{1}^2 \vec{n} + 1477 \vec{p}_{1}\cdot\vec{n} \vec{p}_{1} \nn\\ 
&& + 2081 \vec{p}_{2}\cdot\vec{n} \vec{p}_{1} -6581 ( \vec{p}_{1}\cdot\vec{n})^{2} \vec{n} \big) - \vec{S}_{2}\cdot\vec{n} \vec{S}_{1} \big( 2603 p_{1}^2 -2462 ( \vec{p}_{1}\cdot\vec{n})^{2} \big) \nn\\ 
&& + \vec{p}_{1}\cdot\vec{S}_{2} \vec{S}_{1} \big( 929 \vec{p}_{1}\cdot\vec{n} + 976 \vec{p}_{2}\cdot\vec{n} \big) + \vec{S}_{1}\cdot\vec{n} \vec{S}_{2} \big( 2593 p_{1}^2 \nn\\ 
&& - 4751 \vec{p}_{1}\cdot\vec{p}_{2} -568 \vec{p}_{1}\cdot\vec{n} \vec{p}_{2}\cdot\vec{n} - 4909 ( \vec{p}_{1}\cdot\vec{n})^{2} \big) + \vec{p}_{1}\cdot\vec{S}_{1} \vec{S}_{2} \big( 1238 \vec{p}_{1}\cdot\vec{n} \nn\\ 
&& + 1633 \vec{p}_{2}\cdot\vec{n} \big) \Big] - 	\frac{G^3 m_{1}{}^2}{480 r{}^4} \Big[ 21400 \vec{S}_{1}\cdot\vec{n} \vec{S}_{2}\cdot\vec{n} \vec{n} - 7013 \vec{S}_{1}\cdot\vec{S}_{2} \vec{n} - 8168 \vec{S}_{2}\cdot\vec{n} \vec{S}_{1} \Big] \nn\\ 
&& + 	\frac{1091 G^3 m_{1} m_{2}}{32 r{}^4} \vec{S}_{2}\cdot\vec{n} \vec{S}_{1} + 	\frac{277 G^3 m_{2}{}^2}{480 r{}^4} \vec{S}_{2}\cdot\vec{n} \vec{S}_{1} .
\eea

\section{COM Hamiltonians}
\label{newcomhams}

As noted in section \ref{spechamsnobs}, the NLO quartic-in-spin Hamiltonians restricted to the COM frame are written as:
\bea
\tilde{H}^{\text{NLO}}_{\text{S}^4} &=&
C_{1\text{ES}^2} \tilde{H}^{\text{NLO}}_{(\text{ES}_1^2 ) \text{S}_1^2} 
+C_{1\text{ES}^2}^2 \tilde{H}^{\text{NLO}}_{(\text{ES}_1^2 )^2} 
+ C_{1\text{BS}^3} \tilde{H}^{\text{NLO}}_{(\text{BS}_1^3) \text{S}_1 } 
+C_{1\text{ES}^4} \tilde{H}^{\text{NLO}}_{\text{ES}_1^4} 
+C_{1\text{E}^2\text{S}^4} \tilde{H}^{\text{NLO}}_{\text{E}^2\text{S}_1^4}\nn\\
&& +\tilde{H}^{\text{NLO}}_{\text{S}_1^3 \text{S}_2}
+C_{1\text{ES}^2} \tilde{H}^{\text{NLO}}_{(\text{ES}_1^2 ) \text{S}_1 \text{S}_2} 
+ C_{1\text{ES}^2}^2 \tilde{H}^{\text{NLO}}_{C_{\text{ES}^2_1}^2 \text{S}_1^3 \text{S}_2 } 
+ C_{1\text{BS}^3} \tilde{H}^{\text{NLO}}_{(\text{BS}_1^3) \text{S}_2 }\nn\\
&& +\tilde{H}^{\text{NLO}}_{\text{S}_1^2 \text{S}_2^2}
+C_{1\text{ES}^2} \tilde{H}^{\text{NLO}}_{(\text{ES}_1^2 )  \text{S}_2^2} 
+C_{1\text{ES}^2} C_{2\text{ES}^2} \tilde{H}^{\text{NLO}}_{(\text{ES}_1^2 )(\text{ES}_2^2 ) }
+  (1 \leftrightarrow 2),
\eea
with
\bea
\tilde{H}^{\text{NLO}}_{(\text{ES}_1^2 ) \text{S}_1^2}&=& \frac{\nu^2 \tilde{S}_1^4 }{\tilde{r}^6} 
\left[
\frac{3}{2}-\frac{3 \nu }{2} 
+ \tilde{p}_r^2 \tilde{r} \left(
\frac{45}{16}-\frac{21 \nu }{8}
\right) 
+ \frac{1}{ \nu q} \left(
-\frac{3 \nu ^2}{2}+\frac{9 \nu }{2}-\frac{3}{2}
\right.\right.\nn\\
&& \left.\left.
+ \tilde{p}_r^2 \tilde{r} \left(
-\frac{39 \nu ^2}{16}+\frac{33 \nu }{4}-\frac{45}{16}
\right) \right)  \right] 
+\frac{\nu^2 \tilde{S}_1^2 (\tilde{\vec{S}}_1 \cdot \tilde{\vec{L}} )^2}{\tilde{r}^7} \left[
-\frac{3 \nu }{4}
+  \frac{1}{  q} \left(\frac{3}{4}-\frac{3 \nu }{2} \right)  \right]\nn\\
&&+ \frac{\nu^2 \tilde{S}_1^2 (\tilde{\vec{S}}_1 \cdot \vec{n})^2 }{\tilde{r}^6} \left[
9 \nu -9
+\frac{\tilde{L}^2}{\tilde{r}} \left(
\frac{69 \nu }{8}-\frac{135}{16}
\right) + \tilde{p}_r^2 \tilde{r} \left(
\frac{63 \nu }{4}-\frac{135}{8}
\right) \right.\nn\\
&&    + \frac{1}{ \nu q} \left(
9 \nu ^2-27 \nu +9
+\frac{\tilde{L}^2}{\tilde{r}} \left(
\frac{141 \nu ^2}{16}-\frac{51 \nu }{2}+\frac{135}{16}
\right)  \right.\nn\\
&&\left.\left. + \tilde{p}_r^2 \tilde{r} \left(
\frac{117 \nu ^2}{8}-\frac{99 \nu }{2}+\frac{135}{8}
\right) \right)  \right] \nn\\
&&+ \frac{\nu^2  (\tilde{\vec{S}}_1 \cdot \vec{n})^4 }{\tilde{r}^6} \left[
\frac{15}{2}-\frac{15 \nu }{2}
+\frac{\tilde{L}^2}{\tilde{r}} \left(
\frac{195}{16}-\frac{105 \nu }{8}
\right) + \tilde{p}_r^2 \tilde{r} \left(
\frac{225}{16}-\frac{105 \nu }{8}
\right) \right.\nn\\
&&    + \frac{1}{ \nu q} \left(
-\frac{15 \nu ^2}{2}+\frac{45 \nu }{2}-\frac{15}{2}
+\frac{\tilde{L}^2}{\tilde{r}} \left(
-\frac{225 \nu ^2}{16}+\frac{75 \nu }{2}-\frac{195}{16}
\right)  \right.\nn\\
&&\left.\left. + \tilde{p}_r^2 \tilde{r} \left(
-\frac{195 \nu ^2}{16}+\frac{165 \nu }{4}-\frac{225}{16}
\right) \right)  \right]\nn\\
&&+\frac{\nu^2 (\tilde{\vec{S}}_1 \cdot \vec{n})^2 (\tilde{\vec{S}}_1 \cdot \tilde{\vec{L}} )^2}{\tilde{r}^7} \left[
\frac{45}{8}
+  \frac{1}{ \nu q} \left(
\frac{45 \nu ^2}{8}+\frac{45 \nu }{4}-\frac{45}{8}
\right)  \right]\nn\\
&&+\frac{\nu^2 \tilde{p}_r \tilde{S}_1^2 \tilde{\vec{S}}_1 \cdot \vec{n} \tilde{\vec{S}}_1 \times \tilde{\vec{L}} \cdot \vec{n} }{\tilde{r}^6} \left[
\frac{45}{8}-6 \nu
+  \frac{1}{ \nu q} \left(
-\frac{51 \nu ^2}{8}+\frac{69 \nu }{4}-\frac{45}{8}
\right)  \right]\nn\\
&&+\frac{\nu^2 \tilde{p}_r  (\tilde{\vec{S}}_1 \cdot \vec{n})^3 \tilde{\vec{S}}_1 \times \tilde{\vec{L}} \cdot \vec{n} }{\tilde{r}^6} \left[
\frac{15}{8}
+  \frac{1}{ \nu q} \left(
\frac{15 \nu ^2}{8}+\frac{15 \nu }{4}-\frac{15}{8}
\right)  \right],
\eea
\bea
\tilde{H}^{\text{NLO}}_{(\text{ES}_1^2 )^2}&=& \frac{\nu^3 \tilde{S}_1^4 }{\tilde{r}^6} \left[
-\frac{1}{8}
+ \frac{1}{ \nu q} \left(
\frac{1}{8}-\frac{\nu }{8}
\right)  \right] \nn\\
&&+ \frac{\nu^2 \tilde{S}_1^2 (\tilde{\vec{S}}_1 \cdot \vec{n})^2 }{\tilde{r}^6} \left[
\frac{21 \nu }{4}-\frac{9}{2}
+ \frac{1}{ \nu q} \left(
\frac{21 \nu ^2}{4}-\frac{57 \nu }{4}+\frac{9}{2}
\right)  \right] \nn\\
&&+ \frac{\nu^2  (\tilde{\vec{S}}_1 \cdot \vec{n})^4 }{\tilde{r}^6} \left[
\frac{15}{2}-\frac{69 \nu }{8}
+ \frac{1}{ \nu q} \left(
-\frac{69 \nu ^2}{8}+\frac{189 \nu }{8}-\frac{15}{2}
\right)  \right],
\eea
\bea
\tilde{H}^{\text{NLO}}_{(\text{BS}_1^3) \text{S}_1 } &=&  \frac{\nu^2 \tilde{S}_1^4 }{\tilde{r}^6} \left[
-\frac{\nu }{2}-\frac{3}{2}
+ \tilde{p}_r^2 \tilde{r} \left(
\frac{3 \nu }{2}-\frac{3}{2}
\right) 
+ \frac{1}{ \nu q} \left(
-\frac{\nu ^2}{2}-\frac{5 \nu }{2}+\frac{3}{2}
\right.\right.\nn\\
&& \left.\left.
+ \tilde{p}_r^2 \tilde{r} \left(
\frac{3 \nu ^2}{2}-\frac{9 \nu }{2}+\frac{3}{2}
\right) \right)  \right] 
+\frac{\nu^2 \tilde{S}_1^2 (\tilde{\vec{S}}_1 \cdot \tilde{\vec{L}} )^2}{\tilde{r}^7} \left[
\frac{3 \nu }{2}-\frac{3}{2}
+  \frac{1}{\nu q} \left(\frac{3 \nu ^2}{2}-\frac{9 \nu }{2}+\frac{3}{2} \right)  \right]\nn\\
&&+ \frac{\nu^2 \tilde{S}_1^2 (\tilde{\vec{S}}_1 \cdot \vec{n})^2 }{\tilde{r}^6} \left[
6 \nu +9
+\frac{\tilde{L}^2}{\tilde{r}} \left(
6-6 \nu
\right) + \tilde{p}_r^2 \tilde{r} \left(
9-9 \nu
\right) \right.\nn\\
&&  \left.  + \frac{1}{ \nu q} \left(
6 \nu ^2+12 \nu -9
+\frac{\tilde{L}^2}{\tilde{r}} \left(
-6 \nu ^2+18 \nu -6
\right)   + \tilde{p}_r^2 \tilde{r} \left(
-9 \nu ^2+27 \nu -9
\right) \right)  \right] \nn\\
&&+ \frac{\nu^2  (\tilde{\vec{S}}_1 \cdot \vec{n})^4 }{\tilde{r}^6} \left[
-\frac{15 \nu }{2}-\frac{15}{2}
+\frac{\tilde{L}^2}{\tilde{r}} \left(
10 \nu -10
\right) + \tilde{p}_r^2 \tilde{r} \left(
\frac{15 \nu }{2}-\frac{15}{2}
\right) \right.\nn\\
&&    + \frac{1}{ \nu q} \left(
-\frac{15 \nu ^2}{2}-\frac{15 \nu }{2}+\frac{15}{2}
+\frac{\tilde{L}^2}{\tilde{r}} \left(
10 \nu ^2-30 \nu +10
\right)  \right.\nn\\
&&\left.\left. + \tilde{p}_r^2 \tilde{r} \left(
\frac{15 \nu ^2}{2}-\frac{45 \nu }{2}+\frac{15}{2}
\right) \right)  \right]\nn\\
&&+\frac{\nu^2 (\tilde{\vec{S}}_1 \cdot \vec{n})^2 (\tilde{\vec{S}}_1 \cdot \tilde{\vec{L}} )^2}{\tilde{r}^7} \left[
\frac{15}{2}-\frac{15 \nu }{2}
+  \frac{1}{ \nu q} \left(
-\frac{15 \nu ^2}{2}+\frac{45 \nu }{2}-\frac{15}{2}
\right)  \right]\nn\\
&&+\frac{\nu^2 \tilde{p}_r \tilde{S}_1^2 \tilde{\vec{S}}_1 \cdot \vec{n} \tilde{\vec{S}}_1 \times \tilde{\vec{L}} \cdot \vec{n} }{\tilde{r}^6} \left[
\frac{9 \nu }{2}-\frac{9}{2}
+  \frac{1}{ \nu q} \left(
\frac{9 \nu ^2}{2}-\frac{27 \nu }{2}+\frac{9}{2}
\right)  \right]\nn\\
&&+\frac{\nu^2 \tilde{p}_r  (\tilde{\vec{S}}_1 \cdot \vec{n})^3 \tilde{\vec{S}}_1 \times \tilde{\vec{L}} \cdot \vec{n} }{\tilde{r}^6} \left[
\frac{5}{2}-\frac{5 \nu }{2}
+  \frac{1}{ \nu q} \left(
-\frac{5 \nu ^2}{2}+\frac{15 \nu }{2}-\frac{5}{2}
\right)  \right],
\eea
\bea
\tilde{H}^{\text{NLO}}_{\text{ES}_1^4} &=& \frac{\nu^2 \tilde{S}_1^4 }{\tilde{r}^6} \left[
\frac{5 \nu }{2}-\frac{23}{8}
+\frac{\tilde{L}^2}{\tilde{r}} \left(
\frac{15 \nu }{16}-\frac{3}{16}
\right)  + \tilde{p}_r^2 \tilde{r} \left(
\frac{9 \nu }{8}+\frac{9}{16}
\right) \right.\nn\\.
&& \left.  + \frac{1}{ \nu q} \left(
\frac{5 \nu ^2}{2}-\frac{33 \nu }{4}+\frac{23}{8}
+\frac{\tilde{L}^2}{\tilde{r}} \left(
\frac{9 \nu ^2}{8}-\frac{21 \nu }{16}+\frac{3}{16}
\right) + \tilde{p}_r^2 \tilde{r} \left(
\frac{9 \nu ^2}{4}-\frac{9}{16}
\right) \right)  \right] \nn\\
&&+\frac{\nu^2 \tilde{S}_1^2 (\tilde{\vec{S}}_1 \cdot \tilde{\vec{L}} )^2}{\tilde{r}^7} \left[
\frac{3}{4}-\frac{3 \nu }{4}
+  \frac{1}{ \nu q} \left( 
-\frac{3 \nu ^2}{4}+\frac{9 \nu }{4}-\frac{3}{4} \right)  \right]\nn\\
&&+ \frac{\nu^2 \tilde{S}_1^2 (\tilde{\vec{S}}_1 \cdot \vec{n})^2 }{\tilde{r}^6} \left[
30-\frac{105 \nu }{4}
+\frac{\tilde{L}^2}{\tilde{r}} \left(
-\frac{51 \nu }{8}-\frac{9}{8}
\right) + \tilde{p}_r^2 \tilde{r} \left(
-\frac{57 \nu }{4}-\frac{21}{8}
\right) \right.\nn\\
&&    + \frac{1}{ \nu q} \left(
-\frac{105 \nu ^2}{4}+\frac{345 \nu }{4}-30
+\frac{\tilde{L}^2}{\tilde{r}} \left(
-\frac{33 \nu ^2}{4}+\frac{33 \nu }{8}+\frac{9}{8}
\right)  \right.\nn\\
&&\left.\left. + \tilde{p}_r^2 \tilde{r} \left(
-\frac{51 \nu ^2}{2}+9 \nu +\frac{21}{8}
\right) \right)  \right] \nn\\
&&+ \frac{\nu^2  (\tilde{\vec{S}}_1 \cdot \vec{n})^4 }{\tilde{r}^6} \left[
\frac{125 \nu }{4}-\frac{285}{8}
+\frac{\tilde{L}^2}{\tilde{r}} \left(
\frac{95 \nu }{16}+\frac{45}{16}
\right) + \tilde{p}_r^2 \tilde{r} \left(
\frac{145 \nu }{8}+\frac{25}{16}
\right) \right.\nn\\
&&    + \frac{1}{ \nu q} \left(
\frac{125 \nu ^2}{4}-\frac{205 \nu }{2}+\frac{285}{8}
+\frac{\tilde{L}^2}{\tilde{r}} \left(
\frac{65 \nu ^2}{8}-\frac{5 \nu }{16}-\frac{45}{16}
\right)  \right.\nn\\
&&\left.\left. + \tilde{p}_r^2 \tilde{r} \left(
\frac{125 \nu ^2}{4}-15 \nu -\frac{25}{16}
\right) \right)  \right]\nn\\
&&+\frac{\nu^2 (\tilde{\vec{S}}_1 \cdot \vec{n})^2 (\tilde{\vec{S}}_1 \cdot \tilde{\vec{L}} )^2}{\tilde{r}^7} \left[
\frac{15 \nu }{4}-\frac{15}{4}
+  \frac{1}{ \nu q} \left(
\frac{15 \nu ^2}{4}-\frac{45 \nu }{4}+\frac{15}{4}
\right)  \right]\nn\\
&&+\frac{\nu^2 \tilde{p}_r \tilde{S}_1^2 \tilde{\vec{S}}_1 \cdot \vec{n} \tilde{\vec{S}}_1 \times \tilde{\vec{L}} \cdot \vec{n} }{\tilde{r}^6} \left[
\frac{3 \nu }{2}+\frac{9}{4}
+  \frac{1}{ \nu q} \left(
\frac{21 \nu ^2}{4}+3 \nu -\frac{9}{4}
\right)  \right]\nn\\
&&+\frac{\nu^2 \tilde{p}_r  (\tilde{\vec{S}}_1 \cdot \vec{n})^3 \tilde{\vec{S}}_1 \times \tilde{\vec{L}} \cdot \vec{n} }{\tilde{r}^6} \left[
-\frac{15 \nu }{2}-\frac{5}{4}
+  \frac{1}{ \nu q} \left(
-\frac{65 \nu ^2}{4}+5 \nu +\frac{5}{4}
\right)  \right],
\eea
\bea
\tilde{H}^{\text{NLO}}_{\text{E}^2\text{S}_1^4}&=& \frac{27\nu^2 \tilde{S}_1^4 }{35\tilde{r}^6} \left[
\frac{1}{8}-\frac{\nu }{8}  
+ \frac{1}{ \nu q} \left(
-\frac{\nu ^2}{8}+\frac{3\nu }{8}-\frac{1}{8}
\right)  \right] \nn\\
&&+ \frac{9\nu^2 \tilde{S}_1^2 (\tilde{\vec{S}}_1 \cdot \vec{n})^2 }{7\tilde{r}^6} \left[
\frac{3\nu }{4}-\frac{3}{4}
+ \frac{1}{ \nu q} \left(
\frac{3\nu ^2}{4}-\frac{9 \nu }{4}+\frac{3}{4}
\right)  \right] \nn\\
&&+ \frac{\nu^2  (\tilde{\vec{S}}_1 \cdot \vec{n})^4 }{\tilde{r}^6} \left[
\frac{9}{8}-\frac{9 \nu }{8} 
+ \frac{1}{ \nu q} \left(
-\frac{9 \nu ^2}{8}+\frac{27 \nu }{8}-\frac{9}{8}
\right)  \right],
\eea
\bea
\tilde{H}^{\text{NLO}}_{\text{S}_1^3 \text{S}_2}&=&\frac{\nu^3 \tilde{S}_1^2  \tilde{\vec{S}}_1 \cdot \tilde{\vec{S}}_2 }{\tilde{r}^6}\left[
3 
+ \frac{45  }{8} \tilde{p}_r^2 \tilde{r} 
+ \frac{1}{\nu q} \left(
3 \nu -3
+ \tilde{p}_r^2 \tilde{r} \left(
\frac{21 \nu }{4}-\frac{45}{8}
\right)\right) \right]\nn\\
&&+\frac{\nu^3 \tilde{S}_1^2  \tilde{\vec{S}}_1 \cdot \tilde{\vec{L} }  \tilde{\vec{S}}_2 \cdot \tilde{\vec{L} }  }{\tilde{r}^7}\left[
\frac{3  }{2}
+ \frac{1}{\nu q} \left(
\frac{3 \nu }{2}-\frac{3}{2}
\right) \right] 
+\frac{\nu^3   (\tilde{\vec{S}}_1 \cdot \tilde{\vec{L} })^2  \tilde{\vec{S}}_1 \cdot \tilde{\vec{S} }_2   }{\tilde{r}^7}\left[
-\frac{3  }{2}
+\frac{3}{2} \frac{1}{\nu q}   \right]\nn\\
&&+\frac{\nu^3 \tilde{S}_1^2   \tilde{\vec{S}}_1 \cdot \vec{n}  \tilde{\vec{S}}_2 \cdot \vec{n} }{\tilde{r}^6}\left[
-12 
-6  \frac{\tilde{L}^2}{\tilde{r}}
-\frac{177  }{8} \tilde{p}_r^2 \tilde{r} \right.\nn\\
&&\left.+ \frac{1}{\nu q} \left(
12-12 \nu
+ \frac{\tilde{L}^2}{\tilde{r}} \left(
6-6 \nu
\right)+ \tilde{p}_r^2 \tilde{r} \left(
\frac{177}{8}-\frac{81 \nu }{4}
\right)\right) \right]\nn\\
&&+\frac{\nu^3    (\tilde{\vec{S}}_1 \cdot \vec{n})^2  \tilde{\vec{S}}_1 \cdot \tilde{\vec{S}}_2 }{\tilde{r}^6}\left[
-6 
-\frac{87  }{8} \frac{\tilde{L}^2}{\tilde{r}}
-\frac{93  }{8} \tilde{p}_r^2 \tilde{r} \right.\nn\\
&&\left.+ \frac{1}{\nu q} \left(
6-6 \nu
+ \frac{\tilde{L}^2}{\tilde{r}} \left(
\frac{87}{8}-\frac{45 \nu }{4}
\right)+ \tilde{p}_r^2 \tilde{r} \left(
\frac{93}{8}-\frac{45 \nu }{4}
\right)\right) \right]\nn\\
&&+\frac{\nu^3    (\tilde{\vec{S}}_1 \cdot \vec{n})^3  \tilde{\vec{S}}_2 \cdot  \vec{n} }{\tilde{r}^6}\left[
15
+\frac{195}{8} \frac{\tilde{L}^2}{\tilde{r}}
+\frac{225}{8} \tilde{p}_r^2 \tilde{r} \right.\nn\\
&&\left.+ \frac{1}{\nu q} \left(
15 \nu -15
+ \frac{\tilde{L}^2}{\tilde{r}} \left(
\frac{105 \nu }{4}-\frac{195}{8}
\right)+ \tilde{p}_r^2 \tilde{r} \left(
\frac{105 \nu }{4}-\frac{225}{8}
\right)\right) \right]\nn\\
&&+\frac{\nu^3   (\tilde{\vec{S}}_1 \cdot \vec{n})^2 \tilde{\vec{S}}_1 \cdot \tilde{\vec{L} } \tilde{\vec{S}}_2 \cdot \tilde{\vec{L} }      }{\tilde{r}^7}\left[
\frac{15}{4}
-\frac{15}{4} \frac{1}{\nu q}   \right]
+\frac{\nu^3   \tilde{\vec{S}}_1 \cdot \vec{n} \tilde{\vec{S}}_2 \cdot \vec{n} (\tilde{\vec{S}}_1 \cdot \tilde{\vec{L} }  )^2 }{\tilde{r}^7}\left[
\frac{15}{2}
-\frac{15}{2} \frac{1}{\nu q}   \right]
\nn\\
&&+\frac{\nu^3 \tilde{p}_r \tilde{S}_1^2  \tilde{\vec{S}}_1 \cdot \vec{n}    \tilde{\vec{S}}_2 \times \tilde{\vec{L} }   \cdot \vec{n}   }{\tilde{r}^6}\left[
\frac{39}{4}
+ \frac{1}{\nu q}\left(
6 \nu -\frac{39}{4}
\right)   \right]\nn \\
&&+\frac{\nu^3 \tilde{p}_r  (\tilde{\vec{S}}_1 \cdot \vec{n})^3   \tilde{\vec{S}}_2 \times \tilde{\vec{L} }   \cdot \vec{n}   }{\tilde{r}^6}\left[
-\frac{15}{4}
+\frac{15}{4} \frac{1}{\nu q}
\right]\nn\\
&& +\frac{\nu^3 \tilde{p}_r \tilde{S}_1^2  \tilde{\vec{S}}_2 \cdot \vec{n}    \tilde{\vec{S}}_1 \times \tilde{\vec{L} }   \cdot \vec{n}   }{\tilde{r}^6}\left[
6
+ \frac{1}{\nu q}\left(
6 \nu -6
\right)   \right]\nn \\
&&+\frac{\nu^3 \tilde{p}_r \tilde{\vec{S}}_1  \cdot\tilde{\vec{S}}_2  \tilde{\vec{S}}_1 \cdot \vec{n}    \tilde{\vec{S}}_1 \times \tilde{\vec{L} }   \cdot \vec{n}   }{\tilde{r}^6}\left[
-\frac{9}{2}
+\frac{9}{2} \frac{1}{\nu q}
\right]\nn \\
&&+\frac{\nu^3 \tilde{p}_r (\tilde{\vec{S}}_1  \cdot \vec{n})^2  \tilde{\vec{S}}_2 \cdot \vec{n}    \tilde{\vec{S}}_1 \times \tilde{\vec{L} }   \cdot \vec{n}   }{\tilde{r}^6}\left[
\frac{15}{2}
-\frac{15}{2} \frac{1}{\nu q}
\right],
\eea
\bea
\tilde{H}^{\text{NLO}}_{(\text{ES}_1^2 ) \text{S}_1 \text{S}_2} &=&\frac{\nu^3 \tilde{S}_1^2  \tilde{\vec{S}}_1 \cdot \tilde{\vec{S}}_2 }{\tilde{r}^6}\left[
\frac{5}{2}
-\frac{9}{2} \tilde{p}_r^2 \tilde{r} 
+ \frac{1}{\nu q} \left(
\frac{5 \nu }{2}+6
+ \tilde{p}_r^2 \tilde{r} \left(
\frac{9}{2}-\frac{33 \nu }{8}
\right)\right) \right]\nn\\
&&+\frac{\nu^3 \tilde{S}_1^2  \tilde{\vec{S}}_1 \cdot \tilde{\vec{L} }  \tilde{\vec{S}}_2 \cdot \tilde{\vec{L} }  }{\tilde{r}^7}\left[
-\frac{3}{2}
+ \frac{1}{\nu q} \left(
\frac{3}{2}-3 \nu
\right) \right] \nn\\
&& +\frac{\nu^3   (\tilde{\vec{S}}_1 \cdot \tilde{\vec{L} })^2  \tilde{\vec{S}}_1 \cdot \tilde{\vec{S} }_2   }{\tilde{r}^7}\left[
-3
+ \frac{1}{\nu q} \left(
3-3 \nu
\right)  \right]\nn\\
&&+\frac{\nu^3 \tilde{S}_1^2   \tilde{\vec{S}}_1 \cdot \vec{n}  \tilde{\vec{S}}_2 \cdot \vec{n} }{\tilde{r}^6}\left[
-\frac{33}{2}
+6  \frac{\tilde{L}^2}{\tilde{r}}
+\frac{33}{2} \tilde{p}_r^2 \tilde{r} \right.\nn\\
&&\left.+ \frac{1}{\nu q} \left(
-\frac{33 \nu }{2}-12
+ \frac{\tilde{L}^2}{\tilde{r}} \left(
\frac{51 \nu }{8}-6
\right)+ \tilde{p}_r^2 \tilde{r} \left(
\frac{129 \nu }{8}-\frac{33}{2}
\right)\right) \right]\nn\\
&&+\frac{\nu^3    (\tilde{\vec{S}}_1 \cdot \vec{n})^2  \tilde{\vec{S}}_1 \cdot \tilde{\vec{S}}_2 }{\tilde{r}^6}\left[
-\frac{33}{2}
+12 \frac{\tilde{L}^2}{\tilde{r}}
+\frac{21}{2} \tilde{p}_r^2 \tilde{r} \right.\nn\\
&&\left.+ \frac{1}{\nu q} \left(
-\frac{33 \nu }{2}-15
+ \frac{\tilde{L}^2}{\tilde{r}} \left(
12 \nu -12
\right)+ \tilde{p}_r^2 \tilde{r} \left(
\frac{69 \nu }{8}-\frac{21}{2}
\right)\right) \right]\nn\\
&&+\frac{\nu^3    (\tilde{\vec{S}}_1 \cdot \vec{n})^3  \tilde{\vec{S}}_2 \cdot  \vec{n} }{\tilde{r}^6}\left[
\frac{93}{2}
-30 \frac{\tilde{L}^2}{\tilde{r}}
-\frac{45}{2} \tilde{p}_r^2 \tilde{r} \right.\nn\\
&&\left.+ \frac{1}{\nu q} \left(
\frac{93 \nu }{2}+15
+ \frac{\tilde{L}^2}{\tilde{r}} \left(
30-\frac{255 \nu }{8}
\right)+ \tilde{p}_r^2 \tilde{r} \left(
\frac{45}{2}-\frac{165 \nu }{8}
\right)\right) \right]\nn\\
&&+\frac{\nu^3   (\tilde{\vec{S}}_1 \cdot \vec{n})^2 \tilde{\vec{S}}_1 \cdot \tilde{\vec{L} } \tilde{\vec{S}}_2 \cdot \tilde{\vec{L} }      }{\tilde{r}^7}\left[
\frac{15}{2}
+ \frac{1}{\nu q} \left(
15 \nu -\frac{15}{2}
\right)  \right]\nn\\
&&+\frac{\nu^3   \tilde{\vec{S}}_1 \cdot \vec{n} \tilde{\vec{S}}_2 \cdot \vec{n}( \tilde{\vec{S}}_1 \cdot \tilde{\vec{L} } )^2   }{\tilde{r}^7}\left[
15
+\frac{1}{\nu q} \left(
\frac{75 \nu }{4}-15
\right)   \right]
\nn\\
&&+\frac{\nu^3 \tilde{p}_r \tilde{S}_1^2  \tilde{\vec{S}}_1 \cdot \vec{n}    \tilde{\vec{S}}_2 \times \tilde{\vec{L} }   \cdot \vec{n}   }{\tilde{r}^6}\left[
\frac{3}{2}
+ \frac{1}{\nu q}\left(
-\frac{21 \nu }{8}-\frac{3}{2}
\right)   \right]\nn \\
&&+\frac{\nu^3  \tilde{p}_r (\tilde{\vec{S}}_1 \cdot \vec{n})^3   \tilde{\vec{S}}_2 \times \tilde{\vec{L} }   \cdot \vec{n}   }{\tilde{r}^6}\left[
-\frac{15}{2}
+ \frac{1}{\nu q}\left(
\frac{15}{2}-\frac{15 \nu }{8}
\right)
\right]\nn\\
&&+\frac{\nu^3 \tilde{p}_r \tilde{S}_1^2  \tilde{\vec{S}}_2 \cdot \vec{n}    \tilde{\vec{S}}_1 \times \tilde{\vec{L} }   \cdot \vec{n}   }{\tilde{r}^6}\left[
-6
+ \frac{1}{\nu q}\left(
6-\frac{51 \nu }{8}
\right)   \right]\nn \\
&&+\frac{\nu^3 \tilde{p}_r \tilde{\vec{S}}_1  \cdot\tilde{\vec{S}}_2  \tilde{\vec{S}}_1 \cdot \vec{n}    \tilde{\vec{S}}_1 \times \tilde{\vec{L} }   \cdot \vec{n}   }{\tilde{r}^6}\left[
-9
+ \frac{1}{\nu q} \left(
9-\frac{21 \nu }{4}
\right)
\right]\nn \\
&&+\frac{\nu^3\tilde{p}_r (\tilde{\vec{S}}_1  \cdot \vec{n})^2  \tilde{\vec{S}}_2 \cdot \vec{n}    \tilde{\vec{S}}_1 \times \tilde{\vec{L} }   \cdot \vec{n}   }{\tilde{r}^6}\left[
15
+ \frac{1}{\nu q} \left(
\frac{105 \nu }{8}-15
\right)
\right],
\eea
\bea
\label{squadeform}
\tilde{H}^{\text{NLO}}_{C_{\text{ES}^2_1}^2 \text{S}_1^3 \text{S}_2 } &=& \frac{\nu^3  (  \tilde{\vec{S}}_1 \cdot \vec{n} )^2 \tilde{\vec{S}}_1 \cdot \tilde{\vec{S}}_2 }{\tilde{r}^6}\left[ 
9
+ \frac{1}{\nu q} \left( 9 \nu -9 \right) \right] +\frac{\nu^3    (\tilde{\vec{S}}_1 \cdot \vec{n})^3  \tilde{\vec{S}}_2 \cdot  \vec{n} }{\tilde{r}^6}\left[ -9 + \frac{1}{\nu q} \left( 9-9 \nu \right) \right], \nn \\
\eea
\bea
\tilde{H}^{\text{NLO}}_{(\text{BS}_1^3) \text{S}_2 }&=&\frac{\nu^3 \tilde{S}_1^2  \tilde{\vec{S}}_1 \cdot \tilde{\vec{S}}_2 }{\tilde{r}^6}\left[
-6
-\frac{3}{2}\frac{\tilde{L}^2}{\tilde{r}}
+ \frac{3}{2} \tilde{p}_r^2 \tilde{r} \right.\nn\\
&&\left.+ \frac{1}{\nu q} \left(
\frac{23}{2}-6 \nu
+\frac{\tilde{L}^2}{\tilde{r}}\left(
\frac{3}{2}-\frac{9 \nu }{4}
\right)
+ \tilde{p}_r^2 \tilde{r} \left(
\frac{7}{4}-3 \nu
\right)\right) \right]\nn\\
&&+\frac{\nu^3 \tilde{S}_1^2  \tilde{\vec{S}}_1 \cdot \tilde{\vec{L} }  \tilde{\vec{S}}_2 \cdot \tilde{\vec{L} }  }{\tilde{r}^7}\left[
\frac{5}{2}
+ \frac{1}{\nu q} \left(
\frac{5 \nu }{2}-\frac{17}{4}
\right) \right] \nn\\
&& +\frac{\nu^3   (\tilde{\vec{S}}_1 \cdot \tilde{\vec{L} })^2  \tilde{\vec{S}}_1 \cdot \tilde{\vec{S} }_2   }{\tilde{r}^7}\left[
\frac{1}{2}
+ \frac{1}{\nu q} \left(
\frac{\nu }{2}-\frac{1}{2}
\right)  \right]\nn\\
&&+\frac{\nu^3 \tilde{S}_1^2   \tilde{\vec{S}}_1 \cdot \vec{n}  \tilde{\vec{S}}_2 \cdot \vec{n} }{\tilde{r}^6}\left[
34
-\frac{5}{2}  \frac{\tilde{L}^2}{\tilde{r}}
+\frac{5}{2} \tilde{p}_r^2 \tilde{r} \right.\nn\\
&&\left.+ \frac{1}{\nu q} \left(
34 \nu -\frac{109}{2}
+ \frac{\tilde{L}^2}{\tilde{r}} \left(
\frac{5 \nu }{4}-\frac{11}{2}
\right)+ \tilde{p}_r^2 \tilde{r} \left(
25 \nu -\frac{23}{4}
\right)\right) \right]\nn\\
&&+\frac{\nu^3    (\tilde{\vec{S}}_1 \cdot \vec{n})^2  \tilde{\vec{S}}_1 \cdot \tilde{\vec{S}}_2 }{\tilde{r}^6}\left[
\frac{61}{2}
+\frac{11}{2} \frac{\tilde{L}^2}{\tilde{r}}
-\frac{11}{2} \tilde{p}_r^2 \tilde{r} \right.\nn\\
&&\left.+ \frac{1}{\nu q} \left(
\frac{61 \nu }{2}-58
+ \frac{\tilde{L}^2}{\tilde{r}} \left(
\frac{37 \nu }{4}-\frac{11}{2}
\right)+ \tilde{p}_r^2 \tilde{r} \left(
17 \nu -\frac{43}{4}
\right)\right) \right]\nn\\
&&+\frac{\nu^3    (\tilde{\vec{S}}_1 \cdot \vec{n})^3  \tilde{\vec{S}}_2 \cdot  \vec{n} }{\tilde{r}^6}\left[
-\frac{155}{2}
+\frac{5}{2}\frac{\tilde{L}^2}{\tilde{r}}
-\frac{5}{2} \tilde{p}_r^2 \tilde{r} \right.\nn\\
&&\left.+ \frac{1}{\nu q} \left(
130-\frac{155 \nu }{2}
+ \frac{\tilde{L}^2}{\tilde{r}} \left(
\frac{15}{2}-\frac{25 \nu }{4}
\right)+ \tilde{p}_r^2 \tilde{r} \left(
\frac{75}{4}-55 \nu
\right)\right) \right]\nn\\
&&+\frac{\nu^3   (\tilde{\vec{S}}_1 \cdot \vec{n})^2 \tilde{\vec{S}}_1 \cdot \tilde{\vec{L} } \tilde{\vec{S}}_2 \cdot \tilde{\vec{L} }      }{\tilde{r}^7}\left[
-\frac{25}{2}
+ \frac{1}{\nu q} \left(
\frac{85}{4}-\frac{25 \nu }{2}
\right)  \right]\nn\\
&&+\frac{\nu^3   \tilde{\vec{S}}_1 \cdot \vec{n} \tilde{\vec{S}}_2 \cdot \vec{n}( \tilde{\vec{S}}_1 \cdot \tilde{\vec{L} } )^2   }{\tilde{r}^7}\left[
-\frac{5}{2}
+\frac{1}{\nu q} \left(
\frac{25}{2}-\frac{5 \nu }{2}
\right)   \right]
\nn\\
&&+\frac{\nu^3 \tilde{p}_r  \tilde{S}_1^2  \tilde{\vec{S}}_1 \cdot \vec{n}    \tilde{\vec{S}}_2 \times \tilde{\vec{L} }   \cdot \vec{n}   }{\tilde{r}^6}\left[
\frac{35}{4}
+ \frac{1}{\nu q}\left(
5 \nu -\frac{43}{4}
\right)   \right]\nn \\
&&+\frac{\nu^3 \tilde{p}_r  (\tilde{\vec{S}}_1 \cdot \vec{n})^3   \tilde{\vec{S}}_2 \times \tilde{\vec{L} }   \cdot \vec{n}   }{\tilde{r}^6}\left[
-\frac{55}{4}
+ \frac{1}{\nu q}\left(
\frac{55}{4}-5 \nu
\right)
\right]\nn\\
&&+\frac{\nu^3 \tilde{p}_r \tilde{S}_1^2  \tilde{\vec{S}}_2 \cdot \vec{n}    \tilde{\vec{S}}_1 \times \tilde{\vec{L} }   \cdot \vec{n}   }{\tilde{r}^6}\left[
-\frac{5}{4}
+ \frac{1}{\nu q}\left(
-5 \nu -2
\right)   \right]\nn \\
&&+\frac{\nu^3 \tilde{p}_r \tilde{\vec{S}}_1  \cdot\tilde{\vec{S}}_2  \tilde{\vec{S}}_1 \cdot \vec{n}    \tilde{\vec{S}}_1 \times \tilde{\vec{L} }   \cdot \vec{n}   }{\tilde{r}^6}\left[
\frac{3}{2}
+ \frac{1}{\nu q} \left(
\frac{17}{2}-6 \nu
\right)
\right]\nn \\
&&+\frac{\nu^3 \tilde{p}_r (\tilde{\vec{S}}_1  \cdot \vec{n})^2  \tilde{\vec{S}}_2 \cdot \vec{n}    \tilde{\vec{S}}_1 \times \tilde{\vec{L} }   \cdot \vec{n}   }{\tilde{r}^6}\left[
\frac{35}{4}
+ \frac{1}{\nu q} \left(
35 \nu -\frac{5}{2}
\right)
\right],
\eea
\bea
\tilde{H}^{\text{NLO}}_{\text{S}_1^2 \text{S}_2^2}&=&\frac{\nu^3 \tilde{S}_1^2   \tilde{\vec{S}}_2^2 }{2\tilde{r}^6}\left[
-\frac{27}{4}\frac{\tilde{L}^2}{\tilde{r}}
-\frac{3}{4} \tilde{p}_r^2 \tilde{r}  \right]
+\frac{\nu^2 (\tilde{\vec{S}}_1  \cdot  \tilde{\vec{S}}_2 )^2  }{2\tilde{r}^6}\left[
3
+\frac{27 \nu }{4} \frac{\tilde{L}^2}{\tilde{r}}
+ \frac{3 \nu }{2} \tilde{p}_r^2 \tilde{r} \right] \nn\\
&& +\frac{27\nu^3  \tilde{S}_1^2 (\tilde{\vec{S}}_2 \cdot \tilde{\vec{L} })^2     }{4\tilde{r}^7}
-\frac{33\nu^3   \tilde{\vec{S}}_1 \cdot \tilde{\vec{L}}  \tilde{\vec{S}}_2 \cdot \tilde{\vec{L}} \tilde{\vec{S}}_1 \cdot \tilde{\vec{S}}_2}{4\tilde{r}^7}\nn\\
&&+\frac{\nu^2    \tilde{S}_1^2 (\tilde{\vec{S}}_2 \cdot \vec{n})^2 }{\tilde{r}^6}\left[
3 \nu -3
+\frac{27 \nu }{4} \frac{\tilde{L}^2}{\tilde{r}}
+\frac{3 \nu }{4} \tilde{p}_r^2 \tilde{r} 
+ \frac{3 \nu }{ q}  \right]\nn\\
&&+\frac{\nu^2 \tilde{\vec{S}}_1 \cdot \tilde{\vec{S}}_2   \tilde{\vec{S}}_1 \cdot \vec{n} \tilde{\vec{S}}_2 \cdot  \vec{n} }{2\tilde{r}^6}\left[
-15
-\frac{51 \nu }{4}\frac{\tilde{L}^2}{\tilde{r}}
-6 \nu \tilde{p}_r^2 \tilde{r}  \right]\nn\\
&&+\frac{\nu^2   (\tilde{\vec{S}}_1 \cdot \vec{n})^2 (\tilde{\vec{S}}_2 \cdot \vec{n})^2     }{2\tilde{r}^6}\left[
15
-\frac{15 \nu }{4}\frac{\tilde{L}^2}{\tilde{r}}
+\frac{15 \nu }{4} \tilde{p}_r^2 \tilde{r}  \right]
-\frac{15\nu^3    (\tilde{\vec{S}}_1 \cdot \vec{n} )^2  (\tilde{\vec{S}}_2 \cdot \tilde{\vec{L}})^2}{4\tilde{r}^7}\nn\\
&&+\frac{15\nu^3    \tilde{\vec{S}}_1 \cdot \vec{n}  \tilde{\vec{S}}_2 \cdot \vec{n}  \tilde{\vec{S}}_1 \cdot \tilde{\vec{L}} \tilde{\vec{S}}_2 \cdot \tilde{\vec{L}}}{\tilde{r}^7}
-\frac{9\nu^3 \tilde{p}_r  \tilde{S}_1^2  \tilde{\vec{S}}_2 \cdot \vec{n}    \tilde{\vec{S}}_2 \times \tilde{\vec{L} }   \cdot \vec{n}   }{4\tilde{r}^6}\nn\\
&&+\frac{3\nu^3 \tilde{p}_r   \tilde{\vec{S}}_1 \cdot \tilde{\vec{S}}_2 \tilde{\vec{S}}_1 \cdot \vec{n}    \tilde{\vec{S}}_2 \times \tilde{\vec{L} }   \cdot \vec{n}   }{2\tilde{r}^6}
+\frac{15\nu^3 \tilde{p}_r   (\tilde{\vec{S}}_1 \cdot \vec{n})^2 \tilde{\vec{S}}_2 \cdot \vec{n}    \tilde{\vec{S}}_2 \times \tilde{\vec{L} }   \cdot \vec{n}   }{4\tilde{r}^6},
\eea
\bea
\tilde{H}^{\text{NLO}}_{(\text{ES}_1^2 )  \text{S}_2^2} &=& \frac{\nu^2 \tilde{S}_1^2   \tilde{\vec{S}}_2^2 }{\tilde{r}^6}\left[
\frac{11 \nu }{2}+3
+\frac{\tilde{L}^2}{\tilde{r}}\left(
\frac{87 \nu }{4}-\frac{201}{8}
\right)+\tilde{p}_r^2 \tilde{r} \left(
\frac{3 \nu }{8}-\frac{15}{16}
\right) \right.\nn\\
&&\left.+ \frac{\nu}{q} \left(
\frac{11 }{2}
+ \frac{201 }{8} \frac{\tilde{L}^2}{\tilde{r}}
+ \frac{15 }{16} \tilde{p}_r^2 \tilde{r} \right)\right]
+\frac{\nu^2 (\tilde{\vec{S}}_1  \cdot  \tilde{\vec{S}}_2 )^2  }{\tilde{r}^6}\left[
-\frac{3 \nu }{2}
+\frac{\tilde{L}^2}{\tilde{r}}\left(
\frac{201}{8}-\frac{87 \nu }{4}
\right)\right.\nn\\
&&\left.+\tilde{p}_r^2 \tilde{r} \left(
\frac{21}{8}-\frac{9 \nu }{4}
\right) 
+ \frac{\nu}{q} \left(
-\frac{3 }{2}
-\frac{201 }{8} \frac{\tilde{L}^2}{\tilde{r}}
-\frac{21 }{8} \tilde{p}_r^2 \tilde{r} \right)\right] \nn\\
&& +\frac{\nu^2  \tilde{S}_1^2 (\tilde{\vec{S}}_2 \cdot \tilde{\vec{L} })^2     }{\tilde{r}^7} \left[\frac{225}{8}-24 \nu -\frac{225 }{8} \frac{\nu}{q} \right] 
+\frac{\nu^2  \tilde{S}_2^2 (\tilde{\vec{S}}_1 \cdot \tilde{\vec{L} })^2     }{\tilde{r}^7} \left[\frac{201}{8}-\frac{87 \nu }{4} -\frac{201}{8} \frac{\nu}{q} \right] 
\nn\\
&&+\frac{\nu^2    \tilde{\vec{S}}_1 \cdot \tilde{\vec{L}}  \tilde{\vec{S}}_2 \cdot \tilde{\vec{L}} \tilde{\vec{S}}_1 \cdot \tilde{\vec{S}}_2}{\tilde{r}^7} \left[
42 \nu -\frac{195}{4} 
+\frac{195}{4} \frac{\nu}{q} \right]\nn\\
&&+\frac{\nu^2    \tilde{S}_1^2 (\tilde{\vec{S}}_2 \cdot \vec{n})^2 }{\tilde{r}^6}\left[
-\frac{21 \nu }{2}-3
+ \frac{\tilde{L}^2}{\tilde{r}} \left(
\frac{345}{16}-\frac{147 \nu }{8}
\right)+ \tilde{p}_r^2 \tilde{r} \left(
\frac{15}{16}-\frac{3 \nu }{8}
\right)\right.\nn\\
&&\left.+ \frac{ \nu }{ q} \left(
-\frac{21}{2}
-\frac{345}{16} \frac{\tilde{L}^2}{\tilde{r}} 
-\frac{15}{16} \tilde{p}_r^2 \tilde{r} \right) \right]\nn\\
&&+\frac{\nu^2    \tilde{S}_2^2 (\tilde{\vec{S}}_1 \cdot \vec{n})^2 }{\tilde{r}^6}\left[
-15 \nu -9
+ \frac{\tilde{L}^2}{\tilde{r}} \left(
\frac{201}{8}-\frac{87 \nu }{4}
\right)+ \tilde{p}_r^2 \tilde{r} \left(
\frac{9 \nu }{8}+\frac{3}{16}
\right)\right.\nn\\
&&\left.+ \frac{ \nu }{ q} \left(
-15
-\frac{201}{8} \frac{\tilde{L}^2}{\tilde{r}} 
-\frac{3}{16}\tilde{p}_r^2 \tilde{r} \right) \right]\nn\\
&&+\frac{\nu^2 \tilde{\vec{S}}_1 \cdot \tilde{\vec{S}}_2   \tilde{\vec{S}}_1 \cdot \vec{n} \tilde{\vec{S}}_2 \cdot  \vec{n} }{\tilde{r}^6}\left[
-21 \nu -6
+ \frac{\tilde{L}^2}{\tilde{r}} \left(
\frac{99 \nu }{2}-\frac{225}{4}
\right) + \tilde{p}_r^2 \tilde{r} \left(
\frac{21 \nu }{2}-\frac{45}{4}
\right)\right.\nn\\
&&\left. + \frac{\nu}{q} \left(
-21
+ \frac{225}{4} \frac{\tilde{L}^2}{\tilde{r}}  
+\frac{45}{4} \tilde{p}_r^2
\right) \right]\nn\\
&&+\frac{\nu^2   (\tilde{\vec{S}}_1 \cdot \vec{n})^2 (\tilde{\vec{S}}_2 \cdot \vec{n})^2     }{\tilde{r}^6}\left[
\frac{105 \nu }{2}+15
+ \frac{\tilde{L}^2}{\tilde{r}} \left(
\frac{285}{16}-\frac{135 \nu }{8}
\right) + \tilde{p}_r^2 \tilde{r}  \left(
\frac{135}{16}-\frac{75 \nu }{8}
\right)\right.\nn\\
&&\left.+ \frac{\nu}{q} \left(
\frac{105}{2}
-\frac{285}{16} \frac{\tilde{L}^2}{\tilde{r}}
-\frac{135}{16} \tilde{p}_r^2 \tilde{r}
\right) \right]
+\frac{\nu^2    (\tilde{\vec{S}}_1 \cdot \vec{n} )^2  (\tilde{\vec{S}}_2 \cdot \tilde{\vec{L}})^2}{\tilde{r}^7} \left[30 \nu -\frac{285}{8} 
+\frac{285}{8} \frac{\nu}{q} \right]\nn\\
&&+\frac{\nu^2    (\tilde{\vec{S}}_2 \cdot \vec{n} )^2  (\tilde{\vec{S}}_1 \cdot \tilde{\vec{L}})^2}{\tilde{r}^7} \left[\frac{45 \nu }{2}-\frac{105}{4}
+\frac{105}{4} \frac{\nu}{q} \right]\nn\\
&&+\frac{\nu^2    \tilde{\vec{S}}_1 \cdot \vec{n}  \tilde{\vec{S}}_2 \cdot \vec{n}  \tilde{\vec{S}}_1 \cdot \tilde{\vec{L}} \tilde{\vec{S}}_2 \cdot \tilde{\vec{L}}}{\tilde{r}^7} \left[
\frac{135}{4}-30 \nu
-\frac{135}{4}\frac{\nu}{q}
\right]\nn\\
&& +\frac{\nu^2 \tilde{p}_r  \tilde{S}_1^2  \tilde{\vec{S}}_2 \cdot \vec{n}    \tilde{\vec{S}}_2 \times \tilde{\vec{L} }   \cdot \vec{n}   }{\tilde{r}^6} \left[
\frac{45}{8}-\frac{9 \nu }{2}
-\frac{45}{8} \frac{\nu}{q}
\right]\nn\\
&&+\frac{\nu^2 \tilde{p}_r   \tilde{\vec{S}}_1 \cdot \tilde{\vec{S}}_2 \tilde{\vec{S}}_1 \cdot \vec{n}    \tilde{\vec{S}}_2 \times \tilde{\vec{L} }   \cdot \vec{n}   }{\tilde{r}^6} \left[
15-12 \nu
-15 \frac{\nu}{q}
\right]\nn\\
&& +\frac{\nu^2 \tilde{p}_r   (\tilde{\vec{S}}_1 \cdot \vec{n})^2 \tilde{\vec{S}}_2 \cdot \vec{n}    \tilde{\vec{S}}_2 \times \tilde{\vec{L} }   \cdot \vec{n}   }{\tilde{r}^6}\left[
\frac{45 \nu }{2}-\frac{225}{8}
+ \frac{225}{8} \frac{\nu}{q}
\right]\nn\\
&& +\frac{\nu^2 \tilde{p}_r  \tilde{S}_2^2  \tilde{\vec{S}}_1 \cdot \vec{n}    \tilde{\vec{S}}_1 \times \tilde{\vec{L} }   \cdot \vec{n}   }{\tilde{r}^6} \left[
6 \nu -9
+9 \frac{\nu}{q}
\right]\nn\\
&&+\frac{\nu^2 \tilde{p}_r   \tilde{\vec{S}}_1 \cdot \tilde{\vec{S}}_2 \tilde{\vec{S}}_2 \cdot \vec{n}    \tilde{\vec{S}}_1 \times \tilde{\vec{L} }   \cdot \vec{n}   }{\tilde{r}^6} \left[
3 \nu -\frac{15}{4}
+\frac{15}{4} \frac{\nu}{q}
\right]\nn\\
&& +\frac{\nu^2 \tilde{p}_r   (\tilde{\vec{S}}_2 \cdot \vec{n})^2 \tilde{\vec{S}}_1 \cdot \vec{n}    \tilde{\vec{S}}_1 \times \tilde{\vec{L} }   \cdot \vec{n}   }{\tilde{r}^6}\left[
\frac{75}{4}-15 \nu
-\frac{75}{4} \frac{\nu}{q}
\right],
\eea
\bea
\tilde{H}^{\text{NLO}}_{(\text{ES}_1^2 )(\text{ES}_2^2 ) }&=&\frac{\nu^3 \tilde{S}_1^2   \tilde{\vec{S}}_2^2 }{2\tilde{r}^6}\left[
\frac{9}{2}
+ \frac{\tilde{L}^2}{\tilde{r}} \left(
-\frac{3 \nu }{8}-\frac{21}{8}
\right)+ \tilde{p}_r^2 \tilde{r} \left( 
\frac{15}{8}-\frac{9 \nu }{4}
\right) \right]\nn\\
&&+\frac{\nu^2 (\tilde{\vec{S}}_1  \cdot  \tilde{\vec{S}}_2 )^2  }{2\tilde{r}^6} \left[
\frac{9}{2}
+ \frac{\tilde{L}^2}{\tilde{r}} \left(
-\frac{3 \nu }{4}-\frac{3}{4}
\right)+ \tilde{p}_r^2 \tilde{r} \left( 
-\frac{9 \nu }{2}-\frac{9}{4}
\right) \right] \nn\\
&& +\frac{\nu^2  \tilde{S}_1^2 (\tilde{\vec{S}}_2 \cdot \tilde{\vec{L} })^2     }{\tilde{r}^7} \left[
\frac{9 \nu }{4}-\frac{3}{4}
+\frac{9}{4}\frac{\nu}{q}
\right]
-\frac{3\nu^2   \tilde{\vec{S}}_1 \cdot \tilde{\vec{L}}  \tilde{\vec{S}}_2 \cdot \tilde{\vec{L}} \tilde{\vec{S}}_1 \cdot \tilde{\vec{S}}_2}{4\tilde{r}^7}\nn\\
&&+\frac{\nu^2    \tilde{S}_1^2 (\tilde{\vec{S}}_2 \cdot \vec{n})^2 }{\tilde{r}^6}\left[
-3 \nu -15
+ \frac{\tilde{L}^2}{\tilde{r}}\left(
\frac{33 \nu }{8}+\frac{39}{8}
\right)+ \tilde{p}_r^2 \tilde{r} \left(
\frac{21 \nu }{4}+\frac{21}{8}
\right)\right.\nn\\
&&\left.+ \frac{3 \nu }{ q} \left(
-3
+ \frac{9}{4} \frac{\tilde{L}^2}{\tilde{r}}
-6 \tilde{p}_r^2 \tilde{r} 
\right) \right]\nn\\
&&+\frac{\nu^2 \tilde{\vec{S}}_1 \cdot \tilde{\vec{S}}_2   \tilde{\vec{S}}_1 \cdot \vec{n} \tilde{\vec{S}}_2 \cdot  \vec{n} }{2\tilde{r}^6}\left[
-48
+ \frac{\tilde{L}^2}{\tilde{r}} \left(
\frac{15 \nu }{2}+\frac{27}{2}
\right) + \tilde{p}_r^2 \tilde{r} \left(
45 \nu +\frac{33}{2}
\right) \right]\nn\\
&&+\frac{\nu^2   (\tilde{\vec{S}}_1 \cdot \vec{n})^2 (\tilde{\vec{S}}_2 \cdot \vec{n})^2     }{2\tilde{r}^6}\left[
93
+\frac{\tilde{L}^2}{\tilde{r}} \left(
-\frac{105 \nu }{8}-\frac{225}{8}
\right) + \tilde{p}_r^2 \tilde{r} \left(
-\frac{315 \nu }{4}-\frac{195}{8}
\right) \right]\nn\\
&&+\frac{\nu^2    (\tilde{\vec{S}}_1 \cdot \vec{n} )^2  (\tilde{\vec{S}}_2 \cdot \tilde{\vec{L}})^2}{\tilde{r}^7}\left[
\frac{15}{4}-\frac{15 \nu }{4}
-\frac{15}{4}\frac{\nu}{q}\right]
\nn\\
&&+\frac{15\nu^2    \tilde{\vec{S}}_1 \cdot \vec{n}  \tilde{\vec{S}}_2 \cdot \vec{n}  \tilde{\vec{S}}_1 \cdot \tilde{\vec{L}} \tilde{\vec{S}}_2 \cdot \tilde{\vec{L}}}{4\tilde{r}^7}
+\frac{\nu^2 \tilde{p}_r  \tilde{S}_1^2  \tilde{\vec{S}}_2 \cdot \vec{n}    \tilde{\vec{S}}_2 \times \tilde{\vec{L} }   \cdot \vec{n}   }{\tilde{r}^6}\left[
\frac{21 \nu }{2}-\frac{9}{4}
+ \frac{57}{4}\frac{\nu}{q}\right]\nn\\
&&+\frac{\nu^2 \tilde{p}_r   \tilde{\vec{S}}_1 \cdot \tilde{\vec{S}}_2 \tilde{\vec{S}}_1 \cdot \vec{n}    \tilde{\vec{S}}_2 \times \tilde{\vec{L} }   \cdot \vec{n}   }{\tilde{r}^6}\left[
-6
+\frac{15}{2}\frac{\nu}{q}\right]\nn\\
&& 
+\frac{\nu^2 \tilde{p}_r   (\tilde{\vec{S}}_1 \cdot \vec{n})^2 \tilde{\vec{S}}_2 \cdot \vec{n}    \tilde{\vec{S}}_2 \times \tilde{\vec{L} }   \cdot \vec{n}   }{\tilde{r}^6}\left[
\frac{15 \nu }{2}+\frac{45}{4}
-\frac{75}{4}\frac{\nu}{q}\right].
\eea

\bibliographystyle{jhep}
\bibliography{gwbibtex}

\end{document}